\pdfoutput=1

\documentclass[pdftex,twocolumn,epjc3]{svjour3}          

\RequirePackage[T1]{fontenc}

\smartqed  

\RequirePackage{graphicx}
\RequirePackage{mathptmx}      
\RequirePackage{flushend}
\RequirePackage[numbers,sort&compress]{natbib}
\RequirePackage[colorlinks,citecolor=blue,urlcolor=blue,linkcolor=blue]{hyperref}
\RequirePackage{lineno}
\usepackage{reviewphysics}
\usepackage{tabularx} 

\journalname{Eur. Phys. J. C}

\begin{document}

\title{Review of single vector boson production in pp collisions at $\sqrt{s} = 7$\,TeV}

\author{Matthias Schott\thanksref{addr1} 
\and
Monica Dunford\thanksref{addr2} 
}

\thankstext{e1}{e-mail: Matthias.Schott@cern.ch}
\thankstext{e2}{e-mail: Monica.Dunford@cern.ch}

\institute{Institut f\"ur Physik, Johannes Gutenberg Universit\"at Mainz\label{addr1}
\and
Kirchhoff Institut f\"ur Physik, Ruprecht-Karls-Universit\"at Heidelberg\label{addr2}
}

\date{Submitted to Eur. Phys. J. C}

\maketitle


\begin{abstract}
This review summarises the main results on the production of single vector bosons in the Standard Model, both inclusively and in 
association with light- and heavy-flavour jets, at the Large Hadron Collider in proton-proton collisions at a center-of-mass energy of $7\,\TeV$. 

The general purpose detectors at this collider, ATLAS and CMS, each recorded an integrated luminosity of 
$\approx 40\,{\rm pb^{-1}}$ and $5\,{\rm fb^{-1}}$ in the years 2010 and 2011, respectively. 
The corresponding data offer the unique possibility to precisely study the properties of the production of heavy vector
 bosons in a new energy regime. The accurate understanding of the Standard Model  is not only crucial for searches of unknown particles and phenomena 
 but also to test predictions of perturbative Quantum-Chromodynamics calculations and for 
 precision measurements of observables in the electroweak sector.

Results from a variety of measurements in which single \Wboson or \Zboson bosons are identified are reviewed. Special 
emphasis in this review is given to interpretations of the experimental results in the context of state-of-the-art predictions.

\end{abstract}

\tableofcontents
\section{Introduction}

The \Wboson and \Zboson bosons, since their discovery at UA1~\cite{Arnison:1983rp, Arnison:1983mk} and 
UA2~\cite{Banner:1983jy, Bagnaia:1983zx} in the early 1980s, have been the subject of detailed measurements at both
electron-positron and hadron colliders. The ALEPH, DELPHI, L3 and OPAL experiments at the large electron-positron collider, LEP, preformed many precision studies 
of these vector bosons, including measurements of the branching ratios~\cite{ALEPHDELPHIL3:2005aa}, the magnetic dipole moment and the electric quadrupole moment
~\cite{ALEPHDELPHIL3200501}, all of which were measured with a precision of better than 1\%. At hadron colliders, single vector boson 
production has been explored at $\sqrt{s}=0.63\,\TeV$ at the CERN S$\bar{p} p$S by UA1 and UA2, and at both $\sqrt{s}=1.8\,\TeV$ and
$\sqrt{s}=1.96\,\TeV$ at the Tevatron by CDF~\cite{Abe:1988ar, Abulencia:2005ix} and D0~\cite{Abbott:1999gn, Abazov:2007ac}. 
The distinct advantage of \Wboson and \Zboson  
production measurements at the hadron colliders is that the number of single vector boson  
events is large, roughly 138,000 $Z \rightarrow ee$ and 470,000 $W \rightarrow e\nu$ candidates 
using 2.2 \ifb of data at CDF~\cite{Aaltonen:2012fi, Aaltonen:2012bp} with relatively low background rates, roughly 0.5\% of \Zboson candidates and 1\% of \Wboson candidates at CDF. 
The major disadvantage is that the parton center-of-mass energy can not be determined for each event because of the uncertainties in the structure of the proton. Despite these challenges, the CDF and D0 experiments have reported 
measurements of the mass of the \Wboson~\cite{Aaltonen:2012bp, Abazov:2012bv} with a precision comparable to the measurements at LEP.  
In addition with the large data samples, measurements of \Wboson and \Zboson production in association with 
jets~\cite{Abe:1993si, Aaltonen:2007ae, Aaltonen:2007ip, Abazov:2013gpa, Abazov:2011rf, Abazov:2006gs} 
and the production of \Wboson and \Zboson production in association with heavy flavour quarks~\cite{Aaltonen:2012wn, Aaltonen:2009qi, Aaltonen:2008mt, Abazov:2008qz, D0:2012qt, Abazov:2013uza} were preformed. 
Measurements of single vector boson production at the S$\bar{p} p$S and the Tevatron have been especially important for the development 
of leading-order and next-to-leading order theoretical predictions, most of which are used today for comparisons to data at the Large Hadron Collider (LHC). Finally \Wboson and \Zboson production has also been observed at heavy-ion colliders at RHIC at $\sqrt{s}=0.5\,\TeV$~\cite{STAR:2011aa, Adare:2010xa} 
and the LHC at $\sqrt{s}=2.76$ \TeV~\cite{Chatrchyan:2012nt, Chatrchyan:2011ua, Aad:2012ew}. For a detailed review of the measurements preformed at LEP and S$\bar{p} p$S see~\cite{Ellis:2004gu} and~\cite{Altarelli:1989ub} respectively. 

Today, the focus of measurements of \Wboson and \Zboson production at the LHC is to test the theory
of perturbative Quantum-Chromodynamics 
(QCD) in a new energy regime, to provide better constraints on the parton distribution functions, 
and to improve electroweak precision measurements, such as the mass of the \Wboson and $\sin^2\theta$.
As \Wboson and \Zboson production are dominant backgrounds to Higgs boson measurements and searches for physics beyond the Standard Model, 
these new measurements also provide insight to these studies. 

For tests of the predictions of perturbative QCD, the be\-nefit of the increase in energy at the LHC can readily be seen in Figure~\ref{fig:TIntroX1X2}, where the Bjorken $x$-values of the interacting partons for a given process, e.g. the production of a \Zboson boson, is shown. The reach in the low-$x$ region has been 
increased by more than two orders of magnitude compared to that of the SPS and the Tevatron. 
As a matter of fact, these new measurements not only benefit from the higher center-of-mass energy but also from  
improved statistical and systematic uncertainties. At the LHC, copious amounts of \Wboson and \Zboson boson events, more than a million 
$Z \rightarrow ee$ events at each of the ATLAS and CMS experiments during the 2011 $\sqrt{s}=7\,\TeV$ run,
were detected, with an improved experimental precision. For example, the uncertainty on the jet energy scale 
is almost a factor of three better~\cite{Bhatti:2005ai, Aad:2013tea} compared to that at the Tevatron experiments. Furthermore, the detectors have been designed to have an increased rapidity 
acceptance and can measure electrons for some measurements to $|\eta | < 4.9$ and jets to $|\eta | < 4.4$. As a result, a large fraction of these low-$x$ events 
shown in Figure~\ref{fig:TIntroX1X2} can be reconstructed by the LHC detectors.

The theoretical predictions used for comparison to these measurements have been extended and improved. 
For inclusive \Wboson and \Zboson production, 
theoretical predictions at next-to-next-to-leading order in perturbation theory are available. For measurements of single vector boson production in association with jets
predictions at next-to-leading order for up to five additional partons in the final state exist. 
The magnitude of the theoretical uncertainties in these calculations are comparable to those of the experimental uncertainties. 
In addition several advanced leading-order predictions exist which simulate the entire event process
from the hard scatter to the parton showering and the fragmentation. Although many of these predictions have been vetted by measurements at previous
hadron colliders, the LHC measurements will test these predictions in previously unexplored regions of the phase space. 

\begin{figure}[h]
\resizebox{0.5\textwidth}{!}{\includegraphics{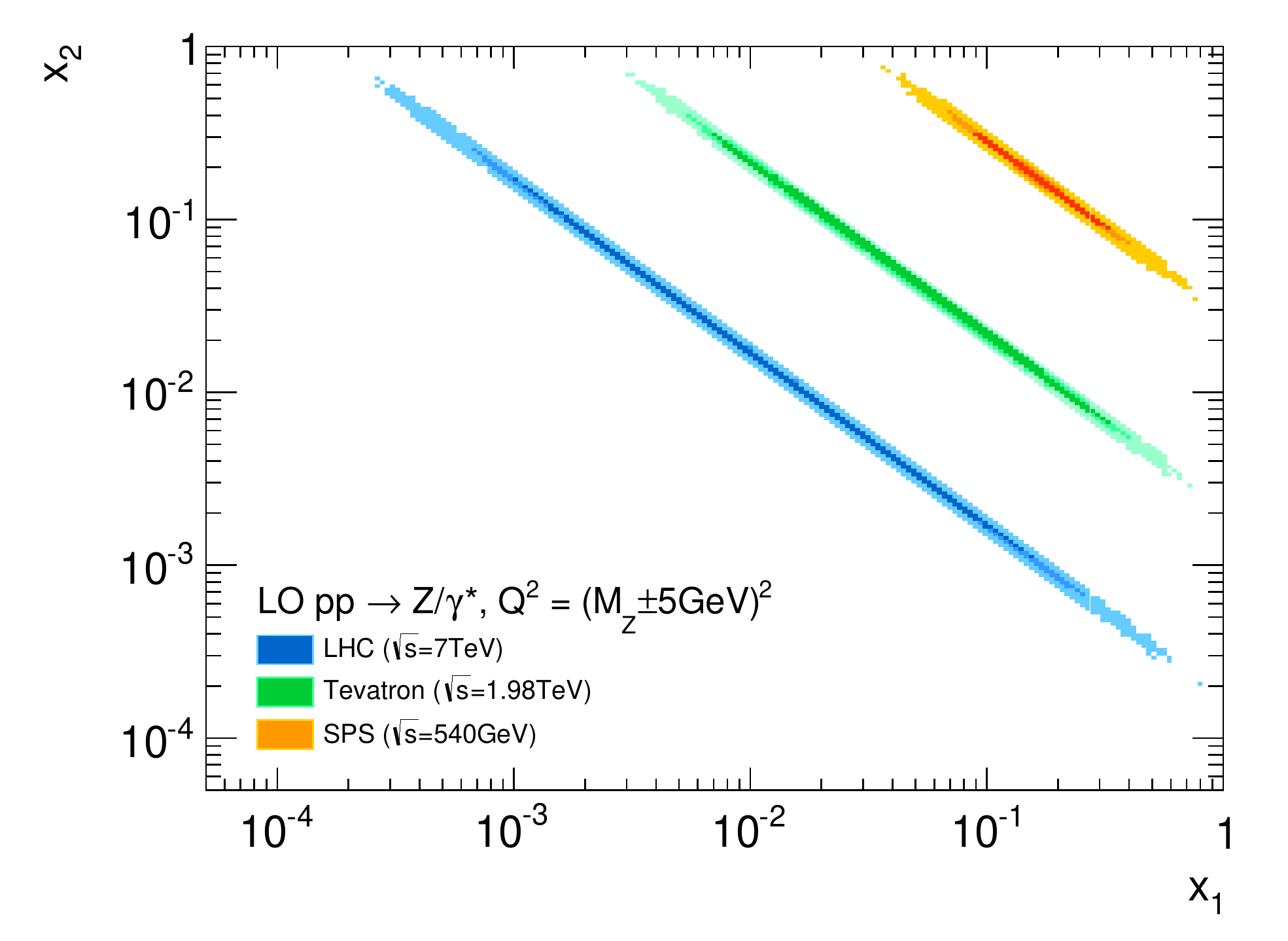}}
\caption{\label{fig:TIntroX1X2} Correlation for different hadron colliders between the Bjorken $x$ values of the two interacting partons at leading order in the reaction $pp\rightarrow Z/\gamma^*$ for LHC and SPS and $p\bar p \rightarrow Z/\gamma^*$ for Tevatron, respectively.}
\end{figure}

The structure functions of the proton, which are a dominant source of uncertainties in electroweak precision 
measurements at hadron colliders, can also be constrained through studies of the differential cross-sections of \Wboson 
or \Zboson bosons production. This is illustrated in Figure \ref{fig:TIntroXQ2}, where the kinematic plane as a function of the 
Bjorken $x$ and $Q^2$ for Drell-Yan scattering at the Tevatron, the LHC and the corresponding 
deep inelastic scattering experiments are compared. Similarly, measurements of the \Wboson production in association with a charm quark 
test the contributions in the proton from the strange quarks at $x \approx 0.01$ as well as any $s-\bar{s}$ asymmetries. 

\begin{figure}[h]
\resizebox{0.5\textwidth}{!}{\includegraphics{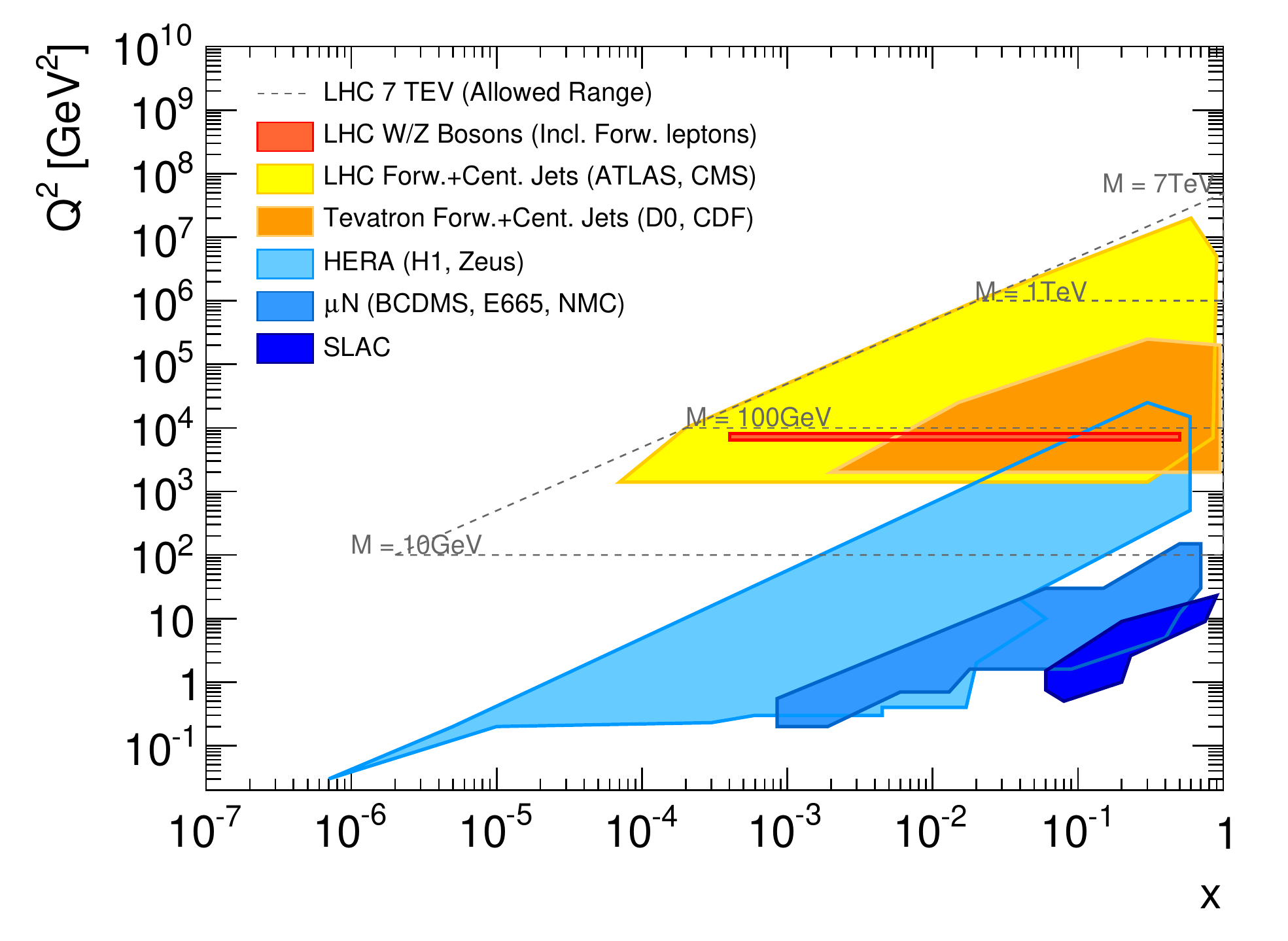}}
\caption{\label{fig:TIntroXQ2} Illustration of kinematic plane in bins of Bjorken $x$ and $Q^2$ for Drell-Yan scattering at the Tevatron and the LHC and the corresponding deep inelastic scattering experiments.}
\end{figure}

This review article summarises the major results of the single \Wboson and \Zboson production in the proton-proton 
collision data at $\sqrt{s}=7\,\TeV$ recorded in the years 2010 and 2011 at the LHC by the two general purpose experiments, ATLAS and CMS. The article is organised as follows. 
First, in Section~\ref{sec:VecBosProduction}, we review the basic theory 
behind single vector boson production. In this section, we pay special attention to some of the basic elements of 
cross-section calculations such as the matrix element calculations, the parton shower modelling and parton 
distribution functions. We also summarise here the theoretical predictions used in this review. In Section~\ref{sec:DectAndData}, 
we describe the ATLAS and CMS detectors at the LHC and discuss in a general 
manner the basic principles of lepton and jet reconstruction. Section~\ref{sec:CrossMeasPhil} delineates how cross-sections are measured at the LHC, 
while Section~\ref{sec:SelAndBackground} highlights the event 
selection and the background estimates for the measurements presented here. Finally we present the results for 
inclusive single vector boson production in Section~\ref{sec:InclusiveAndDiffMeasurements} as well as the 
production in association with jets in Section~\ref{sec:VjetsMeasurements}. In Section~\ref{sec:SummaryOutlook}, 
we conclude and provide thoughts on future measurements.

\section {\label{sec:VecBosProduction} Vector Boson Production in the Standard Model}

The electroweak Lagrangian of the Standard Model after electroweak symmetry breaking, i.e. after the Higgs Boson has acquired a vacuum expectation value, can be written as \cite{Beringer:1900zz}

\begin{eqnarray}
\label{EQN:SMLagrangian}
\nonumber
\LC_{EW} &=& \LC_{K} + \LC_{N} + \LC_{C} + \LC_{WWV} + \LC_{WWVV} \\
		& &  + \LC_{H} + \LC_{HV} + \LC_{Y}.
\end{eqnarray}
where the terms in Equation \ref{EQN:SMLagrangian} are schematically illustrated as tree level Feynman graphs in Figure \ref{fig:TSMLagrangian}. The kinetic term, $\LC_{K}$, describes the free movement all fermions and bosons. It involves quadratic terms and the respective masses. The term $\LC_{N}$ describes the interaction of the photon and the \Zboson boson to fermions, while $\LC_{C}$ describes the interaction of the W-Boson to left-handed particles or right-handed anti-particles. The self interaction of gauge bosons is a direct consequence of the $SU(2)_L$ group structure and is described by $\LC_{WWV}$ and $\LC_{WWVV}$, representing three-point and four-point interactions, respectively. The three- and four-point self-interaction of the Higgs boson is described by $\LC_{H}$, while the interaction of the Higgs boson to the gauge bosons is represented in $\LC_{HV}$. The last term in Equation \ref{EQN:SMLagrangian}, $\LC_{Y}$ characterises the Yukawa couplings between the massive fermions of the Standard Model and the Higgs field.

\begin{figure}[h]
\resizebox{0.5\textwidth}{!}{\includegraphics{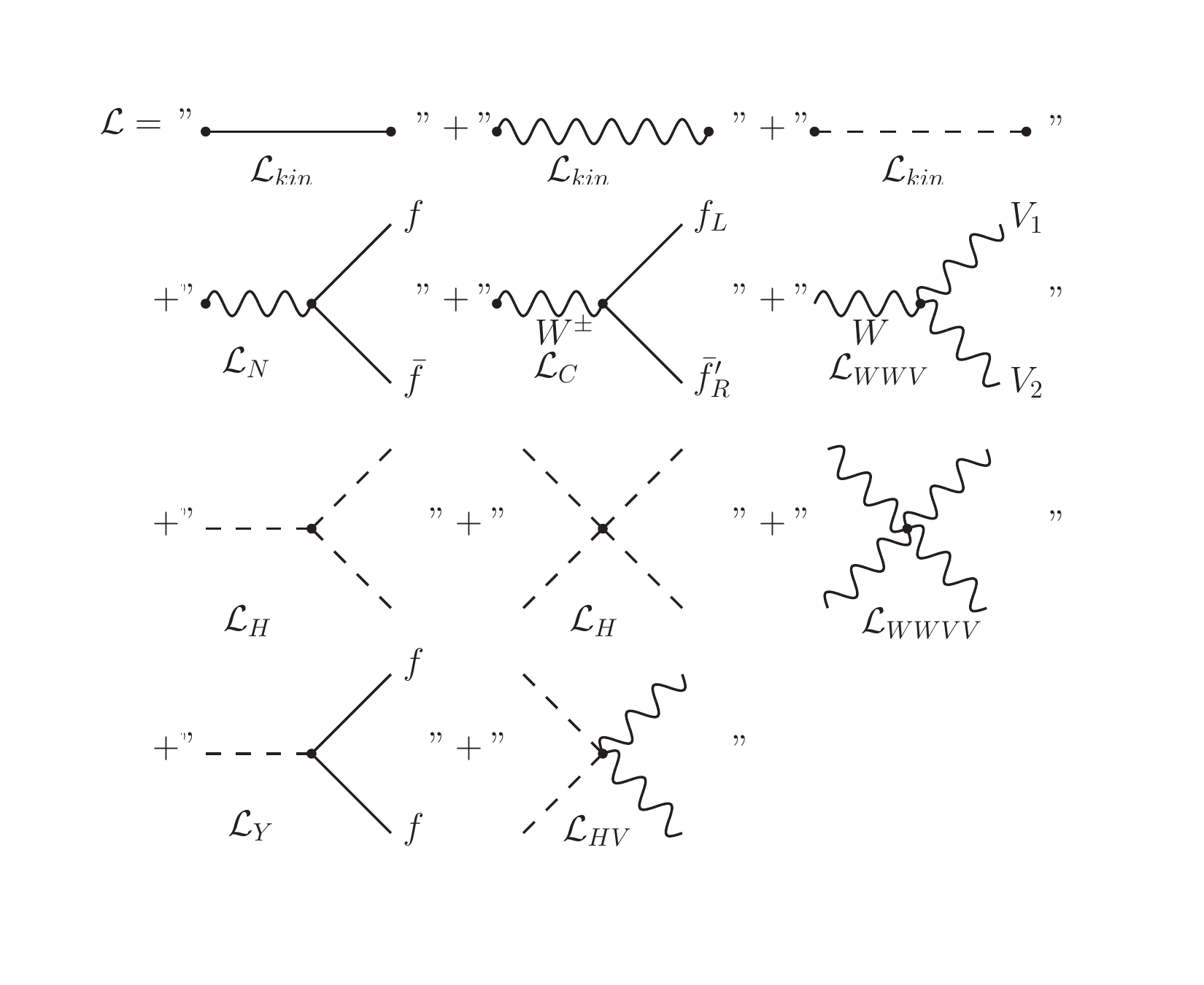}}
\caption{\label{fig:TSMLagrangian} Schematic illustration of the Lagrangian describing the electroweak sector of the Standard Model (See Equation \ref{EQN:SMLagrangian}).}
\end{figure}

For the single \Zboson boson production at hadron colliders, $\LC_{N}$ is 

\begin{equation}
\label{EQN:LN}
\LC_N = e J_{\mu}^{em}A^\mu + \frac{g}{\cos\theta _W} \left( J_\mu^3 - sin^2\theta_W J_\mu^{em}\right) Z^\mu,
\end{equation}
where $J_\mu^{em}$ describes the electromagnetic current, i.e. the sum over all fermion fields weighted by their electromagnetic charges, and $J_\mu^3$ represents the weak current, involving only left-handed particles and right-handed anti-particles with their respective weak isospins. The weak mixing angle, $\theta_W$, describes the relative contribution of the weak and electromagnetic part of the interaction. The production of the single \Wboson bosons is described by

\begin{eqnarray}
\label{EQN:LC}
\nonumber
\LC_C &=& - \frac{g}{\sqrt{2}} \left[\bar u_i \gamma^\mu \frac{1-\gamma^5}{2} V_{ij}^{CKM} d_j + \bar v_i \gamma ^\mu \frac{1-\gamma^5}{2} e_i \right] W_\mu^+\\
&& +h.c.,
\end{eqnarray}
where only the terms for first generation are explicitly shown. The quark and lepton spinor fields are denoted by $u_i, d_j$ and $v_i, e_i$. The term $(1-\gamma^5)$ acts as a projector for the left-handed components of the spinors, meaning that the charge current acts exclusively on left-handed particles and right-handed anti-particles, while for the neutral current all spinor components play a role due to the electromagnetic part of the interaction term. The Cabbibo-Kobayashi-Maskawa (CKM) matrix is denoted as $V_{ij}^{CKM}$ \cite{Kobayashi:1973fv}. In this review article, we will concentrate on the terms of the Electroweak Lagrangian represented in Equation \ref{EQN:LN} and \ref{EQN:LC}, as they describe the single vector boson production within the Standard Model.

Before discussing the experimental results, we first review the central parts of the theoretical predictions
of the gauge boson production cross-sections at the LHC. In Section \ref{sec:XsecCalc} the calculation of the cross-section is 
defined, which is shown to consist of two main parts; the matrix-element 
term describing the parton interactions and the parton distribution functions describing the proton. 
As the lowest-order matrix-element term for \Wboson and \Zboson 
production is a frequent example in many particle physics textbooks, we
extend the formalism in Section \ref{sec:MECalc} by discussing higher-order corrections from the QCD and Electroweak 
theories and emphasise why these are important to the experimental measurements. The hard scatter process, which at high energy scales can be connected to the lower energies scales via 
parton showering models is discussed in Section \ref{sec:PartonShower}. Finally, the second part of the cross-section calculation, the parton distribution functions, is briefly reviewed in Section \ref{sec:PDF}. In addition, we discuss some critical inputs which are needed to perform these calculations. 
This includes the available models for hadronisation of final state particles (section \ref{sec:Hadron}) and the description 
of multiple particle interactions (section \ref{sec:MPI}). 
A discussion of the available computing codes, 
which are used to compare the latest LHC measurements with the predictions of 
the Standard Model, can be found in Section \ref{sec:ComputerCodes}. Our discussion ends with an
overview of the definition and interpretation of some observables which are important for understanding QCD dynamics (section \ref{sec:QCDDynamics}). 

Several introductory articles on the production of vector bosons in hadron collisions are available. We summarise here the essential aspects along the lines of \cite{Nunnemann:2007qs}, \cite{Buckley:2011ms}. Related overview articles on parton density functions at next-to-leading order at hadron colliders and subsequently jet physics in electron-proton can be found in \cite{Carli:2010rw} and \cite{SchornerSadenius:2012de}, respectively.

\subsection {\label{sec:XsecCalc} Cross-section calculations}

The calculation of production cross-sections in 
proton-proton collisions at the LHC is, in general, a combination of two energy regimes: the short-distance or high-energy 
regime and the long-distance or low-energy regime. 
By the factorisation theorem, the production cross-section can therefore be expressed as a product of two terms:
one describing the parton-parton cross-section, $\hat \sigma_{q\bar q \rightarrow n}$ at short-distances 
and another describing the complicated internal structure of protons at long distances. For large momentum transfers of the 
interacting partons in the short-distance term, the parton-parton interaction can be evaluated using perturbative QCD calculations.
However, in the long-distance term, where perturbative calculations are no longer applicable, parton density functions (PDFs) are
used to describe the proton structure in a phenomenological way. These functions are written as 
$f_{a/A} (x,Q^2)$ for the parton $a$ in the proton A where $x=\frac{p_a}{p_A}$ is the relative momentum 
 of the parton in direction of the proton's momentum and $Q^2$ is this energy scale 
of the scattering process. The scale at which the long-distance physics of the PDF description 
and the short-distance physics of parton-parton interaction separate is called the factorisation scale and is defined as $\mu_F = Q$. 
For the production of a vector boson via quark-fusion, the energy scale is set to the mass of the 
vector boson, which in turn can be expressed by 
the invariant mass of final state fermions $f$, i.e. $Q^2=m_V^2=m_{f\bar f}$. 

\begin{figure}[h]
\resizebox{0.5\textwidth}{!}{\includegraphics{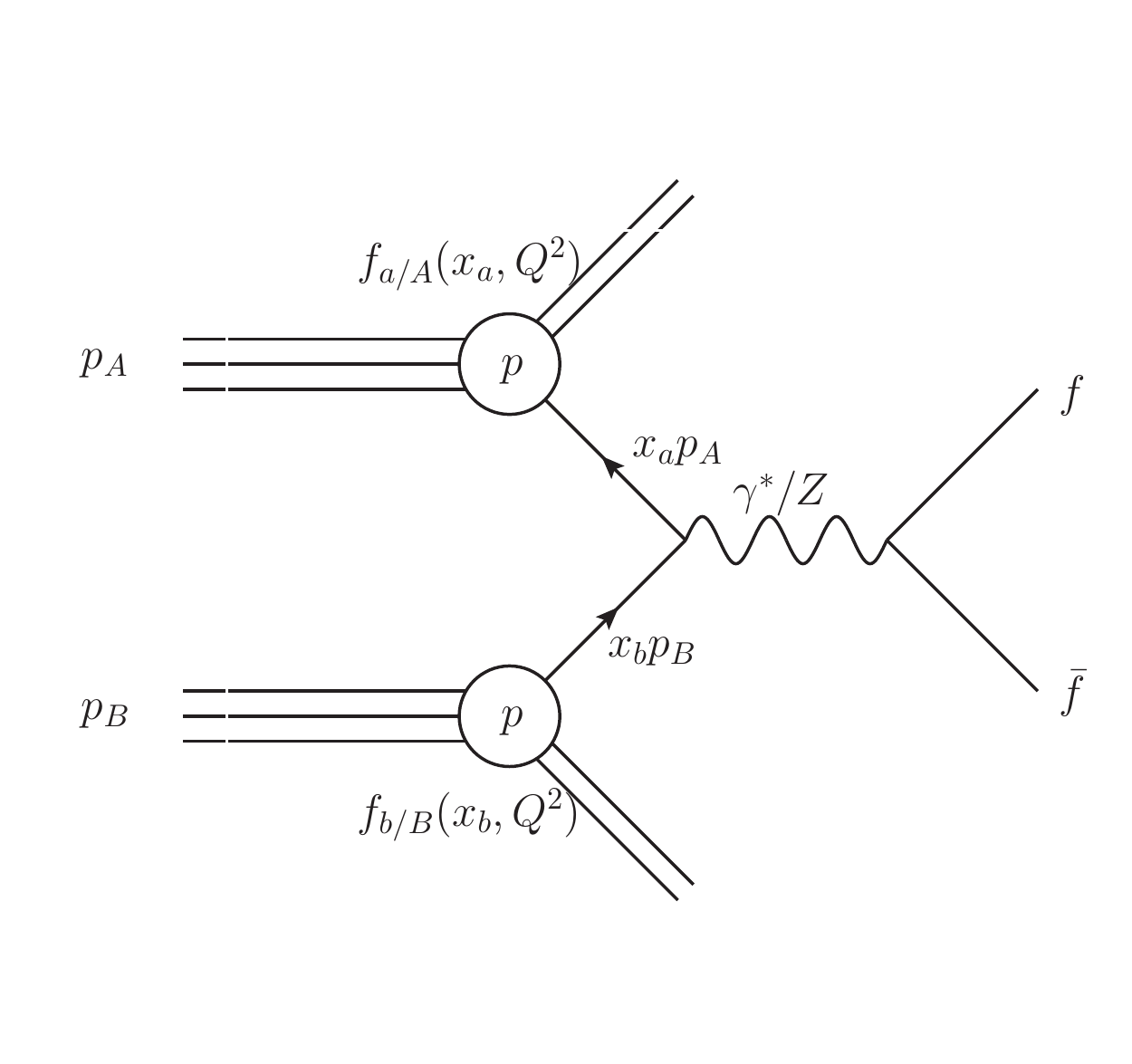}}
\caption{Illustration of cross-section calculation in a proton-proton collision at the LHC.}
\label{fig:TCrossSectionAtLHC}
\end{figure}

The proton-proton cross-section is thereby expressed as

\begin{eqnarray}
\sigma_{p_A p_B \rightarrow n} = \sum _q \int dx_a dx_b f_{a/A} (x_a, 
Q^2) f_{b/B} (x_b, Q^2) \hat \sigma _{ab \rightarrow n}  
\end{eqnarray}
and shown graphically for \Zboson production in Figure \ref{fig:TCrossSectionAtLHC}. 
The functions $f_{a/A}$ and $f_{b/B}$ denote the PDFs for the partons $a$ and 
$b$ in protons $A$ and $B$. All quark flavours are accounted for in the sum and the 
integration is performed over $x_a$ and $x_b$, describing the respective momentum fractions of the interacting partons. 
The subset of perturbative corrections from real and virtual gluon emissions, 
which are emitted collinearly to the incoming partons, lead to large logarithms 
that can be absorbed in the PDFs.

Inclusive hard-scattering processes can be described using the factorisation theorem \cite{Bjorken:1969ja}, \cite{Collins:1985ue}.
This approach is also applicable when including the higher order perturbative QCD corrections, which are discussed in 
more detail in Section \ref{sec:QCDCorrections}. When expanding the parton-parton cross-section in terms of $\alpha_s$,
the formula for the cross-section becomes:

\begin{eqnarray}
\nonumber
\label{eqn:prodcross}
\sigma_{p_A p_B \rightarrow n} &=& \sum _q \int dx_a dx_b \int f_{a/A} (x_a,  Q^2) f_{b/B} (x_b, Q^2) \times \\
&& \times [\hat \sigma _0 + \alpha_s(\mu_R^2) \hat \sigma_1 + ....]_{ab \rightarrow n},
\end{eqnarray}
\noindent where $\sigma _0$ is the tree-level parton-parton cross-section and
 $\sigma _1$ is the first order QCD correction to the parton-parton cross-section, etc. The renormalisation scale, $\mu_R$ is the 
reference scale for the running of $\alpha_s(\mu_R^2)$, caused by ultraviolet divergences in finite-order calculations. 

Writing this equation in terms of the matrix elements yields
 
\begin{eqnarray}
\nonumber
\sigma_{p_A p_B \rightarrow n} &=& \sum _q \int dx_a dx_b \int d\phi_n  \times \\
&& \times f_{a/A}(x_a, Q^2) f_{b/B} (x_b, Q^2) \frac{1}{2 \hat s} |m_{q\bar q \rightarrow n}|^2(\phi_n),
\end{eqnarray}
where $1/(2\hat s)$ is the parton flux, $\phi_n$ is the phase space of the 
final state and $|m_{q\bar q \rightarrow n}|$ is the corresponding matrix 
element for a final state $n$, which is produced via the initial state $q \bar 
q$. The matrix element can then be evaluated according to perturbation theory as a 
sum of Feynman diagrams, $m_{q\bar q\rightarrow n} = \sum_i F^{(i)}_{q\bar q 
n}$. The evaluation of these integrals over the full phase space is typically achieved via Monte Carlo 
sampling methods. 

\subsection{\label{sec:MECalc} Matrix-element calculations}

\subsubsection{\label{sec:LOC} Leading-order calculations}

The calculation of the electron-positron annihilation cross-section in pure Quantum Electrodynamics (QED), i.e. $e^+e^- 
\rightarrow \mu^+\mu^-$, is straightforward and can easily be extended to the quark-antiquark annihilation cross-section by including the 
colour factor of $1/3$ and accounting for the charge $Q_q$ of the involved quarks $q$:

\begin{equation} 
\hat \sigma _{q\bar q \rightarrow l \bar l'} = \frac{4 \pi}{9 \hat s} \cdot  \alpha^2_{em} \cdot Q^2_q\, ,
\end{equation}
where $\hat s = (x_A P_A + x_B P_B)^2 = x_A x_B s$ and $\sqrt{s}$ denotes the center-of-mass 
energy of the proton-proton collision. In an electroweak unified theory, the cross-section 
must also include the exchange of a \Zboson boson for larger energies ($\sqrt{s}>40\,\GeV$) and is therefore extended by

\begin{eqnarray} 
\nonumber
&\hat \sigma& _{q\bar q \rightarrow l \bar l'} = \frac{4 \pi}{9 \hat s} \left\{ Q^2_q - Q_q \frac{\sqrt{2} G_F m_Z^2}{4 \pi \alpha} g_V^l g_V^q Re(K(\hat s)) +\right. \\ 
\nonumber
&&+ \left. \frac{G_F^2 m_Z^4}{8 \pi ^2 \alpha_{em} ^2} \cdot ((g_V^l)^2 + (g_A^l)^2) \cdot ((g_V^q)^2 + (g_A^q)^2) |K(\hat s) |^2 \right\}\, .
\end{eqnarray}
The vector and axial couplings of the \Zboson bosons to the leptons and quarks are 
denoted as $g_V=g_L+g_R$ and $g_A=g_L-g_R$ which can be expressed as combinations of left- and right-handed chiral states for the quarks $q$ and leptons $l$. 
The \Zboson boson propagator $K (\hat s)$ can be written as

\begin{equation}  
K(\hat s) = \frac{\hat s}{\hat s - m_Z^2 + i m_Z \Gamma _Z}\,.
\end{equation} 
In the narrow-width approximation, the \Zboson boson is assumed to be a stable particle and the propagator reduces to a $\delta$-function. 
This approximation is based on the fact that the width of the \Zboson boson ( $\Gamma _Z \approx 2.5\, \GeV$) is small 
compared to its mass ($m_Z\approx 91\,\GeV$). Hence the parton-parton cross-section can be expressed as
\begin{equation}  
\hat \sigma _{q\bar q \rightarrow Z} = \frac{\sqrt{2 \pi}}{3} G_F m_Z^2  ((g_V^q)^2 + (g_A^q)^2) \delta(\hat s - m_Z^2),
\end{equation} 
when omitting the interference with the photon-exchange in the s-channel 
\footnote{The interference at $\hat s = m_Z^2$ is at per mille level}. 

The decay of the \Zboson boson into fermion pairs is described by the branching ratio 
$Br(Z\rightarrow f\bar f) = \Gamma(Z\rightarrow f\bar f) / \Gamma_Z$, where the 
partial width $\Gamma(Z\rightarrow f\bar f)$ is given in lowest order by

\begin{equation}
\Gamma(Z\rightarrow f\bar f) = N_C \frac{G_F m_Z^3}{6 \sqrt{2} \pi}  ((g_V^f)^2 + (g_A^f)^2),
\end{equation} 
where the factors $g_V^f$ and $g_A^f$ are again the vector- and axial couplings for 
the respective fermions $f$ to the \Zboson boson. The colour factor $N_C$ is $1$ for leptons and $3$ for quarks.
 This leads to a prediction of 
$\approx70\%$ decays into quark and antiquarks, but only $\approx 3.4\%$ for 
the decay into a single generation of charged leptons. 

The lowest-order cross-section for the \Wboson boson production via quark-antiquark 
fusion can be derived in a similar manner. In contrast to the \Zboson boson 
production, quarks from different generations can couple to the \Wboson boson, while 
the interference with the electromagnetic sector is not present. The 
cross-section in the narrow width approximation is given by

\begin{equation}  
\hat \sigma _{q\bar q' \rightarrow W} = \frac{\sqrt{2} \pi }{3} G_F m_W^2 |V^{CKM}_{q q'}|^2 \delta(\hat s - m_W^2),
\end{equation} 
where the CKM matrix element accounts for the quark-generation mixing. The partial decay width of 
the \Wboson boson at lowest order is

\begin{equation}  
\Gamma(W\rightarrow f\bar f') = N_C \frac{G_F m_W^3}{6 \sqrt{2} \pi},
\end{equation} 
leading to $1/3$ probability for leptonic decays and $2/3$ for decays into quark/antiquark pairs.

\subsubsection{\label{sec:QCDCorrections} Perturbative QCD corrections and jets}

The leading-order calculations of the \Wboson and \Zboson boson production, as shown in Section 
\ref{sec:LOC}, suggest that the momentum of the boson in the transverse plane is zero. 
Yet it is well known from collider experiments that the transverse momentum ($\pT$) of \Wboson and \Zboson bosons 
peaks at few GeV, with a pronounced tail to high values, i.e. $\pT \gg  m_V$ \cite{Abe:1991pu}, \cite{Abe:1991rk}, \cite{Abbott:1998jy}, \cite{Abbott:1999yd}. 
To understand the physical 
origin of this, two different effects have to be taken into account. First, the 
interacting partons are believed to have an intrinsic transverse momentum 
($k_T$) relative to the direction of the proton,
leading to an exponentially decreasing $\pT$ distribution of the vector boson.
The experimentally determined value of the average intrinsic momentum is $<k_T> = 
0.76\,\GeV$, measured in proton-neutron collisions \cite{Ellis:1991qj} and is not large enough
to explain the observed $\pT$ distribution of vector bosons in hadron 
collisions. The second, larger effect arises from higher order QCD corrections to the 
vector boson production, which can lead to the radiation of additional quarks 
and gluons in the final state in the transverse plane. The vector sum of these 
emissions has to be balanced by the transverse momentum of the produced vector 
boson, which in turn acquires a transverse momentum. In the regime where $\alpha_s$ is small, 
these perturbative QCD corrections may be calculated. The two contributing classes of next-to-leading-order (NLO)
corrections, i.e. the virtual loop corrections and the real emissions of gluons/quarks, are 
illustrated in Figure \ref{fig:QCDCorrections}.
The correction terms with virtual loops do not affect the transverse momentum spectrum of the vector boson directly. The real corrections however, imply the existence of 
$2\rightarrow2$ processes, leading to an additional parton in the final state which boosts the \Wboson or \Zboson boson in the transverse plane. 

\begin{figure}[b]
\resizebox{0.5\textwidth}{!}{\includegraphics{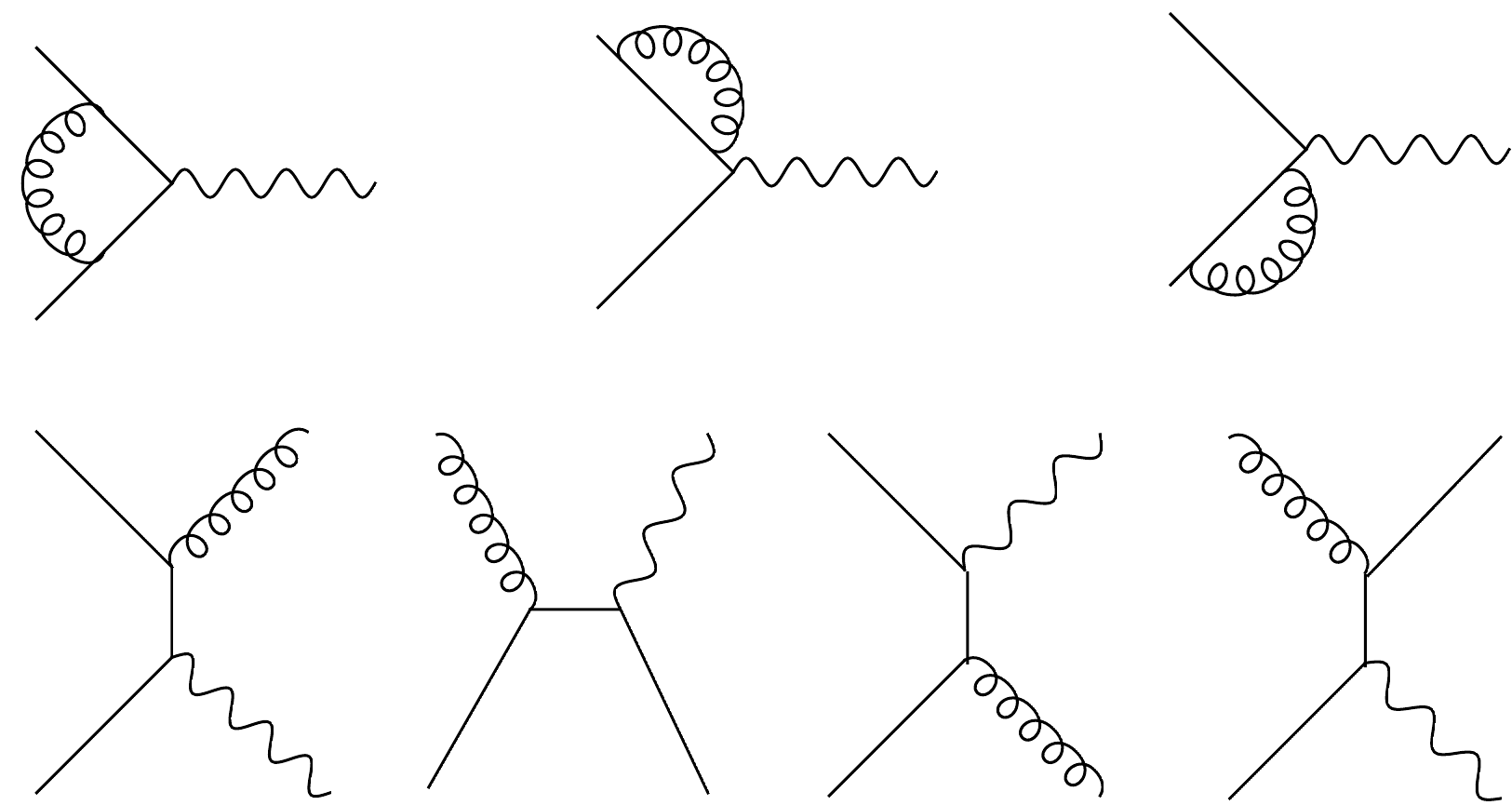}}
\caption{Perturbative QCD corrections}
\label{fig:QCDCorrections}
\end{figure}

The generic form of the production cross-section for the processes 
$q\bar q \rightarrow Vg$ and $qg \rightarrow V q'$, where $V$ stands for a 
vector boson, can be expressed by the Madelstam variables, describing the 
Lorentz invariant kinematics of a $2\rightarrow2$ scattering process. The resulting cross-section at NLO is proportional to

\begin{equation} 
\label{EQN:QCDPT}
\sigma \sim \frac{2\cdot m_V^2 \cdot s + t^2 + u^2}{tu},
\end{equation}
As $t,u\rightarrow 0$, divergencies in Equation \ref{EQN:QCDPT} occur. This can be 
interpreted as final state quarks or gluons with a vanishing transverse momentum, 
i.e. those which are collinear to the incoming parton. Therefore, a minimal 
$\pT$ requirement  of the additional quark or gluon in the final state needs to 
be applied to obtain a finite production cross-section prediction. In the calculation of the 
fully inclusive production cross-section, the divergencies are compensated by the virtual loop corrections.

Two main energy regimes of the transverse momentum 
spectrum of the vector boson production are considered here: A high energy regime, where $\pT\gg m_V$ 
and an intermediate regime, where $k_T<\pT(V)<m_V/2$. 
For very large transverse momenta of the vector bosons ($\pT\gg m_V$), the 
real NLO corrections lead to an expected transverse momentum distribution of

\begin{equation} 
\frac{d^2\sigma}{d^2\pT} \sim \frac{\alpha_s (\pT^2)}{\pT^4}.
\end{equation} 

\noindent The linear dependence of $\alpha_s$ is a consequence of the NLO 
QCD corrections, leading to one additional parton in the final state (V+1 jet production). 

Each additional parton in the final state requires one additional higher order QCD correction and therefore an additional order of $\alpha_s$. 
Examples of leading-order Feynman diagrams for the Z+2 jet production are shown in Figure~\ref{fig:Z2Jets}. 
For QCD corrections with multiple jets, the probability that an additional parton is a radiated gluon
is governed by a Poisson distribution. This implies that the leading-order term for a 
V+n-jet final state, called {\it Poisson scaling}, has the form of
\begin{equation} 
\label{eqn:poisson}
\sigma _{V+n-jet}^{LO} \sim \frac{\bar{n} e^{-\bar{n}}}{n!} \sigma_{tot} \, ,
\end{equation} 
where $\sigma_{tot}$ is the total cross-section, $\bar{n}$ is a expectation value of the Poisson.
This is the expected behaviour at $e^+ e^-$ colliders~\cite{Gerwick:2012hq}, where PDFs do not play a role.
However, at hadron colliders, the experimentally observed V+n-jet final state, called {\it staircase scaling}, has the form of
\begin{equation} 
\label{eqn:starcaise}
\sigma _{V+n-jet}^{LO} \sim \sigma_{0} e^{-a n} \, ,
\end{equation} 
where the coefficients $a$ depend on the exact definition of a jet and $\sigma_0$ is the zero-jet exclusive cross-section.
The ratio of the n-jet and (n+1)-jet cross-sections is then a constant value, $\frac{\sigma_{n+1}}{\sigma_n} = e^{-a}$,
where $e^{-a}$ is a phenomenological parameter which is measurement dependent.
The reason for observed staircase scaling at hadron colliders is two-fold. At small numbers of additional 
partons, the emission of an additional parton is suppressed in the parton density function. 
At larger numbers of additional partons, the probability of gluon radiation no longer follows
a Poisson distribution due to the non-abelian nature of QCD theory, which states that a gluon can radiate 
from another gluon. For large 
jet multiplicities a deviation from the staircase scaling behaviour is expected, as 
the available phase space for each additional jet in the final state decreases. 

Today, several leading-order calculations, such as \cite{Mangano:2002ea} and \cite{Caravaglios:1998yr}, 
are available that describe more than six partons in the final state. The inclusive production cross-section and the associated rapidity distribution for vector bosons is known today to 
next-to-next-to-leading order (NNLO) \cite{Hamberg:1990np}, \cite{vanNeerven:1991gh}, \cite{Anastasiou:2003ds}.


\begin{figure}[htb]
\resizebox{0.5\textwidth}{!}{\includegraphics{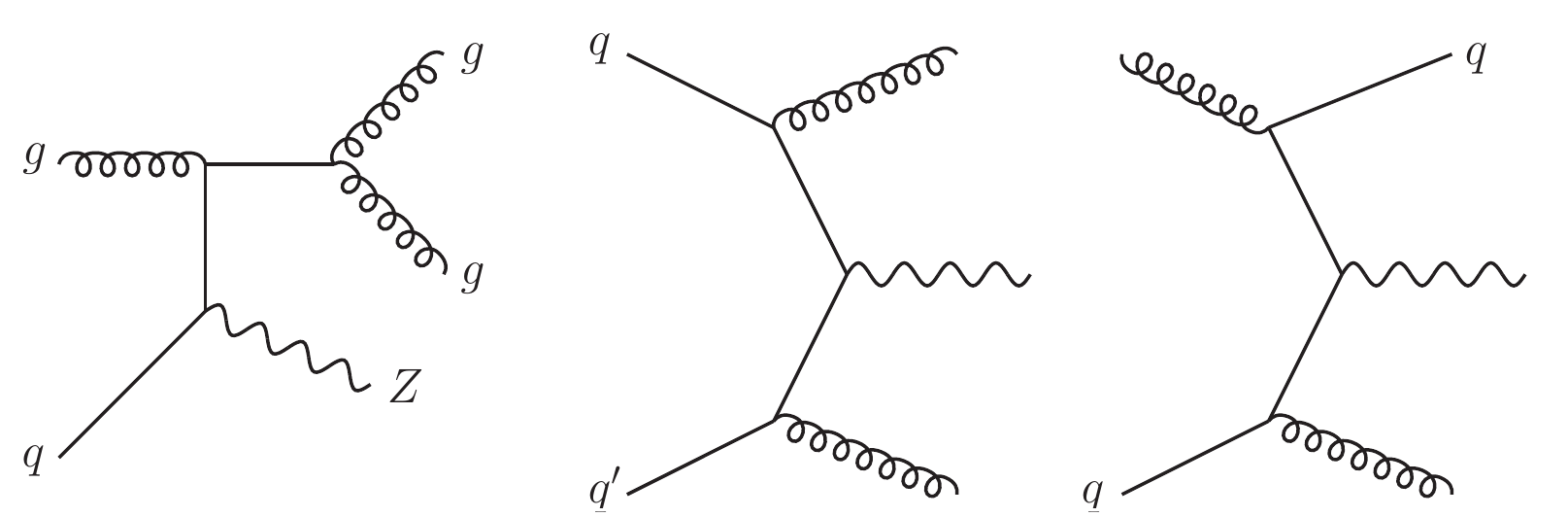}}
\caption{Leading-order Feynman diagram for $Z+2$-jet Production}
\label{fig:Z2Jets}
\end{figure}

The intermediate momentum range of 
$k_T<\pT(V)<m_V/2$ can also be assessed with perturbative calculations.
However, higher-order corrections, manifested as low energetic gluons emitted off the incoming
partons at intermediate energies, must be included for a correct description
of the experimental data. This can be most easily seen 
in the limit of $t\rightarrow 0$ and $u\rightarrow 0$; then the final state gluon in 
the $q \bar q' \rightarrow Vg$ process becomes collinear to the incoming parton. 
The corresponding Feynman diagram can be redrawn as initial state 
radiation (ISR) as shown in Figure \ref{fig:ISRRadiation}.

\begin{figure}[htb]
\resizebox{0.5\textwidth}{!}{\includegraphics{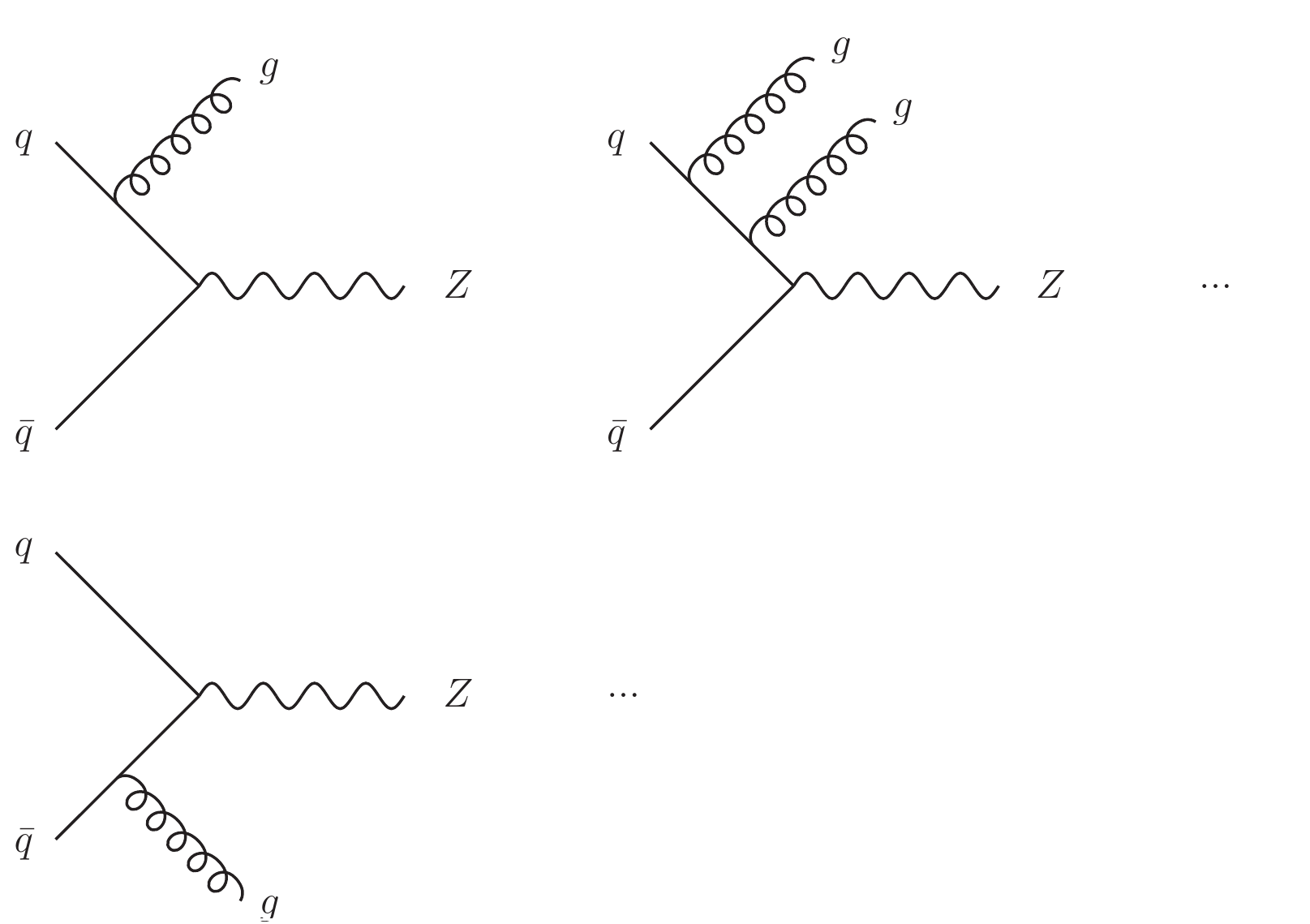}}
\caption{Example Feynman diagrams illustrating initial state radiation of gluons in the leading-order Drell-Yan process.}
\label{fig:ISRRadiation}
\end{figure}

The main contributions of these collinear gluon emissions to the cross-section at the $n$-th order 
are given by
\begin{equation}
\frac{1}{\sigma} \frac{d\sigma}{d\pT^2} \sim \frac{1}{\pT^2} \alpha_s^n ln 
^{2n-1}\frac{m_V^2}{\pT^2}\, .
\end{equation}
Such collinear gluon emissions are also the basis for parton showers, which 
will be discussed in Section \ref{sec:PartonShower}. Summing up the gluon emissions to all 
orders leads to
\begin{equation}
\frac{1}{\sigma} \frac{d\sigma}{d\pT^2} \approx \frac{d}{d\pT^2} e^{ \left( - \frac{\alpha _s}{2 \pi} C_F ln^2 \frac{m_V^2}{\pT^2}\right)},
\end{equation}
where $C_F=4/3$ is the QCD colour factor for gluons. This approach, known as \textit{Resummation} \cite{Laenen:1991af},  has been 
significantly improved and extended in recent years and provides currently the 
most precise predictions for the transverse momentum distribution of vector 
bosons in the low energy regime.

\subsubsection{Electroweak corrections}

So far, only QCD corrections to \Wboson and \Zboson boson production have been 
discussed. The virtual one loop QED corrections and the real photon radiation 
corrections are illustrated via Feynman diagrams in Figure \ref{fig:QEDCorrections}. The NLO 
corrections to the charged and neutral currents are well known \cite{Baur:2001ze}, \cite{Balossini:2008cs}. In 
particular, the full $\mathcal{O}(\alpha_{em})$ corrections to the $pp\rightarrow \gamma/Z
\rightarrow l^+l^-$ process with $\mathcal{O}(g^4\mT^2/m_W^2)$ corrections to the effective 
mixing angle $sin^2\theta _{eff} ^2$ and $m_W$ are available \cite{Arbuzov:2005ma, Bardin:1999yd, Akhundov:1985fc, Bardin:1986fi, Degrassi:1995mc}.

\begin{figure}[htb]
\resizebox{0.5\textwidth}{!}{\includegraphics{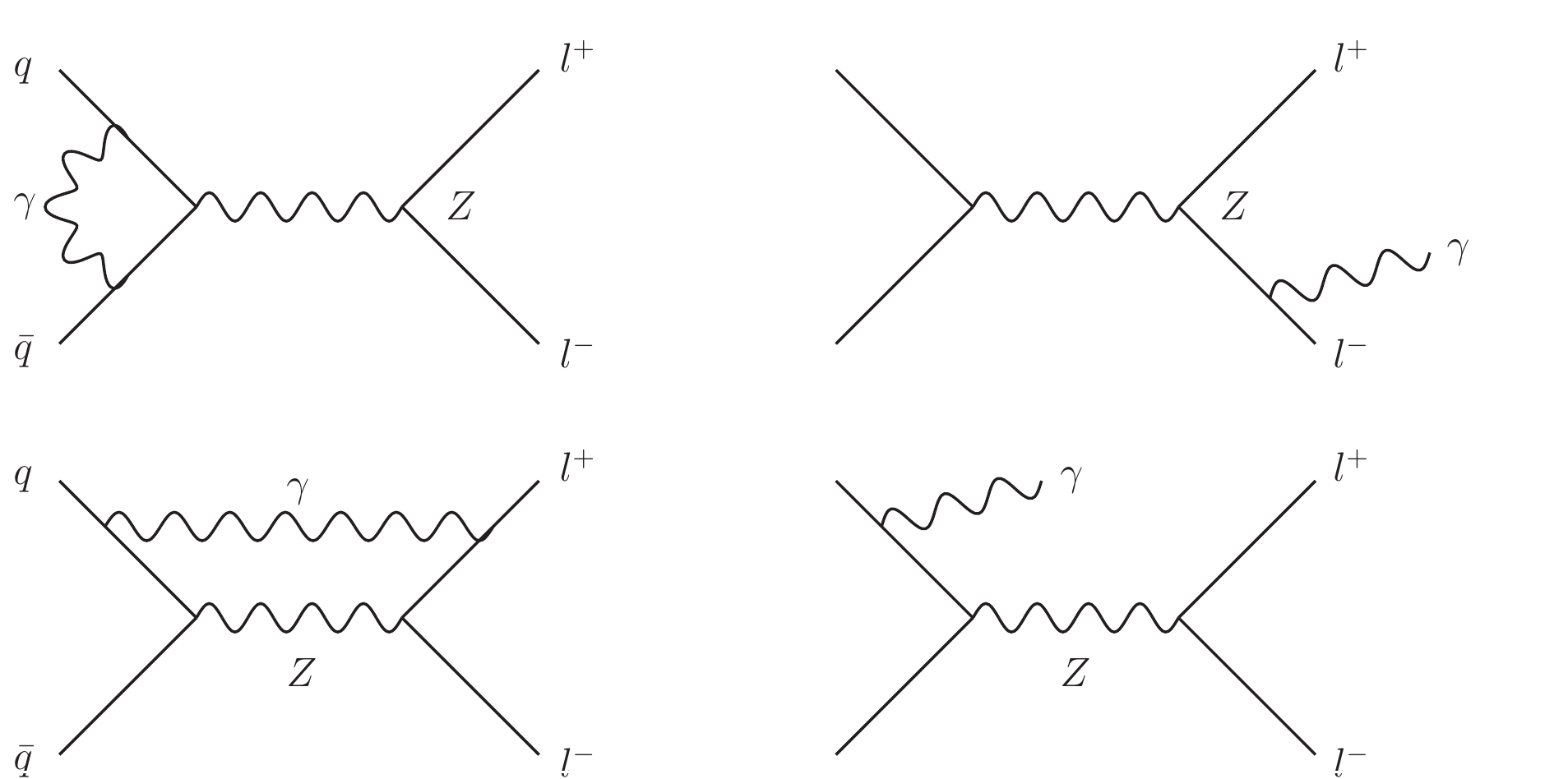}}
\caption{Examples of leading-order QED corrections}
\label{fig:QEDCorrections}
\end{figure}

Even though the electroweak corrections to the vector boson production cross-section are 
small compared to the higher-order QCD corrections, they lead to a significant distortion of the 
line shape of the invariant mass and subsequently to the transverse momentum 
spectrum of the decay leptons. The comparison between the corresponding 
distributions with and without electroweak corrections is illustrated in Figure \ref{fig:TheoryQEDCorImpact}. 

\begin{figure}[htb]
\resizebox{0.5\textwidth}{!}{\includegraphics{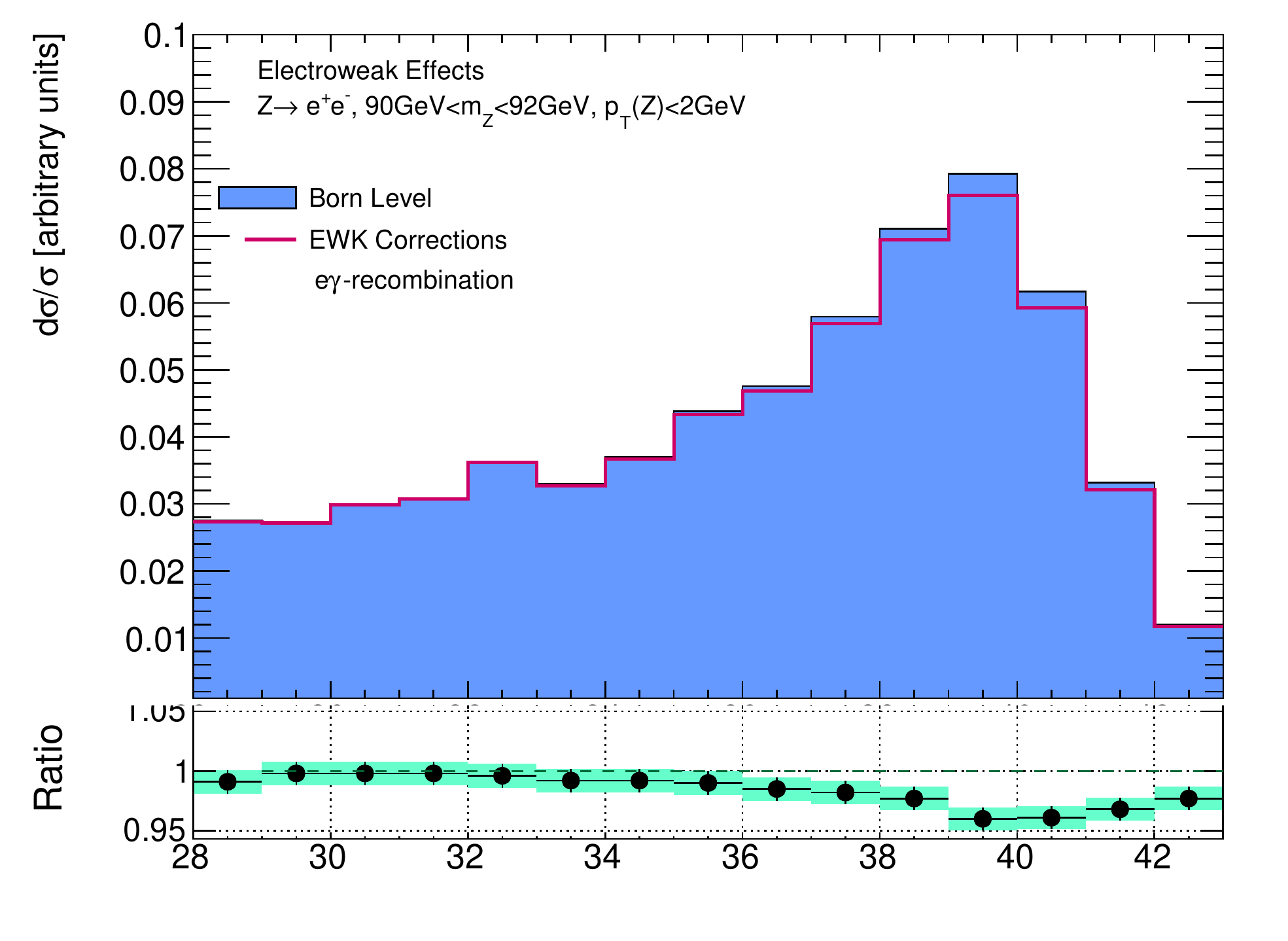}}
\caption{Electroweak (EWK) corrections to the lepton transverse momentum for the neutral current Drell-Yan process at the LHC  \cite{Brensing:2007qm}. Results are presented for bare electrons and electrons employing electron-photon recombination, respectively. }
\label{fig:TheoryQEDCorImpact}
\end{figure}

In general, electroweak corrections at moderate energies are dominated by final-state 
radiations (FSR) of photons, which is indicated by the upper right diagram in Figure \ref{fig:QEDCorrections}. 
Certain measurements, like the determination of the \Wboson boson mass from the decay lepton $\pT$ distribution, 
are sensitive to these corrections. For this measurement, this effect can induce a shift 
of up to $10\,\MeV$ on $m_W$ in the muon decay channel. In contrast, the electron decay channel is less effected due
to the nature of electron reconstruction in the detector, where 
the FSR photons are usually reconstructed together with the decay electron. Therefore the relative magnitude of the electroweak corrections
with respect to the QCD corrections must be considered individually for each measurement. 
It should be noted that there are also diagrams with photons in the initial state and hence the parton density functions evolve with combined QCD and QED evolution equations. That can lead to sizable corrections in the high and low mass Drell-Yan production \cite{Ball:2013hta}.

\subsection{\label{sec:PartonShower}Parton shower models}

As discussed in the previous section, matrix-element calculations at fixed-order 
provide cross-sections assuming that the final state partons have large momenta and are 
well separated from each other. Parton shower models provide a relation between the 
partons from the hard interaction ($Q^2\gg \Lambda _{QCD}$) to partons near 
the energy scale of $\Lambda_{QCD}$. Here $\Lambda_{QCD}$ is defined as the transition energy
between the high-energy and low-energy regions. A commonly used approach for parton shower models
is the leading-log approximation, where showers are modelled as a sequence of splittings of a
parton $a$ to two partons $b$ and $c$. QCD theory allows for three types of 
possible branchings, $q\rightarrow qg$, $g\rightarrow gg$ and $g\rightarrow q\bar q$, 
while only two branchings exist in QED theory, namely $q\rightarrow q\gamma$ and 
$l\rightarrow l \gamma$.

The differential probability $dP_{a}$ for a branching for QCD emissions is 
given by the Dokshitzer-Gribov-Lipatov-Altarelli-Parisi (DGLAP) evolution equations \cite{Gribov:1972ri, Dokshitzer:1977sg, Altarelli:1977zs} 

\begin{equation}
\label{EQN:DGLPA1}
dP_a = \sum_{b,c} \frac{\alpha_s(t)}{2 \pi} P_{a\rightarrow b,c}(z) dt dz\, ,
\end{equation}
with the evolution time $t$ defined as $t=ln(Q^2/\Lambda^2_{QCD})$. $Q^2$ denotes the 
momentum scale at the branching and $z$ the momentum fraction of the parton $b$ 
compared to parton $a$. The sum runs over all possible branchings and  
$P_{a\rightarrow b,c}$ denotes the corresponding DGLAP splitting kernel. 

This relation also holds for the QED branchings, where the coupling constant in Equation~\ref{EQN:DGLPA1}
is replaced with $\alpha_{em}$. For final state radiation, the emission of particles due 
to subsequent branchings of a mother parton is evolved from $t=Q^2_{hard}$ 
at the hard interaction to the non-perturbative regime $t\approx \Lambda_{QCD}$. 
Initial-state radiation can be ordered by an increasing time, i.e. going 
from a low energy scale to the hard interaction. This can be
interpreted as a probabilistic evolution process connection two different scales: the initial 
scale $Q_0^2$ of the interaction to the scale of the hard interaction scale $Q_{hard}^2$.
During this evolution, all possible configurations of branchings, leading to a defined set of partons 
taking part in the hard interaction, are considered.

The implementation of parton showers is achieved with Monte Carlo techniques. 
They are used to calculate the step-length $t_0$ to $t$ where the virtuality 
of the parton decreases with no emissions. At the evolution time $t$, a 
branching into two partons occurs where the resulting sub-partons have 
smaller virtuality than the initial parton. This procedure is then repeated 
for the sub-partons, starting at the new evolution time $t_0=t$. 

Therefore the probability for the parton 
$a$ at the scale $t_0$ not to have branched when it is found at the scale $t$ has to be determined. 
This probability $P_{no-branching}(t_0,t)$ is given by the Sudakov form factor \cite{Sudakov:1954sw},
which can be expressed as

\begin{equation}
\label{EQN:SUFAKOV}
S_a(t) = \exp \left( - \int _{t_0}^t dt' \sum_{b,c} I_{a\rightarrow b,c}(t') \right),
\end{equation}
where

\begin{equation}
I_{a\rightarrow b,c}(t)= \int_{z_-(-t)}^{z_+(t)} dz \frac{\alpha_s(t)}{2}\cdot P_{a\rightarrow b,c} (z)
\end{equation}
is the differential branching probability for a given evolution time $t$ with 
respect to the differential range $dt'$. The latter relation follows directly 
from 
Equation \ref{EQN:DGLPA1}, by integrating over the allowed momentum 
distributions $z$. The probability for a branching of a given parton $a$ at 
scale $t$ can then be expressed by the derivative of the Sudakov form factor 
$S_a(t)$:

\begin{equation}
\frac{dP_a(t)}{dt} = \left(\sum_{b,c} I_{a\rightarrow b,c} \right) S_a(t)\, .
\end{equation}
This relation describes the effect known as Sudakov suppression: The first 
factor $I_{a\rightarrow b,c}$, which describes the branching probability at a 
given time $t$, is suppressed by the Sudakov form factor $S_a(t)$, i.e. by taking into account the 
possibility of branchings before reaching the actual scale $t$.

The branching of the initial-state partons during the parton shower therefore leads to
the emission of gluons or quarks, which in turn may add an additional jet to the 
event. The final-state partons predicted within the leading-log approximation 
are dominated by soft and collinear radiations and hence large momentum jets 
that are not expected to be described correctly within this approximation. The kinematics of 
the missing hard scatter components are predicted by the corresponding higher-order diagrams, which have
been discussed in Section \ref{sec:QCDCorrections}. 

The main advantage of the parton shower approach is its simplicity compared to matrix-element calculations which increase 
in complexity when considering more independent partons in the initial and final states.
However, there is an important difference between the soft-gluon emission described by parton shower
and the emission of a gluon calculated by NLO matrix element. While the full matrix-element calculation 
includes the spin-1 nature of the gluon and hence induced polarisation effects on the intermediate
gauge boson, the parton shower algorithm does not take into account spin effects.

\subsection{\label{sec:PDF}Parton distribution functions and scale dependencies}

The PDFs play a central role not only in the calculation of the cross-section in 
Equation \ref{eqn:prodcross}, but also in the modelling of parton showers and hadronisation 
effects. A generic PDF $f_i(x,\mu _F, \mu_R)$ describes at lowest order the 
probability of finding a parton of type $i$ with a momentum fraction $x$ when a 
proton is probed at the scale $\mu _F$. The factorisation and renormalisation scale parameters $\mu_F$ 
and $\mu_R$ in the PDF definition act as cut-off parameters to prohibit infrared and 
ultraviolet divergences. If a cross-section could be calculated to 
all orders in perturbation theory, the calculation would be independent from the choice of the 
scale parameters, since the additional terms would lead to an exact cancellation of 
the explicit scale dependence of the coupling constant and the PDFs. 
Both scales are usually chosen to be on the 
order of the typical momentum scales of the hard scattering process, to avoid terms with large logarithms appearing in the perturbative corrections. 
For the Drell-Yan process at leading order this implies $\mu_F=\mu_R=m_Z$. It should be noted that both 
scales are usually assumed to be equal, even though there is no reason from 
first principles for this choice. The dependence of the predicted
cross-section on $\mu_F$ and $\mu_R$ is thus a consequence of the 
missing/unknown higher-order corrections. The dependence is therefore reduced 
when including higher orders in the perturbation series. The uncertainty on the 
cross-section prediction due to scale uncertainties is usually estimated by 
varying both scales simultaneously within $0.5\cdot Q<\mu_F,\mu_R<2\cdot Q$, where Q is 
the typical momentum scale of the hard process studied. However, this 
evaluation procedure sometimes provides results that are too optimistic and the differences between the leading-order and NLO calculations are
not always covered by the above procedure.

As the actual form of $f_i(x,\mu_F)$ cannot be predicted with perturbative QCD theory, a parameterised functional form has to be 
fitted to experimental data. The available data for the PDF determination comes 
mainly from deep inelastic scattering experiments at HERA, neutrino data, 
as well as Drell-Yan and jet production at the Tevatron and LHC colliders. Note that the 
scale dependence of $f_i$ is predicted by the DGLAP evolution equations.

In order to fit PDFs to data, a starting scale, where perturbative QCD 
predictions can be made, has to be chosen and a certain functional form of the 
PDFs has the be assumed. A typical parametrisation of $f_i(x,\mu_F)$ takes the 
form

\begin{equation}
f_i(x,\mu _F) = a_0 x^{a_1} (1-x)^{a_2} P(x, a_3, a_4, ...)\, ,
\end{equation}
where P is a polynomial function and $a_j$ are free fit parameters which cannot be predicted from perturbative QCD calculations, but can only be
 determined by experiment. In a second step, a factorisation scheme, i.e. a model for the 
handling of heavy quarks, and an order of perturbation theory has to be chosen. 
The DGLAP evolution equations can then be solved in order to evolve the chosen PDF 
parametrisation to the scale of the data. The measured observables can then be  
computed and fitted to the data. The PDF fits are currently performed and 
published for leading-order, NLO and NNLO calculations. Even though most matrix elements 
are known to NLO order in QCD theory, some parton shower models are still based on leading-order 
considerations and therefore leading-order PDF sets are still widely used.

PDF fitting is performed by several groups. The \CTEQ-TEA \cite{Nadolsky:2008zw}, \MSTW~
\cite{Martin:2009iq}, ABKM \cite{Alekhin:2009ni}, GJR \cite{Gluck:2007ck} and \NNPDF~\cite{Ball:2011us} collaborations include all available data for their 
fits, but face the problem of possible incompatibilities of the input 
data, caused by inconsistent measurement results from different experiments.
These results differ in the treatment of the parametrisation assumptions 
of $f_i$. The \HeraPDF~\cite{CooperSarkar:2011aa} group bases their PDF fits on a subset of the 
available data, where only the HERA measurements have been chosen as input and therefore 
possible inconsistencies in the fit are reduced. 

It should be highlighted that the PDF approach and fitting is subject to several assumptions and model uncertainties. The actual form of the input distributions is arbitrary and hence the choice of the analytical function implies a model uncertainty. The approach of the \NNPDF\,group is an exception as the parametrisation is chosen to be handled by a flexible neural network approach. In addition it is commonly assumed that the strange-quark content follows $s = \bar s = (\bar u + \bar d)/4$. The suppression of $s$ and $\bar s$-quark content is due to their larger masses compared to the $\bar u$ and $\bar d$ quarks, but a rigorous argument of the chosen suppression factor cannot be derived from first principles. Similar is the situation for heavy-flavour ($c,b,t$-quarks) contributions to the proton structure.  Their contribution is 0 below the $Q^2$-threshold and is evolved according to the DGLAP-equations above.

\begin{figure}[htb]
\resizebox{0.5\textwidth}{!}{\includegraphics{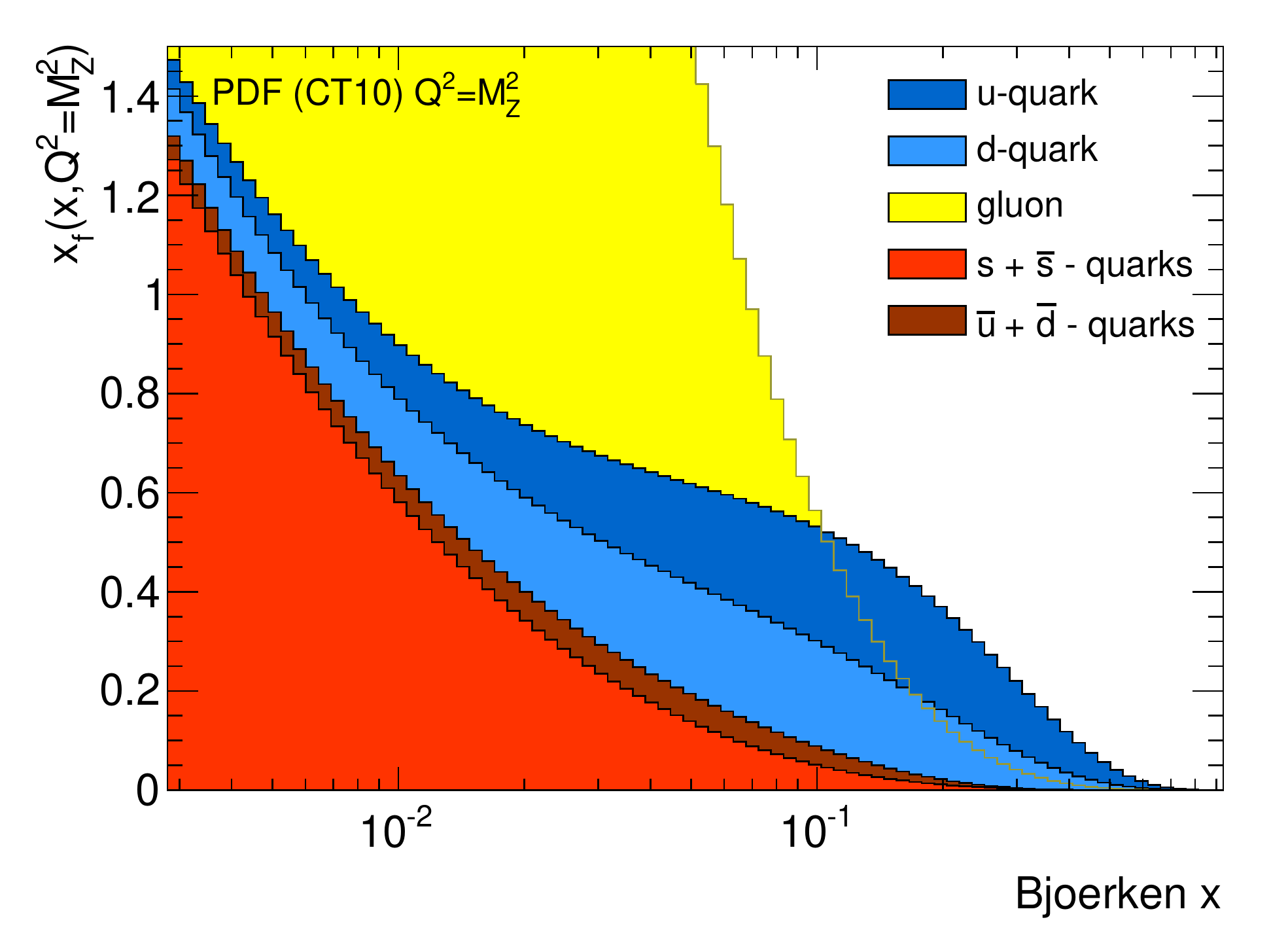}}
\caption{PDF distribution of the CT10 PDF set at $Q^2=m_Z^2$.}
\label{fig:CTEQPDFX}
\end{figure}

\begin{figure}[htb]
\resizebox{0.5\textwidth}{!}{\includegraphics{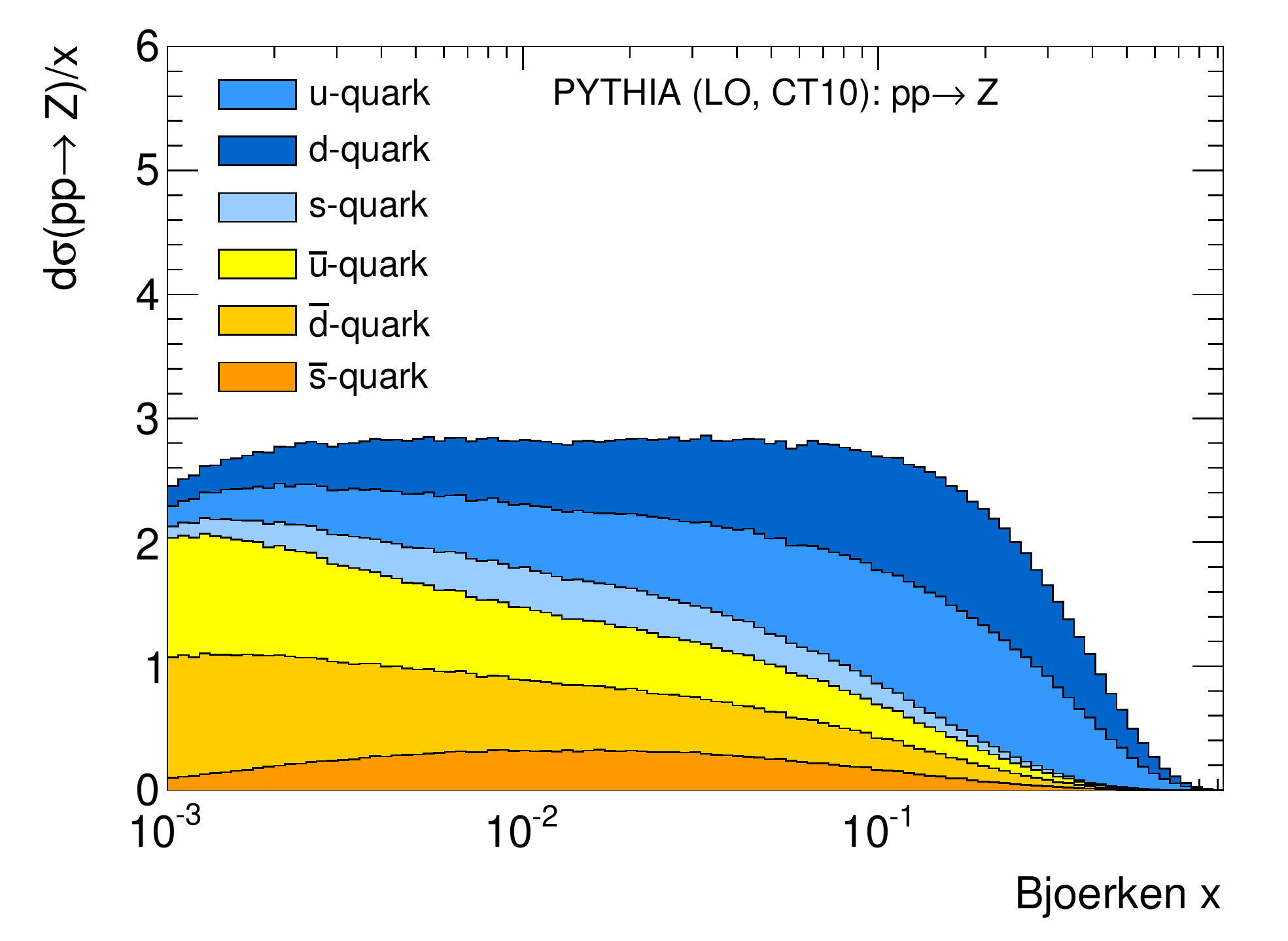}}
\caption{Distribution of Bjorken x-values of partons that are involved in the leading-order production of \Zboson 
bosons at $7\,\TeV$ pp collisions as a stacked histogram. It should be noted that 
each participating valence-quark has to be matched with a corresponding sea-quark.}
\label{fig:CTEQZProd}
\end{figure}

The results presented in this paper rely mainly on the \CTEQ-10 and \MRST~PDF 
sets \cite{Nadolsky:2008zw}, \cite{Martin:2009iq}. The PDF set for $Q^2=m_Z^2$ from the 
CTEQ collaboration are illustrated in Figure \ref{fig:CTEQPDFX}. The Bjorken x-values 
 of partons, which are involved in the leading-order production of \Zboson bosons at $7\,\TeV$ pp collisions, 
 are illustrated in Figure \ref{fig:CTEQZProd}, where the $x$-values of both interacting quarks
 per events have been used.

The associated uncertainties of a given PDF set are 
based on the Hessian method \cite{Pumplin:2001ct}, where a diagonal error matrix 
with corresponding PDF eigenvectors is calculated. The matrix and the PDF 
eigenvectors are functions of the fit-parameters $a_i$. Non-symmetric 
dependencies are accounted for by using two PDF errors for each eigenvector in 
each direction. The PDF uncertainties on a cross-section is then given by

\begin{equation}
\Delta \sigma = \frac{1}{2} \sqrt{\sum_{k=1}^n [\sigma(a_i^+)-\sigma(a_i^-)]^2},
\end{equation}
where $a^\pm_i$ labels the corresponding eigenvectors of a chosen PDF error set. This approach assesses only the PDF fit 
uncertainties within a certain framework, i.e. for a chosen parametrisation 
and various base assumption. Hence, usually also the difference between two 
different PDF sets from two independent groups, e.g. \CTEQ~and \MRST, are taken 
as an additional uncertainty. The same procedure for the impact of PDF related 
uncertainties can be applied for any observable and is not restricted to inclusive 
cross-sections. 

\subsection{\label{sec:Hadron} Hadronisation}

The process of how hadrons are formed from the final state partons is call hadronisation. The scale
at which the hadronisation is modelled is $Q^2 = \Lambda^2 _{QCD}$ .
Since this process is complex, phenomenological models must be used. A detailed discussion can be found elsewhere \cite{Sjostrand:1987su}.

The first models of hadronisation were proposed in the 70s 
\cite{Field:1977fa} and today, two models are widely 
in use. The so-called \textit{string model} \cite{Andersson:1983ia} is based on the assumption of a linear confinement and provides a 
direct transformation from a parton to hadron description. It accurately predicts 
the energy and momentum distributions of primary produced hadrons, 
but requires numerous parameters for the description of flavour properties, 
which have been tuned using data. The second approach for the description of 
hadronisation is known as \textit{cluster model} \cite{Field:1982dg}, \cite{Webber:1983if}, which is based on the 
pre-confinement property of parton showers \cite{Amati:1979fg}. It involves an 
additional step before the actual hadronisation, where colour-singlet subsystems 
of partons (denoted as \textit{clusters}) are formed. The mass spectrum of these 
clusters depend only on a given scale $Q_0$ and $\Lambda _{QCD}$, but not on a 
starting scale Q, with $Q\gg Q_0>\Lambda _{QCD}$. The cluster model has therefore 
fewer parameters than the string model, however, the description of data is 
in general less accurate.

The subsequent decay of primary hadrons is either directly implemented in the 
computing codes for hadronisation, or in more sophisticated libraries such as 
EVTGEN\cite{Lange:2001uf}. Special software libraries can be used for the 
description of the $\tau$-lepton decay, e.g. TAUOLA \cite{Jadach:1993hs}, correctly taking 
into account all branching ratios and spin correlations. Since hadronisation 
effects usually have only a small impact on the relevant observables discussed in this 
article, we refer to \cite{Sjostrand:1987su} for a detailed discussion.

\subsection{\label{sec:MPI} Multiple-parton interactions}

Equation \ref{eqn:prodcross} describes only a single parton-parton interaction within a 
proton-proton collision. However, in reality, several parton-parton interactions 
can occur within the same collision event. This phenomenon is known as 
multiple-parton interactions (MPI). Most of the MPI lead to soft 
additional jets in the event which cannot be reconstructed in the detector
due to their small energies. Hence they contribute only 
as additional energy deposits in the calorimeters. However, a hard perturbative tail of 
the MPI, following $\sim d\pT^2/\pT^4$, where $\pT$ is the transverse momentum 
of the additional jets, can lead to additional jets in the experimental 
data. These effects must be taken into account for the study of vector boson 
production in association with jets. Further information about the current 
available models for MPI can be found in \cite{Bartalini:2010su}. Dedicated studies
of MPI have been done at the LHC using \Wboson events with two associated jets~\cite{Aad:2013bjm}.
The fraction of events arising from MPI is $0.08 \pm 0.01 (\textrm{stat.}) \pm 0.02 (\textrm{sys.})$ 
for jets with a $\pT > 20\, \GeV$ and a rapidity $|y| < 2.5$. This fraction decreases when the \pT~requirements
on the jets increases. 

\subsection{\label{sec:ComputerCodes} Available computing codes}

\subsubsection{Multiple purpose event generators}

Multiple purpose event generators include all aspects of the proton-proton 
collisions: the description of the proton via an interface to PDF sets, 
initial-state shower models, the hard scattering process and the subsequent 
resonance decays, the simulation of final-state showering, MPI, the
hadronisation modelling and further particle decays. Some frequently used generator in the 
following analyses are \Pythia6 \cite{Sjostrand:2006za}, \Pythia8 \cite{Sjostrand:2007gs}, \Herwig\ \cite{Corcella:2000bw}, \HerwigPP\ 
\cite{Bahr:2008pv} and \Sherpa\ \cite{Gleisberg:2008ta}. All of these generators contain an extensive list of 
Standard Model and new physics processes, calculated with fixed-order 
tree-level matrix elements. Some important processes, such as the vector boson production
are also matched to NLO cross-sections.

The \Pythia\ generator family is a long established set of multiple purpose 
event generators. While \Pythia6, developed in Fortran in 1982 is still 
in use, the new \Pythia8 generator was coded afresh in C++. 
The showering in \Pythia6 used in this review is implemented with a \pT-ordered showering 
scheme, whereas the new version used here is based on a dipole showering approach. The 
hadronisation modelling in both versions is realised via the Lund string model. MPI are 
 internally simulated in addition.

Similar to \Pythia, the \Herwig~generator was originally developed in Fortan and 
is now superseded by \HerwigPP, written in C++. Both versions use an 
angular-ordered parton shower model and the cluster model for hadronisation. The 
\Jimmy~library \cite{Butterworth:1996zw} is used for the simulation of MPI.

The \Sherpa~generator was developed in C++ from the beginning and uses the 
dipole approach for the parton showering. The hadronisation is realised with 
the cluster model. The MPI are described with a model that is similar to the 
one used in \Pythia. For the analyses described here, \Sherpa~is generated with up to five additional partons
in the final state.

\subsubsection{Leading-order and NLO matrix-element calculations}

Several programs such as  \Alpgen~\cite{Mangano:2002ea} and \MadGraph~\cite{Alwall:2011uj}
calculate matrix elements for leading-order and some NLO processes, but 
do not provide a full event generation including parton shower or 
hadronisation modelling. These generators as well as \Sherpa~are important because they contain matrix-element calculations 
for the production of vector bosons in 
association with additional partons. \Alpgen~is a leading-order matrix-element generator and 
includes predictions up to six additional partons in the final state. This is achieved by adding real emissions to the leading-order diagrams
before the parton shower modelling. In this way, although the process is calculated at leading-order, tree-level diagrams corresponding to higher jet multiplicities
can be included. Some of the virtual corrections are then added when a parton shower model is used. 
\MadGraph~for the analyses presented here
follows a similar method and produces predictions up to four additional partons. The subsequent event generation, 
starting from the final parton configuration, is then performed by \Pythia~or \Herwig~for \Alpgen~and \Pythia~for \MadGraph. 

\subsubsection{\label{sec:PSMatching}Parton shower matching}

There is significant overlap between the phase space of NLO or n-parton final-state QCD matrix-element 
calculations and the application to parton showers with respect to their initial- 
and final-state partons, as both lead to associated jets. To avoid a potential double counting, 
matching schemes have been developed that 
allow matrix-element calculations for different parton multiplicities in the initial state and final state
to be combined with parton shower models. The main strategies 
are based on re-weighting methods and veto-algorithms. 
The Catani-Krauss-Kuler-Webber (CKKW) matching scheme \cite{Catani:2001cc}, \cite{Krauss:2002up} and the Mangano 
(MLM) scheme \cite{Mangano:2001xp} are widely used for tree-level generators. 
For example, the \Alpgen~generator uses the MLM scheme, whereas 
\Sherpa~uses CKKW matching for leading-order matrix-element calculation. 
A detailed discussion can be found in the references given. 

An alternative, less generalised approach to matching schemes are merging strategies. Here the parton 
showers are reweighted by weights calculated by matrix-element calculations. In the \Pythia~
generator only the first branching is corrected, while \Herwig~modifies all 
emissions which could be in principle the hardest. These approaches model correctly one
additional jet, but fail for higher jet multiplicities.

\subsubsection{NLO generators}

While matrix-element calculations give both a good description for the hard 
emission of jets in the final states and handle inferences of initial 
and final states correctly, they are not NLO calculations. A combined NLO 
calculation with parton shower models therefore is much desired. 
However, the above described methods work only for leading-order matrix-element calculations. For the 
matching between NLO matrix element and parton shower models more sophisticated methods have to be used. 
The \MCAtNLO~approach \cite{Frixione:2002ik} was the first available prescriptions to match NLO 
QCD matrix elements to the parton shower framework of the \Herwig~generator. The 
basic idea is to remove all terms of the NLO matrix-element expression which are generated 
by the subsequent parton shower. Due to this removal approach, negative event 
weights occur during the event generation. The \aMCAtNLO~generator follows a similar approach for NLO calculations. 
The second approach is the \Powheg\ procedure \cite{Frixione:2007vw}, which is currently implemented in the 
\PowhegBox\ framework \cite{Alioli:2010xd}. This framework allows for an automated matching of a 
generic NLO matrix element to the parton shower. The \Powheg\ procedure foresees that the hardest 
emission is generated first. The subsequent softer radiations are passed to the showering generator. 
In contrast to the \MCAtNLO~approach, only positive 
weights appear and in addition the procedure can be interfaced to other event 
generators apart from \Herwig. 
\Pythia8 also includes possibilities to match to NLO matrix element using the \Powheg\ scheme.

\subsubsection{NLO calculations and non-perturbative corrections}

\MCFM~\cite{Campbell:2003hd} and \Blackhat~\cite{Berger:2008sj} provide NLO calculations up to two 
and five additional partons respectively. These calculations differ from NLO generators as they do not provide any
modelling of the parton shower. These calculations compute both the virtual and real emission corrections for higher jet
multiplicities. For the virtual corrections, the calculation is achieved by evaluating one-loop corrections to the tree-level
diagrams, while the real emission corrections are obtained by matrix-element calculations which include an additional emitted parton.
Several different techniques are used for these calculations, see for example~\cite{Berger:2008sj}.

Since \MCFM~and \Blackhat~are not matched to a parton shower model, they can not be directly compared to data or simulations
which have parton showering and hadronisation applied to the final-state particles. To mimic the effects of both
the parton shower and the hadronisation, \textit{non-perturbative corrections} are estimated using a multiple purpose generator such as 
\Pythia. These corrections are derived by comparing \Pythia~with and without the parton shower and hadronisation models and
applied directly to prediction cross-sections. The non-perturbative corrections are on the order of 7\% for jets of $\pT < 50\, \GeV$ 
and reduce to zero at higher \pT~values. 

Finally, the inclusive \Wboson and \Zboson boson production cross-sections 
in proton-proton collisions are also known to NNLO precision in $\alpha_s$ 
and can be calculated with the \FEWZ~generator \cite{Gavin:2010az}. This generator allows 
also the prediction of several observables of the final-state objects, such 
as the rapidity distribution of the produced vector bosons.

\subsubsection{\label{sec:TheoResum}Calculations based on resummation}

There are specific programs available, such as \ResBos~\cite{Balazs:1997xd}, 
which are based on resummed calculations and therefore are suited to describe 
the transverse momentum spectrum of vector boson production. \ResBos~provides a 
fully differential cross-section versus the rapidity, the invariant mass and the 
transverse momentum of the vector boson as intermediate state of a 
proton-proton collision. The resummation is performed to NNLL approximation and 
matched to NNLO perturbative QCD calculations at large boson momenta.

\subsubsection{\label{sec:CrossPred}Overview and predicted inclusive cross-sections}

A summary of all Monte Carlo (MC) generators used to describe the relevant signal processes in this work is shown in 
Table \ref{tab:Generators}. The order of perturbation theory, the parton shower matching algorithms and 
the corresponding physics processes are also stated. 

\begin{table*}
\centering
\caption{Monte Carlo programs which are used for the analyses described in this review article. The information on the order of \OAlphaS in the matrix-element calculation, the generator functionality, the possibility to match matrix-element calculations with parton showers and the functionality within the analyses are given. The structure of the table is based on \cite{Nunnemann:2007qs}.}
\label{tab:Generators}       
\begin{tabular*}{\textwidth}{@{\extracolsep{\fill}} lllll }
\hline
\hline
Program 		& Matrix-Element 	& Full Event  				& Merging/Matching 		& Functionality w.r.t.  \\
 			& \OAlphaS 		& Generator 				& 					& \Wboson/\Zboson~Production  \\
\hline
\Pythia		& LO				& yes					& matrix-element correction		& inclusive production \\
			& 				& 						& for first branching		& 					 \\
\Herwig 		& LO				& yes					& matrix-element correction		& inclusive production \\
			& 				& 						& for hardest branching	& 					 \\
\MCAtNLO	& NLO			& yes (interface	 to \Herwig )	& PS matching			& inclusive production \\
			& 				& 						& 					& 					 \\
\aMCAtNLO	& NLO			& yes (interface	 to \Herwig )	& PS matching			& inclusive production \\
			& 				& 						& 					& 					 \\
\PowhegBox	& NLO			& yes (interface to \Pythia~or 	& PS matching			& inclusive production \\
			& 				& \Herwig )				& 					& 					 \\
\Alpgen		& LO				& no (but interface to			& MLM (all			& \Wboson/\Zboson+Jets \\
			& 				& \Pythia/\Herwig )			& parton multiplicities)		& (incl. large. multipl.) \\
\MadGraph	& LO				& no (but interface to			& n.a. (all				& \Wboson/\Zboson+Jets \\
			& 				& \Pythia)					& parton multiplicities)	& (incl. large multipl.) \\
\Sherpa		& LO				& yes 					& CKKW (all 			&  \Wboson/\Zboson+Jets \\
			& 				& 						& parton multiplicities)	&  (incl. large multipl.) \\
\Blackhat		& NLO			& no	(only					& n.a.				&  \Wboson/\Zboson+Jets \\
			& 				& parton level)				& 					&  (incl. large multipl.)\\
			& 				& 						& 					&  \\
\ResBos		& Resummation	& no (only 				& n.a. 				&  $\pT$ spectrum  \\
			& 				& boson kinematics)			& 					&  of \Wboson/\Zboson~bosons \\			
\MCFM		& NLO			& no (only 				& n.a. 				&  NLO corrections to \\
			& 				& parton level)				& 					&  integral rates and shapes\\
\FEWZ		& NNLO			& no (only 				& n.a. 				&  NNLO corrections to \\
			& 				& boson kinematics)			& 					&  integral rates and shapes\\
\hline
\hline
\end{tabular*}
\end{table*}

Table \ref{tab:Predictions2} summarises several predictions for different generators and 
PDF sets for production cross-sections of selected final states in 
proton-proton collisions at $\sqrt{s} = 7\, \TeV$. Uncertainties due to scale and PDF variations are shown in addition. As indicated in the table, 
the increase of the cross-section from leading-order to NNLO predictions is more than $15\%$. The difference between different PDF sets is in 
the order of  $1.5\%$ and covered by the associated PDF uncertainties. 

\begin{table*}
\centering
\caption{Prediction of the cross-sections of \Wboson and \Zboson boson ($66\,\GeV<m_{ll}<116\,\GeV$) production in $\sqrt{s}=7\,\TeV$ pp-collisions at leading order, NLO and NNLO in $\alpha_s$ calculated by the FEWZ generator. The given uncertainty is calculated for the NNLO cross-section and includes PDF- and scale-uncertainties.}
\label{tab:Predictions2}       
\begin{tabular*}{\textwidth}{@{\extracolsep{\fill}}llllll}
\hline
Process 		& LO in \OAlphaS		 	&	NLO in \OAlphaS				&	NNLO in \OAlphaS			& NNLO in \OAlphaS   	& Uncertainty\\
PDF-Set		& MSTW2008LO					&	MSTW2008NLO				& 	MSTW2008NNLO			& CT10 		& \\	
\hline
$\sigma(pp\rightarrow Z+X) \times BR(Z\rightarrow l^+l^-)$ [nb]				&	$0.753$	&	$0.931$		&	$0.960$		& 	$0.991$		& 0.05	\\
\hline
$\sigma(pp\rightarrow W^+ + X) \times BR(W^+\rightarrow l^+\nu)$ 	[nb]			&	$4.80$		&	$5.80$	&	$5.98$		&	$6.16$		& 0.3		\\
$\sigma(pp\rightarrow W^- + X) \times BR(W^-\rightarrow l^- \nu)$ 	[nb]			&	$3.27$		&	$4.06$	&	$4.20$		&	$4.30$		& 0.2		\\
$\sigma(pp\rightarrow W^\pm + X) \times BR(W^\pm \rightarrow l^\pm \nu)$[nb] 	&	$8.11$		&	$9.86$	&	$10.18$		& 	$10.46$		& 0.3		\\


\hline
\end{tabular*}
\end{table*}

\subsection{\label{sec:QCDDynamics} QCD dynamics and angular coefficients}

To leading order, the angular distribution of the decay products in the process $e^+e^-\rightarrow \mu^+\mu-$
can easily be calculated and exhibits a $(1+\cos^2\theta)$ dependence, where $\theta$ is the angle between the incoming electron
and the outgoing positive charged muon. A similar angular dependence is derived for the quark/antiquark annihilation including
a \Zboson boson exchange in the corresponding s-channel diagram. However, the coupling and gauge structure of the weak interaction as well as 
higher-order corrections leads to new angular-dependent terms in the  differential production cross-sections. The measurement of these terms therefore provides 
not only an important test of perturbative QCD but also of the fundamental properties of the electroweak sector, as described in more detail in 
the following paragraphs. The discussion starts with two definitions of rest frames that allow for the 
definition of the angle $\theta$ in proton-proton collisions (Section \ref{sec:frames}). Then the general form of the differential Drell-Yan cross-section is
introduced in Section \ref{sec:DiffDrellYan}, while the interpretation of the corresponding angular coefficients is discussed in Section \ref{ref:interpretationCoeff}.

\subsubsection{\label{sec:frames} Collins-Soper and helicity frame}

The direction of the incoming particles and anti-particles 
is known in an e$^+$e$^-$ collider and hence the reference axis for the definition 
of the angle $\theta$ can be defined in a straightforward way. The situation is 
very different at a proton-proton collider such as the LHC. 
Since the vector boson originates from a $q\bar q$ annihilation, a natural 
choice would be the direction the incoming quark but this can not be done for two reasons.
First, the direction of the incoming quark is not know at the LHC. Second, the incoming partons 
are subject to initial-state radiation, which leads to a non-negligible transverse momenta, $\pT$, 
of the vector boson when annihilating that can not be determined for the two interacting partons. 

To overcome these problems, the rest frame of the vector boson is 
typically chosen as rest frame in which the angular distributions of the 
decay leptons is measured. However, the definition of the axes is this 
rest frame is still ambiguous. 
To minimise the effect due to the lack of information about the kinematics of the 
incoming partons, the polar axis can be 
defined in the rest frame of the vector boson, such that it is bisecting the 
angle between the momentum of the incoming protons. The y-axis can then be 
defined as the normal vector to the plane spanned by the two incoming protons 
and the x-axis is chosen such that a right-handed Cartesian coordinate system 
is defined (Figure \ref{fig:frames}). The resulting reference frame is called 
Collins-Soper (CS) frame~\cite{Collins:1977iv}.

When measuring $\theta$ in an analogous way as in an e$^+$e$^-$ collision, 
the direction of the incoming quark and antiquark must also be known but cannot be inferred on an 
event-by-event basis. However, this can be addressed on a statistical basis. Vector 
bosons with a longitudinal momentum $p_z(V)$ have been produced by partons which 
have significantly different Bjorken-$x$ values. Figure \ref{fig:CTEQPDFX} illustrates that 
large $x$-values enhance the probability for having valence quarks in the 
interaction and therefore the corresponding antiquark can be associated to the 
smaller $x$-values. Hence the measurement of $p_z(V)$ allows us to assign the 
longitudinal quark and antiquark directions on a statistical basis. It should 
be noted that large $p_z(Z)$ values also imply large rapidities and therefore 
the statistical precision for the correct quark/antiquark assignment is 
enhanced for \Zboson bosons in the forward region.

In summary, the angle $\theta$ can be expressed in the CS frames as

\begin{equation}
\cos\theta^*_{CS} = \frac{p_z(V)}{|p_z(V)|} \cdot \frac{2(p_1^+ 
p_2^- - p_1^- p_2^+)}{m_{ll}\sqrt{m_{ll}^2 + \pT(ll)^2}}\, ,
\end{equation}
with $p_{1/2}^\pm = 1/\sqrt{2} \cdot (E_{1/2} \pm p_{z,1/2})$, where $E$ and 
$p_Z$ are the energy and longitudinal momenta of the first and second lepton. The first 
term of this equation defines the sign and hence the direction of the incoming 
quark. As previously discussed, large rapidities enhance the probability for a 
correction assignment of the direction. The second term of the equation 
corrects for the measured boost due to the hadronic recoil of the event and 
defines an average angle between the decay leptons and the quarks. 

While angular measurements of the \Zboson boson decays are usually done 
in the CS frame, measurements of the \Wboson boson are traditionally performed in the so-called 
helicity frame. The helicity frame is also chosen to be the rest frame of the 
vector boson. The z-axis is defined along the \Wboson laboratory direction of flight 
and the x-axis is defined orthogonal in the event plane, defined by the two 
protons, in the hemisphere opposite to the recoil system. The y-axis is 
then chosen to form a right-handed Cartesian coordinate system as shown in 
Figure \ref{fig:frames}.

\begin{figure*}
\resizebox{1.00\textwidth}{!}{\includegraphics{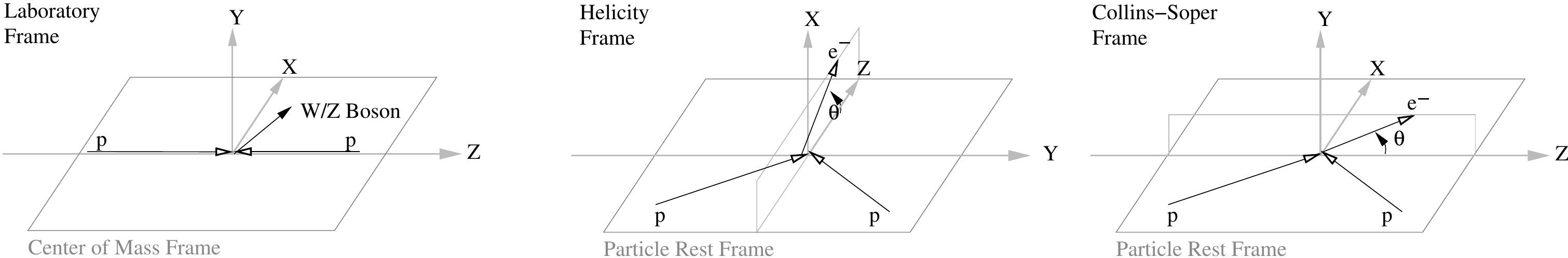}}
\caption{Center-of-mass frame (left), helicity frame (middle) and Collins-Soper frame (right)}
\label{fig:frames}
\end{figure*}

\subsubsection{\label{sec:DiffDrellYan}Differential cross-section of the Drell-Yan process}

The general form of the differential cross-section of the Drell-Yan process 
$pp\rightarrow Z(W) + X\rightarrow l^+ l^- (l \nu) + X$ can be decomposed as \cite{Mirkes:1994eb}, \cite{Mirkes:1994dp}

\begin{eqnarray}
\label{EQN:DECOMP}
\frac{d\sigma}{d\pT^2 dy d\cos\theta d\phi} &=&  \frac{d\sigma_{unpol}}{d\pT^2 dy} \cdot \left( (1+\cos^2\theta) \right. \\
\nonumber
&& + A_0 \frac{1}{2} (1-3 \cos^2(\theta)) \\
\nonumber
&& + A_1 \sin(2\theta) \cos(\phi)  \\
\nonumber
&& + A_2 \frac{1}{2} \sin^2(\theta) \cos(2\phi)  \\
\nonumber
&& + A_3 \sin(\theta) \cos(\phi)  \\
\nonumber
&& + A_4(\cos\theta)  \\
\nonumber
&& + A_5 \sin^2(\theta) \sin(2\phi)  \\
\nonumber
&& + A_6 \sin(2\theta) \sin(\phi)  \\
\nonumber
&& + A_7 \sin(\theta) \sin(\phi) ),
\end{eqnarray}
where $\theta$ and $\phi$ are the polar and azimutal angles of the 
charged lepton in the final state\footnote{For measurements of the \Zboson the negative lepton is used.} 
in the CS frame to the direction of 
the incoming quark/antiquark. This decomposition is valid in the limit of 
massless leptons in a 2-body phase space and helicity conservation in the 
decay. 

While the functional dependence of Equation \ref{EQN:DECOMP} on $\theta$ and $\phi$ is 
independent on the reference frame chosen, the parameters $A_i$ are 
frame dependent. When no cuts on the final-state kinematics are applied, the 
parameters $A_i$ can be transformed from one reference frame to another. 
Due to the limited detector coverage and additional analysis requirements on 
the kinematics, the coefficients exhibit an experiment-dependent kinematic 
behaviour. Hence the optimal choice of the reference frame will differ for 
each analysis. 

The angular coefficients $A_i$ are functions of the vector boson 
kinematics, i.e. its transverse momentum, $\pT(V)$, its rapidity, $Y_V$, and $m_V$ and contain information about the 
underlying QCD dynamics. They are subject to modifications from higher-order 
perturbative and non-perturbative corrections, structure functions and 
renormalisation and factorisation scale uncertainties. Since the PDFs of the 
proton impact the vector boson kinematics, the coefficients $A_i$ also depend 
indirectly on the PDFs themselves. The 1-dimensional angular distributions 
can be obtained by integrating either over $\cos\theta$ or $\phi$, leading to

\begin{equation} 
\frac{d\sigma}{\cos\theta} \sim (1+\cos^2\theta) + \frac{1}{2} A_0 
(1-3\cos^2\theta) + A_4 \cos\theta
\end{equation}

\begin{eqnarray} 
\nonumber
\frac{d\sigma}{\cos\phi} &\sim & 1 + \frac{2\pi}{16} A_3 (\cos^2\phi) + \frac{1}{4} A_2 (\cos(2\phi)) + \\
&& + \frac{3\pi}{16} A_7 \sin\phi + \frac{1}{4} A_5 \sin(2\phi)
\end{eqnarray}
which can be used to extracted several coefficients independently in the case of small data samples.

\subsubsection{\label{ref:interpretationCoeff}Interpretation of coefficients}

The $(1+\cos^2\theta)$ term in Equation \ref{EQN:DECOMP} comes from the pure leading-order 
calculation of the vector boson production and decay. The terms corresponding 
to the coefficients $A_0, A_1, A_2$ are parity conserving, while the terms 
$A_3$ to $A_7$ are parity violating. The $A_0$ to $A_4$ coefficients receive 
contributions from the QCD theory at leading and all higher orders, while the parameters 
$A_5, A_6$ and $A_7$ appear only in NLO QCD calculations and are typically 
small. Several studies have been published which discuss and predict these 
coefficients for hadron colliders \cite{Mirkes:1992hu}. 

It should be noted that all terms except of $A_4$ are symmetric in $\cos\theta$. 
In the case of the $Z/\gamma^*$ exchange, $A_4$ appears also in leading-order calculations as 
it is directly connected to the forward backward asymmetry $A_{fb}$ via

\begin{equation}
A_{fb}= \frac{3}{8}A_4
\end{equation}
which will be discussed in more detail in Section \ref{sec:AFB}. 

The $A_4$ parameter also plays a special role in the case of \Wboson boson polarisation 
measurements. As discussed in the previous section, vector 
bosons tend to be boosted in the direction of the initial quark. In the 
massless quark approximation, the quark must be left-handed in the case of the \Wboson 
boson production and as a result \Wboson bosons with large rapidities are 
expected to be purely left-handed. For more centrally produced \Wboson bosons, there 
is an increasing probability that the antiquark carries a larger momentum 
fraction and hence the helicity state of the \Wboson bosons becomes a mixture of 
left- and right-handed states. The respective proportionals are labelled with 
$f_L$ and $f_R$. For \Wboson bosons with a larger transverse momenta, the production 
via a gluon in the initial or final state becomes relevant, e.g. via $u\bar d 
\rightarrow W^+ g$. Hence the vector nature of the gluon has to be taken into 
account in the prediction of the production mechanisms. For high transverse 
momenta, also polarisations in the longitudinal state of the \Wboson bosons can 
appear. This fraction is denoted by $f_0$ and is directly connected to the 
massive character of the gauge bosons. The helicity fractions $f_L$, $f_R$ and 
$f_0$ can be directly connected to the coefficients $A_0$ and $A_4$ via

\begin{eqnarray}
f_L(Y_W, \pT^W) = \frac{1}{4} (2-A_0(Y_W, \pT^W)) \mp A_4(Y_W, \pT^W) \\
f_R(Y_W, \pT^W) = \frac{1}{4} (2-A_0(Y_W, \pT^W)) \pm A_4(Y_W, \pT^W) \\
f_0(Y_W, \pT^W) = \frac{1}{2} A_0(Y_W, \pT^W),
\end{eqnarray}
where the upper (lower) signs correspond to $W^+$ ($W^-$). In particular, the 
difference of $f_L$ and $f_R$ depends only on $A_4$ as

\begin{equation}
f_R - f_L = \pm \frac{1}{2} A_4\, .
\end{equation}
The coefficients $A_0$ and $A_2$ also play a particular role in the angular 
decay distributions, as they are related via the Lam-Tung relation \cite{Lam:1980uc}. This 
relation states that $A_0(\pT)$ and $A_2(\pT)$ are identical for all $\pT$ if 
the spin of the gluon is one. In case of a scalar gluon, this relation would be 
broken. It should be noted that the test of this relation is therefore not a 
test of QCD theory, but a consequence of the rotational invariance of decay angles and 
the properties of the quark-coupling to $Z/\gamma^*$ and the $W$ boson. At the 
Z-pole, the leading-order predictions of the $\pT$ dependence of $A_{0/2}$ for 
a gluon of spin one are given by \cite{Chaichian:1981va, Lindfors:1979rc, Lam:1978zr} 

\begin{equation}
A_{0,2} = \frac{\pT^2}{\pT^2 + m^2_Z}
\end{equation}
for the process $q\bar q \rightarrow \Zboson g$ and by

\begin{equation}
A_{0,2} = \frac{5\cdot \pT^2}{5\cdot  \pT^2 + m^2_Z}
\end{equation}
for the process $q g \rightarrow \Zboson q$. NLO order corrections do not impact 
$A_0$ significantly, while $A_2$ receives contributions up to $20\%$.

\section{Detectors and Data}
\label{sec:DectAndData}

\subsection{The LHC and the data collected at $\sqrt{s}=7\, \TeV$}

From March 2010 to October 2011, the Large Hadron Collider \cite{Evans:2008zzb} delivered 
proton-proton collisions at a center-of-mass energy of $\sqrt{s} = 7 \, \TeV$ to 
its four main experiments ATLAS \cite{Aad:2008zzm}, CMS \cite{Chatrchyan:2008aa}, LHCb \cite{Alves:2008zz} and ALICE \cite{Aamodt:2008zz}. The primary LHC 
machine parameters at the end of the data taking in 2010 and 2011 are given in 
Table \ref{tab:LHCParameter}. From 2010 to 2011, the number of circulating proton bunches was 
increased by a factor of $3.8$, the spacing between two bunches was decreased 
from $150\,\ns$ to $50\,\ns$ and the beam-focus parameter $\beta^*$ was reduced by a 
factor of $3.5$. This resulted in a significant increase of instantaneous 
luminosity from $L = 2\times10^{32}\cm^{-2}s^{-1}$ in 2010 to $L = 3.7\times10^{33}\cm^{-2}s^{-1}$ in 2011 \cite{Fournier:2012np}. 

\begin{table}
\caption{Parameters of the LHC at the end of 2010 and the end of 2011 including an estimate of the average number of interactions per bunch-crossing \cite{Fournier:2012np}.}
\label{tab:LHCParameter}       
\begin{tabular}{lll}
\hline\noalign{\smallskip}
Parameter 		& 2010 				& 	2011 \\
\noalign{\smallskip}\hline\noalign{\smallskip}
$\sqrt{s}$			& 7 TeV				&	7 TeV	\\
$N(10^{11}p/b)$	& 1.2					&	1.5	\\
$k (n_{bunches})$	& 368				&	1380 \\
Bunch Spacing (ns)	& 150				&	50 \\
L ($cm^{-2}s^{-1}$)	& $2\cdot 10^{32}$		&	$3.6\cdot 10^{33}$ \\
Average pp-interactions per bunch-crossing&$\approx 1.2$		&	$\approx 10-15$\\
\noalign{\smallskip}\hline
\end{tabular}
\end{table}

The total integrated luminosity delivered to the experiments was 
$L\approx44\,\ipb$ in 2010 and $L\approx6.1\,\ifb$ in 2011. The data taking 
efficiency of ATLAS and CMS, when the detector and 
data-acquisition systems were fully operational, was above $90\%$ for both years. The recorded 
integrated luminosity, which was used as the data samples for the published physics 
analyses for ATLAS and CMS in 2010 and 2011, is shown in Table \ref{tab:IntLumi} together 
with their respective relative uncertainties. 

\begin{table}
\caption{Overview of recorded integrated luminosity in 2010 and 2011 by the ATLAS and CMS experiments. Also shown is the integrated luminosity which is used for physics analyses.}
\label{tab:IntLumi}       
\begin{tabular}{lllll}
\hline\noalign{\smallskip}
Experiment 	&  \multicolumn{2}{c}{\IntLumi (2010)}   	&  \multicolumn{2}{c}{\IntLumi (2010)}  \\
			& recorded 		& used 			& recorded 		& used 		         \\
\noalign{\smallskip}\hline\noalign{\smallskip}
ATLAS 		& 45\ipb 			& 35\ipb			& 5.08\ifb 			& 4.6\ifb			\\
CMS 		& 40.8\ipb 		& 36\ipb			& 5.55\ifb 			& 4.5-4.8\ifb			\\
\noalign{\smallskip}\hline
\end{tabular}
\end{table}

The precise knowledge of the recorded integrated luminosity is a crucial 
aspect for all cross-section measurements. The Van der 
Meer methods \cite{VanDerMeer:1}, \cite{Aad:VanDerMeer} was applied in total three times in 2010 and 2011 to 
determine the luminosity for ATLAS and CMS, leading to relative uncertainties 
below $2\%$. It should be noted that the luminosity determination is highly 
correlated between ATLAS and CMS, leading to correlated uncertainties in the 
corresponding cross-section measurements. 

The change in the machine settings from 2010 to 2011 leads to an increase of 
\textit{pile-up} noise, which is the occurrence of several independent, inelastic 
proton-proton collisions during one or more subsequent proton-proton bunch 
crossings. These additional collisions can lead to a significant performance 
degradation of some observables which are used in physics analysis. The \textit{in-time} pile-up, 
i.e. the additional collisions occurring within the same bunch crossing, can be 
described by the number of reconstructed collision vertices $N_{vtx}$ in one 
event. The \textit{out-of-time} pile-up is due to additional collisions from previous 
bunch crossings, that can still affect the response of the detector, in 
particular calorimeters, whose response time is larger than two subsequent 
bunch crossings. The number of interactions per crossing is denoted as $\mu$ and 
can be used to quantify the overall pile-up conditions. On average, there is
roughly a linear relationship between $\mu$ and $N_{vtx}$, i.e. $<N_{vtx}> \approx 
0.6 <\mu>$. In 2010, the average number of interactions per collision was 
$\mu=2$. The first \IntLumi$\approx 1\ifb$ in 2011 had $<\mu>\approx6$, 
while $<\mu>$ of greater than 15 was reached by the end of 2011. This 
affects several systematic uncertainties related to 
precision measurements at the LHC.

\subsection{Coordinate system}

The coordinate system of the ATLAS and CMS detectors are orientated 
such that the z-axis is in the beam direction, the x-axis points to the center of the 
LHC ring and the y-axis points vertically upwards (Figure \ref{fig:TCoordinate}). The radial coordinate in the 
x-y plane is denoted by $r$, the azimuthal angle $\phi$ is measured from the x-axis. 
The pseudorapidity $\eta$ for particles coming from the primary vertex is defined as 
$\eta = - log \frac{\theta}{2}$, where  $\theta$ is the polar angle of the particle direction 
measured from the positive z-axis. The transverse momentum $\pT$ is defined as the 
transverse momentum component to the beam direction, i.e. in the x-y-plane. 
The transverse energy is defined as $\ET = E \sin \theta$.

\begin{figure}[h]
\resizebox{0.5\textwidth}{!}{\includegraphics{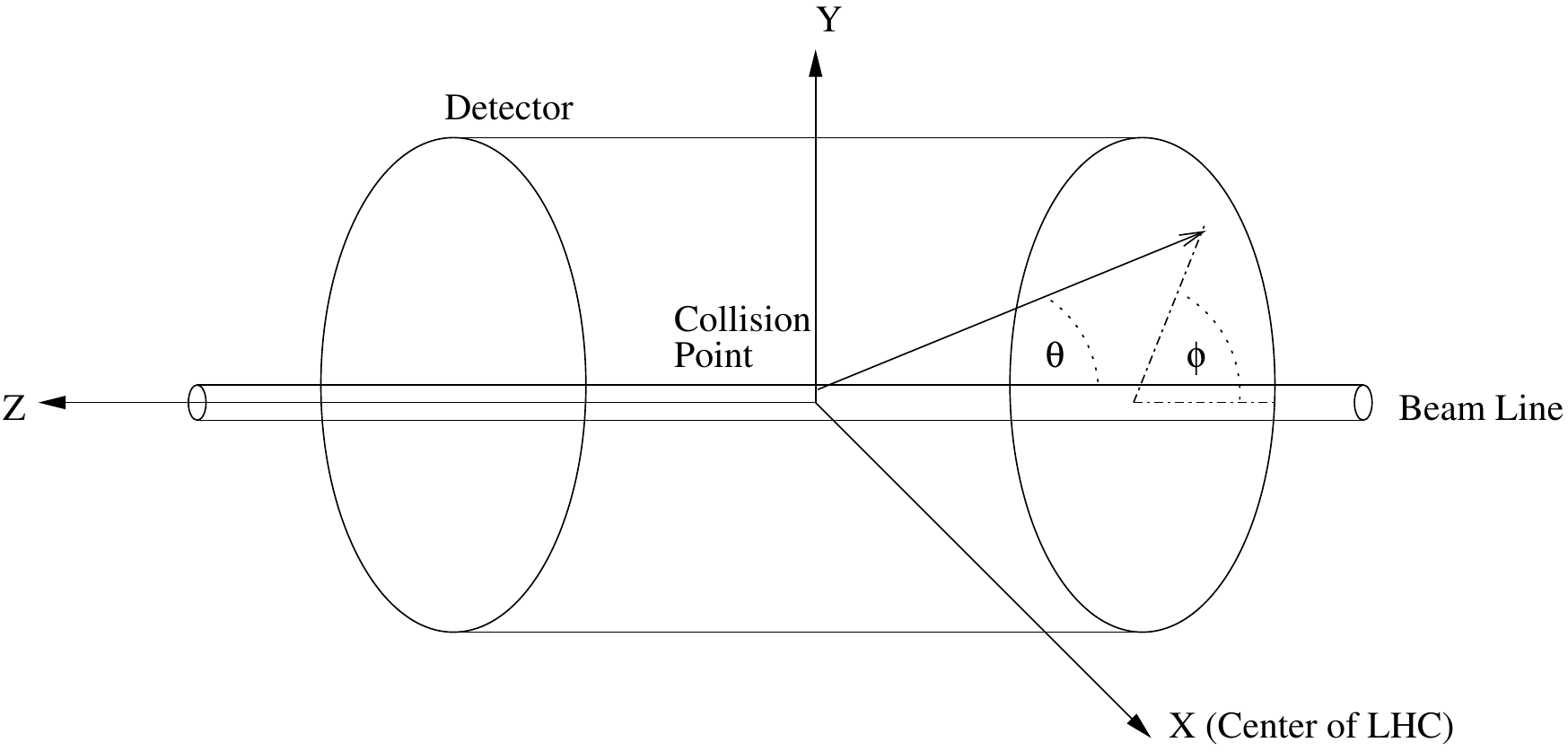}}
\caption{Illustration of the ATLAS and CMS coordinate system.}
\label{fig:TCoordinate}
\end{figure}

\subsection{The ATLAS detector}

The "A Toroidal LHC ApparatuS" (ATLAS) detector is one of the two general 
purpose detector at the LHC. It has a symmetric cylinder 
shape with nearly $4\pi$ coverage (Figure \ref{fig:ATLAS}). ATLAS has a 
length of 45m, a diameter of 25m and weighs about 7,000 tons. It can be 
grouped into three regions: the barrel region in the center of the detector and 
two end-cap regions which provide coverage in the forward and backward direction 
with respect to the beam-pipe. ATLAS consists of one tracking, two calorimeter and one muon system, 
which are briefly described below. A detailed review can be 
found in \cite{Aad:2008zzm}. 

\begin{figure*}
\begin{minipage}{0.5\textwidth}
\resizebox{1.0\textwidth}{!}{\includegraphics{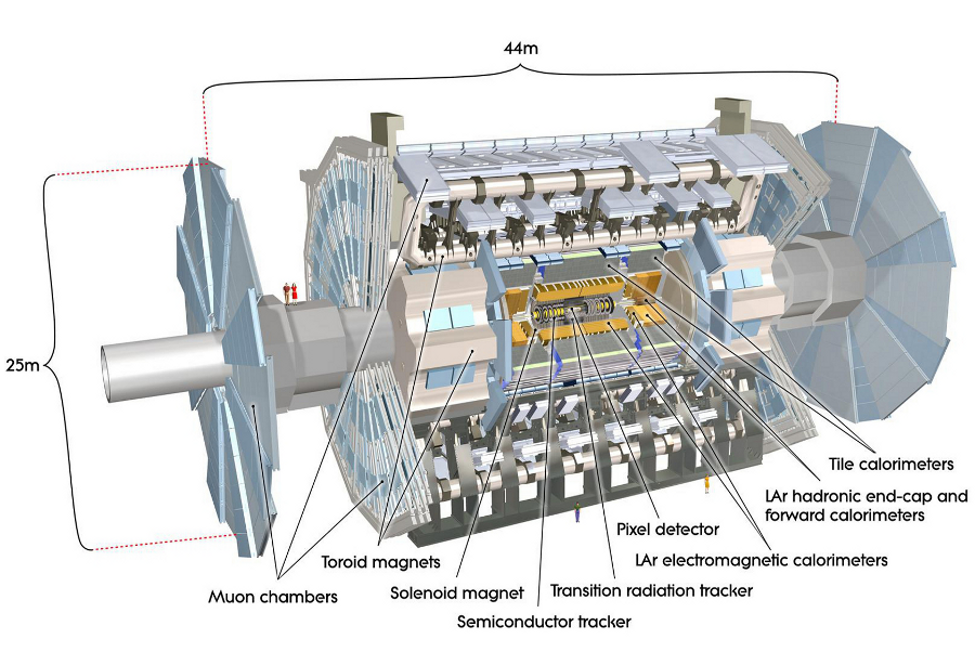}}
\caption{\label{fig:ATLAS} ATLAS Experiment}
\end{minipage}
\begin{minipage}{0.5\textwidth}
\resizebox{1.0\textwidth}{!}{\includegraphics{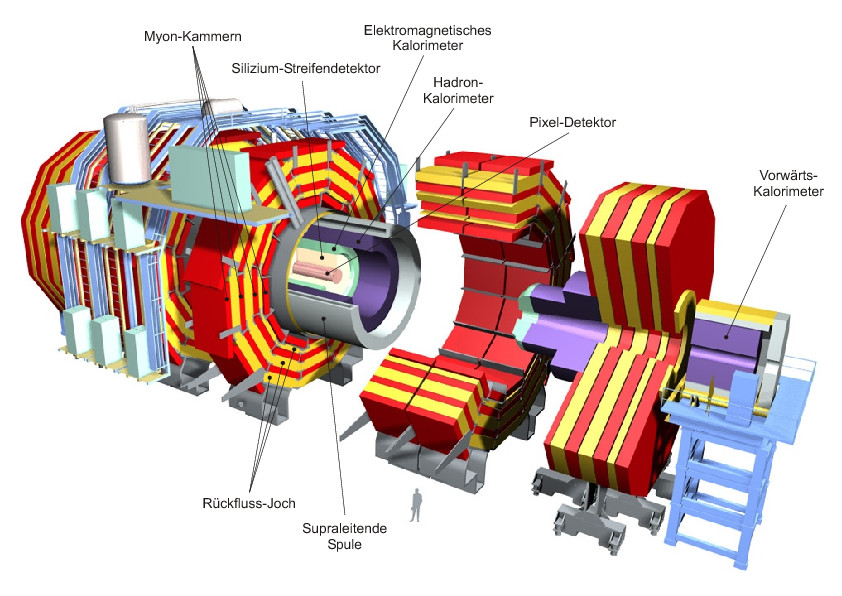}}
\caption{\label{fig:CMS} CMS Experiment}
\end{minipage}
\end{figure*}

The tracking detector is the closest to the LHC beam pipe 
and extends from an inner radius of $5\cm$ to an outer radius of $1.2\meter$. It
measures tracking information of charged particles in a $2\,$T axial magnetic 
field provided by a superconducting solenoid magnet system. In addition, the tracking detector
provides vertex information, which can be used to identify the interaction 
point of proton-proton collision and the decay of short-lived 
particles. Three technologies are used. The innermost part of the tracking detector consists 
of three silicon pixel detector layers. Each pixel has a size of $50 \times 400 
\umeter$, leading in total to 80 million readout channels. The pixel detector 
provides tracking information up to a pseudorapidity of $|\eta|=2.5$. The same 
region is also covered by the semi-conductor tracker, which 
surrounds the pixel detector. It consists of narrow silicon strips in the 
size of $80\umeter \times 12\,\cm$, which are ordered in four double 
layers. The outermost part of the tracking detector is the transition radiation tracker 
which uses straw detectors and covers an area up to $|\eta|=2.0$. It provides up to 36 additional 
measurement points of charged particles with a spatial resolution of $200\,\mu 
m$. In addition, the produced transition radiation can be used for 
electron identification. 

In the electromagnetic (EM) calorimeter of ATLAS, the energy of incoming electrons and photons is 
fully deposited to the detector materials and can be precisely determined. 
Moreover, the ATLAS calorimeter can measure the location of the 
deposited energy to a precision of $0.025\rad$. Liquid argon is used as active 
material, while lead plates act as absorbers. The absorbers are arranged in an 
accordion shape which ensures a fast and uniform response of the calorimeter. 
The barrel region covers a range up to $|\eta|<1.475$, the two endcaps provide 
coverage for $1.375<|\eta|<3.2$. A presampler detector is installed in the region 
up to $|\eta|<1.8$, which is used to correct the energy loss of 
electrons and photons in front of the calorimeter.

The hadronic calorimeter ranges from $r=2.28\,\meter$ to $r=4.23\,\meter$ and measures 
the full energy deposition of all remaining hadrons. The barrel part is the so-called 
tile calorimeter and covers a region up to $|\eta| < 1.0$. An extended barrel 
detector is used for the region $0.8 < |\eta| < 1.7$. Scintillating plastic 
tiles are used as active medium. Iron plates act not only as absorber material, 
but also as the return yoke for the solenoid magnetic field of the tracking detector. The 
granularity of $\Delta \phi \times \Delta \eta = 0.1 \times 0.1$ determines the 
position information of the measured energy deposits, which is roughly 
$0.1\,\rad$. The hadronic endcap calorimeter covers a pseudorapidity range from 
$1.5 < |\eta| < 3.5$, where liquid argon is used as the ionisation material and 
copper as the absorber. The very forward region from $3.1 < |\eta| < 4.5$ is 
covered by the forward calorimeters, which also uses liquid argon with copper and tungsten as absorbers. 
Electrons and photons are also detected in the forward calorimeters, as no dedicated electromagnetic 
calorimeter is present in that region.

The muon spectrometer is not only the largest part of the ATLAS Experiment, 
ranging from $r=4.25\meter$ to $r=11.0\meter$, but also its namesake. Three 
air-core toroidal magnets provide a toroidal magnetic field with an average 
field strength of $0.5\,$T. Muons with an energy above $\approx 6 \, \GeV$ that 
enter the toroidal magnetic field will be deflected. This deflection is 
measured in three layers of tracking chambers. In the barrel-region 
($|\eta|<1.0$) and partly in the endcaps up to $|\eta|<2.0$. Monitored 
drift-tube chambers provide the precise tracking information of 
incoming muons. For $2.0<|\eta|<2.7$, cathode strip chambers with a 
higher granularity are used. The trigger-system of the muon spectrometer is 
based on resistive plate chambers in the barrel region and by thin gap 
chambers in the endcap. Since the ATLAS muon system is filled with air, 
effects from multiple scattering are minimised. In addition, the long bending path of 
the muons provides an accurate measurement of their momentum. 

The trigger system of the ATLAS detector has three levels. The first level is a 
hardware based trigger, which uses a reduced granularity information of the 
calorimeters and the muon system. It defines so-called regions-of-interest, 
in which possible interesting objects have been detected, and reduces 
the event rate to $\approx$ 75\,kHz. The second level trigger is 
software based and has the full granularity information within the region-of-interest and the 
inner detector information. By this system, the rates are reduced to 1\,kHz. The 
last trigger level has access to the full event information with full granularity and uses 
reconstruction algorithms that are the same or similar to those used in the offline reconstruction. 
The final output rate is $\approx$ 400\,Hz.

\subsection{The CMS detector}

The Compact Muon Solenoid (CMS) detector is the second general purpose detector 
at the LHC with a similar design as the ATLAS detector. It offers also a nearly 
full $4\pi$ coverage which is achieved via one barrel and two endcap sections (Figure 
\ref{fig:CMS}). CMS is 25\,m long, has a diameter of 15\,m and weights 12500 tons. Most 
of its weight is due to its name-giving solenoid magnet, which provides a 
3.8\,T magnetic field. The magnet is 12.5\,m long with a diameter of 6\,m and consists 
of four layers of superconducting niobium-titanium at an operating temperature 
of 4.6\,K. The CMS tracking system as well as the calorimeters are within the 
solenoid, while the muon system is placed within the iron return yoke. We 
briefly discuss the four main detector systems of CMS below; a 
detailed description can be found in \cite{Chatrchyan:2008aa}.

The inner tracking system of CMS is used for the reconstruction of charged 
particle tracks and is fully based on silicon semi-conductor technology. The 
detector layout is arranged in 13 layers in the central region and 14 layers 
in each endcap. The first three layers up to a radius of 11cm 
consist of pixel-detectors with a size of $100\umeter \times 150\umeter$. The 
remaining layers up to a radius of $1.1$\,m consist of silicon strips with 
dimensions $100\umeter \times 10\cm$ and $100\umeter \times 25\cm$. In total, the 
CMS inner detector consists of 66 million readout-channels of pixels and 96 
million readout channels of strips, covering an $\eta$-range of up to $2.5$.

The CMS electromagnetic calorimeter is constructed from crystals of lead tungstate (PbWO$_4$). The 
crystalline form together with oxygen components in PbWO$_4$ provide a highly 
transparent material which acts as a scintillator. The readout of the 
crystals in achieved by silicon avalanche photodiodes. The barrel part of the 
EM calorimeter extends to $r=1.29\,\meter$ and consists of 61200 crystals (360 in $\phi$ and 
170 in $\eta$), covering a range of $|\eta| < 1.479$. The EM calorimeter endcaps 
are placed at $z=\pm3.154\,\meter$ and cover an $\eta$-range up to $3.0$ with 7324 
crystals on each side. A pre-shower detector 
is installed in order to discriminate between pions and photons.

The hadronic calorimeter of the CMS detector is a sampling calorimeter which 
consists of layers of brass or steel as passive material, interleaved with 
tiles of plastic scintillator. It is split in four 
parts. The barrel part ($|\eta|$<1.3) consists of 14 brass absorbers and two 
stainless steel absorbers as the innermost and outermost plates. The granularity is 
$0.087\times0.087$ in the $\eta, \phi$-plane. Due to the space limitations 
from the solenoid, an outer calorimeter has been installed. It 
consists of two scintillators at $r=3.82\,\meter$ and $r=4.07\,\meter$ with $12.5\,\cm$ steel in 
between. The endcap calorimeters cover $1.3 < |\eta| < 3.0$ and 
are made of 18 layers of $77\,$mm brass plates interleaved by $9\,$mm scintillators. 
The $\eta$-region from $3.0 < |\eta| < 5.0$ is covered by forward calorimeters, 
positioned at $z=\pm11\,\meter$. They will also register the energy deposits of 
electrons and photons in this rapidity range.

The barrel and endcap parts of the CMS muon system consist of four layers of 
precision tracking chambers. The barrel part covers a range up to $|\eta|=1.3$ 
and drift-tube chambers are used for the tracking. The tracking information 
in both endcaps ($0.9<|\eta|<2.4$) is provided by cathode strip 
chambers. The muon triggers are based on resistive place chambers, 
similar to the ATLAS experiment \cite{Aad:2012xs}. 

The CMS trigger system has two levels. The first level trigger is hardware 
based and uses coarsely segmented information from the calorimeters and the 
muon system. It reduces the rate to $100$\,kHz. The second level trigger, 
called the high-level trigger, is a software-based trigger system which is based on fast 
reconstruction algorithms. It reduces the final rate for 
data-recording down to $400$\,Hz.

\subsection{Reconstructed objects}
\label{sec:reco-objects}

Measurements of single vector boson production using ATLAS and CMS data involve in general five primary physics 
\textit{objects}. These objects are electrons, photons, muons, neutrinos, whose energy can only be inferred, and particle jets, which
originate from hadronised quarks and gluons. 
An overview of the ATLAS and CMS detector performance for several physics 
objects is summarised in Table \ref{tab:PhysPerform}.

\subsubsection{Electron, photon and muon reconstruction}
\label{sec:reco-lep}

Electrons candidates are identified by requiring that significant energy deposits in the 
EM calorimeter, which are grouped to so-called electromagnetic clusters, exist and that
there is an associated track in the tracking detector. The transverse momenta of the electrons are calculated from 
the energy measurement in the EM calorimeter and the track-direction information is taken from the 
tracking detector. A series of \textit{quality cuts} are defined to select electron candidates. These cuts include 
cuts on the shower-shape distributions in the calorimeter, track-quality requirements and 
the track-matching quality to the clusters. Stringent cuts on these quantities 
ensure a good rejection of non-electron objects, such as particle jets faking electron signatures in 
the detector. ATLAS has three different quality definitions for electrons, named
\textit{loose}, \textit{medium} and \textit{tight} \cite{Aad:2011mk} and CMS analyses use two definitions, called \textit{loose} and \textit{tight} \cite{Chatrchyan:2013dga}. 

For some analyses in both ATLAS and CMS, electron clusters in the transition 
region between the barrel and endcap sections are rejected,  as cables 
and services in this detector region lead to a lower quality of  
reconstructed clusters. These regions are defined as $1.37<|\eta|<1.52$ and 
$1.44<|\eta|<1.57$ in ATLAS and CMS respectively. Electron candidates in the 
forward region from $2.5<|\eta|<4.9$ (used by some ATLAS analyses) have no 
associated track information and therefore their identification is based solely 
on the shower-shape information. 

Photons candidates are reconstructed by clustered energy deposits on the EM 
calorimeter in a range of $|\eta| < 2.37$ and $|\eta|<2.5$ for ATLAS \cite{Aad:2011mk} 
and CMS \cite{Chatrchyan:2013dga}, respectively, as well as 
specific shower shape cuts. If no reconstructed track in the tracking detector can be associate 
to the electromagnetic cluster, then the photon candidate is marked as an \textit{unconverted photon 
candidate}. If the EM cluster can be associated to two tracks, which are 
consistent with a reconstructed conversion vertex, the candidate is defined as a
\textit{converted photon candidate}. 

Muon candidates are identified by one reconstructed track in the muon spectrometer. \textit{Combined 
muons} are required to have in addition an associated track in the tracking detector. The 
measured 4-momenta, in particular the transverse momenta, of combined muons 
are based on a statistical combination of the independent measurements 
within the tracking and muon detectors or a complete refit of all available parameters. 
For the measurements presented here, the momentum resolution for muons is 
dominated by the information from the tracking detector for both experiments. CMS can 
reconstruct muons within $|\eta|<2.4$ \cite{Chatrchyan:2012xi}, while the ATLAS muon spectrometer reaches $|\eta|<2.7$ \cite{ATLAS:Muon}. 
However, muons that are reconstructed beyond $|\eta|>2.5$, have no associated 
information from the tracking detector available and therefore only kinematic 
information from the muon spectrometer can be used. ATLAS analyses therefore often restrict the muon range to $|\eta|<2.4$. 


In many single vector boson measurements, the leptons are required to be isolated 
meaning that there is no significant energy deposited around the lepton itself. 
Requiring isolation greatly reduces the number of particle jets which are misreconstructed
as electron, photons or muons.  

Isolation can be defined in several ways. First, a
tracking-based isolation can be used that is defined as
\begin{equation}
\pT^{iso} = \sum _i ^{\Delta R(\eta,\phi)<0,3} \pT^i / \pT^{lepton},
\end{equation}
\noindent where $i$ indicates the sum over all reconstructed tracks in the tracking detector
with an energy above a given threshold and within a 
cone-radius of 0.3 in the $(\eta,\phi)$-plane. The track from the lepton candidate itself is not considered. 
This quantity can be normalised by the transverse momentum 
of the lepton candidate, which ensures a more stable isolation definition 
for larger transverse momenta.  A similar definition 
can be made using the EM calorimeter, i.e. 
\begin{equation}
\ET^{iso} = \sum _i ^{\Delta R(\eta,\phi)<0.3} E^i_T / \ET^{lepton},
\end{equation}
where $i$ runs over all EM clusters within $\Delta R<0.3$ that are not associated to the 
reconstructed lepton. ATLAS uses both, tracking- and calorimeter-based isolation criteria as defined above. CMS uses similar isolation 
variables, but in addition applies  non-normalised isolation definitions based on tracks or energies in the hadronic 
calorimeter.

\subsubsection{Hadronic jets and missing energy reconstruction}
\label{sec:reco-jet}

Hadronised partons are detected as particle jets in the EM and hadronic calorimeters. 
To reconstruct particle jets in ATLAS, the energy deposits are merged to 
topological clusters in a pseudorapidity range of $|\eta|<4.9$. Clusters are seeded by 
calorimeter cell deposits with a four sigma deviation from the noise level. An 
anti-$k_T$ algorithm \cite{Catani:1993hr} is then used to reconstruct the clusters into jets. 
The typical ATLAS distance parameter of the jet definition is 
$R=0.4$. In addition it is often required that the distance between leptons 
and jets in the ($\eta, \phi$)-plane of the detector satisfies the condition $\Delta R 
(l, jet)>0.3$ to avoid double counting \cite{Aad:2011he}. The jet energy and direction 
is corrected for effects like additional dead material in the detector, the difference in energy response in the calorimeter
to hadronic versus EM interactions, the loss of energy outside the jet radius and the presence of energy from pile-up interactions~\cite{Aad:2011he}.
For CMS analyses, where the detector features a superior tracking system but offers a less precise calorimetry system, 
the so-called particle flow technique 
\cite{CMS:2009nxa} is used. This method combines information from all detector systems, in 
particular the calorimeter and the tracking detector and aims to identify all particles in mutual exclusive 
categories: muons, electrons, including the identification of bremsstrahlung 
photons, converted and unconverted photons, charged and uncharged hadrons. 
Thus, a full event description of each particle is attempted. This event 
description is used as input to an anti-$k_T$ algorithm with a typical distance parameter
of $R=0.5$. The jet energy and direction is also corrected for effects like the presence of pile-up and the non-compensation of the calorimeter~\cite{Chatrchyan:2011ds}.

Both experiments have dedicated algorithms to identify 
particle jets originating from $b$- and $c$-quarks. These algorithms combine 
information about the impact parameter significance of tracks with the expected 
topology of semileptonic $b$- or $c$-decays~\cite{Chatrchyan:2012jua}.

The energy of neutrinos, which leave the detector unseen, must be inferred though \textit{missing energy}. While the 
initial collision energy in beam direction of the partons that are involved in 
the hard scattering process is not known at hadron colliders, the vector sum 
of all transverse momenta and energies in the initial state must be zero. Due to 
energy and momentum conservation, this must also hold for all final state 
objects in the transverse plane, defined as
\begin{equation}
\vec 0 = \sum_{i} \vec E^{calo}_{T,i} + \sum_{i} \vec p_{t,i}^\mu + \sum_{i} \vec p_{T,i}^\nu\, ,
\end{equation}
where the first term corresponds to the vector sum of all transverse energy 
deposits in the calorimetric system, the second term corresponds to the 
transverse momenta of the muons reconstructed by the muon systems and the last 
term corresponds to the transverse momentum sum of all neutrinos in the final 
state. The latter term is called the missing energy, $\vec \ETMiss= - \sum_{i} \vec E^{calo}_{T,i} - \sum_{i} \vec p_{t,i}^\mu$ and its absolute value is 
called missing transverse energy $\ETMiss$. 

The ATLAS measurement of $\vec \ETMiss$ uses all electromagnetic and hadronic energy clusters up to 
$|\eta|=4.9$. Cells which are associated with a particle jet are calibrated with 
a hadronic energy scale correction, while cells associated to electromagnetic 
showers are calibrated via the electromagnetic energy scale \cite{Aad:2011he}. The CMS measurement of 
$\ETMiss$ follows similar lines, but again uses the information provided by the 
particle flow algorithms to improve the measurement \cite{CMS:2009nxa}.  


\begin{table*}
\centering
\caption{Reconstruction performance of electrons, muons and particle jets of the ATLAS and CMS experiment. The reconstruction efficiency and momentum/energy resolutions are shown for the kinematic ranges defined. Further details can be found in the references given.}
\label{tab:PhysPerform}       
\begin{tabular*}{\textwidth}{@{\extracolsep{\fill}}llllll}
\hline
\multicolumn{6}{c}{ATLAS}\\
Object 	& definition and algorithm 	& kinematic range 			& reconstruction efficiency 	& $\pT$-resolution 	& Reference \\
\hline
Electron	& medium	 quality definition 	& $|\eta|<2.4$ 				&$94\%-98\%$				& $\approx 2\%$	& \cite{Aad:2011mk}	\\
		& 					 	& $20\,\GeV<\ET<40\,\GeV$	& 						&				&			\\
Muon	& combined tracking+muon 	& $|\eta|<2.5$ 				&$\approx95\%$			& $\approx 2\%$	& \cite{ATLAS:Muon}\\
		& 					 	& $20\,\GeV<\pT<40\,\GeV$	& 						&				&			\\
Jet		& anti-$k_T$ ($\Delta R=0.4$)	& $|\eta|<0.8$, $\ET=100\, \GeV$					& 100\%					& $\approx 10\*$	& \cite{Aad:2011he}		\\
\hline
\multicolumn{6}{c}{CMS}\\
Object 	& definition and algorithm & kinematic range 	& reconstruction efficiency 	& $\pT$-resolution 	& Reference \\
\hline
Electron	& medium	, multivariate 		& $|\eta|<1.479$ 			&$75-85\%$				& $\approx 3-4\%$	& \cite{CMS:ECAL}	\\
		& 					 	& $20\,\GeV<\ET<40\,\GeV$	& 						&				&			\\
Muon	& combined tracking+muon 	& $|\eta|<1.2$ 				&$\approx95\%$			& $\approx 2\%$	& \cite{Chatrchyan:2012xi}\\
		& 					 	& $20\,\GeV<\pT<40\,\GeV$	& 						&				&			\\
Jet		& anti-$k_T$ ($\Delta R=0.5$)	& $|\eta|<0.5$, $\ET=100\, \GeV$& 100\%					& $\approx 10\*$	& \cite{CMS:2009nxa}		\\
\hline
\end{tabular*}
\end{table*}


\subsection{\label{sec:DetCalib}Detector simulation and calibration}

A detailed simulation of the ATLAS and CMS detector response has 
been developed over recent years. Both simulations are based on the GEANT4 
package \cite{Agostinelli:2002hh}, which offers the possibility to describe the interaction of all 
final-state particles with the detectors at a microscopic level. In a second 
step, the digitisation of the simulated detector interactions is performed and 
the nominal data reconstruction algorithms are applied. 

Several methods are used to calibrate the detector and to compare data to the simulated events. 
One important calibration for lepton and jet reconstruction is based on the study of the 
leptonic decays of the \Zboson boson, which will be briefly summarised in the following. More details on lepton and 
jet calibration can be found in~\cite{Chatrchyan:2011ds},~\cite{Aad:2011he}.

The lepton reconstruction and identification efficiencies can be determined in 
data via the \textit{tag-and-probe  method}. This method makes use of well known decay 
properties of a resonance, e.g. the \Zboson boson, into two well identified 
particles. One particle is selected with a strict selection (the \textit{tag}) to obtain
a low background rate. The second particle (the \textit{probe}) is required to only pass 
loose selection cuts and can then be used to determine the selection efficiency 
for tighter requirements. 

A simple example is the reconstruction efficiency of 
muons in the muon spectrometer. The \Zboson boson decays into two muons, 
resulting in two oppositely charged tracks in the tracking and muon detectors. The tag object is 
required to have a track in both detector systems. The probe object is 
required to have only a track in the tracking detector which forms an invariant mass close to 
the \Zboson boson mass. This ensures a rather clean sample of \Zboson boson events in the 
muon decay channel. The corresponding reconstruction efficiency can be 
determined by testing if a matching track to the probe 
can be found.

Since the \Zboson boson mass and width is precisely known from the LEP 
experiments, it can also be used to calibrate the energy scale and resolution of 
leptons. Here, the invariant mass spectra of the leptonic \Zboson boson decays are 
compared in data and simulations. The peak of the mass distribution is sensitive to 
the lepton energy/momentum scale, while the width of the distributions gives a handle 
on the energy and momentum resolutions.

The production of the \Zboson boson also offers the possibility to calibrate the 
energy scale of particle jets and the hadronic activity in the 
calorimeters. \Zboson bosons which are produced with large transverse momenta must 
balance this momenta with additional partons in the final state. The transverse 
momenta of the \Zboson boson can be reconstructed rather precisely by the 4-momentum 
measurements of its decay leptons. This transverse momentum must be balanced by 
reconstructed particle jets, or to be more general, by the total measured hadronic 
activity. Hence the energy scale and the resolution of particle jets can 
be calibrated in data.

It should be noted that these methods rely on the available statistics 
of the corresponding control samples, e.g. on the available number of recorded 
Z boson events in the leptonic decay channel. While the uncertainties on the 
detector calibration are usually treated as systematic uncertainties in the 
physics analyses, they have a significant statistical component which can be 
reduced by studying more data. Analyses which are based on the 2010 data have 
therefore significant larger uncertainties due to the limited statistics of the 
calibration samples.

\section{\label{sec:CrossMeasPhil} Production Cross Section Measurements at the LHC}


For experimental measurements, 
the production cross-section is calculated via the following equation
\begin{equation}
\label{EQN:CrossSectionExp}
\sigma_{V}^{incl} = \frac{N_{signal}}{\epsilon \cdot BR \cdot \IntLumi}\,.
\end{equation}
The number of signal events is determined as $N_{signal} = N_{data}-N_{bkg}$, 
where $N_{data}$ is the number selected events in data and
$N_{bkg}$ is the number of background events surviving the signal selection.
The factor $\epsilon$ is the efficiency of the signal events passing the
signal selection criteria. To correct the cross-section for
the choice of a specific decay channel, a branching ratio factor, $BR$ is applied. These ratios are known
to a high accuracy for the gauge bosons from LEP experiments \cite{Beringer:1900zz}. 
Finally, \IntLumi is the integrated luminosity, which is a measure
of the size of the data sample used.


The efficiency correction factor $\epsilon$ is usually estimated with 
simulations of the signal process. These simulations include both a detailed description of 
the object reconstruction in the detector, called the \textit{reconstruction level}, and the final-state particle information 
of the generator calculations, called the \textit{generator level}. The same signal 
selection cuts as applied to the data can be applied to
the simulated events at reconstruction level.  However, the
simulation do not model the data perfectly and these
differences are corrected in the estimation of $\epsilon$, following the methods
described in Section \ref{sec:DetCalib}. In addition, basic signal selection cuts, such as minimal $\pT$ cut,
can also be applied to the final-state particles at the generator level. The object value for the final-state particles though differs from the reconstructed quantity.
Following these definitions, $\epsilon$ can be defined as the ratio of all events which pass the
signal selection on reconstruction level $N^{\textrm{selected}}_{\textrm{reco.}}$ over the number of all generated
events $N^{\textrm{all}}_{\textrm{gen.}}$. 

The efficiency correction $\epsilon$ can further be decomposed as the product of a
fiducial acceptance, $A$, and a detector induced- correction factor, $C$, i.e. $\epsilon =
A \cdot C$. The fiducial acceptance is the ratio of the number of
events that pass the geometrical and kinematic cuts in the analysis on generator level
($N^{\textrm{selected}}_{\textrm{gen.}}$)  over the total number of generated events in 
a simulated sample of signal process ($N^{\textrm{all}}_{\textrm{gen.}}$). These selection cuts on generator level usually require geometrical
and kinematic constraints close to the cuts applied on the 
reconstructed objects, e.g. leptons in the final state should fulfil
$\pT>20\,\GeV$ and $|\eta | < 2.5$. The dominant uncertainties on the fiducial acceptance are the scale and PDF uncertainties.

The detector correction factor $C$ is defined as the number of selected
events in simulated sample ($N^{\textrm{selected}}_{\textrm{reco.}}$), which now includes a detailed simulation of the detector response, over the number of
events in the fiducial phase space at generator level ($N^{\textrm{selected}}_{\textrm{gen.}}$). Hence the product of $A\cdot C$ can be written as

\begin{equation}
\epsilon = C\cdot A =  
\frac{N^{\textrm{selected}}_{\textrm{reco.}} }{N^{\textrm{selected}}_{\textrm{gen.}} } 
\cdot 
\frac{N^{\textrm{selected}}_{\textrm{gen.}}}{N^{\textrm{all}}_{\textrm{gen.}}} 
= \frac{N^{\textrm{selected}}_{\textrm{reco.}} }{N^{\textrm{all}}_{\textrm{gen.}} }\, .
\end{equation}

The uncertainties associated with the detector correction factor
are dominated by experimental sources, such as limited knowledge
of reconstruction or cut efficiencies and the accuracy of the energy/momentum
measurements. This factor can be larger than unity
due to migration effects from outside the fiducial region 
into the reconstructed sample. However, in practice this is rarely the case,
as detector inefficiencies and the selection criteria on reconstructed objects reduce the number of events.

Defining $\epsilon$ as $A \cdot C$ is convenient because if the definition of the fiducial volume used for $N^{\textrm{selected}}_{\textrm{gen.}}$ is close to the cuts applied to the data, this factorisation allows for a separation of
theoretical and experimental uncertainties. The fiducial acceptance, $A$, is completely independent of the detector response whereas the detector correction factor, $C$, is largely independent of theoretical modelling uncertainties. 

In many experimental measurements, the fiducial cross-section, defined as
\begin{equation}
\label{EQN:CrossSectionFid}
\sigma_{V}^{fid.} = \frac{N_{data} - N_{bkg}}{C \cdot BR \cdot
\IntLumi} = \sigma_{V}^{incl.} \cdot A\, ,
\end{equation}
is therefore used, as this definition is only affected to a small extend by theoretical uncertainties. Using fiducial cross-sections has the 
added benefit that experimental data can be more easily compared to future theory predictions 
with improved theoretical uncertainties. On the 
other hand, it should be noted that the theoretical predictions for fiducial cross-sections are also subject to sizeable
PDF and scale uncertainties.

In addition to measurements of the inclusive cross-section, measurements of the cross-section as a function of 
one or more observables can also be made. For a given range of a single observable, this is expressed as
\begin{equation}
\label{EQN:CrossSectionDiff}
\frac{d\sigma_{V}^{fid.}}{dx} = \frac{N(\Delta x)_{data} - N(\Delta x)_{bkg}}{C(\Delta x) \cdot BR \cdot
\IntLumi} 
\end{equation}
\noindent where $x$ is the observable being measured, $N(\Delta x)_{data}$, $N(\Delta x)_{bkg}$ and $C(\Delta x)$ are the same as defined above but for a specific range of $x$. Differential cross-section measurements allow for a comparison of distributions of the theoretical predictions to the data. 

The challenge of measuring the differential cross-section is the transformation of the 
measured distribution, which is distorted by the limited resolution and efficiencies of the 
detector, to the underlying or true distribution. One possibility to infer the true distribution
from the measurement is to directly use Equation \ref{EQN:CrossSectionDiff}, which is known as \textit{bin-by-bin unfolding}. 
However, this method is only a valid approach if the
\textit{purity}, defined as the ratio of events which fall in the same range of $\Delta x$ in both the reconstructed- and generator-level
selections over the total number of events that have been generated in the range $\Delta x$, of the underlying distribution is high, typically above $90\%$. When the purity is low, $C(\Delta x)$ can have large theoretical uncertainties, 
since simulations are the only means to estimate this migration between bins. To reduce these uncertainties, advanced
unfolding methods have to be used. One widely used approach
 is Bayesian unfolding \cite{D'Agostini:1994zf}. Here, the experimental
detector effects are represented in a response matrix, whose elements are
the probability of an event in the $i$-th bin at generator level to be
reconstructed in the $j$-th bin at reconstruction level. The
bin size is chosen to be wider than the detector resolution
effects, aiming at a purity of $>60\%$. In the first iteration, the response
matrix is derived from simulations. It is then multiplied to the measured
spectrum, resulting in a first unfolded spectrum of the data. For the $n$-th iteration, the response
matrix is reweighted to the unfolded spectrum of step $n-1$ in order to minimise
the bias of the initial prediction. Thus the unfolded spectrum becomes insensitive
to the original prior.  Other unfolding techniques, such as 
matrix inversion or single value decomposition \cite{Hocker:1995kb}, are also used.

\section{Event Selection and Background Estimates}
\label{sec:SelAndBackground}

The event selection of vector bosons is similar for all 
studies that are discussed in this article. Hence we introduce a general 
approach for the signal selection and background estimation for \Wboson and \Zboson 
bosons. The discussion is mainly based on the published
inclusive cross-section analyses based on the 2010 data sample  \cite{Aad:2011dm}, \cite{CMS:2011aa}.
However, important differences in the signal selection for other analyses, such as 
the production of vector bosons in association with jets, are also highlighted. 
As the event selection is rather technical matter in nature, this section should be understood as an introduction to the basic concepts. Experienced 
readers might find the relevant information summarised in Tables \ref{tab:SignalSelectionCuts}, \ref{tab:DataAndBackgroundZ} 
and \ref{tab:DataBackgroundW}.

\begin{figure*}
\begin{minipage}{0.49\textwidth}
\resizebox{1.0\textwidth}{!}{\includegraphics{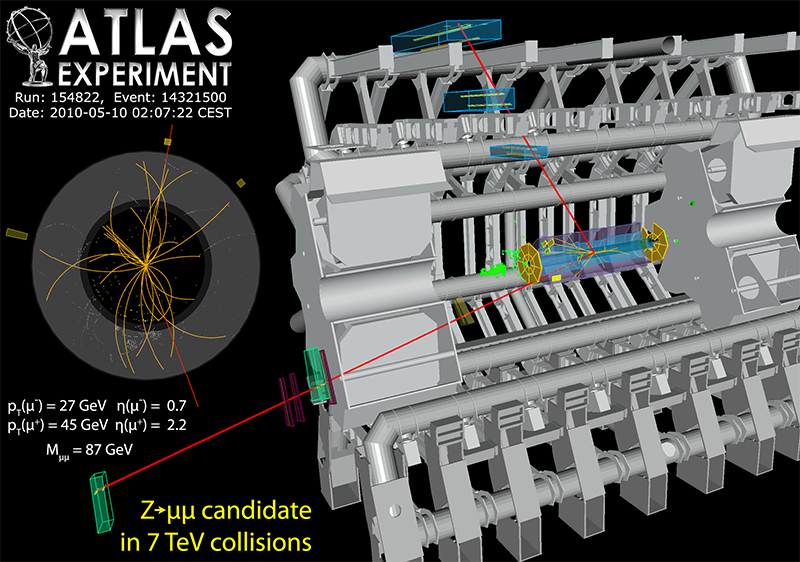}}
\caption{\label{Fig:ATLASZEvent} Event display of a typical $Z\rightarrow \mu \mu$ event candidate, recorded by the ATLAS detector. The reconstructed muon tracks in the barrel  and endcap region are indicated as red lines.}
\end{minipage}
\hspace{0.5cm}
\begin{minipage}{0.49\textwidth}
\resizebox{1.0\textwidth}{!}{\includegraphics{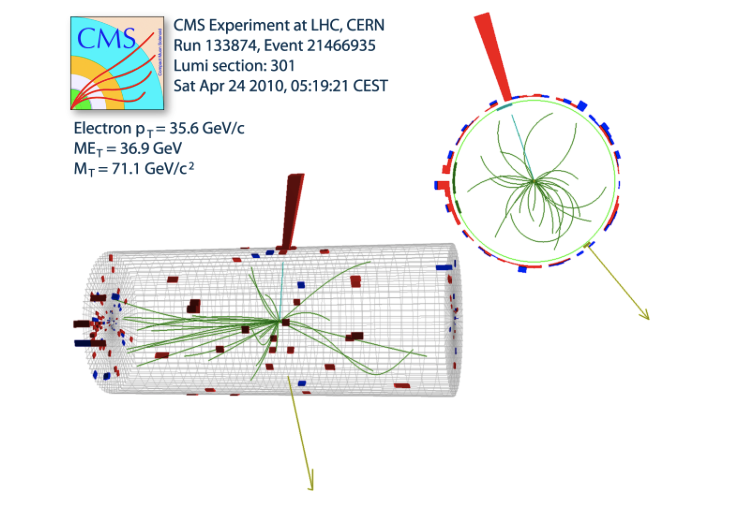}}
\caption{\label{Fig:CMSWEvent} Event display of a typical $W\rightarrow e \nu$ event candidate, recorded by the CMS detector. The reconstructed cluster in the electromagnetic calorimeter is shown in red, the   missing transverse energy in yellow.}
\end{minipage}
\end{figure*}

\subsection{\label{sec:SignalZ}Signal selection and background estimations of $Z/\gamma^*$ events}

The experimental signature of \Zboson bosons in the leptonic decay channel are 
two oppositely charged, isolated and energetic leptons. These leptons 
stem from the same vertex and form an invariant mass close to the \Zboson boson 
mass of $m_Z=91.2\,\GeV$. An event display of the typical $Z\rightarrow \mu \mu$ event candidate, 
recorded by the ATLAS detector, is shown in Figure \ref{Fig:ATLASZEvent}. It should 
be noted that the di-lepton final state 
contains contributions from both \Zboson boson and virtual photons ($\gamma^*$) 
exchange as well as interference. 
Therefore, the measured production cross-sections 
are usually given in terms of a combined $Z/\gamma^*$ exchange. 

In most ATLAS anal\-yses, the generic $Z/\gamma^* \rightarrow l^+l^-$ selection 
requires two oppositely charged leptons with an invariant mass between 
$66\,\GeV<m_{ll}<116\,\GeV$. Muons are required to have a reconstructed track in both the 
tracking and muon detectors within $|\eta|<2.4$ and a minimal transverse momentum of 
$\pT>20\,\GeV$. In addition, the muons are required to pass a relative tracking-based isolation 
requirement based on a cone radius of $\Delta R = 0.2$. Electrons are required 
to fulfil $|\eta|<1.37$ or $1.52<|\eta|<2.4$ with a mi\-ni\-mum transverse energy 
of $\ET>20\,\GeV$. In addition, the medium identification criteria have to be 
satisfied. ATLAS also uses forward electrons for some analyses, i.e electrons 
within $2.4<|\eta|<4.9$ 
which have no associated track in the tracking system. For those electrons 
tight identification criteria have to be fulfilled and at least one 
of the two signal electrons must be within $|\eta|<2.5$ and have a 
corresponding track in the tracking detector.

CMS selects $Z\rightarrow\mu\mu$ events by requiring two reconstructed, oppositely 
charged muons within $|\eta|<2.1$ coming from the same vertex and a transverse momenta 
requirement of $\pT>20\,\GeV$. Both muons have to fulfil a relative tracking-based isolation requirement 
within a cone radius of $\Delta R = 0.3$ and must match to the corresponding 
di-lepton trigger objects. 
Later CMS analyses, extend the $|\eta|$-re\-quire\-ment to 
$2.4$ and lower the cuts on the muon transverse momenta: the 
leading and subleading muon have to fulfil $\pT>14\,\GeV$ and $\pT>9\,\GeV$, respectively. 
For the electron decay channel of $Z\rightarrow ee$, at least two reconstructed electrons within $|\eta|<1.44$ or 
$1.57<|\eta|<2.5$ and $E_T>20\,\GeV$ are required. In later CMS analyses, this was relaxed to require the 
leading electron to have $\pT>20\,\GeV$, while the sub-leading electron must fulfil 
$\pT>10\,\GeV$. While it is not required that both electrons have oppositely reconstructed charge, they must be matched to 
the corresponding trigger objects. The electrons must also satisfy a 
relative tracking-based isolation requirements within a cone radius of $\Delta R<0.3$. 
The \Zboson 
boson mass range for both channels is defined as $60\,\GeV<m_{ll}<120\,\GeV$. 
The chosen mass range of the CMS experiment is larger compared to the ATLAS 
definition. This leads to an increase of the available phase space by a factor 
of $1.015$ for CMS. The signal selection cuts are summarised in Table \ref{tab:SignalSelectionCuts}.

For measurements of $Z/\gamma^*$ production in association with jets, the jets are required to have $\pT>30\,\GeV$. For ATLAS 
measurements, the jets at high rapidities are used, $|y|<4.4$, whereas CMS uses jets within the acceptance of the tracker, $|\eta|
<2.4$. Since ATLAS and CMS use jet algorithms with different size parameters (see Section~\ref{sec:reco-objects}) the jet energies 
can not be directly compared. Jets within an $\Delta R < 0.3$ or $\Delta R < 0.5$ of an electron or muon are not counted for CMS 
and ATLAS analyses, respectively.

\begin{table*}
\centering
\caption{Summary of the kinematic cuts used by the ATLAS \cite{Aad:2011dm} and CMS analysis \cite{CMS:2011aa} on leptons and 
their invariant quantities for the electron and muon decay channel of \Zboson and \Wboson Bosons respectively.}
\label{tab:SignalSelectionCuts}       
\begin{tabular*}{\textwidth}{@{\extracolsep{\fill}}l  ll  ll}
\hline
				&	 \multicolumn{2}{c}{ATLAS}		& \multicolumn{2}{c}{CMS}								
\\ 
				&	$Z\rightarrow l^+l^-$ 			&	 $W^\pm\rightarrow l^\pm \nu 	$    			& $Z\rightarrow l^
+l^-$ 				& $W^\pm\rightarrow l^\pm \nu 	$ 	\\
\hline
Electron-Channel	&	$E_{T}(e^+)>20\,\GeV$			&	$p_{T}(e^\pm)>20\,\GeV$					&$E_{T}(e^
+)>25\,\GeV$				& one $e^\pm$ with	$\ET>25\,\GeV$	\\
				&	$E_{T}(e^-)>20\,\GeV$			&	$p_{T}(\nu)>25\,\GeV$					&$E_{T}
(e^-)>25\,\GeV$				& $|\eta_{e^\pm}|<1.44$ or			\\
				&	$|\eta_{e^\pm}|<1.37$ or			&	$|\eta_{e^\pm}|<1.37$ or					&$|\eta_{e^
\pm}|<1.44$ or			&$1.57<|\eta_{e^\pm}|<2.5$			\\
				&	$1.47<|\eta_{e^\pm}|<2.47$		&	$1.47<|\eta_{e^\pm}|<2.47$				&$1.57<|
\eta_{e^\pm}|<2.5$			&								\\
				&	$66\,\GeV<m_{ee}<116\,\GeV$		&	$\mT>40\,\GeV$						&$60\,
\GeV<m_{ee}<120\,\GeV$		&								\\
\hline
Muon-Channel		&	$p_{T}(\mu^+)>20\,\GeV$			&	$p_{T}(\mu^\pm)>20\,\GeV$				&$p_{T}(\mu^
+)>25\,\GeV$			&one $e^\pm$ with	$\pT>25\,\GeV$	\\
				&	$p_{T}(\mu^-)>20\,\GeV$			&	$p_{T}(\nu)>25\,\GeV$					&$p_{T}
(\mu^-)>25\,\GeV$			&$|\eta_{\mu^\pm}|<2.1$				\\
				&	$|\eta_{\mu^\pm}|<2.4$			&	$|\eta_{\mu^\pm}|<2.4$					&$|\eta_{\mu^
\pm}|<2.1$				&								\\
				&	$66\,\GeV<m_{\mu\mu}<116\,\GeV$	&	$\mT>40\,\GeV$						&$60\,
\GeV<m_{\mu\mu}<120\,\GeV$	&								\\
\hline
\end{tabular*}
\end{table*}

For $Z/\gamma^* \rightarrow l^+l^-$ events, the main background contributions stem from $Z\rightarrow\tau\tau$ events, di-boson 
events, \ttbar~decays and QCD multi-jet events. $Z\rightarrow\tau\tau$ events can pass the signal selection 
when the $\tau$-leptons decay into electrons or muons. 
Di-boson production such as $WZ\rightarrow l^\pm \nu l^+l^-$ and top-quark pair production such as  ($t\bar t \rightarrow W^+ b W^- 
\bar b \rightarrow l^+ \nu b l^- \nu$) both 
have signatures with two energetic and isolated leptons. With the exception of the 
$WZ$ di-boson process, these processes though
do not peak at $m_Z$ and are largely removed by the mass cut. The QCD multi-jet events do not necessarily have a lepton in the final state and are 
discussed in more detail below.

To estimate the backgrounds, CMS often uses a data-driven approach, which exploits the 
fact that most of the mentioned background processes have an $e\mu$ decay 
channel, while the signal has two same flavour leptons in the final 
state. By requiring opposite flavour leptons, the background can be directly 
estimated after correcting for differences in the lepton reconstruction. For most ATLAS analyses, simulations are used for these 
estimates, since these processes are theoretical well understood in both the absolute background contribution as well as the 
predictions of the kinematic distributions. In $Z/\gamma^*$ production in association with jets, the background from 
\ttbar~production becomes more significant for larger jet multiplicities. In this case, ATLAS analyses use a data-driven approach 
similar to the CMS method. 

The QCD multi-jet background cannot be predicted precisely and must be 
estimated with data-driven methods. QCD multi-jet events pass the signal selection cuts in 
one of two ways: a jet is misreconstructed in the calorimeter and \textit{fakes} an electron signature or the jet contains a heavy-flavour 
quark or kaon which decays into an electron or muon. In the first case, jets can fake an electron signal without a real electron 
 in the jet itself whereas in the second case a real lepton is present. The main difference between the lepton signatures for 
QCD multi-jets versus those from $Z/\gamma^*$ events is the isolation 
properties and - in the case of electrons - the calorimeter shower-shapes. While the leptons in signal events appear very isolated in the 
detector, jets contain a significant number of adjacent
particles. Similarly \Wboson boson production in association with jets can also mimic this signature, where one lepton comes from 
the leptonic \Wboson boson decay, and the second lepton originates from or is faked by the accompanying jet. 

To estimate these backgrounds, both experiments use similar data-driven approaches. 
A control region in data dominated by the QCD multi-jet events is used to 
define the kinematic distributions of the background. For the muon decay channel, this 
is achieved by inverting the isolation cut of one of the muons. The control region for 
the electron channel is obtained by requiring a non-isolated electron which 
only passes the loose electron identification cuts. The uncertainties of the 
predicted background distributions can be cross-checked by comparing the spectra to 
same-sign, isolated di-lepton events, which is also expected to be dominated by 
QCD multi-jet background. The absolute normalisation of the QCD multi-jet background is 
then achieved by adjusting the sum of the expected signal and other background 
template to the data as a function of the invariant mass. For $Z/\gamma^*$ production in association with jets, the normalisation of 
the QCD multi-jet background is determined for each jet multiplicity separately. 

Some CMS analyses extract the signal yield together with the lepton trigger and 
reconstruction efficiencies by using a simultaneous fit to the measured invariant mass spectra in several di-lepton candidate categories, e.g. two combined muons or one combined muon and one inner 
detector track. The shapes of the signal and background distributions are taken from MC predictions or data-driven approaches as 
described above.

For measurements of a \Zboson boson in association with $b$-jets, backgrounds from \Zboson events in association with light ($u$, $d$, 
and gluon) jets and $c$-quark jets dominate. To determine the number of Z+$b$-jet events, first di-boson, single-top, \ttbar, and $W \rightarrow \tau \nu$ 
or $Z \rightarrow \tau \nu$ backgrounds are removed. For CMS analyses, the dominant background of \ttbar~is normalised 
to the data at large values of the \Zboson boson mass peak, whereas for the ATLAS analysis, simulations are used to subtract the 
background. To extract the number of \Zboson+$b$-jets events from the light- and $c$-jet events, both ATLAS and CMS use a similar 
approach; a maximum likelihood fit is preformed using a b-tagging observable. For the CMS analysis, this observable used is the invariant mass of the secondary vertex, 
estimated from the b-tagging algorithm.
For ATLAS measurements, the observable used is one of the outputs of the b-tagging 
algorithm's neural network. In both cases, these observables are chosen because they give good separation between light-, $c$- and 
$b$-jet events. The number of Z+$b$-jets is then determined from the fit.


The selected data for the inclusive \Zboson boson production measurements based on the 2010 data sample for ATLAS and CMS 
together with their expected signal and 
background contributions, as well as the respective uncertainties, are summarised in Table 
\ref{tab:DataAndBackgroundZ}. Similar background contaminations and associated uncertainties are seen in the analyses which are 
based on the 2011 data. The invariant mass distributions and the $\pT$ spectra of the decay leptons for 
the selected data samples and the signal MC predictions are shown for 
ATLAS and CMS in Figure \ref{fig:ATLASDilepton} and \ref{fig:CMSDilepton}, respectively. All detector correction effects have been 
applied. 
The background contribution is a few percent, making the leptonic \Zboson boson decay channel one of the cleanest signatures at 
the LHC. Hence it is an ideal channel for precision measurements of the Standard Model as well as for the detector calibration. 
Overall, excellent agreement between data and the predictions can be seen.

\begin{table*}
\centering
\caption{Data sample and background estimations of the ATLAS and CMS inclusive analyses for the process $Z\rightarrow l^+l^-$, 
based on the 2010 data sample.}
\label{tab:DataAndBackgroundZ}       
\begin{tabular*}{\textwidth}{@{\extracolsep{\fill}}l ll ll}
\hline
					&	 \multicolumn{2}{c}{ATLAS}							& \multicolumn{2}{l}{\hspace{2cm}CMS}			\\ 
					&	$Z\rightarrow e^+ e^-$	& $Z\rightarrow \mu^+ \mu^-$		&	$Z\rightarrow e^+ e^-$	& $Z
\rightarrow \mu^+ \mu^-$ \\
\hline
Data (2010)			&	9725					&	11709					&	8452					&13 728					\\
\hline
Total Background		&	$206\pm64$			&	$86\pm32$				&	$35\pm11$			& $60\pm21$				\\
\hline
					& \multicolumn{4}{l}{Percentage of each background compared to the total number of backgrounds}  \\
\hline					
$WW, WZ, ZZ$				&	$10\%$				&	$26\%$					&	$37\%$				& $47\% $					\\
$t\bar t$, $Z\rightarrow\tau\tau$&	$14\%$				&	$22\%$					&	$47\%$				& $50\% $					\\
QCD	multi-jets				&	$76\%$				&	$52\%$					&	$16\%$				& $3\%$					\\
\hline
\hline
\end{tabular*}
\end{table*}

\begin{figure*}
\resizebox{0.245\textwidth}{!}{\includegraphics{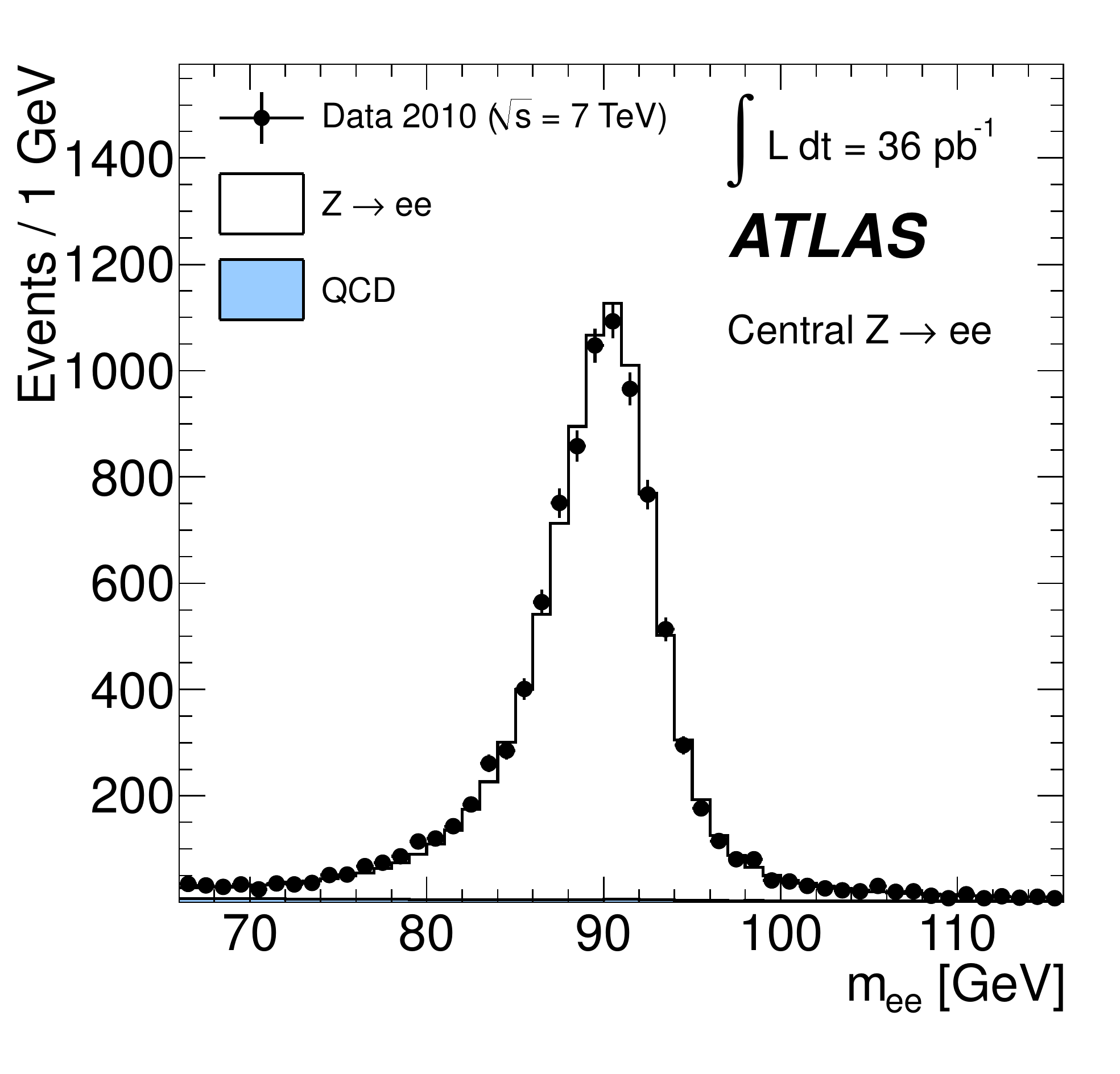}}
\resizebox{0.245\textwidth}{!}{\includegraphics{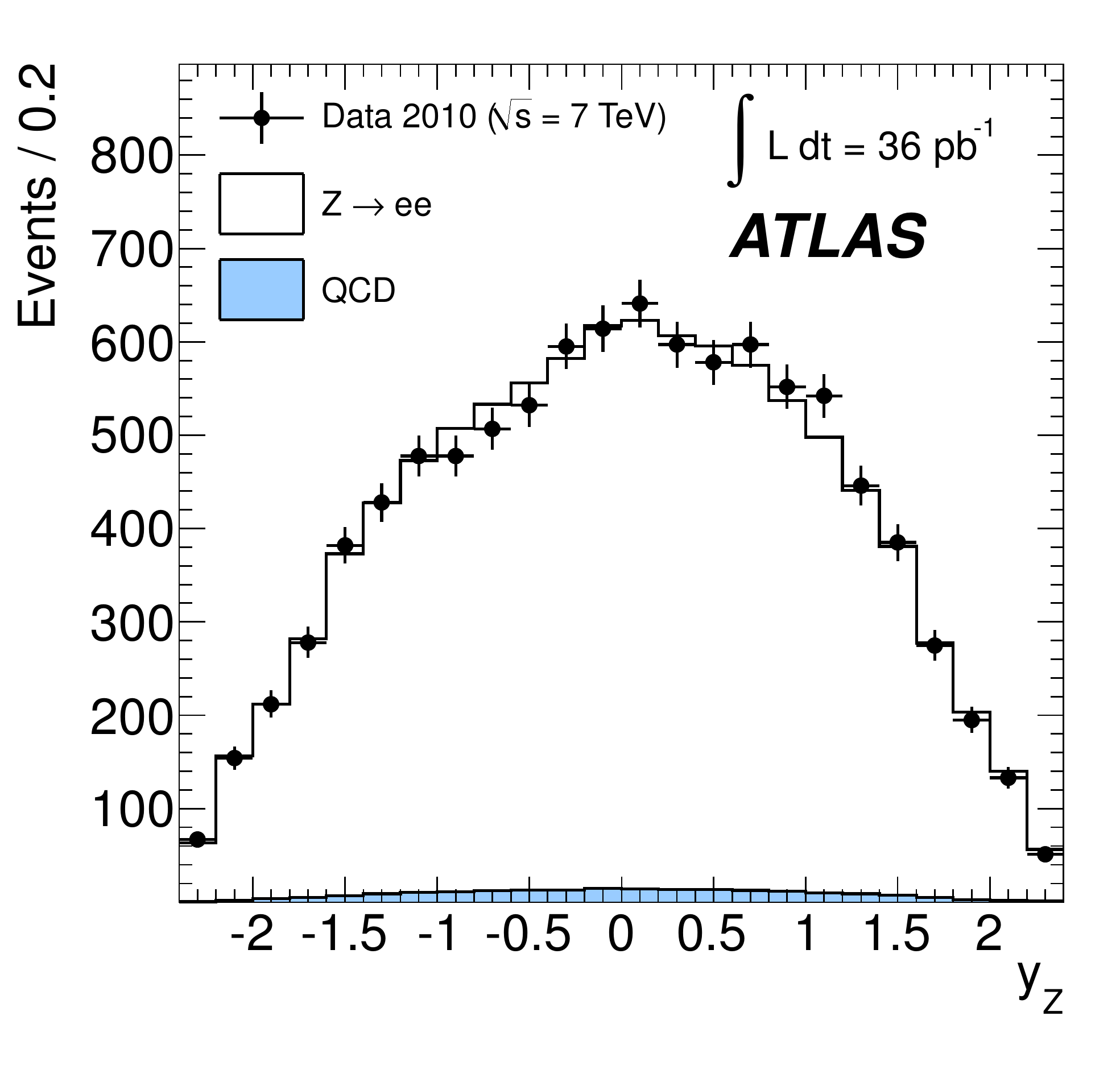}}
\resizebox{0.245\textwidth}{!}{\includegraphics{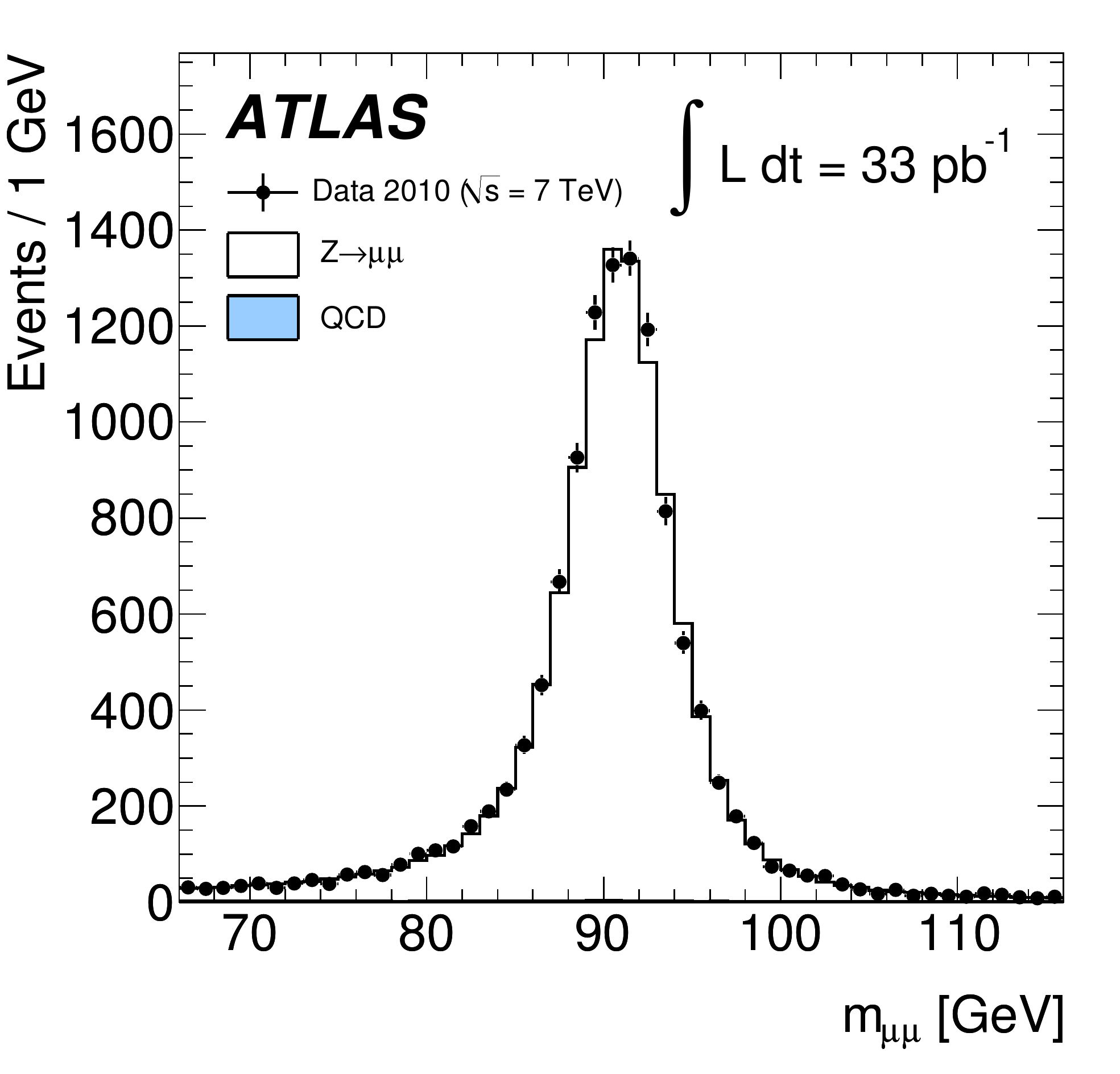}}
\resizebox{0.245\textwidth}{!}{\includegraphics{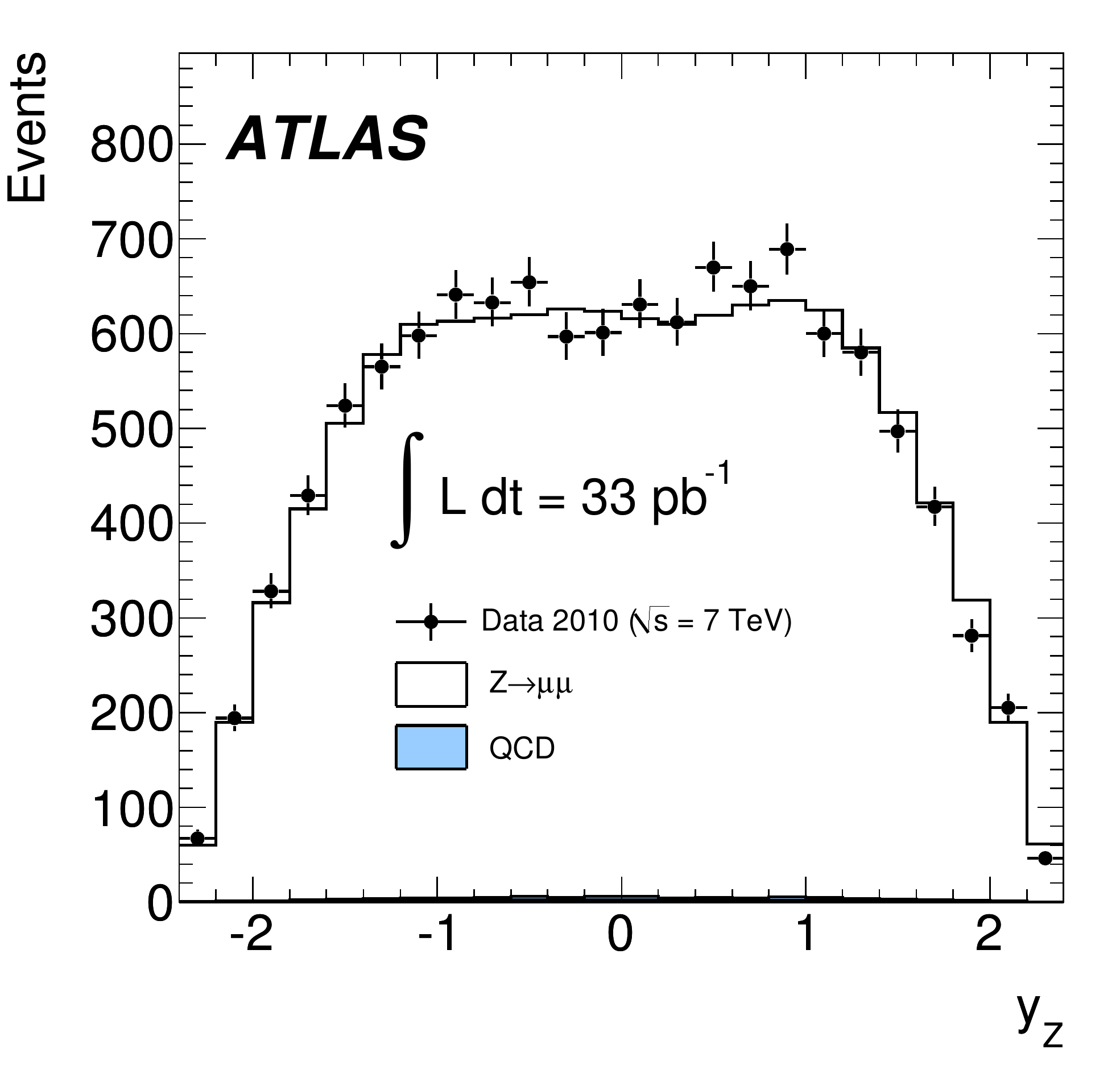}}
\caption{ATLAS \cite{Aad:2011dm}: Di-lepton invariant mass and rapidity $y_Z$-distribution for the central electrons and muons. The simulation is normalised to the data. The QCD multi-jet background shapes have been estimated by data-driven methods.}
\label{fig:ATLASDilepton}
\end{figure*}

\begin{figure*}
\resizebox{0.51\textwidth}{!}{\includegraphics{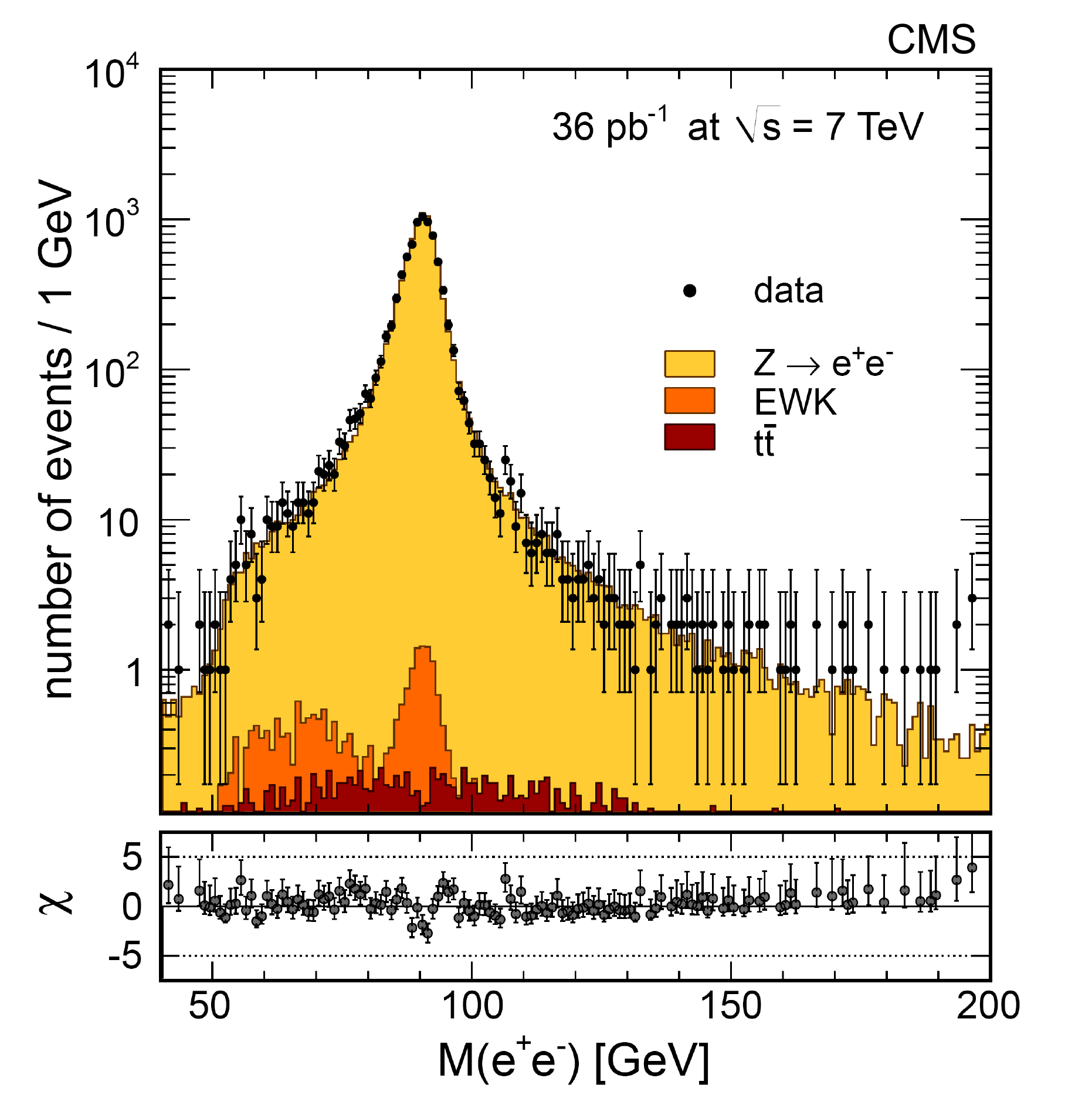}}
\resizebox{0.51\textwidth}{!}{\includegraphics{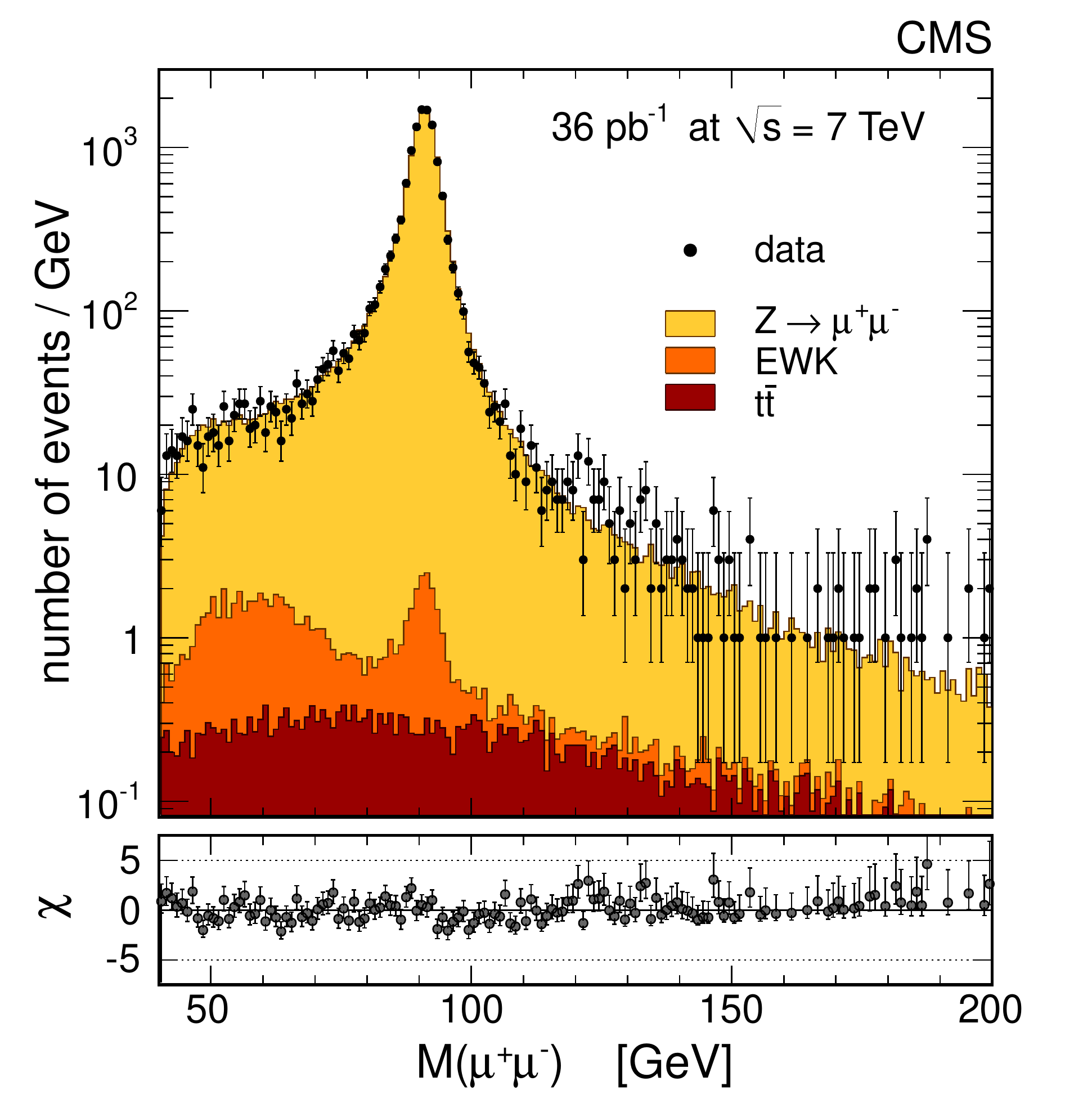}}
\caption{CMS \cite{CMS:2011aa}: Di-lepton invariant mass spectra of selected \Zboson boson events, where all detector corrections have been applied. 
The points with the error bars represent the data. Superimposed are the expected distributions from simulations, normalised to an 
integrated luminosity of 36 \ipb. The expected distributions are the \Zboson signal (yellow, light histogram), other EWK processes 
(orange, medium histogram), and ttbar background (red, dark histogram).}
\label{fig:CMSDilepton}
\end{figure*}

\subsection{\label{sec:SignalW} Signal selection and background estimations of \Wboson boson events}

The leptonic decay of the \Wboson bosons ($W^\pm \rightarrow l^\pm \nu$) leads to an 
isolated and energetic lepton and missing transverse energy. 
An event display of the typical $W\rightarrow e \nu$ event candidate, recorded by the CMS detector, is shown in Figure \ref{Fig:CMSWEvent}.
Since 
no information on the z-component of the missing energy is available, the 
mass of the W-boson cannot be reconstructed. However, the 
invariant mass projection to the transverse plane, defined as
\begin{equation} 
\mT = \sqrt{2 \cdot \pT^l \cdot \pT^\nu \cdot (1-\cos(\phi^l-\phi^\nu))},\,
\end{equation} 
can be reconstructed, since the (x, y) components of the neutrino momentum 
are inferred from the \MET. This observable 
is identical to the \Wboson boson mass when the decay happens purely in the x-y-plane.

The ATLAS analyses select \Wboson boson events by requiring one reconstructed, 
isolated lepton, a minimal \MET~of $25\,\GeV$ and a minimal transverse 
mass of $\mT>40\,\GeV$. For the muon decay channel, one combined 
reconstructed muon with $\pT>20\,\GeV$ within $|\eta|<2.4$ is
required. In the electron decay channel, electrons are required  to 
fulfil $|\eta|<1.37$ or $1.52<|\eta|<2.4$ with a minimum transverse 
energy of $\ET>20\,\GeV$ and medium identification criteria.

The signal selection in the corresponding CMS analysis is significantly 
different. \Wboson boson candidate events are selected by only requiring one 
reconstructed electron with $\ET>25\,\GeV$ and $|\eta|<2.5$ or one 
reconstructed muon with $\pT>25\,\GeV$ and $|\eta|<2.1$. In later analyses, 
the $\eta$ requirement has been relaxed to $|\eta|<2.4$ for muons but the 
threshold for electrons raised to $\ET>35\,\GeV$. Events 
with a second reconstructed lepton with $\pT>15\,\GeV$ are vetoed. No
additional cuts on \MET and $\mT$ are imposed. For the CMS measurement of 
\Wboson events in association with jets, the selection is slightly modified. The cut on the reconstructed lepton is $\pT>20\,\GeV$ with no 
additional leptons above $\pT>10\,\GeV$. A cut of $\mT>20\,\GeV$ is applied.

Both experiments use the signed curvature of the lepton tracks in the inner 
detector to determine its charge and hence also the charge of the 
 \Wboson boson. While the charge misidentification is rare in muon events, 
a significant fraction of electron charges are mismeasured. Due to 
substantial material in the tracking detector, a large fraction of electrons radiate photons 
which in turn may convert to electron-positron pairs close to the
original electrons, leading to charge misidentifications during the track 
reconstruction. In addition to the tracking information, CMS also uses the 
vertex and cluster position of the calorimeter for the charge identification. 

For \Wboson production in association with jets, the jet selection is the same as described in Section~\ref{sec:SignalZ}.

For both experiments, the major sources of backgrounds for the signal selection are $Z/\gamma^*$ production, the $\tau$-lepton
decay channel of the W-boson, di-boson production, QCD multi-jet events and top-pair production.

The $Z/\gamma^*$ process can pass the signal signature when one
lepton is not reconstructed, e.g. by being outside of the detector acceptance, thereby creating significant amounts of \MET. This 
background is theoretically understood to high precision and therefore the kinematic distributions can be predicted directly from 
simulations. The normalisation is either taken from a control sample 
(by requiring two reconstructed leptons), or also from simulations.
Similarly the $W\rightarrow\tau\nu$ background, where the $\tau$
lepton decays further into electrons or muons and di-boson production, where one or both of the bosons decays to leptons, is also 
theoretically well 
understood and modelled to a sufficient precision by simulations. 

As discussed in Section~\ref{sec:SignalZ}, the QCD multi-jet background must be estimated using data-driven techniques. In ATLAS 
analyses, the QCD multi-jet control region in the muon channel is defined by reversing the isolation and removing the cut on \MET. 
For the electron channel, the control sample is defined by inverting some electron identification criteria and not applying an \MET~
requirement. The normalisation of the QCD multi-jet background 
is determined from data using a fit of the \MET distribution, the results of which can seen in Figures~\ref{fig:ATLASMETW} 
and ~\ref{fig:CMSMETW}. For \Wboson production in association with jets, the normalisation is determined separately for each jet 
multiplicity. 

For measurements of the inclusive \Wboson production, CMS extracts the number of \Wboson signal events with a binned,  
extended maximum likelihood fit to the \MET~distributions. The \MET~distributions for the signal and for the $Z/\gamma^*$, $t\bar t
$ and $W\rightarrow \tau \nu$ backgrounds are based on simulations. The shape of the QCD multi-jet 
background \MET~template is determined in a control region, defined by inverting a 
subset of the electron identification criteria or the muon isolation requirement 
for the $W\rightarrow e\nu$ and $W\rightarrow \mu\nu$-channel, respectively. The fit is performed separately for $W^+$ and $W^-$ signal events.

Similar to $Z/\gamma^*$ and di-boson production, top-pair production is also theoretically well understood but this background is large for 
\Wboson production in association with jets. For inclusive measurements of \Wboson production where the top-pair production is a 
small contribution, simulations are used for the background estimates. For \Wboson events with jets, CMS uses a data-driven 
approach to determine simultaneously the number of both the top-pair events as well as the QCD multi-jet events. This method 
exploits two features about $t\bar t$ and QCD multi-jet events. First, since $t\bar t$ events contain a semileptonic decay of the W, 
these events also peak in $\mT$ at the \Wboson mass. In contrast, QCD multi-jet events do not peak and have a falling $\mT$ 
spectra. Second, $t\bar t$ events also contain jets from b-quarks which can be selected via b-tagging. To determine the 
normalisation for $t\bar t$ and QCD multi-jet events, a 2-dimensional fit in $\mT$ and the number of b-tagged jets is performed in 
each jet multiplicity bin. For the ATLAS measurements, the number of $t\bar t$ events is determined using a 1-dimensional fit in 
the rapidity of the lepton as well as the mass of the W-jet system for each jet multiplicity. The fitted number of $t\bar t$ events is 
consistent with those from the simulations but has a large statistical uncertainty. For this reason, the ATLAS measurements use 
$t\bar t$ simulations for the background estimates. 

For the measurement of $W+b$-jets from ATLAS where the \ttbar~background is 
kinematically very similar to the signal events, control regions with four jets are used to constrain the normalisation for the \ttbar~events. In 
addition the normalisation of single-top events is constrained by fitting the invariant mass of the $W$ boson and $b$-jet system. 
The extraction of the $b$-quark events from the light- and $c$-jet background uses a similar approach 
as outline in Section~\ref{sec:SignalZ}.

Again, the 2010 analyses for ATLAS \cite{Aad:2011dm} and for CMS \cite{CMS:2011aa} are chosen as an example of the 
expected background contributions. The resulting number of selected events of ATLAS and CMS together with their expected signal 
and background contributions and their uncertainties are summarised in Table 
\ref{tab:DataBackgroundW}. The corresponding $\MET$ and $\mT$ distributions of selected W-boson events is shown for the ATLAS and CMS
experiments in Figure~\ref{fig:ATLASMETW} and~\ref{fig:CMSMETW} respectively.   

\begin{table*}
\centering
\caption{Data sample and background estimations of the ATLAS and CMS inclusive analyses for the process $W^\pm\rightarrow l^
\pm\nu$, based on the 2010 data sample.}
\label{tab:DataBackgroundW}       
\begin{tabular*}{\textwidth}{@{\extracolsep{\fill}}l ll ll}
\hline
						&	 \multicolumn{2}{c}{ATLAS}							& \multicolumn{2}{c}{CMS}			\\ 
						& $W^\pm\rightarrow e^\pm \nu$	& $W^\pm\rightarrow \mu^\pm \nu$	&	$W^\pm\rightarrow 
e^\pm \nu$	& $W^\pm\rightarrow \mu^\pm \nu$ \\
\hline
Data (2010)				&	130741				&	139748					&	235687				&	166457 				\\
\hline
Total Background			&	$9610\pm 590$		&	$12300\pm 1100$			&	$99684\pm388$		&	$25700\pm383$			\\
\hline
					& \multicolumn{4}{l}{Percentage of each background compared to the total number of backgrounds}  \\
\hline					
$W\rightarrow\tau\nu$		&	34\%					&	34\%						&	4\%					&	16\%					\\
Top 						&	5\%					&	4\%						&	1\%					&	2\%					\\
$Z\rightarrow l^+l^-$, $WW, WZ, ZZ$&	7\%					&	23\%						&	11\%					&	26\%					\\
QCD multi-jet 				&	54\%					&	38\%						&	85\%					&	56\%					\\
\hline
\end{tabular*}
\end{table*}


\begin{figure*}
\resizebox{0.245\textwidth}{!}{\includegraphics{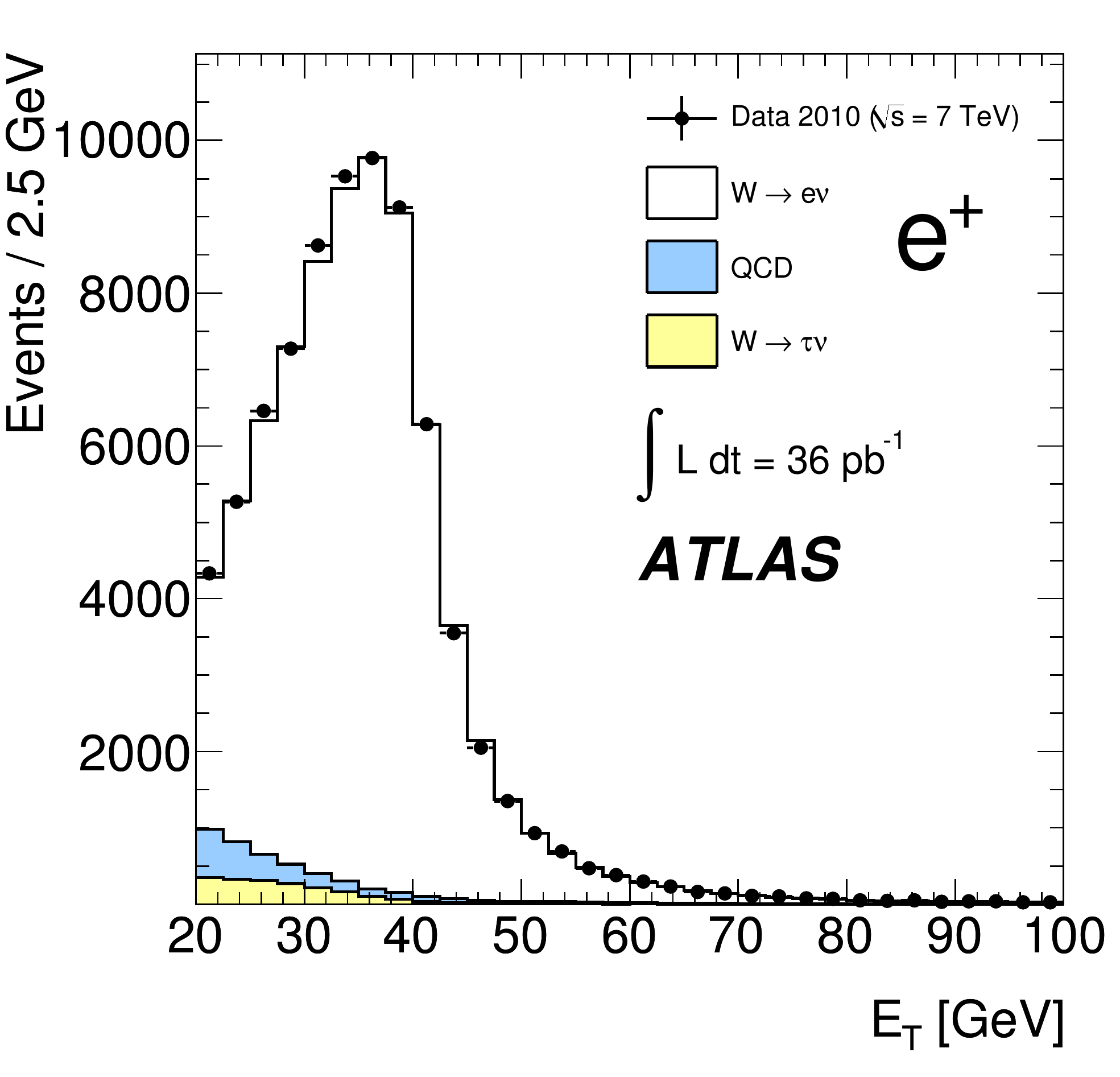}}
\resizebox{0.245\textwidth}{!}{\includegraphics{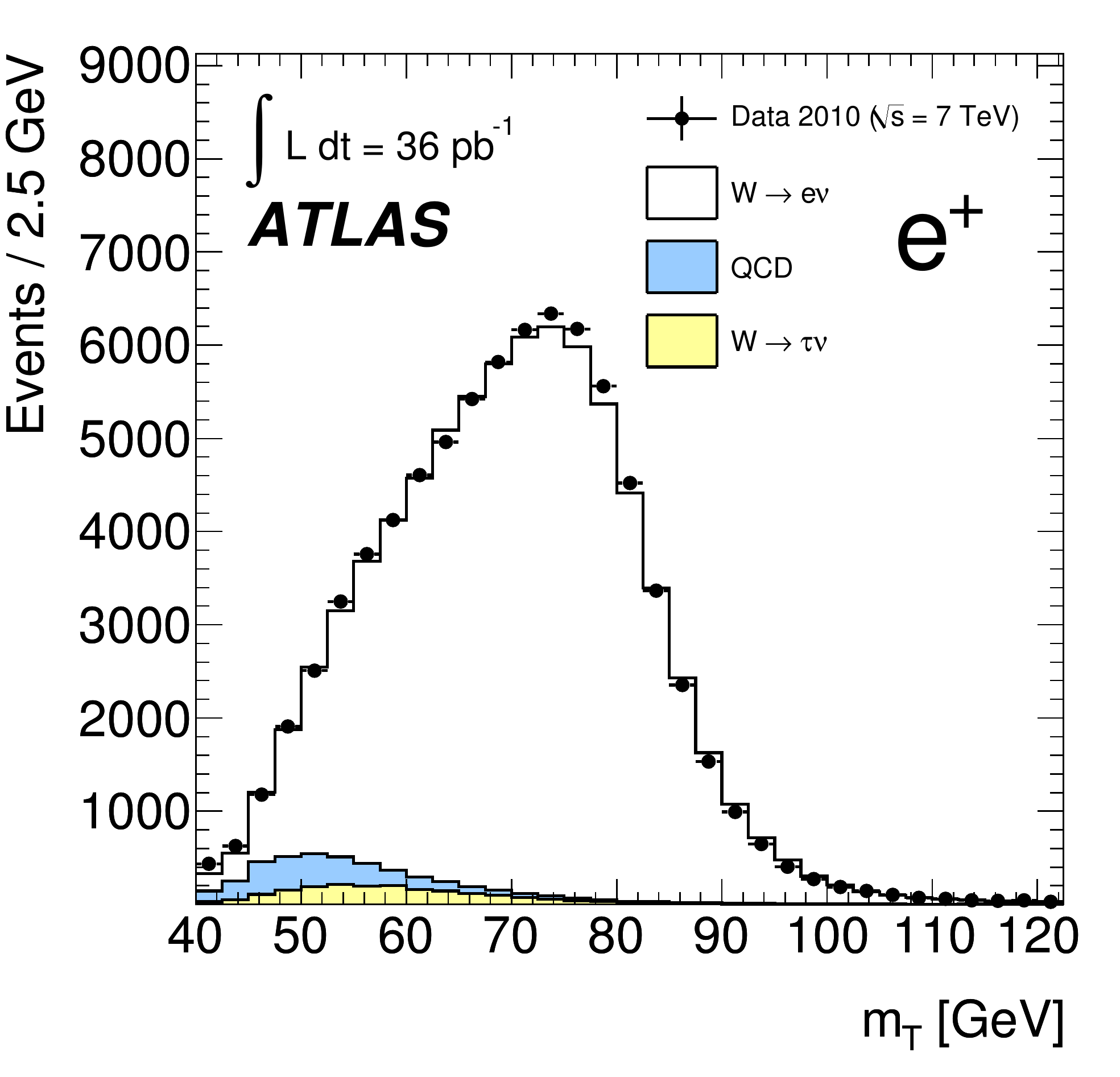}}
\resizebox{0.245\textwidth}{!}{\includegraphics{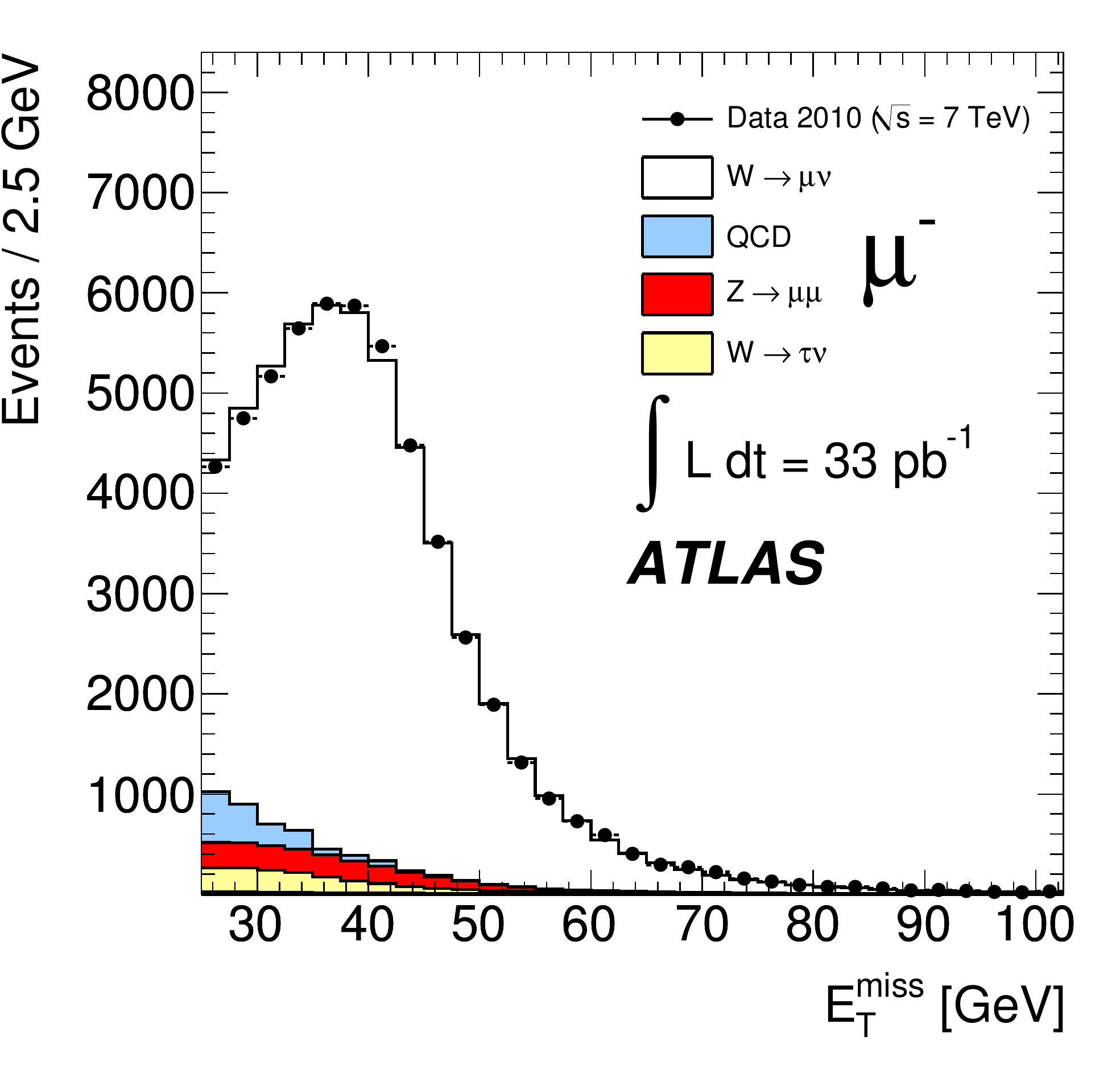}}
\resizebox{0.245\textwidth}{!}{\includegraphics{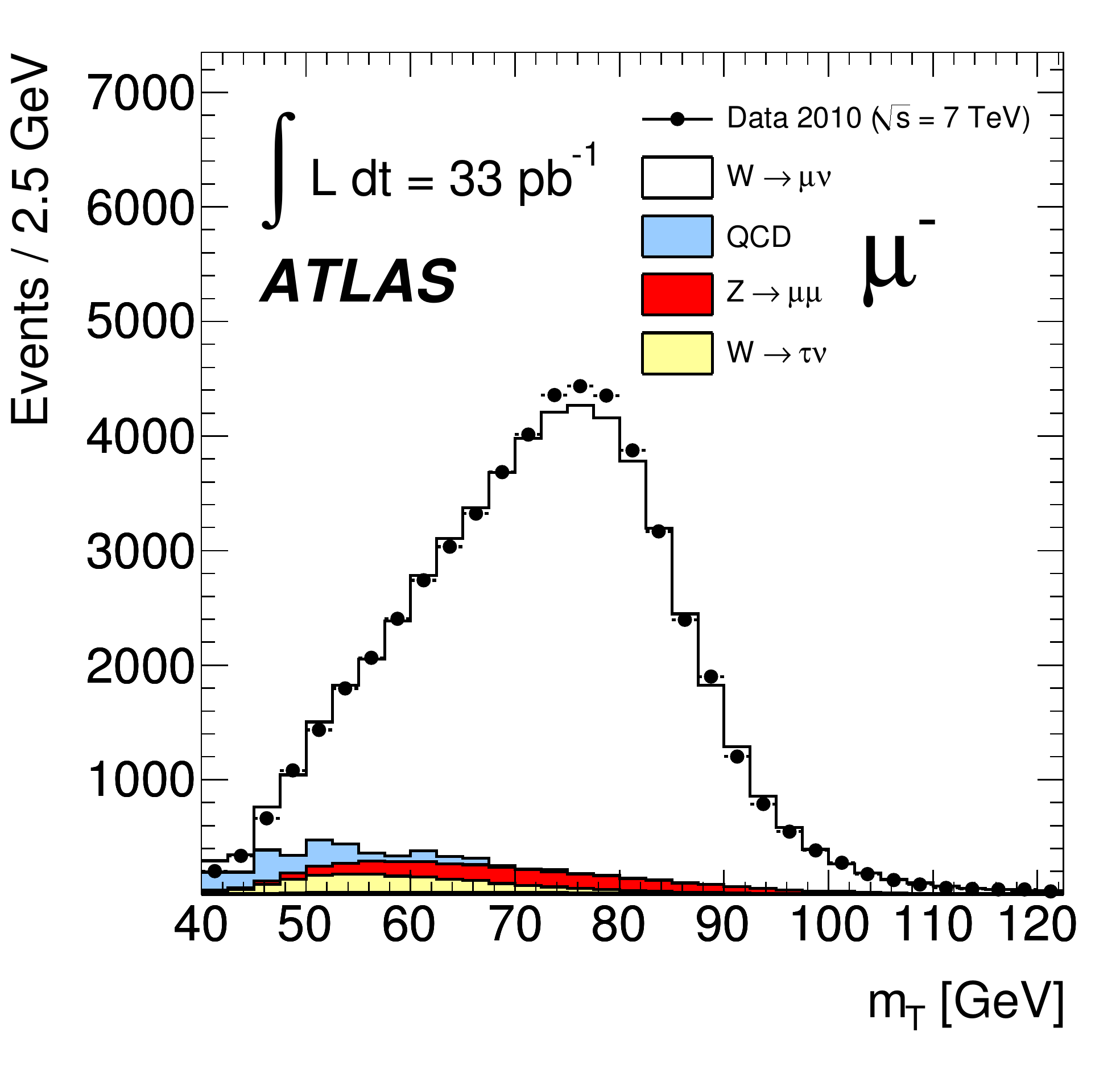}}
\caption{ATLAS \cite{Aad:2011dm}: Distribution of the \MET (left) and $\mT$ (right) in the selected $W^+\rightarrow l^+\nu$ candidate events after all 
cuts for electrons (left two plots) and muons (right two plots). The simulation is normalised to the data. The QCD multi-jet background is 
estimated via data-driven methods.}
\label{fig:ATLASMETW}
\end{figure*}

\begin{figure*}
\resizebox{0.245\textwidth}{!}{\includegraphics{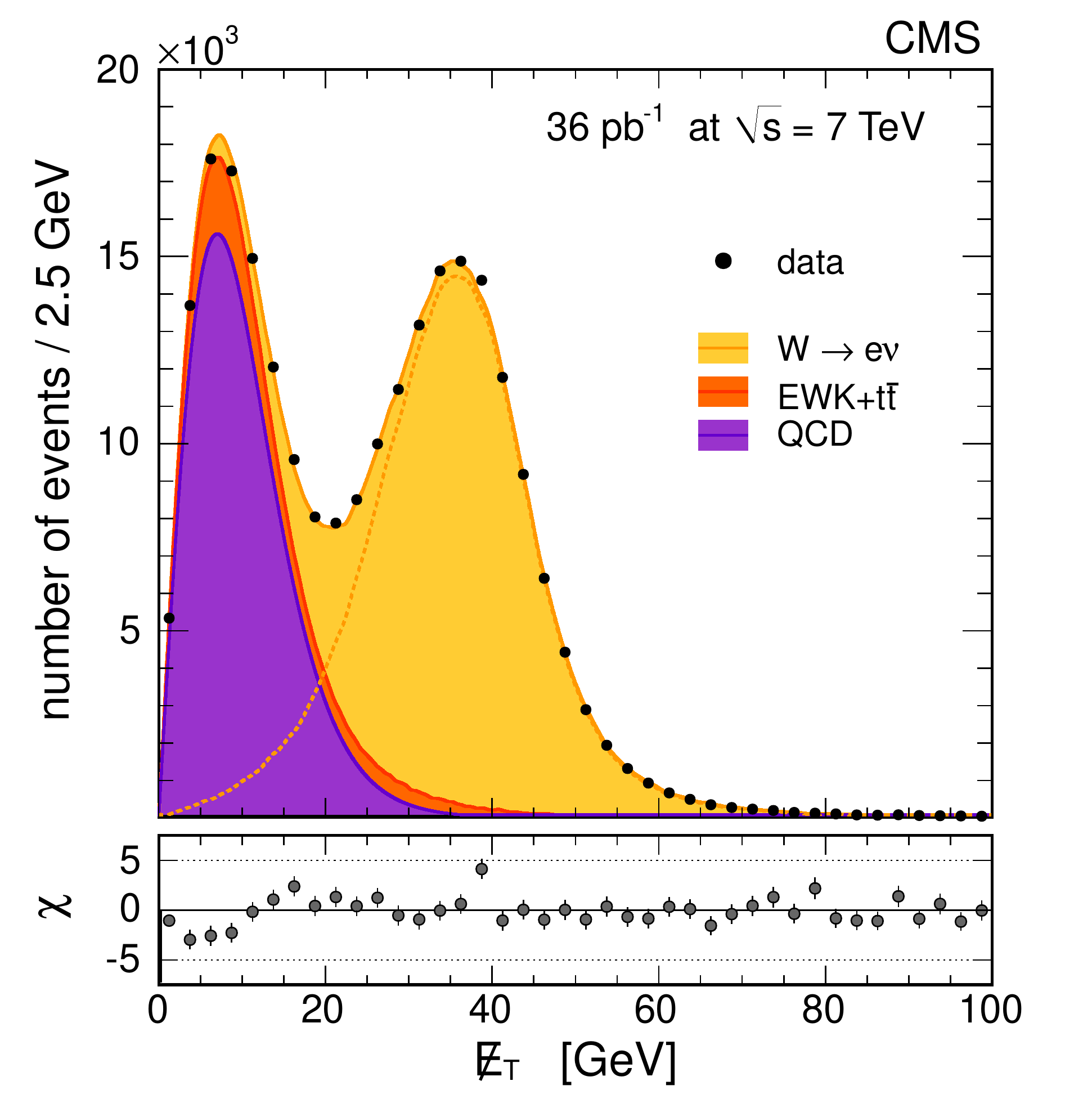}}
\resizebox{0.245\textwidth}{!}{\includegraphics{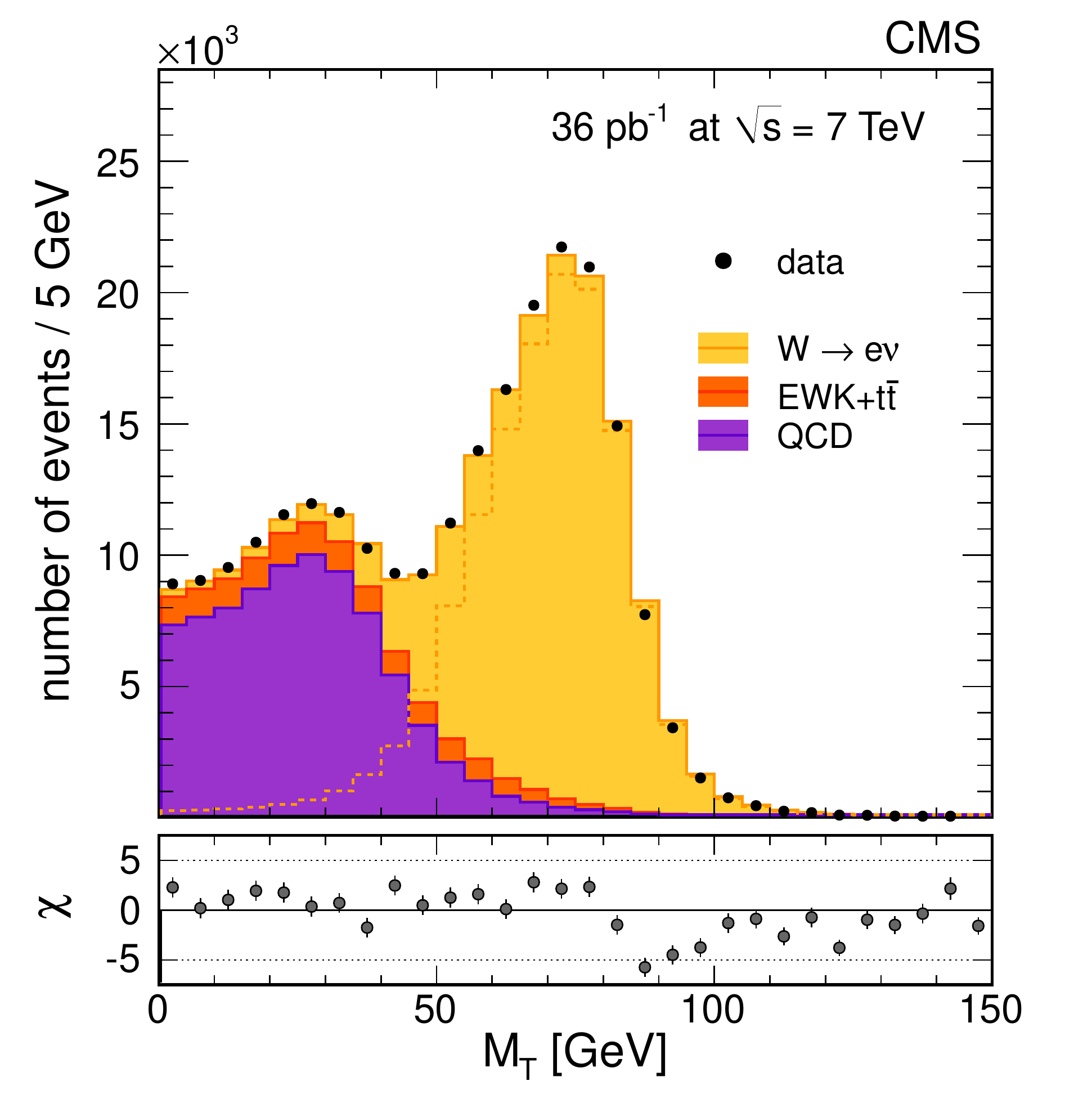}}
\resizebox{0.245\textwidth}{!}{\includegraphics{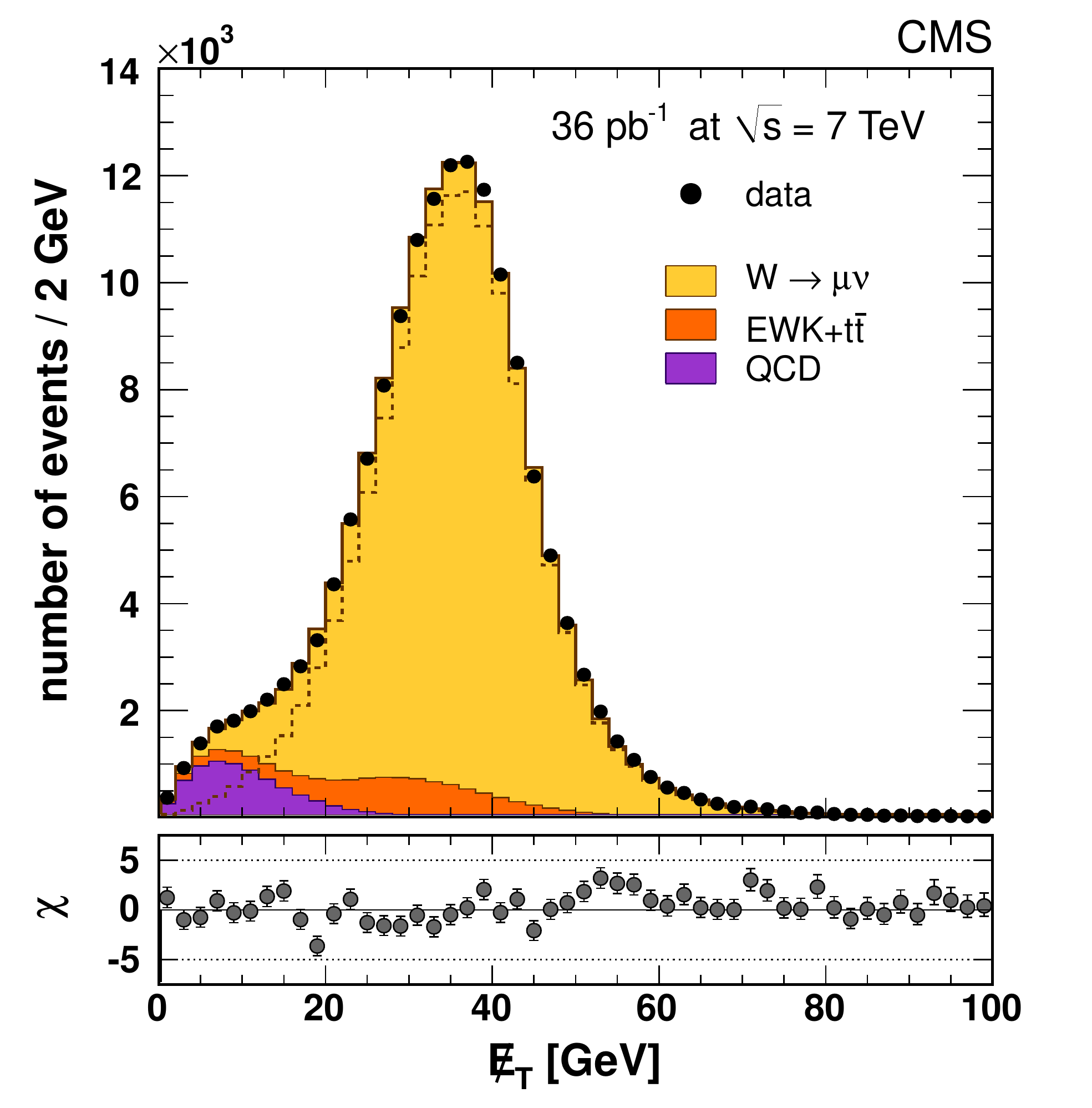}}
\resizebox{0.245\textwidth}{!}{\includegraphics{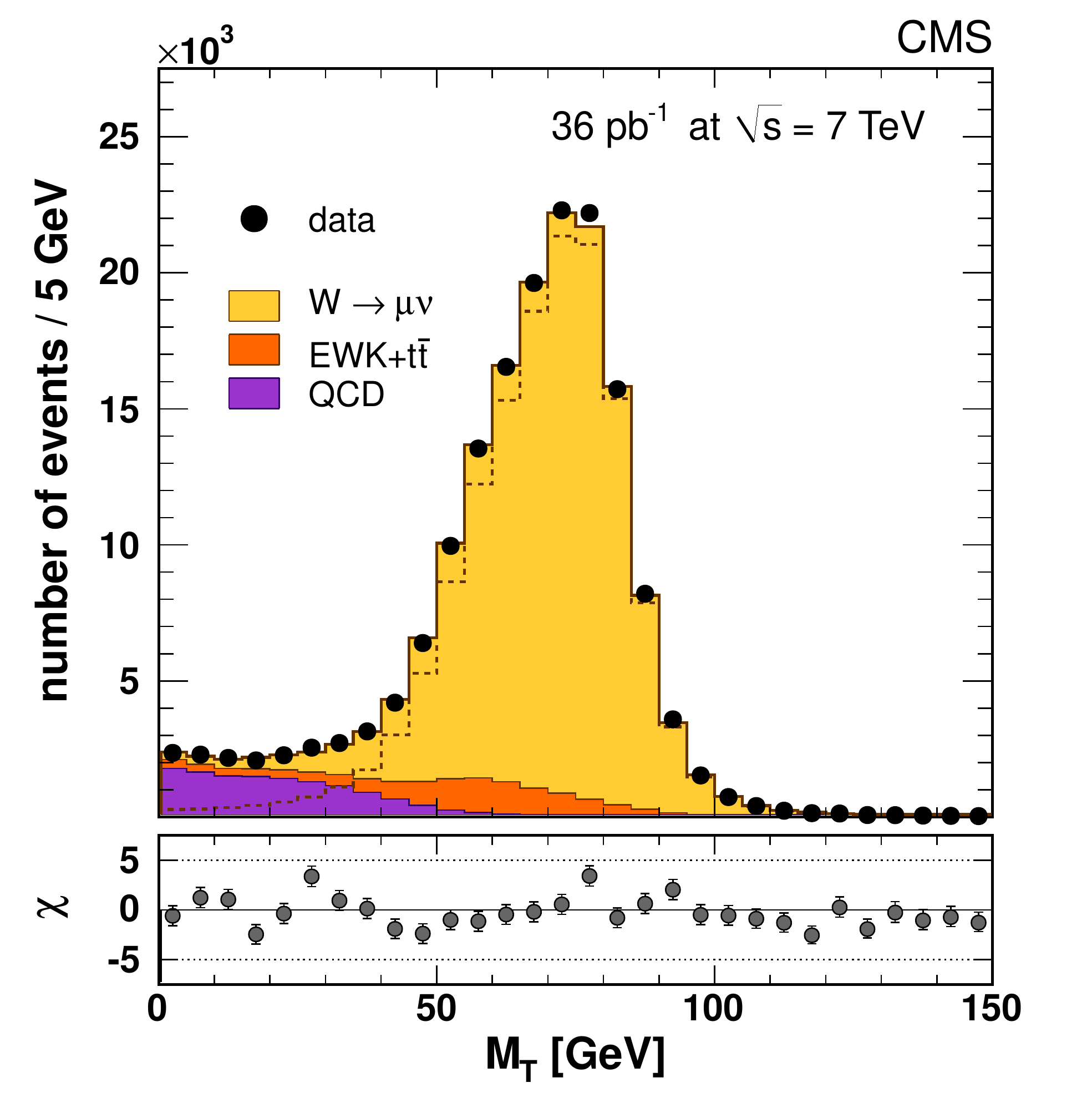}}
\caption{CMS \cite{CMS:2011aa}: The  \MET and $\mT$ distributions for the selected $W^+\rightarrow l^+\nu$ candidate events. The points with the 
error bars represent the data. Superimposed are the contributions obtained with the fit for QCD multi-jet background (violet, dark 
histogram), all other backgrounds (orange, medium histogram), and signal plus background (yellow, light histogram). The orange 
dashed line is the signal contribution.}
\label{fig:CMSMETW}
\end{figure*}

\section{Inclusive and Differential Cross-Section Measurements}
\label{sec:InclusiveAndDiffMeasurements}

The Standard Model predictions of the Drell-Yan processes $pp\rightarrow W^\pm+X$ and $pp\rightarrow Z/\gamma^* + X $ can be tested in a completely new energy regime at the LHC: the study of the Drell-Yan processes provides a unique opportunity to test perturbative QCD 
predictions and improve the knowledge of the proton's PDFs (Section \ref{sec:InclusiveCS}, \ref{sec:DifferentialCS}). 
The measurement of the transverse momentum distribution of vector bosons can be
used to test resummation techniques in addition to higher-order corrections in QCD
calculation. The corresponding measurements are presented in Section \ref{sec:TransverseMomentum}.
Moreover, the measurement of the forward-backward production asymmetry can constrain the 
vector and axial-vector couplings of the \Zboson boson to the quarks, where the latest 
results are presented in Section \ref{sec:AFB} and the measurement of \Wboson boson polarisation can 
test the electroweak properties of the underlying production mechanism, discussed in
Section \ref{sec:WPol}.
Finally, the precise understanding of the Drell-Yan processes is a key-element 
for the search of beyond the Standard Model signatures at the LHC. 
Very similar decay signatures of \Wboson and  \Zboson bosons are for example 
predicted by models of large extra dimensions \cite{Vacavant:2001sd}, 
additional  U(1) gauge groups \cite{Leike:1998wr} 
or quark-lepton compositeness models \cite{Espinosa:2010vn}. Hence, deviations from the predicted production properties
could open the window to new physics.

\subsection{\label{sec:InclusiveCS}Inclusive cross-section measurements}

The inclusive production cross-sections for the Drell-Yan processes for 
the \Wboson boson and $Z/\gamma^*$ exchange are known to NNLO precision in the strong 
coupling constant to a precision of $\approx 2\%$ (Section \ref{sec:CrossPred}). The 
dominating uncertainties are to due the limited knowledge of PDFs, while 
scale uncertainties play only a minor role. 

At tree level, \Zboson bosons are produced by the annihilation of quarks and 
antiquark pairs, i.e. $u \bar u$, $d \bar d$ and to some extent $s\bar s$. 
While the $u$ and $d$ quarks are mainly the valence quarks of the proton, their 
respective antiquarks are always sea-quarks in proton-proton collisions. The 
situation is different for the production of $W^\pm$ bosons, since their 
production mechanism depends on their charge. The dominant processes for $W^+$ 
and $W^-$ are $u \bar d \rightarrow W^+$ and $d \bar u \rightarrow W^-$, 
respectively. Since two $u$-valence quarks are available in the proton, but 
only one $d$-valence quark, more $W^+$ bosons are expected to be produced. The 
ratio of the $W^+$ and $W^-$ production therefore allows for a precise test of 
QCD predictions, as many theoretical and experimental uncertainties cancels in
their ratio (Table \ref{tab:WZRatioResults}).

As one of the first measurements performed at the LHC, the cross-sections times leptonic branching ratios
$\sigma_{W^\pm} \cdot BR(W\rightarrow l^\pm \nu)$ and $\sigma_{Z} \cdot BR(Z\rightarrow l^+l^-)$
of inclusive \Wboson and \Zboson production for electron and muon final states were published by both experiments.  These 
measurements, based on the 2010 data sample with \IntLumi$= 35\,\ipb$, are not limited by their statistical precision \cite{Aad:2011dm}, \cite{CMS:2011aa}, 
but by the knowledge of the integrated luminosity. Hence, the inclusive results, which are discussed in the following, are based solely on the 2010 data sample.

The measurement strategy is based on Equation \ref{EQN:CrossSectionExp}, which was discussed in detail
in Section \ref{sec:CrossMeasPhil}: The number of selected signal events is first corrected
for the expected background contribution and then for detector effects via a factor $C$ within a 
fiducial volume. The division by the integrated luminosity corresponding to the
analysed data sample results in the fiducial cross-section. This can be extrapolated in a second step
to the full inclusive cross-section via the acceptance factor $A$. The detector efficiency factors 
$C$ and the acceptance factors $A$ with their respective uncertainties are shown for 
both experiments and both decay channels in Table \ref{tab:ACAndResults}. The dominating experimental 
uncertainties are due to lepton scales and efficiencies. 

The combined results for the inclusive cross-sections for $W^\pm$ and
$Z/\gamma^*$ for both experiments are also shown in Table \ref{tab:ACAndResults}.
The dominating experimental uncertainties are due to the limited knowledge on the 
integrated luminosity. In fact, by using the theoretical predicted cross-section, the 
integrated luminosity of a data sample can be estimated.

\begin{table*}
\centering
\caption{Summary of the cross-section results of the inclusive \Wboson and \Zboson analyses of ATLAS and CMS based on the 2010 data sample. The combined cross-sections are given with their respective statistical, systematic, acceptance and luminosity uncertainty, respectively.}
\label{tab:ACAndResults}       
\begin{tabular*}{\textwidth}{@{\extracolsep{\fill}}l cc cc cc}
\hline
ATLAS					& $W^+\rightarrow e^+\nu$	& $W^+\rightarrow \mu^+\nu$	& $W^-\rightarrow e^-\nu$	& $W^-\rightarrow \mu^-\nu$	& $Z\rightarrow e^+ e^-$	& $Z\rightarrow \mu^+ \mu^-$ \\
\hline

Acceptance A				& 		$0.479\pm0.008$	&	$0.4595\pm0.008$ 		&	 $0.452\pm0.009$		&	$0.470\pm0.010$ 		&	$0.447\pm0.009$	&	$0.487\pm0.010$				\\
Correction C				& 		$0.693\pm0.012$	&	$0.796\pm0.016$		&	$0.706\pm0.014$		&	$0.779\pm0.015$		&	$0.618\pm0.016$	&	$0.782 ± 0.007$		\\
$\sigma_{incl.}$ [nb]			&		6.06				&	6.06					&	4.15					&	4.20					&	0.952			&	0.935				\\
stat. unc. [nb]				&		0.02				&	0.02					&	0.02					&	0.02					&	0.010			&	0.009				\\
sys. unc. [nb]				&		0.10				&	0.10					&	0.07					&	0.05					&	0.026			&	0.009				\\
lumi. unc. [nb]				&		0.21				&	0.21					&	0.14					&	0.17					&	0.032			&	0.032				\\
theo. unc. [nb]				&		0.10				&	0.10					&	0.09					&	0.07					&	0.019			&	0.019				\\
$\sigma_{incl.}$  (comb.)		&\multicolumn{2}{c}{$6.05\pm0.02\pm0.07\pm0.10\pm0.21$}	&\multicolumn{2}{c}{$4.16\pm0.01\pm0.06\pm0.08\pm0.14$}	&	\multicolumn{2}{c}{$0.937\pm0.006\pm0.009\pm0.016\pm0.032$}\\
\hline
CMS						& $W^+\rightarrow e^+\nu$	& $W^+\rightarrow \mu^+\nu$	& $W^-\rightarrow e^-\nu$	& $W^+\rightarrow \mu^-\nu$	& $Z\rightarrow e^+ e^-$	& $Z\rightarrow \mu^+ \mu^-$ \\
\hline
Acceptance A				& 0.5017 					&	0.4594 				&	0.4808				&	0.4471 				&	0.3876			&	0.3978				\\
Correction C				& $0.737\pm0.01$			&	$0.854\pm0.008$		&	$0.732\pm0.01$		&	$0.841\pm0.008$		&	$0.609\pm0.011$	&	$0.871\pm0.011$		\\
$\sigma_{incl.}$ [nb]			&		6.15				&	5.98					&	4.34					&	4.20					&	0.992			&	0.968				\\
stat. unc. [nb]				&		0.02				&	0.02					&	0.02					&	0.02					&	0.011			&	0.008				\\
sys. unc. [nb]				&		0.10				&	0.07					&	0.07					&	0.05					&	0.018			&	0.007				\\
theo. unc. [nb]				&		0.07				&	0.08					&	0.06					&	0.07					&	0.016			&	0.018				\\
lumi. unc. [nb]				&		0.25				&	0.25					&	0.17					&	0.17					&	0.040			&	0.049				\\
$\sigma_{incl.}$  (comb.)		&\multicolumn{2}{c}{$6.04\pm0.02\pm0.06\pm0.08\pm0.24$}	&\multicolumn{2}{c}{$4.26\pm0.01\pm0.04\pm0.07\pm0.17$}	&	\multicolumn{2}{c}{$0.974\pm0.007\pm0.007\pm0.018\pm0.039$}\\
\hline
\end{tabular*}
\end{table*}

The luminosity uncertainty on the
cross-section measurement cancel in cross-section ratios, as well as some of the 
experimental and theoretical uncertainties. Hence, also cross-section ratios have been
published for ATLAS and CMS. Special focus will be drawn to the 
$\frac{\sigma(W^+)+\sigma(W^-)}{\sigma(Z)}$ ratio, shown in Table \ref{tab:WZRatioResults}, where all correlated uncertainties have 
been taken into account. The NNLO prediction of the ratio is also given.

\begin{table}
\caption{Results of the production cross-section ratio $\sigma(W^+)/\sigma(W^-)$ and $\sigma(W^\pm)/\sigma(Z)$ from the ATLAS and CMS analyses, based on the 2010 data sample. The ATLAS measurement of $\sigma(W^\pm)/\sigma(Z)=10.89$ was extrapolated to the CMS mass-range definition of the Z-Boson. The expected theoretical value is also shown.}
\label{tab:WZRatioResults}       
\begin{tabular}{lccc}
\hline
							& ATLAS 		& CMS 		& Theory (NNLO)  \\
\hline
$\sigma(W^+)/\sigma(W^-)$		& 1.454		& 1.421		& 1.43		\\
Stat. Unc.						& 0.006		& 0.006		& -			\\
Sys. Unc.						& 0.012		& 0.014		& -			\\
Theo. Unc.					& 0.022		& 0.029		& 0.01		\\
\hline
\hline
$\sigma(W^\pm)/\sigma(Z)$		& 10.73		& 10.54		& 10.74		\\
Stat. Unc.						& 00.08		& 00.07		& -			\\
Sys. Unc.						& 00.11		& 00.08		& -			\\
Theo. Unc.					& 00.12		& 00.16		& 0.04		\\
\hline
\end{tabular}
\end{table}

A simple leading-order calculation for the expected cross-section 
ratio for $(W^+ + W^-)/Z$ highlights the dependence on the quark-distribution functions.
Ignoring heavy quark and the $\gamma^*$ contributions, as well as Cabibbo suppressed parts of 
the cross-section, leads to

\begin{eqnarray}
\sigma(W^+)+\sigma(W^-) &=& u_v(x) + \bar d_s(x) + d_v(x) + \bar u_s(x) \\
\nonumber
\sigma(Z) &=& (g_V(u)^2+g_A(u)^2)\cdot u_q(x)) \\
\nonumber
& & + (g_V(d)^2+g_A(d)^2)\cdot v_q(x))
\end{eqnarray}

%

with
\[
u_q(x)=(u_v(x)+\bar u_s(x)),\hspace{0.5cm} v_q(x)=(d_v(x)+\bar d_s(x))
\]
where $u_v(x)$ and $d_v(x)$ are the up- and down-valence quark distributions and 
$u_s(x)$ and $d_s(x)$ the respective sea-quark distributions. When assuming that 
the light sea and antiquark distributions are the same for a given $x$ and considering
that $(g_V(u)^2+g_A(u)^2)\approx (g_V(d)^2+g_A(d)^2)$, this reduces to 

\[\sigma(W^+)+\sigma(W^-) \sim {\sigma(Z)} \]
i.e. only a small dependence on PDFs is expected if the PDFs have been determined 
with the assumption of same light sea and antiquark distributions in the proton, i.e. $\bar q(x)=q(x)$ for $q=u,d,s$. As
this symmetry assumption is inherent for the main PDF fits,\footnote{Small deviations 
are included to account for some light sea quark asymmetry near Bjorken $x\approx 0.1$.} no large 
difference in the theoretical predictions based on different PDF sets are observed. The good
agreement between the measurements (Table \ref{tab:WZRatioResults}) and the predictions are a remarkable 
confirmation of perturbative QCD calculations, but also support strongly 
the assumption of a flavour independent light quark sea at high scales, where x is small compared to 0.1, i.e. $\bar u \approx \bar d \approx \bar c \approx \bar d$ at $Q^2 \approx m_Z^2$. 

The above argument does not hold true for charge-dependent 
cross-section ratios, such as $\sigma(W^+) / \sigma(Z)$, $\sigma(W^-) / \sigma(Z)$
and $\sigma(W^+) / \sigma(W^-)$. They inhibit a significantly larger dependency
on differences in the u- and d-quark distribution functions. However, the largest 
constraints on PDFs do not come from the inclusive cross-section measurements but from 
differential measurements, which are also discussed in the following section.

In summary, the inclusive cross-section measurements were one of the first published measurements at the LHC, that confirmed NNLO perturbative QCD predictions in a new energy regime. By now, also inclusive measurements of the \Wboson and \Zboson cross-section at a center-of-mass energy of $8\,\TeV$ are available \cite{Chatrchyan:2014mua} which are also in very good agreement with the theoretical predictions.

\subsection{\label{sec:DifferentialCS} Differential \Wboson boson and $Z/\gamma^*$ measurements}

In addition to the inclusive production cross-section, the large available statistics 
at the LHC also allows for measurements of differential production cross-sections with high precision.  
Of special importance here is the measurement of the rapidity distribution $y_V$\footnote{The rapidity of a 
particle is defined as $y = \frac{1}{2} \frac{E+p_z}{E-p_z}$, where $E$ is the 
particle's energy and $p_z$ is the longitudinal momentum w.r.t to proton 
direction.} of the vector boson, as it allows for a direct determination of the 
momentum fractions $x_{1/2}$ of the interaction partons, via

\[x_{1/2} = \frac{m_V}{\sqrt{s}} e^{\pm y_V},\]
where $m_V$ is the mass of the vector boson. A center-of-mass energy 
of $\sqrt{s}=7\,\TeV$ allows therefore to reach x-range from $\approx 0.001$ to 
$\approx 0.1$ for the study of \Wboson and \Zboson bosons. 
The boson rapidity distributions are calculated up to NNLO in QCD theory \cite{Gavin:2010az} and 
are dominated by PDF uncertainties. Hence the measurement of the differential 
production cross-section of gauge bosons versus their rapidity 
distribution will provide additional constraints on the proton's PDFs. 
The results of deep inelastic scattering experiments provide constraints on 
the sea quark and gluon distributions at small and medium $x$ values, while the studies of \Wboson and \Zboson production at the Tevatron provided important information on the valence quark distributions. The additional information by measurements at the LHC on the valence quark distribution is therefore expected to be marginal. However, the LHC measurements have a significant impact on the strange-quark PDFs, as well as on the ratio of $u/d$-quark distributions as discussed below. 


\begin{figure*}
\resizebox{0.5\textwidth}{!}{\includegraphics{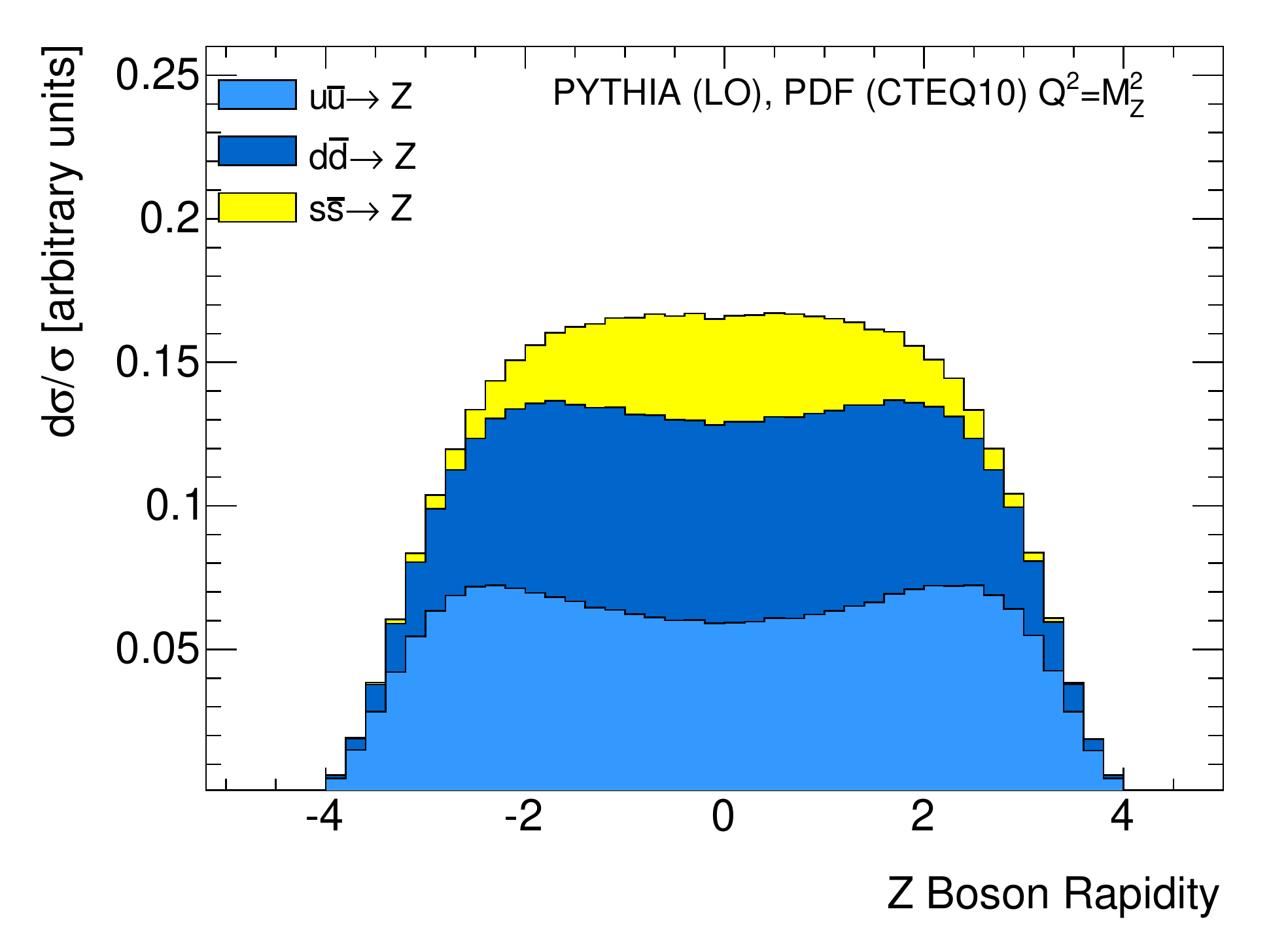}}
\resizebox{0.5\textwidth}{!}{\includegraphics{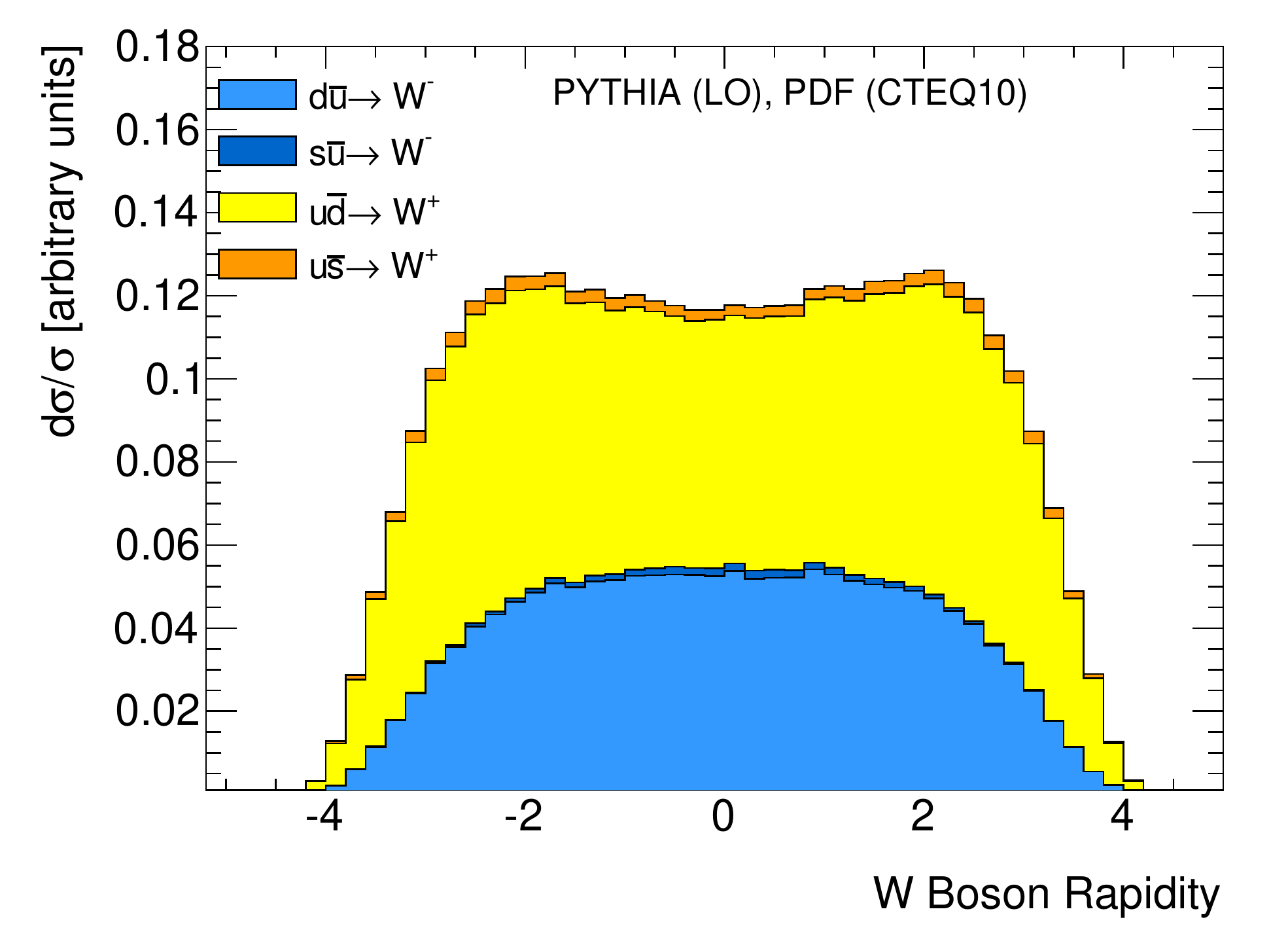}}
\caption{Rapidity Distribution for the leading-order production of \Zboson (left) and \Wboson bosons (right) in $7\,\TeV$ pp collisions. The relative contribution of the different production channels is also shown.}
\label{fig:PDFDecompositionZW}
\end{figure*}

The rapidity distribution of the $Z/\gamma^*\rightarrow l^+l^-$ process can be 
directly inferred from data, since both four-momenta of its decay-leptons can 
be precisely measurement. Hence it will give new information on the $u\bar u$, 
$d\bar d$ and $s\bar s$ PDFs. While the information on $u$ and $d$ quark distribution is already largely determined by previous experiments, in particular the uncertainty on the strange quark content can be improved.
Figure \ref{fig:PDFDecompositionZW} shows the contribution of the 
different quark/antiquark annihilation processes for different $y_Z$ values. 
While the $u\bar u$ annihilation is dominating in the central region, the $d 
\bar d$ annihilation process is expected to have a larger influence for larger 
rapidity values. A precise measurement in the central rapidity region can also 
give additional constraints on the $s \bar s$ PDFs. In addition, the study of 
$y_{Z/\gamma^*}$ for different mass intervals can probe different $x$-regimes, 
e.g. low-mass Drell-Yan events will probe in general small values of $x$ than 
high-mass Drell-Yan events. Such studies can be used to improve the knowledge on the ratio of $u$- and $d$-quark distributions.

ATLAS published a combined differential $d\sigma / d|y_Z|$ cross-section in the fiducial region
\footnote{defined by a  cut on the invariant mass of the di-lepton system of $66\,\GeV< m_{ll}<116\,\GeV$ 
and a minimal requirement of $\pT>20\,\GeV$ and $\eta<2.4$ for both decay leptons.} for 
the electron and muon decay channel of $Z/\gamma^* \rightarrow l^+l^-$ based 
on \IntLumi$\approx 35 \ipb$ \cite{Aad:2011dm}.  Figure \ref{fig:ATLASZRapResults} shows the 
results including NNLO theory predictions with various PDF sets. The largest 
rapidity reached is $y_Z=3.5$ which is due to the inclusion of forward electrons 
in this study. In addition, ATLAS published a differential cross-section of the Drell-Yan process
in the electron decay channel versus the invariant mass of the di-electron pairs, based on the full
2011 data sample \cite{Aad:2013iua}. The comparison of data and NNLO predictions with various PDF sets is shown
in Figure \ref{fig:ATLASZHighMassDrell}.

\begin{figure}
\resizebox{0.5\textwidth}{!}{\includegraphics{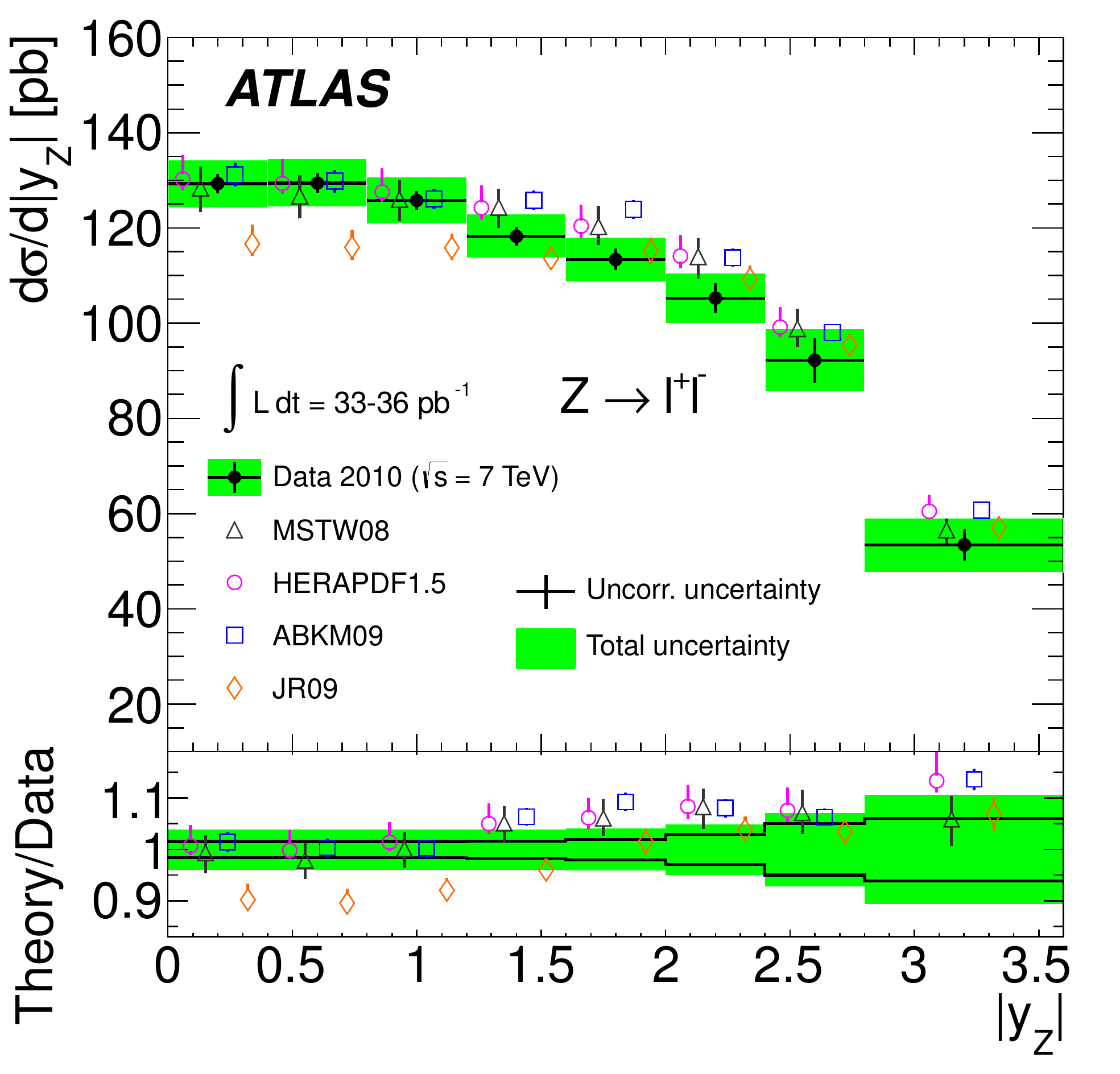}}
\caption{ATLAS \cite{Aad:2011dm}: The combined $d\sigma/d|yZ|$ cross-section measurement for $Z/\gamma^* \rightarrow l^+l^-$ compared to NNLO theory predictions using various PDF sets. The kinematic requirements are $66\,\GeV< m_{ll}<116\,\GeV$ and $\pT^{l}>20\,\GeV$. The ratio of theoretical predictions to data is also shown. Theoretical points are displaced for clarity within each bin.}
\label{fig:ATLASZRapResults}
\end{figure}

\begin{figure}
\resizebox{0.5\textwidth}{!}{\includegraphics{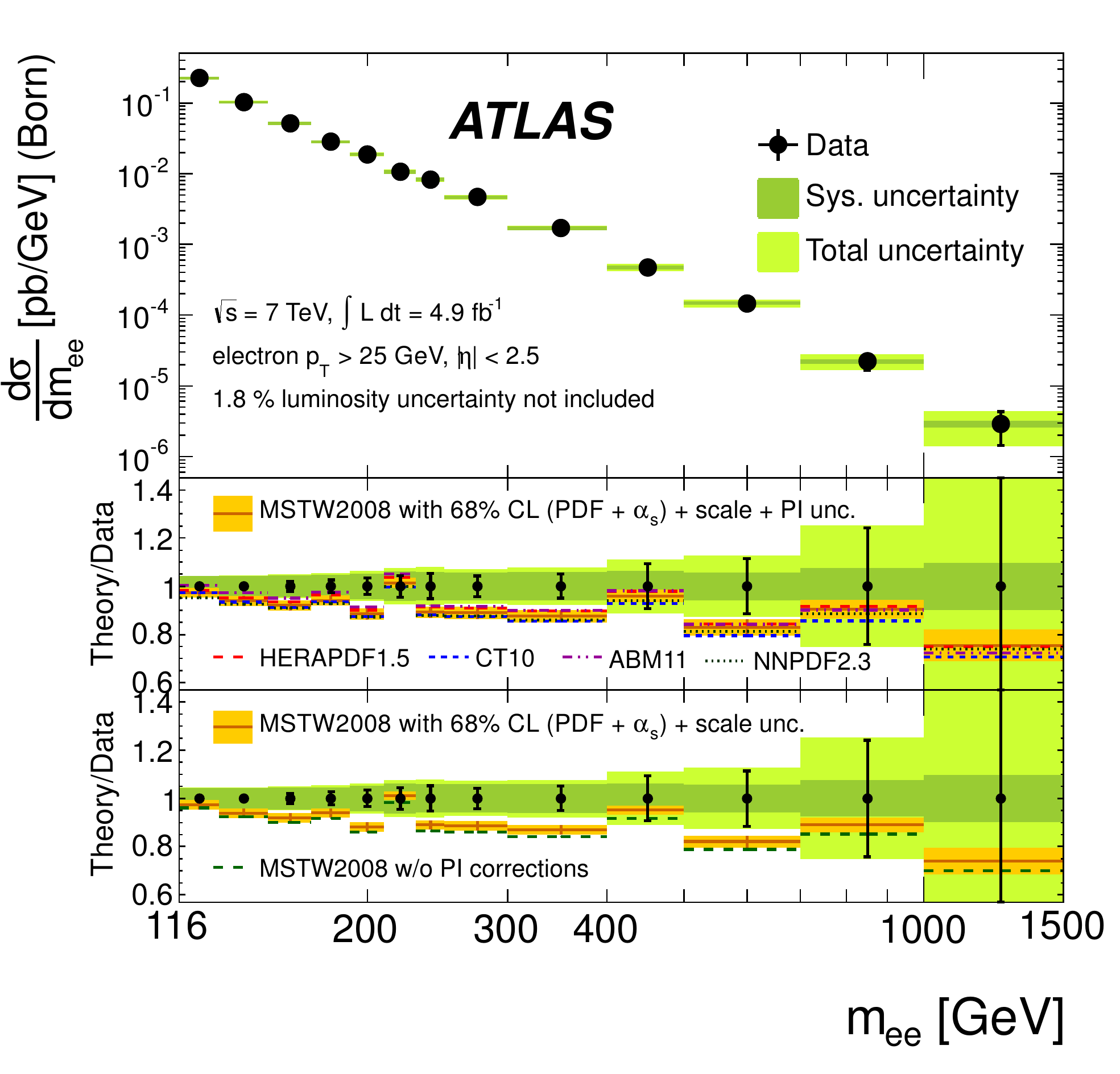}}
\caption{ATLAS \cite{Aad:2013iua}: Measured differential cross-section at the Born level within the fiducial region (electron $\pT > 25\,\GeV$ and $|\eta| < 2.5$) with statistical, systematic, and combined statistical and systematic (total) uncertainties, excluding the 1.8\% uncertainty on the luminosity. The measurement is compared to \FEWZ~ 3.1 calculations at NNLO QCD with NLO electroweak corrections using the $G_mu$ electroweak parameter scheme. }
\label{fig:ATLASZHighMassDrell}
\end{figure}

The latest CMS publication on $Z\rightarrow l^+ l^-$ \cite{Chatrchyan:2013tia} is also
based on an integrated luminosity of \IntLumi$\approx 4.5 \ifb$ and  \IntLumi$\approx 4.8 \ifb$ for the muon and electron channels, respectively. The 
measurement is performed in a double differential way over the mass range of 
$20\,\GeV$ to $1500\,\GeV$ and an absolute di-muon rapidity from $0 < |\eta| < 2.4$. 
The resulting rapidity distributions for three different mass regions 
are shown in Figure \ref{fig:CMSZRapResults}, together with the NNLO prediction for various 
PDF sets. The differential cross-sections have been extrapolated to the full phase space
and normalised to the \Zboson peak cross-section, which is defined in a mass region of $60-120\,\GeV$.
Hence many systematic uncertainties cancel in the ratio. The dominating remaining uncertainties 
are due to the efficiency corrections in the muon channel and energy scale 
uncertainties in the electron channel.

\begin{figure*}
\resizebox{0.3333\textwidth}{!}{\includegraphics{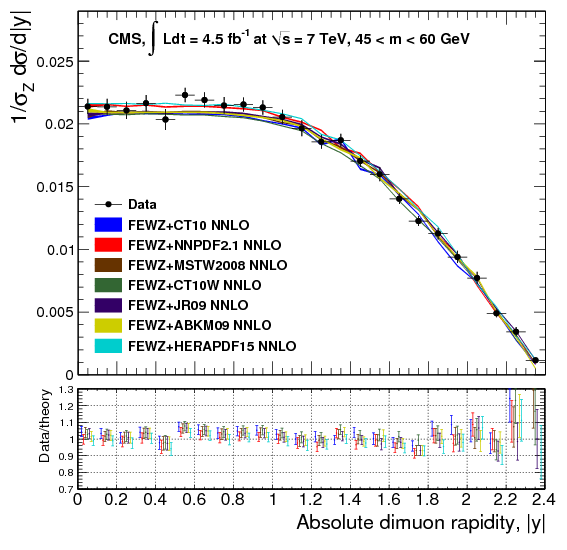}}
\resizebox{0.3333\textwidth}{!}{\includegraphics{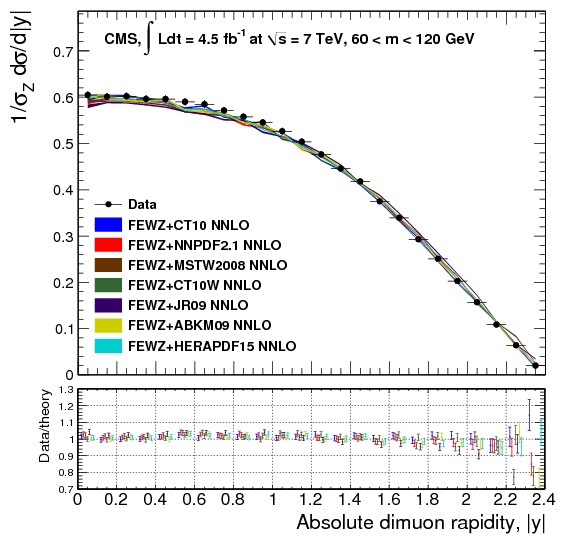}}
\resizebox{0.3333\textwidth}{!}{\includegraphics{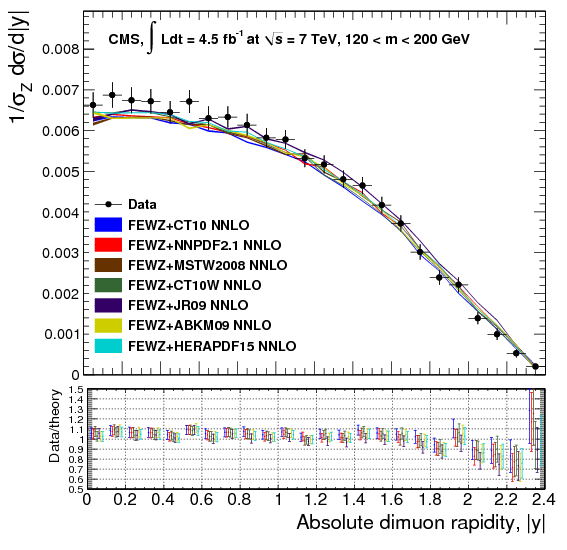}}
\caption{CMS \cite{Chatrchyan:2013tia}: Comparison with theory expectations with various NNLO PDF sets: CT10, HERA, NNPDF2.1, MSTW08, CT10W, JR09, ABKM. The error bands in the theory predictions indicates the statistical calculation error only. The bottom plots show the ratio of data to theory expectation. The error bar is the quadrature sum of experimental uncertainty on the data and statistical calculation error on theory expectation.}
\label{fig:CMSZRapResults}
\end{figure*}

Especially in the low-mass region, sizeable differences between the different 
PDF sets can be seen. The uncertainties of the available data are small enough 
to provide sufficient sensitivity to allow for an improvement over the existing 
PDF sets. The uncertainty on the $u/d$ ratio can be improved by more than 20\%. 

The $W^\pm$ boson rapidity distribution is sensitive to the $u\bar d$- and 
$d\bar u$-quark distribution. Their respective contribution is shown in 
Figure \ref{fig:PDFDecompositionZW}. However, its rapidity distribution in the leptonic decay channel 
is not directly accessible as the longitudinal momentum of the neutrino not 
measured. Therefore only an indirect measurement is possible, which is achieved 
by the measurement of the pseudorapidity of the charged decay leptons $\eta _l$ 
which are correlated to $y_W$. The production and decay of \Wboson boson is 
described by the V-A nature of the weak interaction. It is therefore expected 
that the spin of the \Wboson boson is aligned with the direction of the antiquark 
and the charged decay lepton is preferentially emitted opposite to the boost of 
the decaying boson. The corresponding experimental quantity is the lepton charge 
asymmetry

\begin{equation}
A(\eta) = \frac{d\sigma(W^+\rightarrow l^+\nu)/d\eta - d\sigma(W^-\rightarrow 
l^-\nu)/d\eta}{d\sigma(W^+\rightarrow l^+\nu)/d\eta+d\sigma(W^-\rightarrow 
l^-\nu)/d\eta}\, ,
\end{equation}
where $l$ denotes the lepton and $d\sigma/d\eta$ the differential 
cross-section for charged leptons from the \Wboson events. The definition of $A(\eta)$ has the advantage that 
several systematic uncertainties cancel in its ratio and can constrain 
the $u/d$-quark ratio and the corresponding sea-quark 
densities. Clearly, also the measurement of separate differential 
cross-sections $d\sigma(W^\pm\rightarrow l^\pm\nu)/d\eta$ provides the same 
information when the correlation between the systematic uncertainties is known.

CMS published results on the lepton charge asymmetry in the electron and muon 
decay channels within a fiducial phase space defined by a $\pT>35\,\GeV$ requirement 
for the charged decay leptons.  Since the study of $W\rightarrow e\nu$ \cite{Chatrchyan:2012xt} is based on only 
$\IntLumi = 0.84\ifb$ and the $W\rightarrow \mu \nu$ analyses \cite{Chatrchyan:2012xtmu} uses 
the full available data sample at $\sqrt{s}=7\,\TeV$, we discuss here only the 
latter. The $A(\eta)$ measurement after all corrections is shown in Figure 
\ref{fig:CMSWCharge} for a minimal muon requirement of $\pT>35\,\GeV$. The dominating 
systematic uncertainties are due to efficiencies and scale uncertainties, as 
well as uncertainties on the QCD multi-jet background. Statistical uncertainties are 
small compared to the systematic uncertainties, which range from $0.2\%$ in the 
central region to $0.4\%$ in the forward region. The correlations between 
different $\eta$-bins are small. The results are compared to NLO predictions 
for several PDF sets. 

\begin{figure*}
\resizebox{0.490\textwidth}{!}{\includegraphics{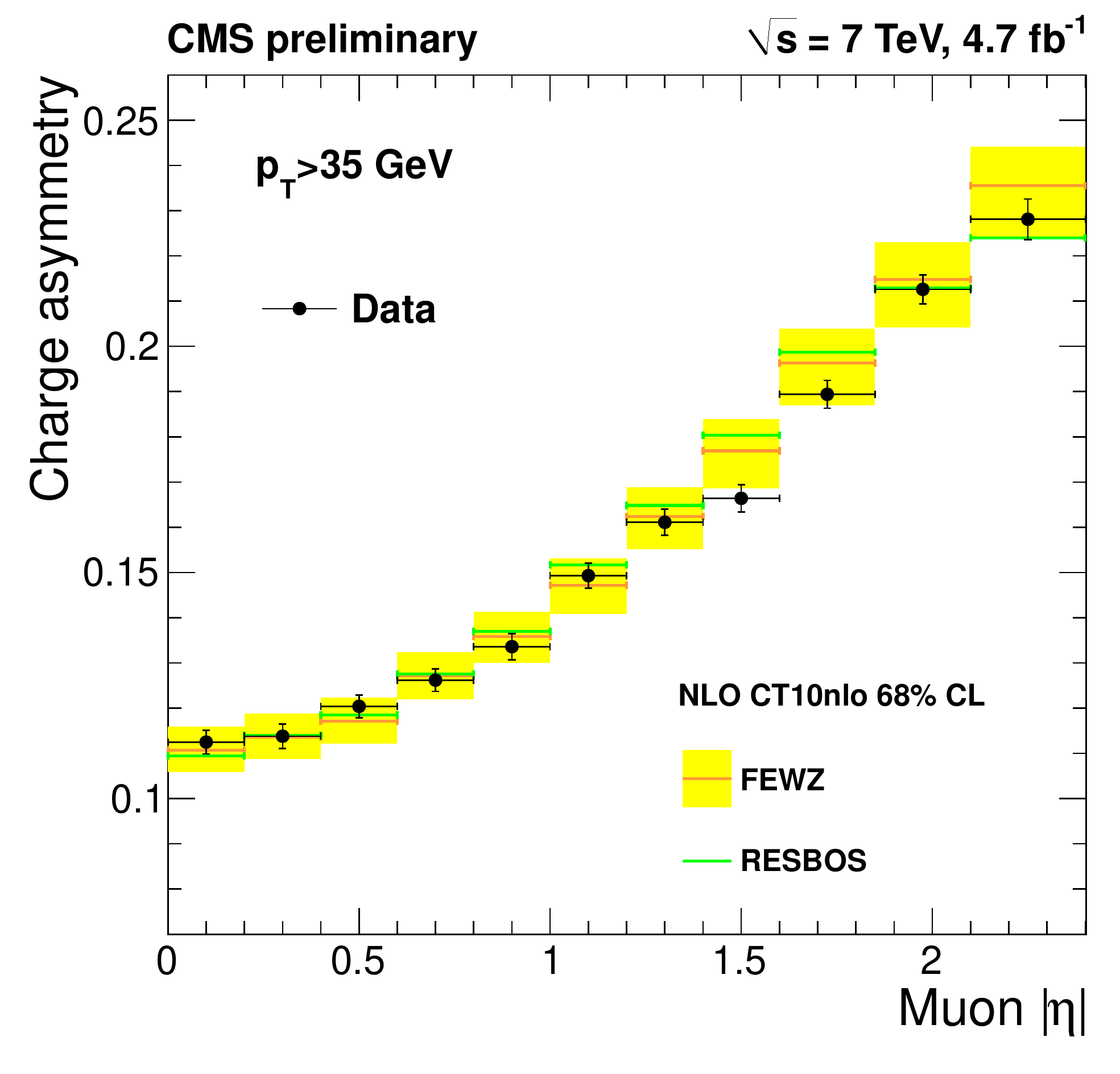}}
\resizebox{0.490\textwidth}{!}{\includegraphics{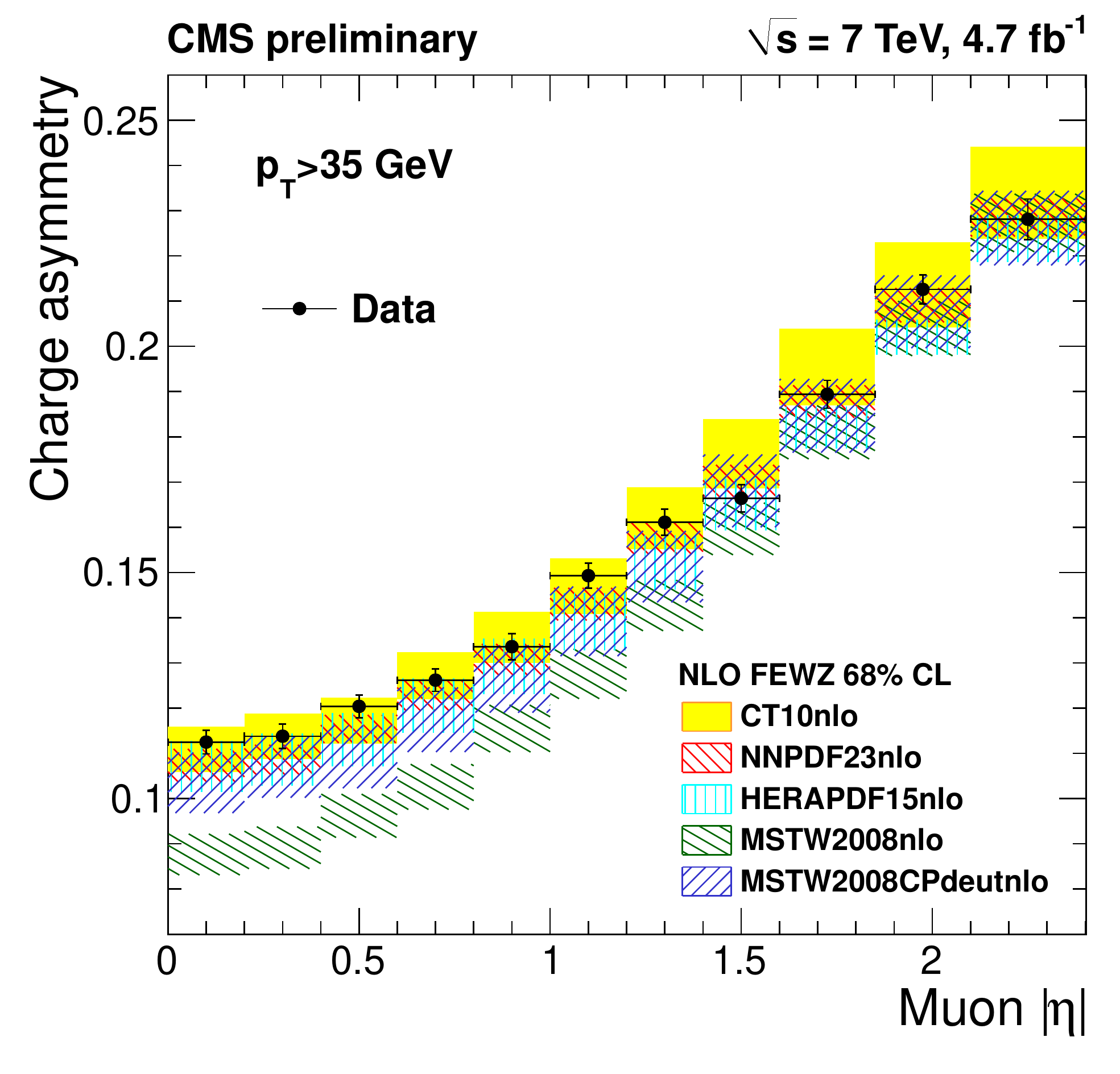}}
\caption{CMS \cite{Chatrchyan:2012xtmu}: Comparison of the measured muon charge asymmetry to theoretical predictions based on \FEWZ~ 3.0 and \ResBos~ calculations, for muons with $\pT > 35 \GeV$ (left). The CT10 NLO PDF is used in both predictions. A comparison of the measured muon charge asymmetries to predictions with CT10, NNPDF2.3, HERAPDF1.5, MSTW2008, and MSTW2008CPdeut NLO PDF models, for muons with $\pT > 35 \GeV$, is shown on the right.}
\label{fig:CMSWCharge}
\end{figure*}

ATLAS has published similar results for the full 2010 data sample in both leptonic 
decay channels within a fiducial phase space, defined by $\pT^{lep}>20\,\GeV$, 
$\pT^{\nu}>25\,\GeV$ and $\mT>40\,\GeV$. In addition to the lepton-charge asymmetry, which is shown 
in Figure \ref{fig:ATLASWCharge}, also the individual lepton charge distributions for $W^+$ and $W^-$ have been derived.
Similar to the CMS results, statistical uncertainties are 
negligible compared to the systematic uncertainties.

\begin{figure*}
\resizebox{0.333\textwidth}{!}{\includegraphics{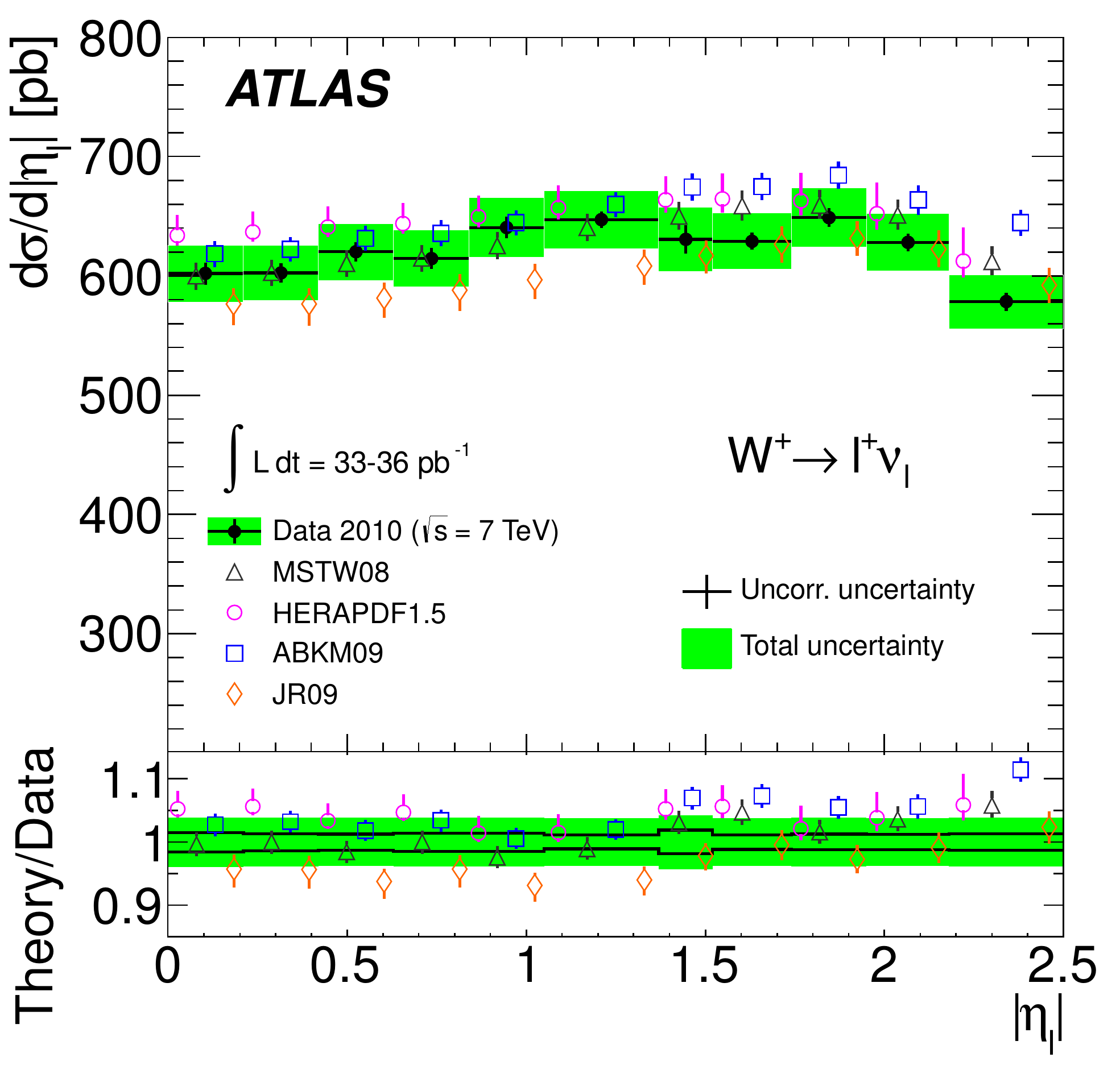}} 
\resizebox{0.333\textwidth}{!}{\includegraphics{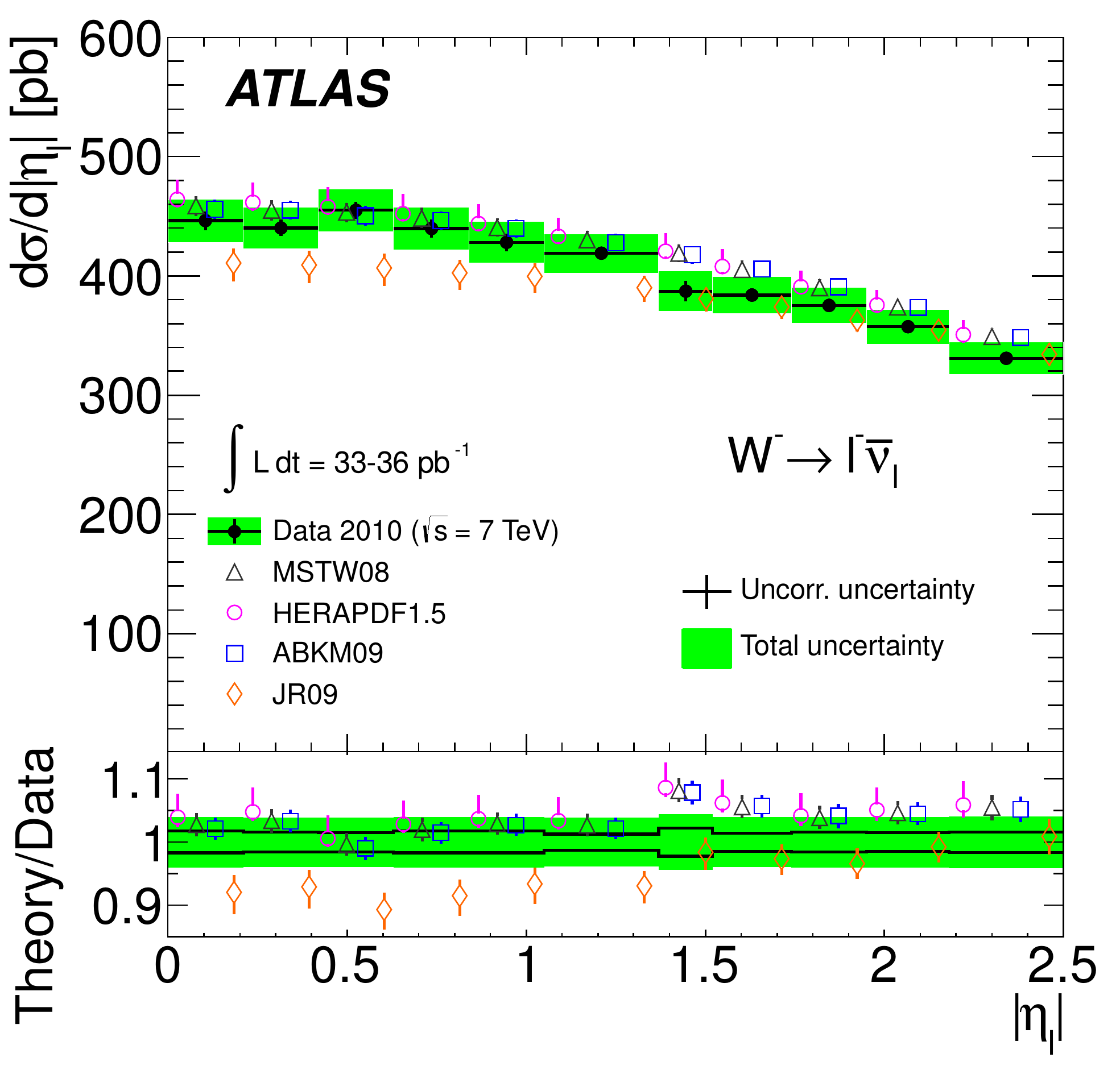}} 
\resizebox{0.333\textwidth}{!}{\includegraphics{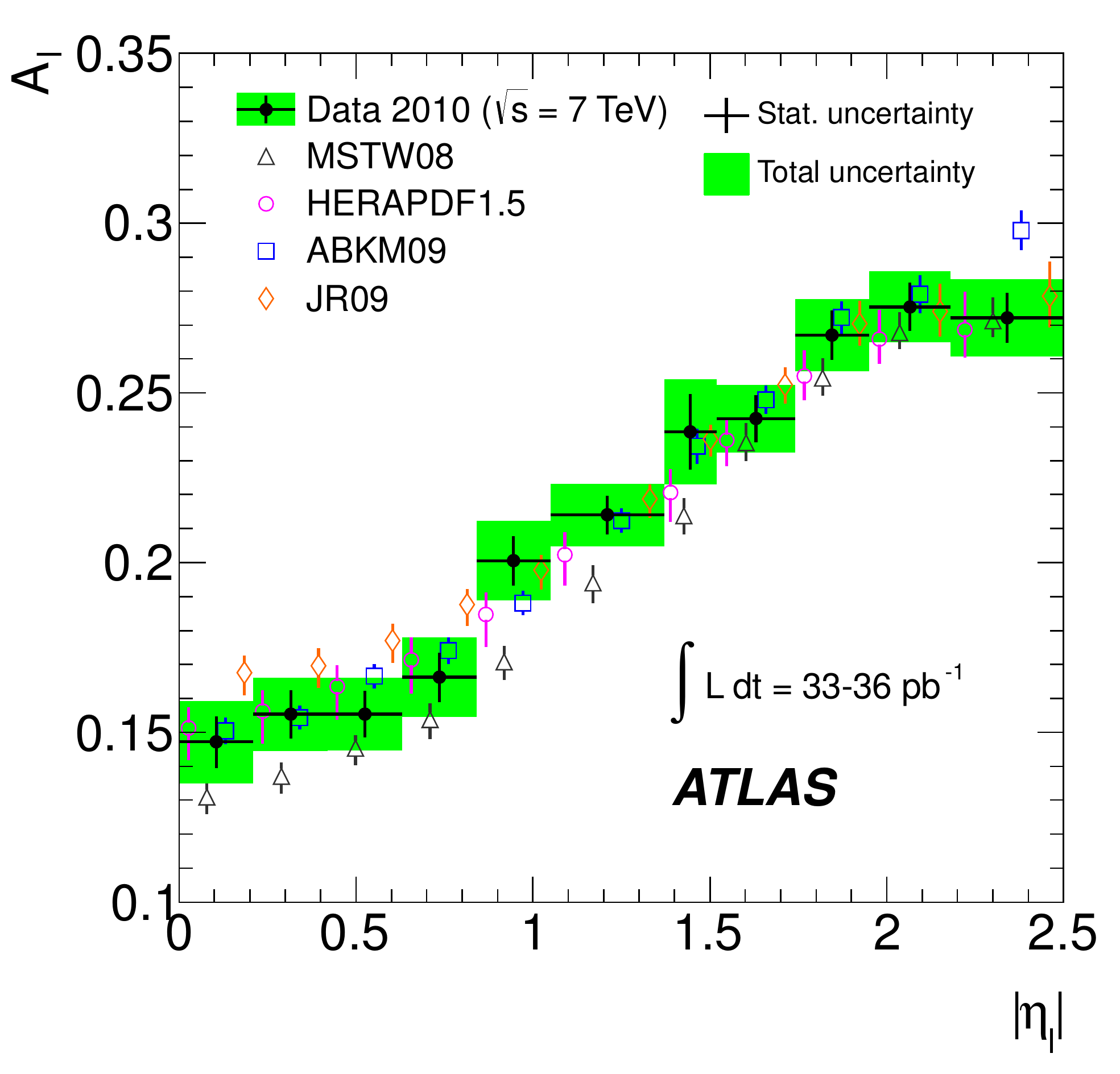}} 
\caption{ATLAS  \cite{Aad:2011dm}: Differential $d\sigma/d|\eta^+$ (left) and $d\sigma/d|\eta^-$ (middle) cross-section measurements within the fiducial volume for $W\rightarrow l\nu $. Measured \Wboson charge asymmetry as a function of lepton pseudorapidity $|\eta|$ is shown on the right. All results are compared to the NNLO theory predictions using various PDF sets.}
\label{fig:ATLASWCharge}
\end{figure*}

While most PDF sets show good agreement with data, the MSTW2008 
PDF parametrisation has a poor agreement  
especially in the region of small rapidities. This is due to a problem in $d$-valence distribution 
which was fixed in the MSTW2008CPdeutnlo set, which is also shown.
Since the uncertainties of the 
measured $A(\eta)$ values are smaller by a factor of 2-3 compared to the 
predicted uncertainties of the studied PDF sets, an improvement of future 
PDF sets is expected. Some preliminary results which make use if the currently published LHC data, can be found for example in \cite{Ball:2012wy}.

Figure \ref{ATLASCMSComparisonZ} and \ref{ATLASCMSComparisonW} show a comparison of the ATLAS and CMS results for the
$Z/\gamma^*$ and $A_W(\eta)$ distributions, respectively. We extrapolate the results 
to a common fiducial volume, defined by $60\,\GeV<m_{ll}<120\,\GeV$
for the $Z/\gamma^*$ process and by $\pT >35\,\GeV$ for the $W^\pm$ boson.
The extrapolation was performed with the \PowhegBox~generator and no additional 
systematic uncertainties have been added to the shown values. Both experiments show 
consistent results. 

\begin{figure*}
\begin{minipage}{0.49\textwidth}
\resizebox{1.0\textwidth}{!}{\includegraphics{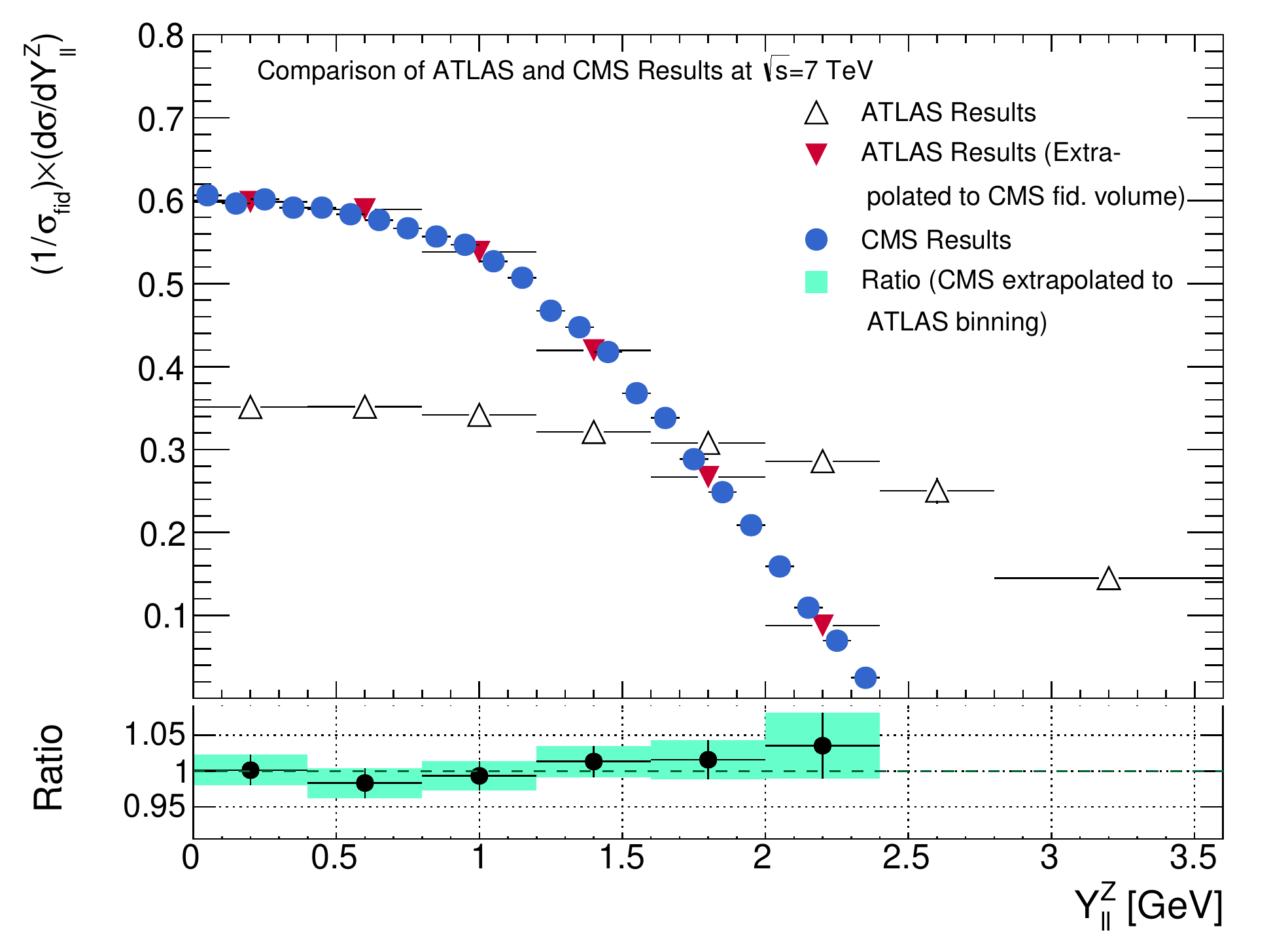}}
\caption{\label{ATLASCMSComparisonZ} Comparison of the ATLAS and CMS: Measurements of the normalised differential cross-section of the Drell-Yan process as a function of the rapidity of the di-lepton system. The ATLAS results have been extrapolated to the CMS fiducial volume, as it is significantly reduced in $Y_{ll}$. For completeness, also the full normalised differential cross-section of the ATLAS experiment is shown.}
\end{minipage}
\hspace{0.5cm}
\begin{minipage}{0.49\textwidth}
\resizebox{1.0\textwidth}{!}{\includegraphics{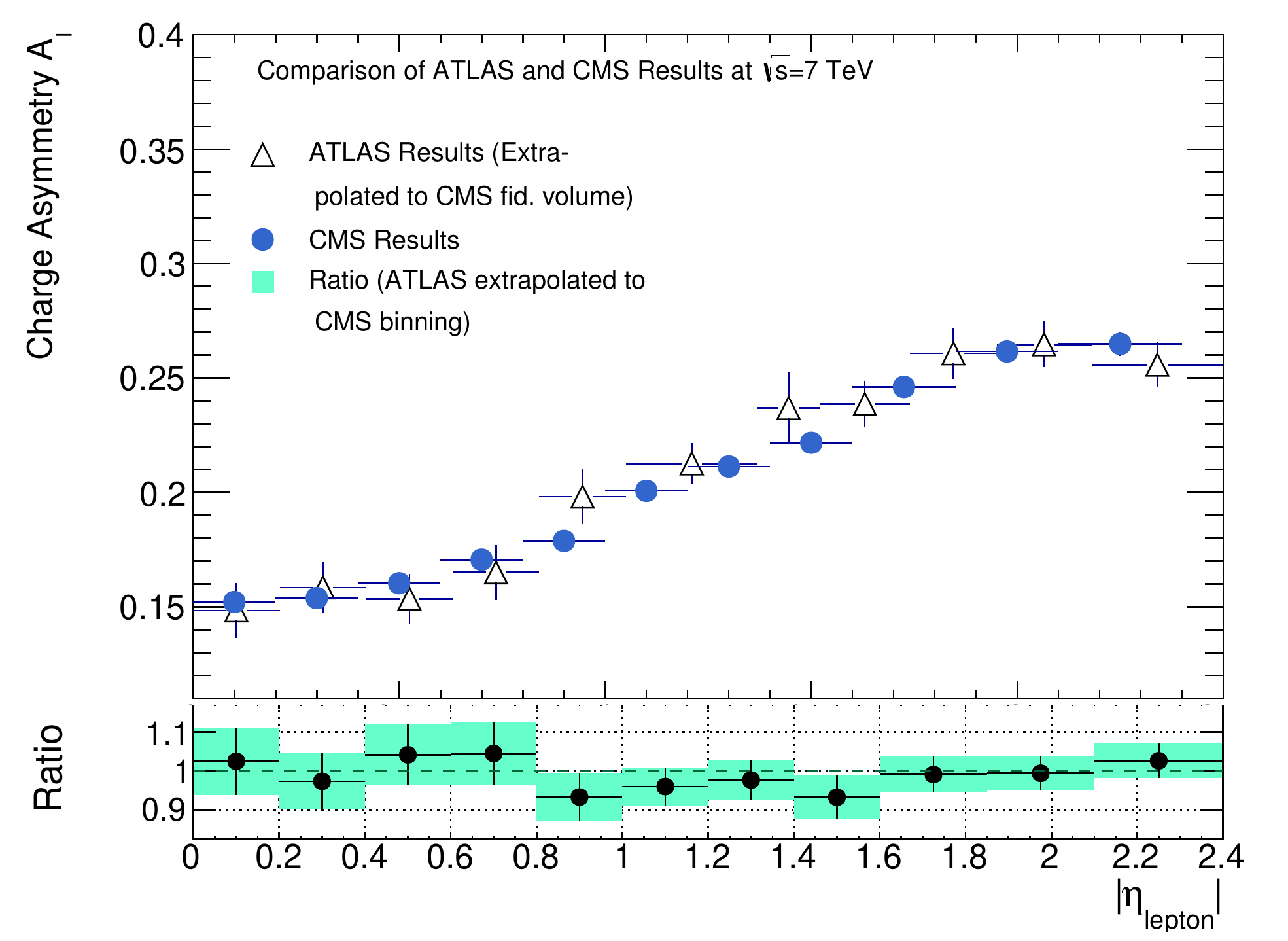}}
\caption{\label{ATLASCMSComparisonW}Comparison of the ATLAS and CMS: Measurements of the \Wboson charge asymmetry, where the ATLAS results have been extrapolated to the CMS fiducial volume.\vspace{1.1cm}}
\end{minipage}
\end{figure*}

ATLAS has studied the impact of their data on the proton PDFs using the HERAFitter 
framework \cite{HeraFitter:2012dc}. Here, especially the $y_Z$ measurement has a large impact on the constraints of the strange-quark PDFs. 
Even with the limited data sample of 2010, the hypothesis of a symmetric composition of the light sea-quarks at 
low $x$ \cite{Aad:2011dm} is supported. Specifically, the ratio of the strange sea-quark content to 
the down sea-quark content at $x=0.023$ was found to be $1^{+0.25}_{-0.28}$ at $Q^2=1.9\,\GeV^2$. This is a remarkable results 
and was confirmed in an improved analyses of the 2011 data sample \cite{Aad:2012sb}. So far it has been assumed in most PDF fitting
approaches that $s = \bar s = \frac{\bar u}{2} = \frac{\bar d}{2}$ due to the mass difference of the quarks at the starting scale, i.e. before the QCD evolution starts. At higher values of $Q^2$, the gluon splitting processes become dominant and lead to a symmetric distribution of sea-quarks. This new results suggests even an equal $\bar u-$, $\bar d$- and $s$-quark content at low $Q^2$ values \cite{Aad:2011dm}. A visualisation of the impacts 
on the strange quark distributions is shown in Figure  \ref{fig:ATLASPDF1}. The inclusion of electrons in the forward-region of the ATLAS detector extends the 
available $y_Z$ regime and should therefore improve the information on 
the valence quark distributions. However, the current experimental uncertainties 
are too large for a significant effect. 

The published results on the lepton charge 
asymmetry for the \Wboson boson production impacts moderately the valence quark 
distribution functions compared to the existing data that include Tevatron results. However, when only using measurements from HERA and LHC, an improvement on the valance quark distributions of more than 30\% for the full x-range can be observed, compared to HERA measurements alone. 

\begin{figure*}
\resizebox{0.495\textwidth}{!}{\includegraphics{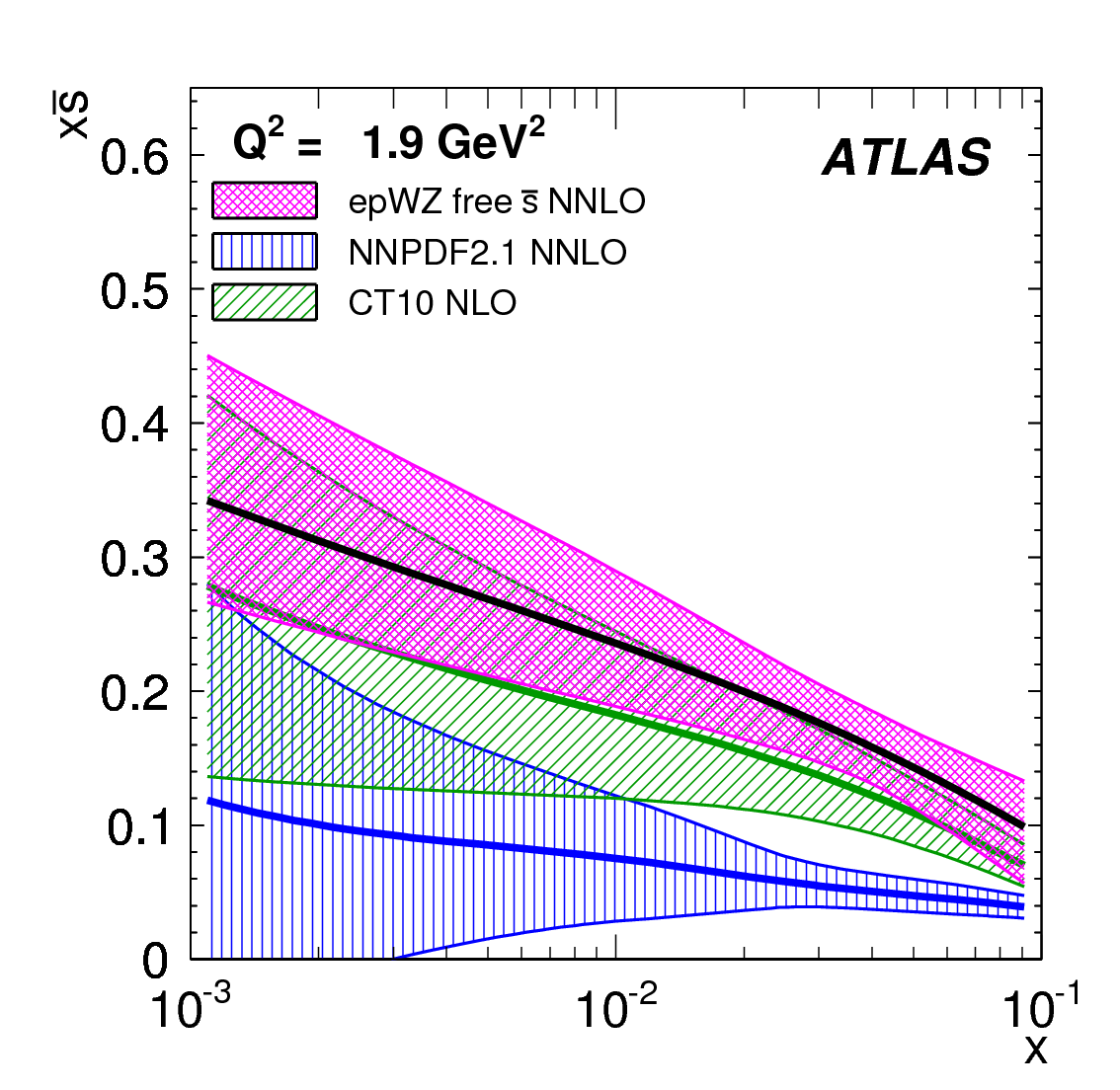}}
\resizebox{0.495\textwidth}{!}{\includegraphics{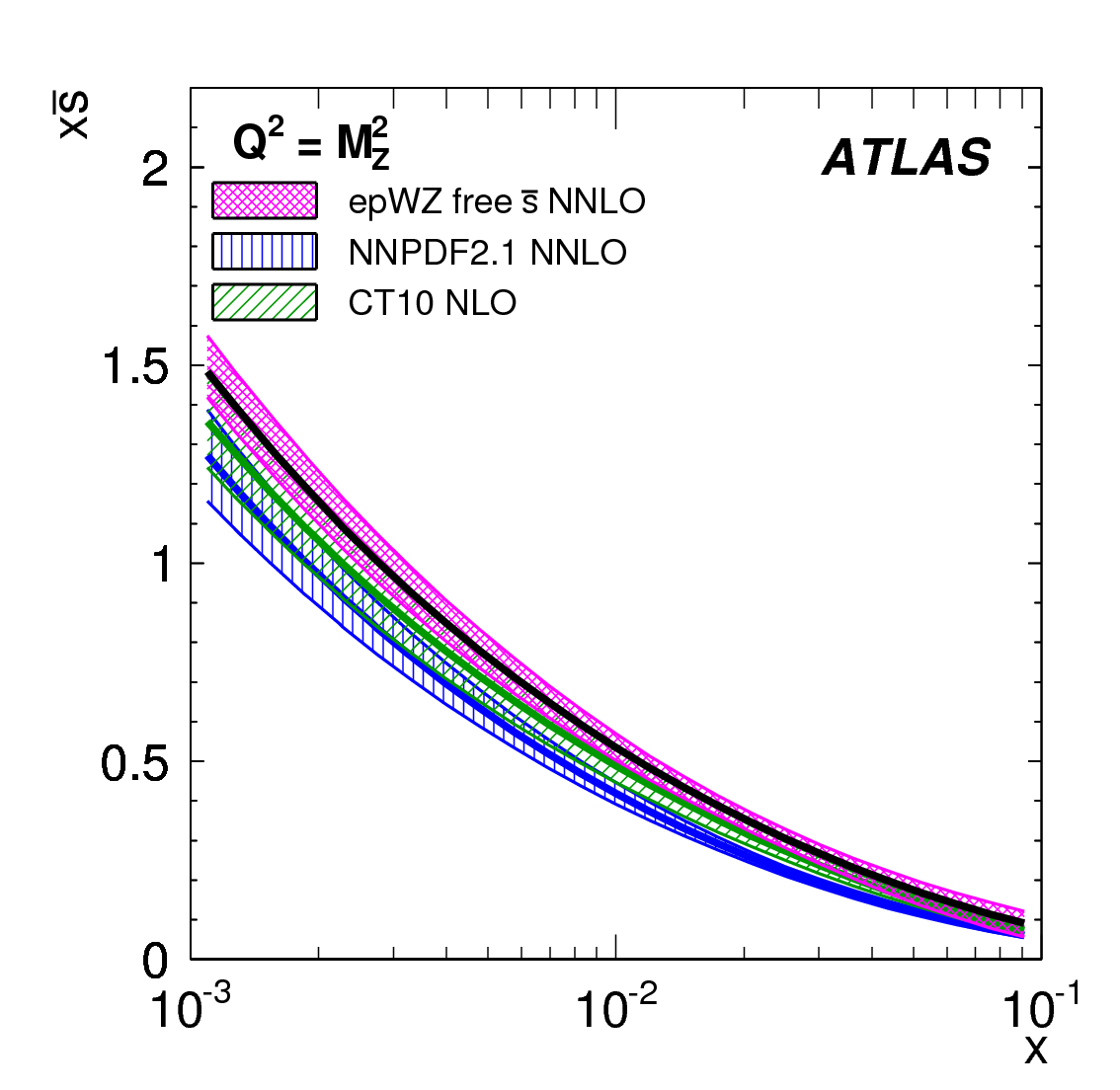}}
\caption{ATLAS \cite{Aad:2011dm}: The strange antiquark density vs $x$ for the ATLAS (denoted as epWZ) free $\bar s$ NNLO fit (magenta band) compared to predictions from NNPDF2.1 (blue hatched) and CT10 (green hatched) at $Q^2= 1.9 \GeV^2$ (left) and $Q^2= 91 \GeV^2$ (right). }
\label{fig:ATLASPDF1}
\end{figure*}

Also the LHCb experiment published a differential cross-section of the $W^\pm$ and $Z$ boson production in the forward 
rapidity region ($2.0<\eta<4.0$), based on \IntLumi$\approx 37 \ipb$, which is consistent with the measurements of ATLAS 
and CMS. A detailed discussion can be found in \cite{Aaij:2012vn}.

In summary, the double-differential cross-section measurements of the \Wboson and \Zboson bosons lead to important constraints to the PDFs of the proton. Figure \ref{fig:PDFNNPDFLHC} shows the improvement of the $\bar u$ and $\bar s$ parton density functions with and without including the current available LHC data based on the NNPDF group \cite{Ball:2012cx}. In particular the ATLAS analysis  \cite{Aad:2011dm} suggests that the strange quark content is comparable to the $\bar u$ and $\bar d$ content even at low scales-
With measurements using the full 2011 data set, the statistical and systematic uncertainties are expected to be decreased significantly and a further improvement $>20\%$ of the strange-quark PDFs and the $u/d$-quark ratio is anticipated. It is not clear how much improvement of a similar study of the full 2012 data set at a center-of-mass energy of $\sqrt{s}=8\,\TeV$ can be expected. Even though the data set is larger by a factor of five, the 2011 results are not statistics limited and the increase of center-of-mass energy is only $10\%$.

\begin{figure*}
\resizebox{0.5\textwidth}{!}{\includegraphics{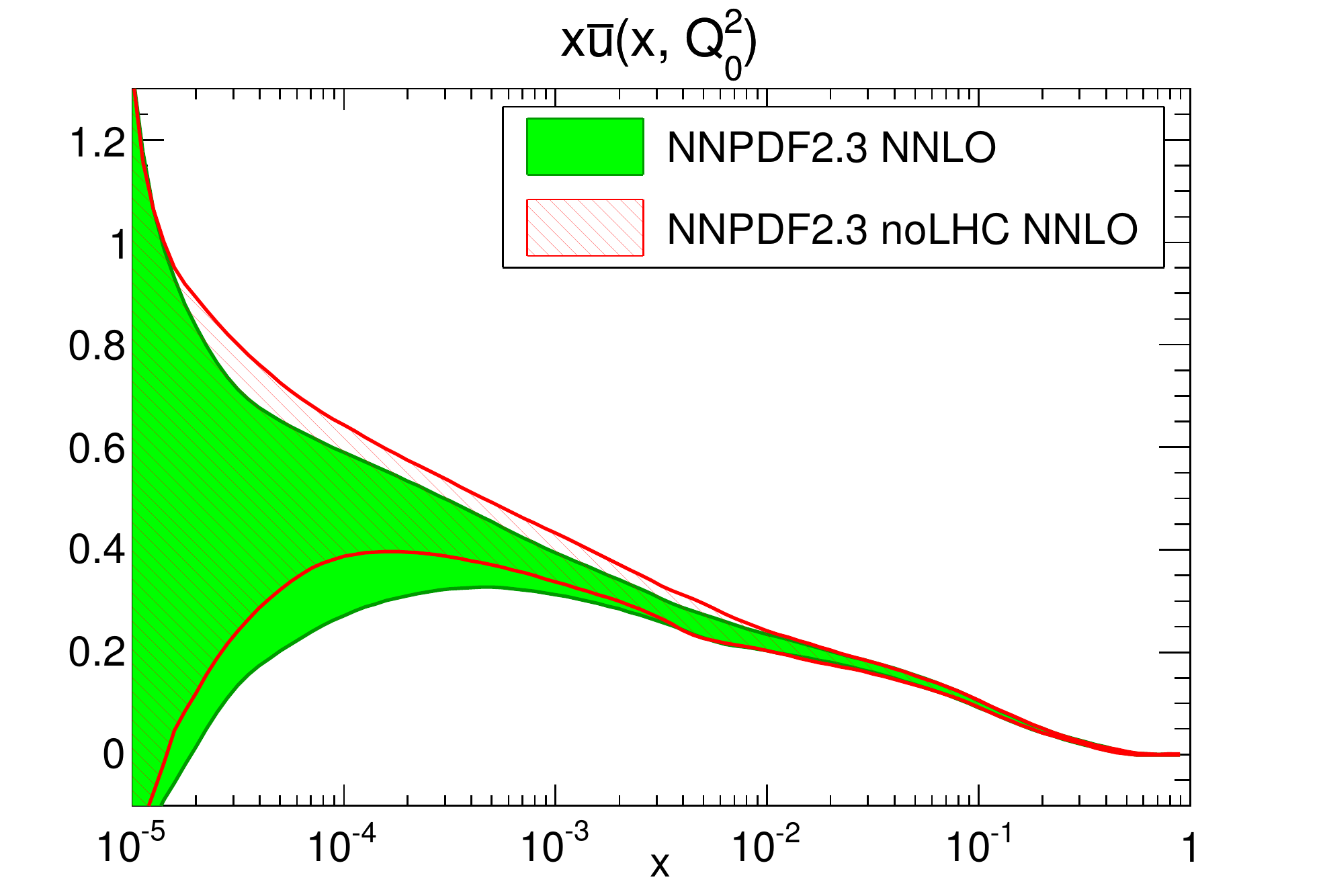}}
\resizebox{0.5\textwidth}{!}{\includegraphics{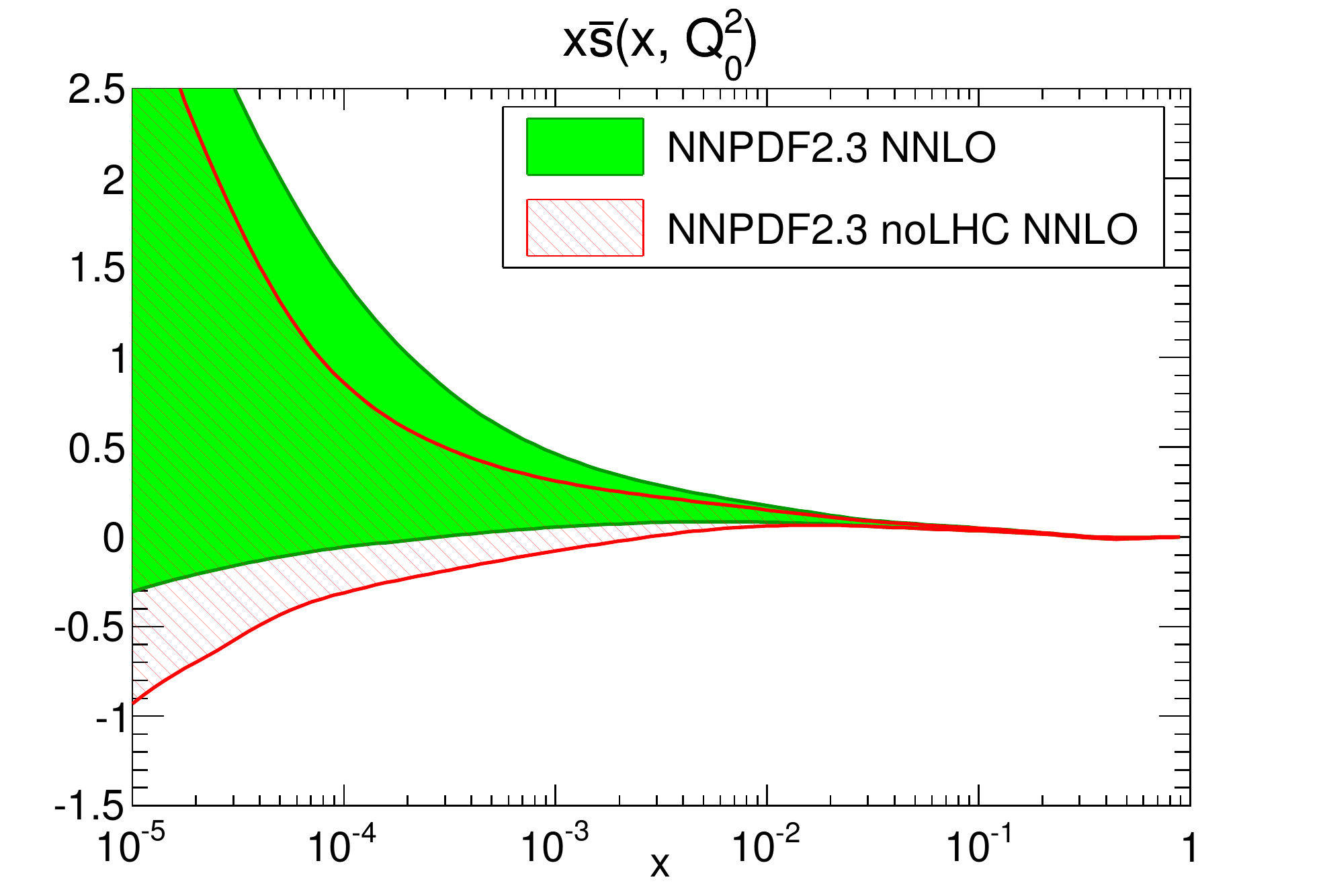}}
\caption{Comparison of $\bar u$-quark (left) and $\bar s$-quark (right) PDFs from the NNPDF2.3 NNLO sets with no LHC data and with LHC data used in the fit. Further comparison plots can be found in \cite{Ball:2012cx} and \cite{Ball:2012wy}.}
\label{fig:PDFNNPDFLHC}
\end{figure*}

\subsection{\label{sec:TransverseMomentum} Transverse momentum measurements of vector bosons}

As discussed in Section \ref{sec:TheoResum}, the transverse momentum distributions 
of vector bosons provides an important test of QCD corrections in the initial  
state of the production process due to the absence of colour flow between the 
initial and final state. In particular, predictions based on resummation 
techniques can be tested, which play an important role for the expected $\pT$ 
spectra between $0$ and $\approx M_V/2\,\GeV$. Only a mild dependence on the 
proton PDFs is expected. Besides the test of QCD calculations, the accurate 
understanding of the vector boson's $\pT$ spectra is essential for the 
measurement of the \Wboson boson mass at the LHC, especially when the $\pT$ spectrum 
of the decay leptons of $W^\pm \rightarrow l^\pm \nu$ is used as a sensitive 
variable for $m_W$. 

The transverse momentum distributions of electron and 
muon pairs from $Z/\gamma^*$ events can be measured directly with the reconstructed four-momentum 
information of the decay leptons. The $\pT(Z)$ momentum resolution for 
$\pT(Z)<40\GeV$ is typically $\approx 3\, \GeV$ for ATLAS and CMS. A finite 
binning with a similar size leads therefore to resolution effects, i.e. 
bin migration, which make a dedicated unfolding procedure necessary (see Section 
\ref{sec:CrossMeasPhil}). 

CMS has published normalised transverse momentum distributions of the $Z/\gamma^*$ process 
based on the 2010 data sample \cite{Chatrchyan:2011wt}. The differential cross-section has 
been normalised to the cross-section integrated over the acceptance region, 
defined by $\pT^l>20\,\GeV$, $|\eta|<2.1$ and $60\,\GeV<m_{ll}<120\,\GeV$. Both 
decay channels have been unfolded using an inverted response matrix and show 
consistent results, as seen in Figure \ref{fig:CMSZPt}.

ATLAS also published an analysis using the full 2010
data sample \cite{Aad:2011gj}, also using both leptonic decay channels. The 
differential cross-sections have been normalised to the fiducial 
cross-section with $66\,\GeV<m_{ll}<116\,\GeV$, $\pT^l>20\,\GeV$ and $|\eta|<2.4$. 
A Bayesian method has been used for the unfolding of the data, where the resulting distribution is shown in 
Figure \ref{fig:ATLASZPt} in comparison with different MC generator predictions\footnote{It should be noted that the differences in predictions of different generators could be due to different scale-parameter settings used.}. The dominating systematic uncertainties are due to uncertainties 
of the momentum scale and resolution uncertainties as well as from the 
unfolding procedure.

\begin{figure*}
\begin{minipage}{0.59\textwidth}
\resizebox{1.0\textwidth}{!}{\includegraphics{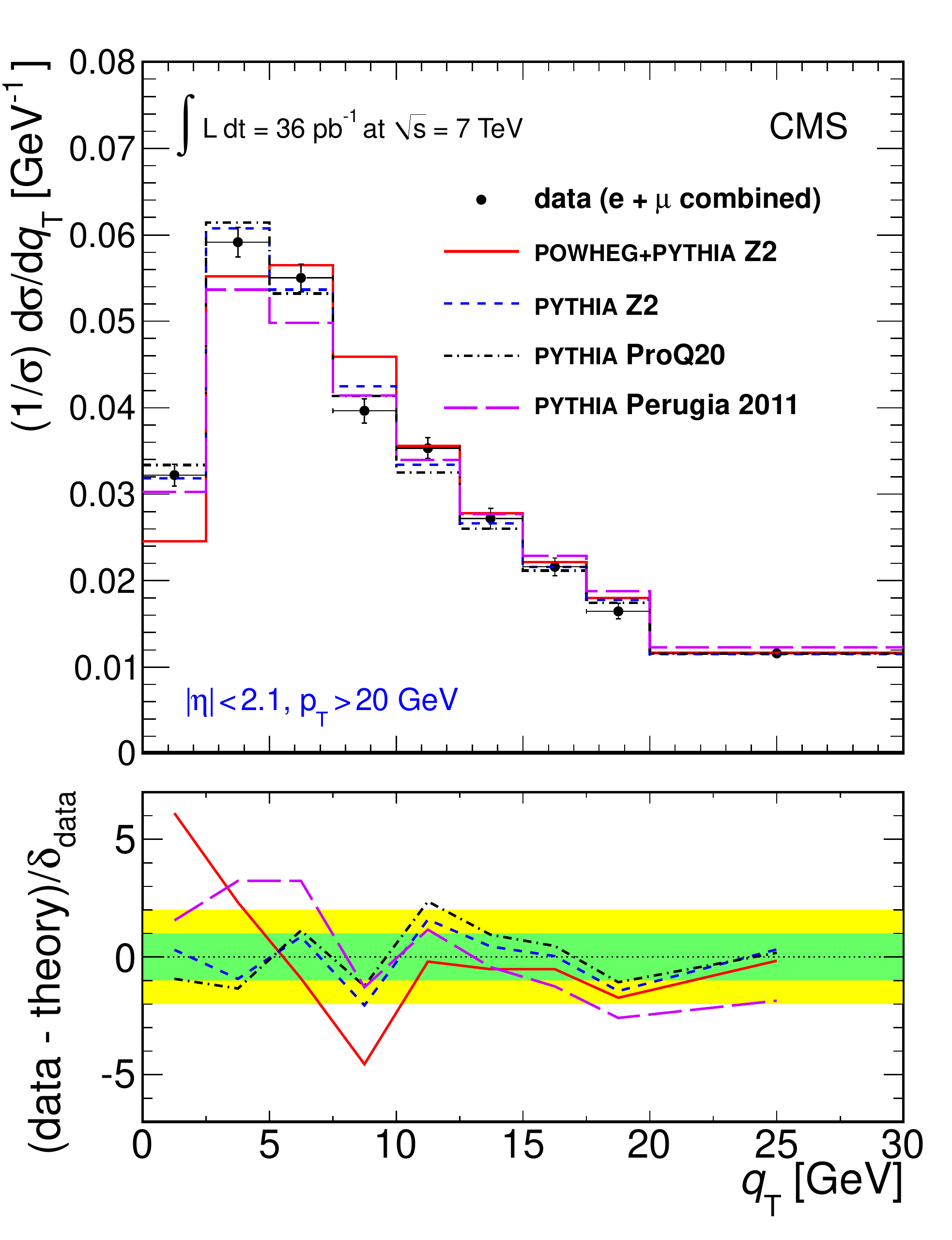}}
\caption{\label{fig:CMSZPt} CMS \cite{Chatrchyan:2011wt}: The combined electron and muon measurement of the \Zboson boson transverse momentum distribution (points) and the predictions of several \Pythia~ tunes and \Powheg~interfaced with \Pythia~ using the Z2 tune (histograms). The error bars on the points represent the sum of the statistical and systematic uncertainties on the data. The lower portion of the figure shows the difference between the data and the simulation predictions divided by the uncertainty delta on the data. The green (inner) and yellow (outer) bands are the $\pm 1 \delta$ and $\pm 2 \delta$  experimental uncertainties.}
\end{minipage}
\hspace{0.2cm}
\begin{minipage}{0.39\textwidth}
\resizebox{1.0\textwidth}{!}{\includegraphics{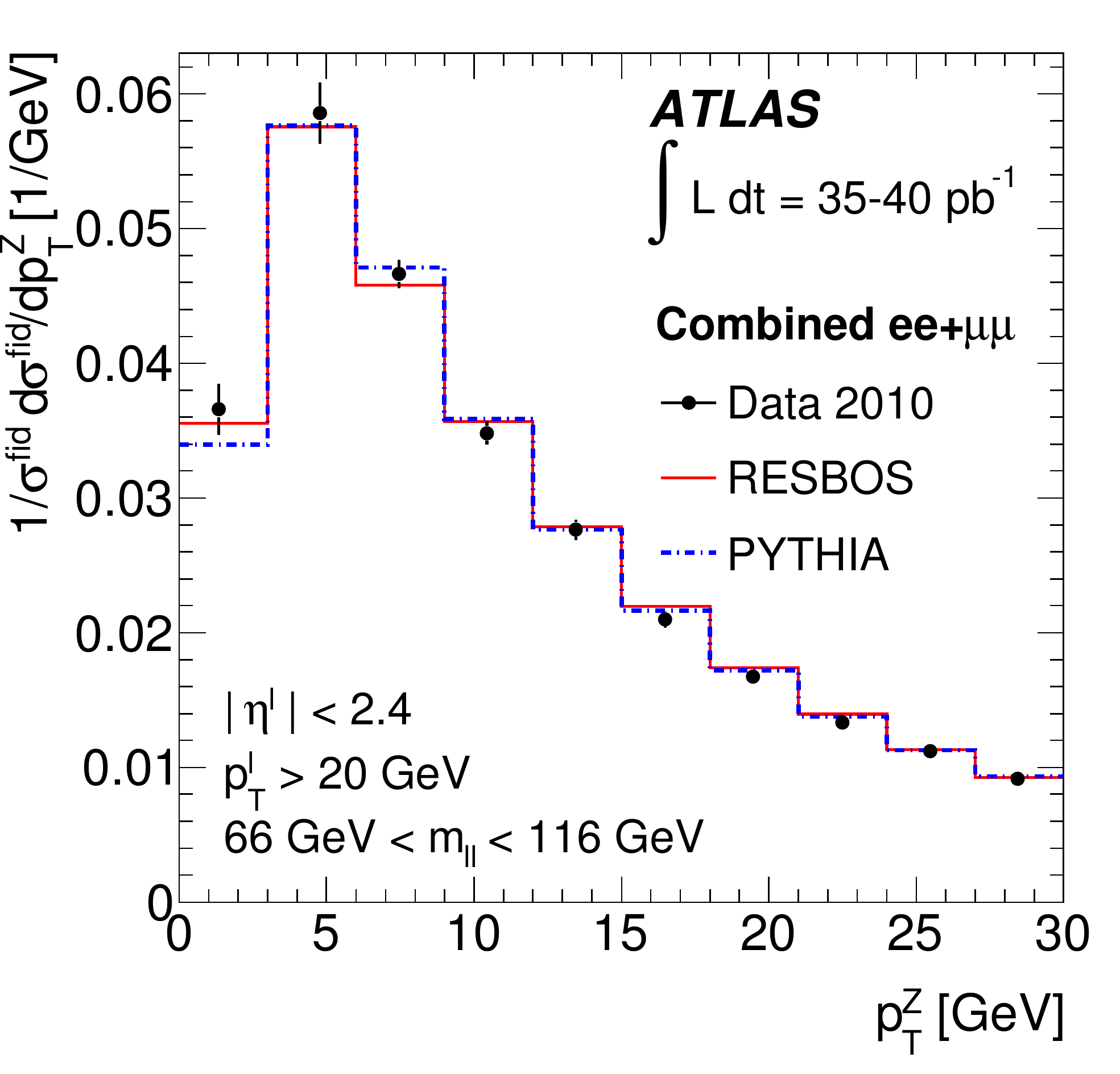}}
\resizebox{1.0\textwidth}{!}{\includegraphics{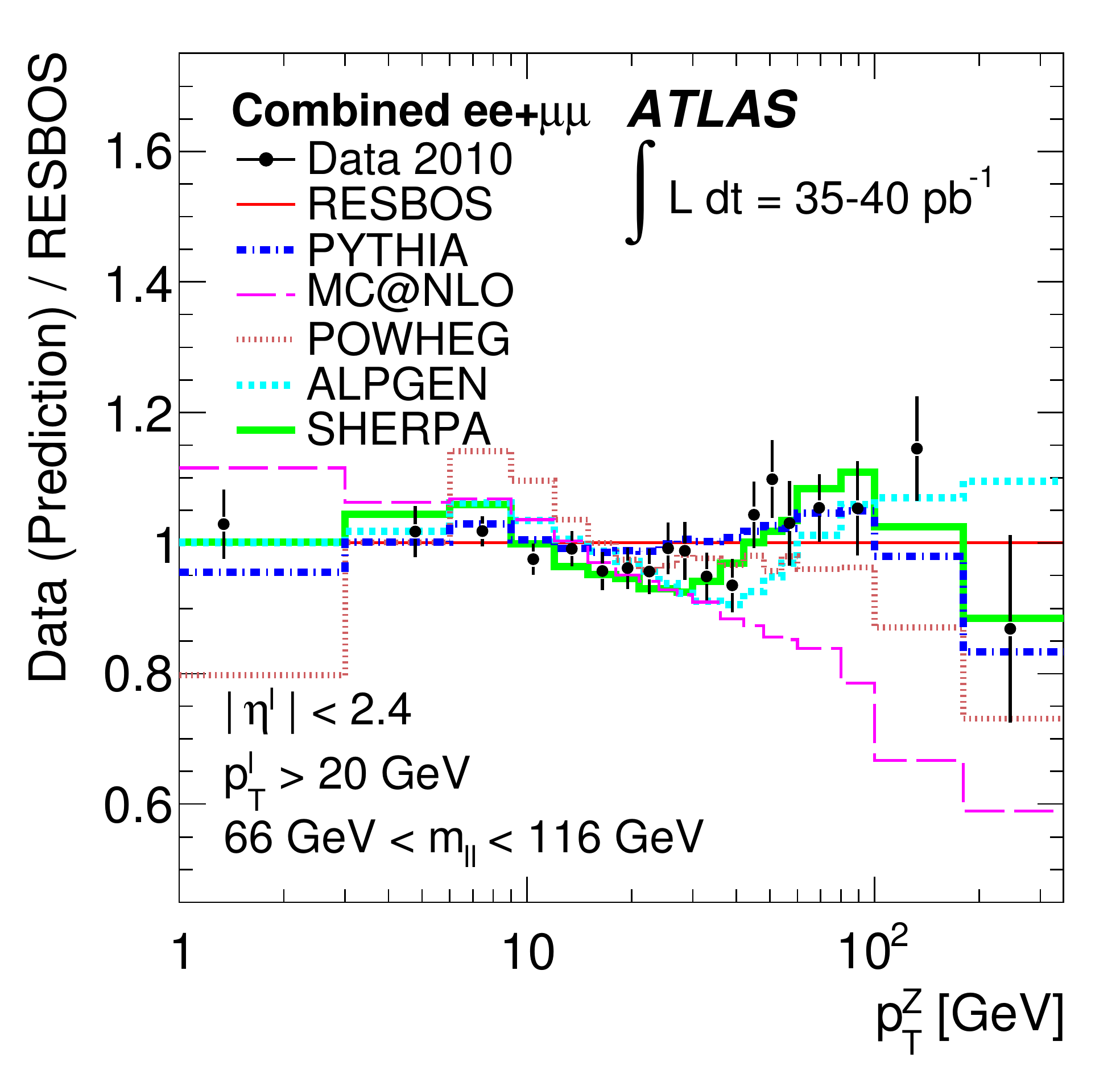}}
\caption{\label{fig:ATLASZPt}ATLAS \cite{Aad:2011gj}: The combined normalised differential cross-section as a function of $\pT(Z)$ for the range $\pT(Z) <30 \GeV$ compared to the predictions of \ResBos~and \Pythia. Ratios of the combined data and predictions from different generators over the \ResBos~prediction for the normalised differential cross-section are shown below. }
\end{minipage}
\end{figure*}

The low $\pT(Z)$ domain of the $Z/\gamma^*$ production can be alternatively 
probed with the $\Phi^*$ observable, defined as

\[\Phi^* = \tan(\frac{\pi-\Delta \Phi}{2}) \cdot sin(\theta^*)\, ,\]

\noindent where $\Delta \Phi$ is the azimuthal opening angle between the two decay 
leptons and 

\[ \cos(\theta^*) = \tanh(\frac{(\eta^- - \eta^+)}{2}) \]

\noindent as the measure of the scattering angle of the positive and negative leptons 
with respect to the beam \cite{Banfi:2010cf}. The $\Phi^*$ observables is highly correlated 
to $\pT(Z)/m_{ll}$; a small $\pT(Z)$ leads to a large opening angle and hence a 
small value of $\Phi^*$, while a large transverse momenta lead to small opening 
angles and therefore larger values of $\Phi^*$. A typical value of 
$\pT(Z)\approx 100\,\GeV$ leads to $\Phi^*\approx 1$. This variable has the 
advantage that it is solely constructed using track-based directions which are known to 
much higher precision than their transverse momenta. ATLAS has published 
unfolded normalised\footnote{The results have been normalised to the same 
fiducial regime as the corresponding $\pT(Z)$ analysis.} $\Phi^*$ distributions 
for three different regions \Zboson boson rapidity regions ($|y_Z|<0.8$, 
$0.8<|y_Z|<1.6$ and $|y_Z|>1.6$) \cite{Aad:2012wfa}. The resulting distribution is shown in 
Figure \ref{fig:ATLASPhiStar} for both decay channels, together with the theoretical prediction 
based on the \ResBos~ generator. Both decay channels lead to consistent results 
and a clear deviation from the theoretical prediction can be observed; these 
are consistent with the published results on $\pT(Z)$. The systematic 
uncertainties are smaller than the associated statistical uncertainties for all 
bins. The statistical precision varies $0.3\%$ for $\Phi^*\approx 0$ to $1.6\%$ 
for $\Phi^*\approx 2.5$.

\begin{figure}
\resizebox{0.5\textwidth}{!}{\includegraphics{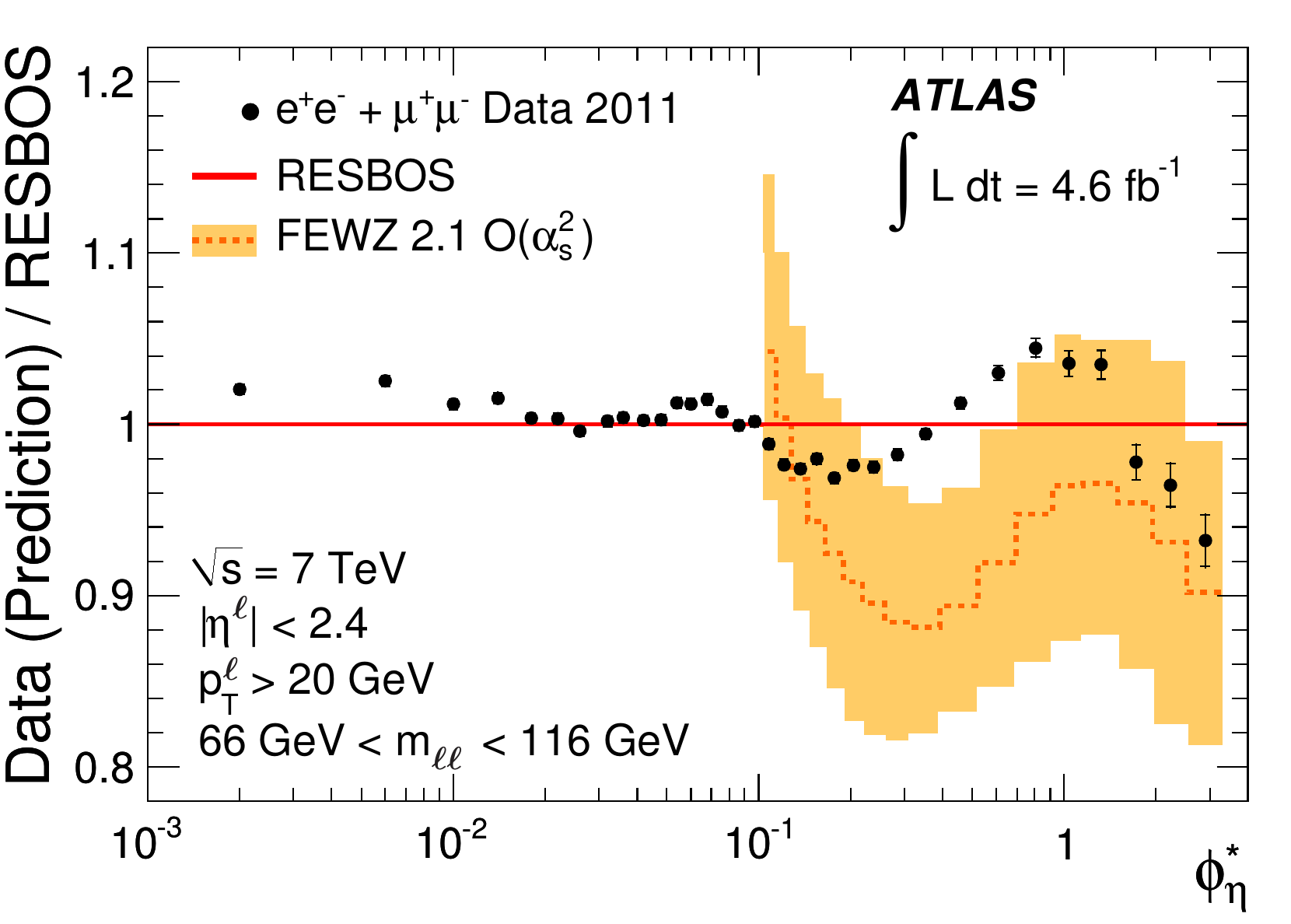}}
\caption{ATLAS: \cite{Aad:2012wfa} The ratio of the combined normalised differential cross-section to \ResBos~ predictions as a function of $\Phi^*$. The inner and outer error bars on the data points represent the statistical and total uncertainties, respectively. The uncertainty due to QED FSR is included in the total uncertainties. The measurements are also compared to predictions, which are represented by a dashed line, from \FEWZ~ 2.1. Uncertainties associated to this calculation are represented by a shaded band. The prediction from \FEWZ~ 2.1 is only presented for $\Phi^*>0.1$.}
\label{fig:ATLASPhiStar}
\end{figure}

The measurements of the $\pT(Z)$ distribution provides important information 
for the tuning of MC generators, which can then be indirectly transferred to 
the prediction of transverse momentum distribution of the \Wboson boson, 
$\vec \pT(W)$. However, an explicit measurement of $\vec \pT(W)$ would allow to 
directly constrain the $\pT$ spectrum of the \Wboson boson's decay leptons and 
therefore estimate uncertainties on an associated $m_W$ measurement.

\begin{figure}
\resizebox{0.5\textwidth}{!}{\includegraphics{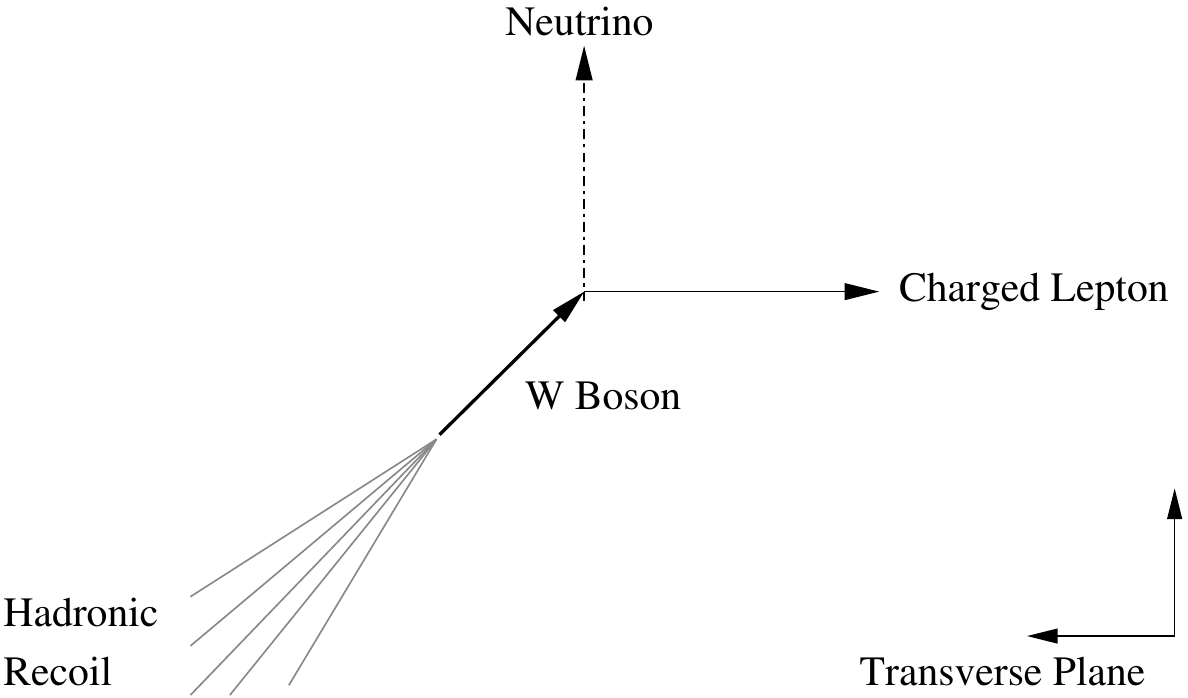}}
\caption{Illustration of the hadronic recoil in $W^\pm \rightarrow l^\pm \nu$ events. While the transverse momenta of the \Zboson boson can be directly interfered by its both decay leptons, the hadronic recoil has to be used in the leptonic decay modes of the \Wboson boson.}
\label{fig:HadRecoil}
\end{figure}

ATLAS also published a measurement of the transverse momentum distribution of 
the \Wboson boson $\vec \pT(W)$ based on the 2010 data sample \cite{Aad:2011fp}. The  $\vec \pT(W)$ 
cannot be directly measured from its decay leptons due to the neutrino. 
However, the $\vec \pT(W)$ must be balanced by the hadronic activity induced by 
QCD corrections in the initial and final state, i.e. 

\[\vec \pT(W) = - \vec \pT(had) = \vec \pT(l^\pm) + \vec \pT(\nu),\]

\noindent where $\pT(had)$ denotes the hadronic recoil (Figure \ref{fig:HadRecoil}).
Hence $\vec \pT(had)$ can be used for the measurement of $\pT(W)$, since it reflects
the underlying hadronic activity from the hard QCD 
interactions. The hadronic recoil has several 
experimental uncertainties (e.g. pile-up) and also a rather poor resolution 
compared to the reconstruction of leptons. Hence a data-driven model of the 
relation between $\vec \pT(had)$ and $\pT(W)$ has been used. This model is 
derived from \Zboson boson events, where the $\pT(Z)$ can be directly 
determined via the decay lepton measurements with a sufficiently good 
resolution. It is then assumed that the dependence of 
the hadronic recoil to the transverse momentum of the vector boson is the same 
in \Wboson and \Zboson boson events. The unfolding is performed with a Bayesian 
approach. The resulting differential cross-section, which has been normalised 
to the fiducial cross-section measured in the phase space defined via 
$\pT^l>20\,\GeV$, $|\eta_l|<2.4$, $\pT^\nu>25\,\GeV$ and $\mT>40\,\GeV$, is 
shown in Figure \ref{fig:ATLASWpt}. It should be noted that the poor resolution of $\vec 
\pT(had)$ implies a significantly larger binning to ensure a stable unfolding 
procedure. The systematic uncertainties of the data-driven modelling of
$\vec \pT(had)$ during the unfolding procedure dominate the overall 
uncertainties up to $\pT(W)<\,75\GeV$. Statistical uncertainties start 
to dominate for larger $\pT(W)$ values. This measurement has not yet been 
repeated for the 2011 data sample, as the increase of pile-up further reduces the 
resolution of $\vec \pT(had)$ and hence complicates the unfolding procedure when 
aiming at a similar binning for the low $\pT(W)$ region.
The comparison to the prediction based on \ResBos~ for both, $\pT(W)$ and $\pT(Z)$, 
(Figure \ref{fig:ATLASZPt}, \ref{fig:ATLASWpt}), shows a significant disagreement  
and hence will allow for an improved tuning of the underlying MC 
generators.

\begin{figure}
\resizebox{0.5\textwidth}{!}{\includegraphics{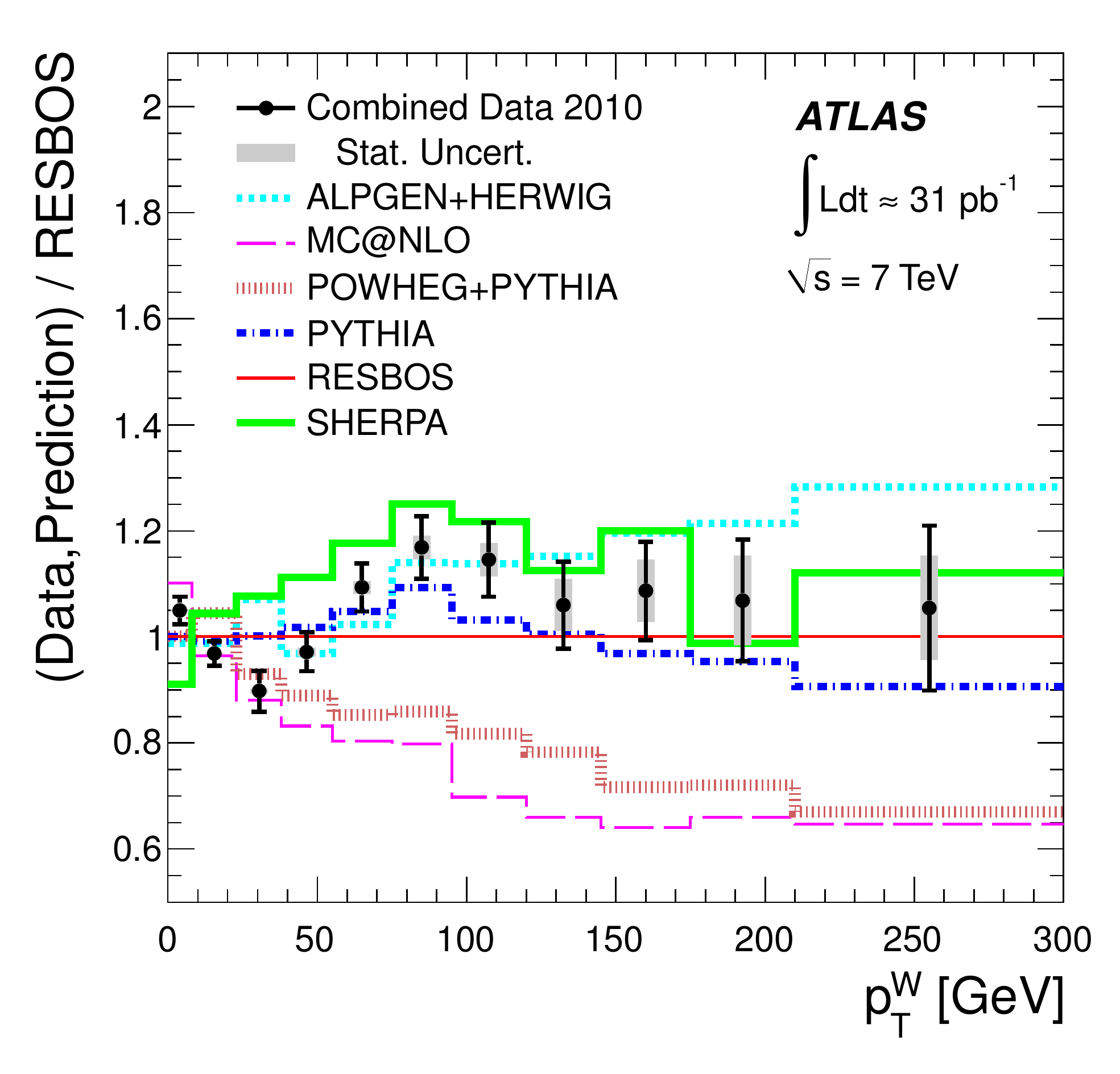}}
\caption{ATLAS \cite{Aad:2011fp}: Ratio of the combined measurement of $\pT(W)$ and various predictions to the \ResBos~ prediction for the normalised differential cross-section from different generators.
}
\label{fig:ATLASWpt}
\end{figure}

A  comparison of the published $\pT(Z)$ distributions of ATLAS and CMS is shown 
in Figure \ref{fig:ATLASCMSWZPTComp}. We extrapolated the ATLAS results to the 
CMS fiducial volume, while the CMS results have been corrected to the ATLAS binning to 
allow for a direct comparison. A slight tension can be observed for $\pT(Z)<10\,\GeV$ which is
not yet significant.

In summary, a new tune of generators like \Pythia~is needed in order to describe the deviations between measurement and simulations. First measurements of the \Zboson boson transverse momentum distribution at $\sqrt{s}=8\,\TeV$ based on a very reduced data set have also become available \cite{CMS-PAS-SMP-12-025}. The expected measurements of the full 2012 data set at $\sqrt{s}=8\,\TeV$ could also allow to test electroweak corrections which are predicted to become sizeable at large transverse momenta. In addition, the higher statistics will allow for the measurement of angular coefficients in the  Drell-Yan production, as introduced in Section \ref{sec:QCDDynamics} (Equation \ref{EQN:DECOMP}), in up 
to three dimensions. Figure \ref{fig:A0A2vaZPT} shows the \Powheg~ and \MCAtNLO~ prediction of 
$A_0$ vs. the transverse momentum of the \Zboson boson.  The observed discrepancy between both generators could be due to the difference in the matching 
scheme of NLO calculations and the underlying parton shower model (see Section \ref{sec:PSMatching}). Therefore a 
precise measurement of the angular correlations with the available 
$\sqrt{s}=8\,\TeV$ data sample is mandatory to test this important aspect in modern QCD calculations.

\begin{figure}
\resizebox{0.5\textwidth}{!}{\includegraphics{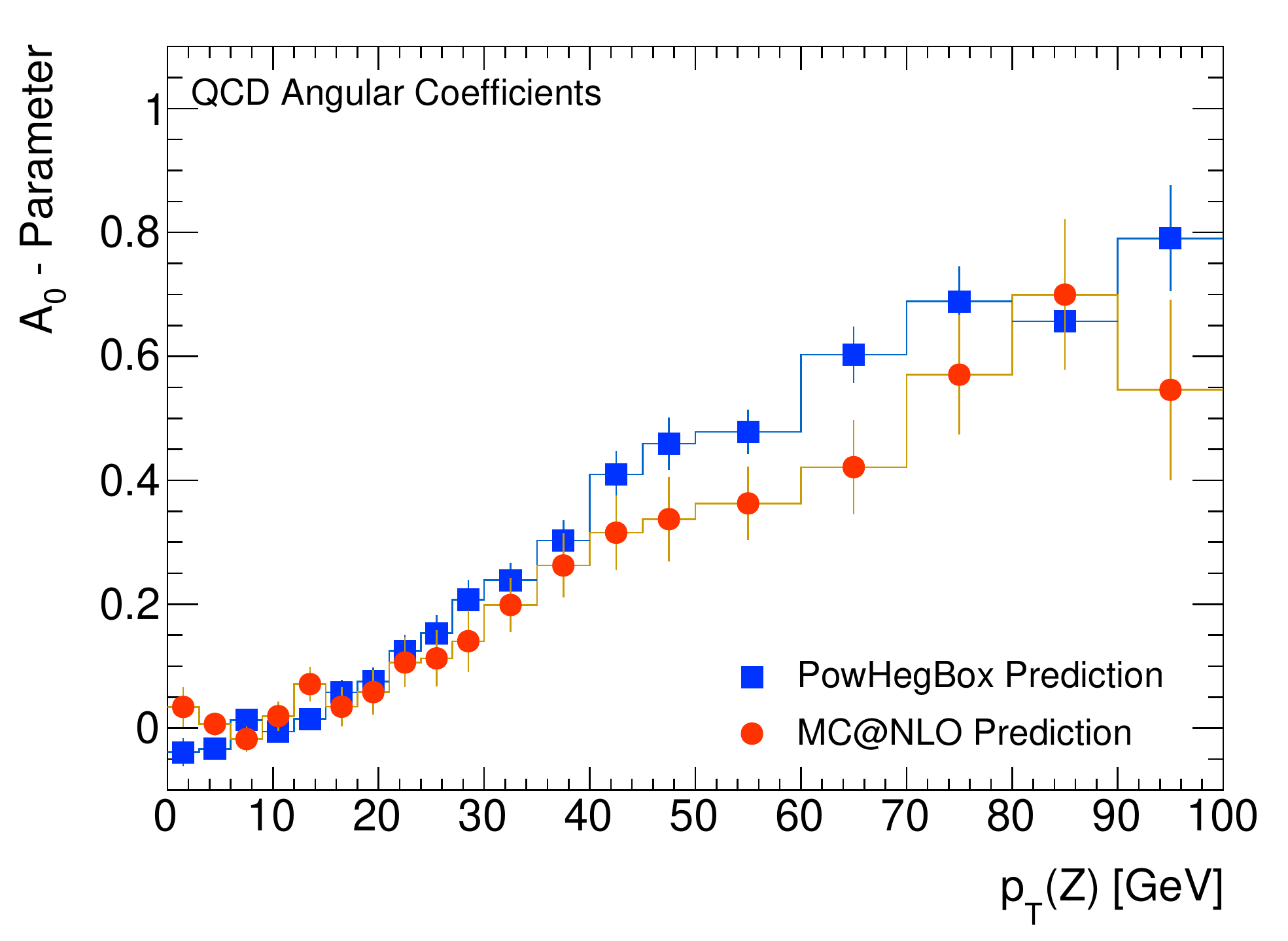}}
\caption{Predictions by \MCAtNLO~ and \PowhegBox~ of the QCD angular parameters $A_0$ (left) and $A_2$ (right) in \Zboson boson events vs. the transverse momentum of the \Zboson boson.}
\label{fig:A0A2vaZPT}
\end{figure}

\begin{figure}
\resizebox{0.5\textwidth}{!}{\includegraphics{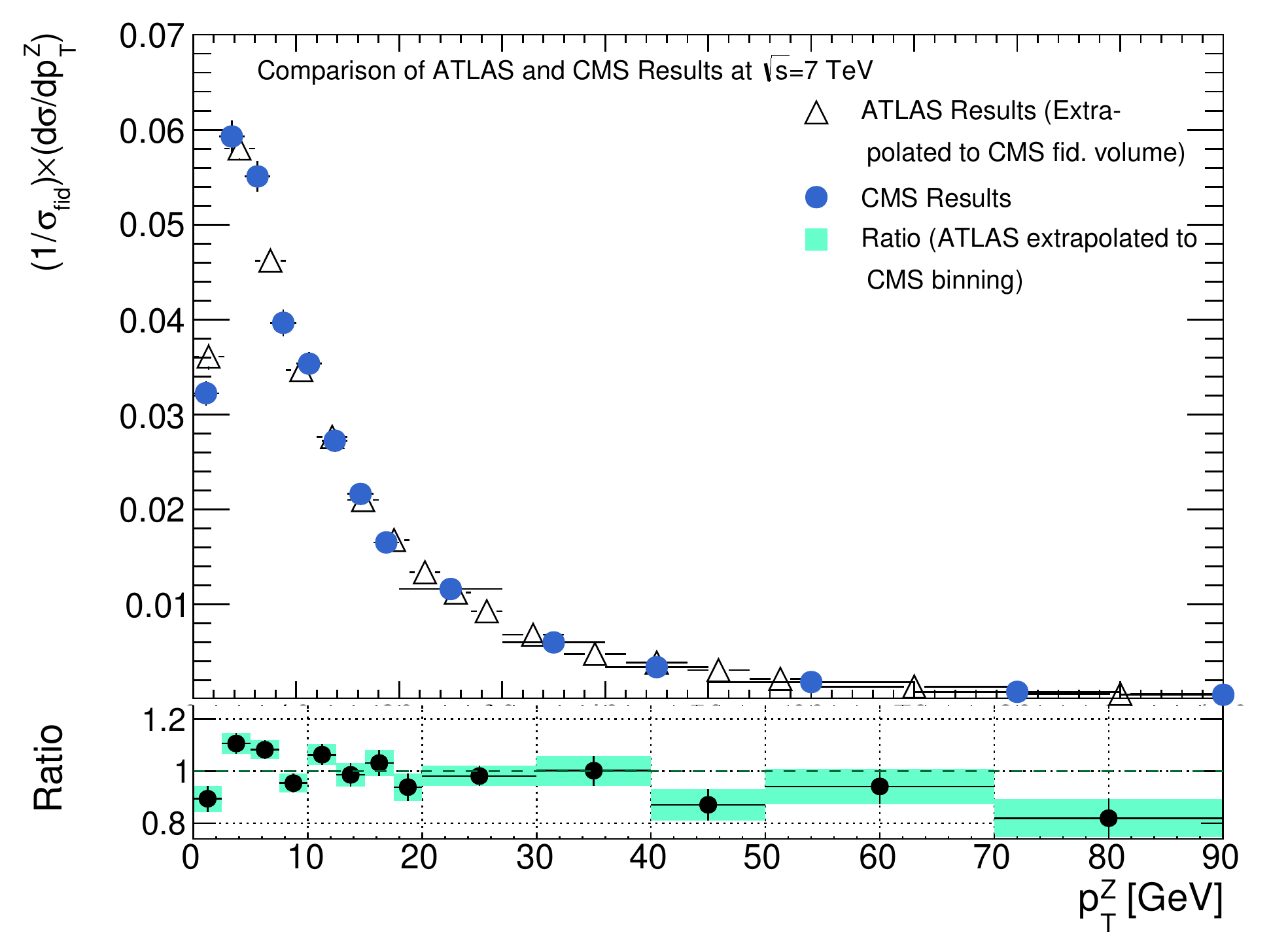}}
\caption{Comparison of the results of the transverse momentum measurement of the \Zboson boson for ATLAS and CMS. The ATLAS results have been extrapolated to the CMS fiducial volume. The CMS results have been corrected to the ATLAS binning for the direct comparison.}
\label{fig:ATLASCMSWZPTComp}
\end{figure}

\subsection{Determination of $\Gamma_W$}

The measurement of the ratio $R$ of leptonic rates for the inclusive production 
of \Wboson and \Zboson bosons at the LHC, as discussed in the previous section, can be 
written as
\begin{eqnarray}
R = \frac{\sigma(pp\rightarrow W+X)}{\sigma(pp\rightarrow Z+X)} \cdot \frac{\Gamma (Z)}{\Gamma (Z\rightarrow l^+ l^-)} \cdot \frac{\Gamma(W^\pm\rightarrow l^\pm \nu)}{\Gamma(W)}.
\end{eqnarray}

The inclusive cross-section ratio is known to NNLO in $\alpha_s$ and has a 
numerical value of $\sigma_W/\sigma_Z= 3.34\pm 0.08$ \cite{Alekhin:2010dd}. The leptonic branching 
ratio for the \Zboson boson $BR(Z\rightarrow l^+l^-)$ is known to be 
$(3.366\pm0.002)\%$ from the LEP and SLC experiments \cite{Beringer:1900zz}. Hence, the measurement 
of $R$ allows for an indirect determination of the leptonic \Wboson boson branching 
ratio $BR(W\rightarrow l^\pm \nu) = \frac{\Gamma(W^\pm\rightarrow l^\pm 
\nu)}{\Gamma(W)}$. 

The CMS analysis \cite{CMS:2011aa} used the measured ratio $R = \sigma(W^\pm \rightarrow l^\pm 
\nu) / \sigma(Z \rightarrow l^+l^-)= 10.54\pm0.19$
\noindent leading to an indirect determination of 
\[
BR(W\rightarrow l^\pm \nu) = 0.106 \pm 0.003
\]
In addition, the partial leptonic decay width of the \Wboson boson can be 
calculated within the Standard Model and is given by
\[
\Gamma(W\rightarrow e \bar \nu_e) = \frac{G_F M_W^3}{6\pi\sqrt{2}} 
(1+\delta_l^{SM}) = 226.2 \pm 0.2 \,\MeV,
\]
\noindent where corresponding electroweak corrections 
$\delta_l^{SM}$ are small, since they are largely absorbed in $G_F$ and $m_W$ 
\cite{Renton:2008ub}. Therefore the \Wboson boson width can be extracted via 
$BR(W\rightarrow l^\pm \nu)$ and results in

\[\Gamma_W = 2144\pm62 \,\MeV,\]
\noindent which is consistent with the current world average $\Gamma_W = 2085\pm42 \,\MeV$.

A combination of the ATLAS and CMS results on the cross-section ratio could reduce the overall uncertainty by $\approx 20\%$, when assuming no correlations between the systematic uncertainties. This would lead to an uncertainty on $\Gamma_W$ of $50\,\MeV$ and hence would lead to a significant reduction of the uncertainty on the current world average.

\subsection{\label{sec:AFB}Forward-backward asymmetries of the \Zboson boson}

The study of the Drell-Yan production at the LHC does also provide 
information on the weak mixing angle. This measurement is complementary to the \Zboson-pole analyses of the LEP experiments \cite{ALEPH:2005ab}, \cite{Haidt:2013rea}. 

The differential cross-section for an 
$f \bar f \rightarrow f' \bar f'$ annihilation process\footnote{with $f\neq 
f'$ since no t-channel contributions should be allowed} at lowest order for a 
\Zboson boson exchange is given by

\begin{eqnarray}
\frac{d\sigma}{d\cos\theta} &=& \frac{N^f_C G_F^2 m_Z^4}{16\pi} \frac{s}{(s-m_Z^2)^2 + \frac{s^2}{m_Z^2} \Gamma_Z^2} \cdot \\
\nonumber
&& [(v_f^2 + a_f^2)(v_{f'}^2 + a_{f'}^2)(1+\cos^2 \theta) + 2 v_f a_f v_{f'}a_{f'}\cdot \cos\theta]
\end{eqnarray}

\begin{equation} \frac{d\sigma}{d\cos\theta} = \kappa [A\cdot(1+\cos^2 \theta) + 
B\cdot \cos\theta]
\end{equation}

\noindent where $\cos\theta$ is the angle between the incoming and outgoing 
fermions\footnote{or between incoming and outgoing anti-fermions}, and $a_f$ 
and $v_f = a_f\cdot(1-4|q_f|sin^2\theta_W)$ are the axial and vector-axial 
couplings to the \Zboson boson and $q_f$ is the fractional charge of the fermion. Several things should be noted. First of all, the 
$(1+\cos^2\theta)$ dependence would also appear in a pure $\gamma$ exchange 
diagram. However, the vector- and axial-vector couplings of the \Zboson boson 
introduce an additional $\cos\theta$ dependence. Secondly, the differential 
cross-section depends only on the weak mixing angle $sin^2\theta_W$, when 
fixing the electric charges, the weak-hypercharges, $m_Z$, and $\Gamma_Z$ for a 
given center-of-mass energy. By defining forward and background cross-sections 
in terms of the angle of the incoming fermions,

\[\sigma_F = \int_0^1 \frac{d\sigma}{d\cos\theta} d\cos\theta, \hspace{0.8cm} \sigma_B = \int_{-1}^0 \frac{d\sigma}{d\cos\theta} d\cos\theta,\]

\noindent where the angle $\theta$ is defined in 
the CS frame as introduced in Section \ref{sec:frames},
a measure for the asymmetry at $\cos\theta=0$ can be defined as 

\[A_{FB} = \frac{\sigma_F-\sigma_B}{\sigma_F + \sigma_B}\]
also known as the forward-backward asymmetry parameter $A_{FB}$. At tree level, 
$A_{FB}$ is given by

\[A_{FB} = \frac{16}{3} 
\cdot \frac{(1-4|q_{f}|\sin^2\theta_W)}{1+(1-4|q_{f}|\sin^2\theta_W)^2} 
\cdot \frac{(1-4|q_{f'}|\sin^2\theta_W)}{1+(1-4|q_{f'}|\sin^2\theta_W)^2}.\]

\noindent The measurement at the \Zboson pole provides the most sensitive measurement as \Zboson 
exchange contributes roughly 100 times more than the $\gamma$ exchange. 
Therefore, only small corrections from the interference and pure $\gamma$ 
exchange terms are expected. It should be noted that electroweak corrections to 
$\sin\theta_W$ can be absorbed by defining an effective weak mixing angle 
$\sin\theta_{eff}$ which is therefore used in the actual measurements.  

The advantage of the $A_{FB}$ measurement is that it reduces to a good 
approximation to a pure counting experiment, e.g. by defining

\[A_{FB} = \frac{N_{\cos\theta>0} - N_{\cos\theta<0}}{N_{\cos\theta>0} + N_{\cos\theta<0}}\]
where the number of events in the forward and backward regions are labelled 
as $N_{\cos\theta>0}$ and $N_{\cos\theta<0}$. 

The measurement of $A_{FB}$ can also be used for the search for new physics. 
While the $A_{FB}$ at the \Zboson boson mass is used for the determination of 
$\sin^2\theta_W$, large invariant masses are governed by virtual photon and \Zboson 
interference terms.  A direct search for a new resonance in the electroweak section via the study of 
the invariant mass spectra of di-lepton events might not show an excess 
if the new resonance has a large width. However, such a new resonance would also 
interfere with the Standard Model amplitudes and hence introduce a structure in the 
measured asymmetries $A_{FB}$ near its mass. 

Both, ATLAS and CMS have published unfolded $A_{FB}$ measurements and also a
determination of the electroweak mixing angle. In addition to the backgrounds from other 
processes, the measured $A_{FB}$ is diluted by a wrong assignment of the incoming quark and antiquark. 
This is accounted for in both analyses using simulations and hence 
relies on a precise knowledge of the PDFs. 

The ATLAS analysis is based on the full 2011 data sample and uses the electron 
and muon decay channels \cite{AtlasAFB}. Since a high acceptance of \Zboson bosons with large 
rapidities reduces the dilution of falsely identified quark-directions, ATLAS 
also includes forward electrons in their analysis. By requiring one electron 
with $|\eta|<2.4$ and allowing a second electron within $|\eta|<4.9$, an 
acceptance for \Zboson boson events up to $|y_Z|<3.6$ is achieved. The $A_{FB}$ is measured 
in the same fiducial region as the \Zboson boson inclusive measurements. 
A Bayesian unfolding technique was used to transform the measured 
raw $\cos\theta_{CS}^*$  distribution in a given mass region to the $\cos\theta_{CS}^*$ at 
parton level. The unfolding does not only remove detector effects, but also effects from QED radiative corrections. The latter lead to deformations of the 
di-lepton invariant mass distribution. In order to account for these corrections, 
the results are unfolded to Born level, i.e. the state, before any emission of 
final-state radiation. The unfolding is based on a \Pythia~MC sample, where NLO 
corrections have been applied. QED final-state radiation was accounted for by interfacing 
the \Photos~generator. The $\pT$ and $y$ distributions of the \Zboson boson have 
been reweighted to NLO QCD predictions. NLO electroweak corrections have been 
estimated with \Horace~\cite{CarloniCalame:2003ux}. The resulting $A_{FB}$ distribution for the electron 
decay channel including one central and one forward electron, is shown in Figure \ref{fig:ATLASAfb}, where the results including forward electrons are denoted as $CF$. 
Dominating systematic uncertainties arise from the limited knowledge
of electron identification efficiencies in the forward region, NLO QCD effects 
and PDF uncertainties.

The effective electroweak mixing angle was not determined from the unfolded 
distributions but measured directly from the raw-data distributions. The 
measured $A_{FB}$ spectra have been compared to MC predictions which have been 
produced by varying initial values for $\sin^2\theta_{eff}$. Each prediction 
was compared to the measured distribution via a $\chi^2$ test.
The minimum of the resulting $\chi^2$ distribution yields then the measured 
$\sin^2\theta_{eff}$ value. The combination for all channels results in

\begin{eqnarray}
\nonumber
\sin^2 \theta_{eff} &=& 0.2297 \pm 0.0004(stat) \pm 0.0009(syst.) \\
\nonumber
&=& 0.2297 \pm 0.001\, ,
\end{eqnarray} 
in agreement with the current world average of $\sin^2\theta_{eff} = 0.23153 \pm 0.00016$. 

The CMS study of the $A_{FB}$ is also based on the full 2011 data sample \cite{Chatrchyan:2012dc} and 
the same fiducial volume as the inclusive measurement in both decay channels. 
The $A_{FB}$ has been measured in four different rapidity regions 
(0 to 1, 1 to 1.25, 1.25 to 1.5 and 1.5 to 2.4) and ten mass regions ranging from $40\,\GeV$ to 
$400\,\GeV$, leading to 40 measurements in total. The mass spectra of forward 
and backward events are unfolded independently for each rapidity region using 
an inverted response matrix approach. The response matrix is based on a NLO
MC predictions from \Powheg~and \Pythia. Also the \Pythia~model for final-state radiation has been 
used. Similar to ATLAS, the unfolding procedure corrects not only for detector 
effects but also gives the number of forward and backward events on Born level. The resulting $A_{FB}$ distribution for the combination of 
both decay channels in the most forward region is shown in Figure \ref{fig:CMSAfb}. 
Dominating systematic uncertainties are due PDFs in the central rapidity regions 
and due final state radiation modelling uncertainties in the forward region. Overall, a good 
agreement with the NLO prediction by \Powheg~is observed for all kinematic 
regions.

\begin{figure*}
\begin{minipage}{0.47\textwidth}
\resizebox{1.0\textwidth}{!}{\includegraphics{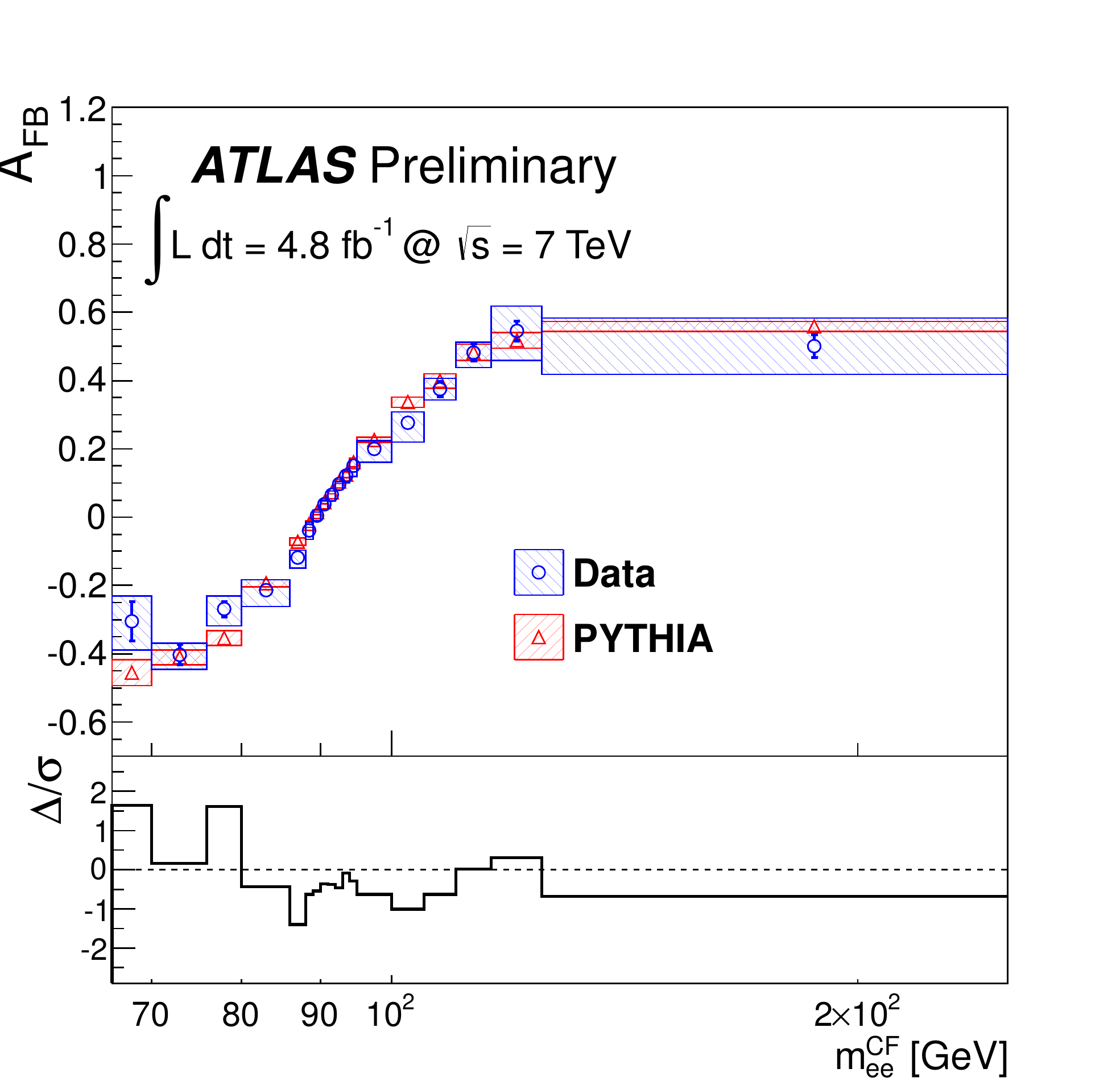}}
\caption{\label{fig:ATLASAfb} ATLAS \cite{AtlasAFB}: $A_{fb}$ unfolded to Born-level central-forward electrons. For the data, the boxed shaded region represents the total (statistical+systematic) uncertainty and the error bars represent the statistical uncertainty. The boxed shaded regions for the \Pythia~predictions represent the statistical uncertainty only. The ratio plots at the bottom display the distribution of pulls.}
\end{minipage}
\hspace{0.2cm}
\begin{minipage}{0.52\textwidth}
\resizebox{1.0\textwidth}{!}{\includegraphics{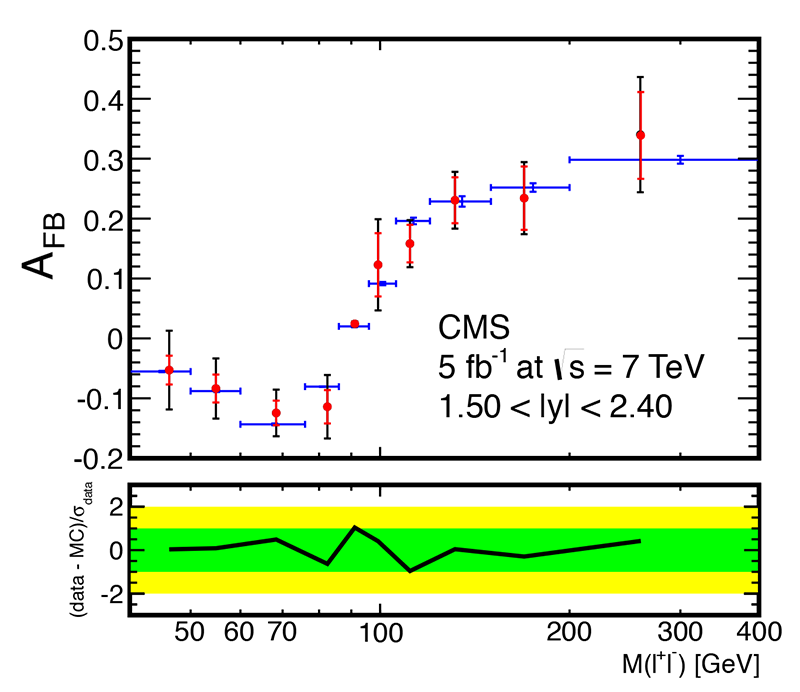}}
\caption{\label{fig:CMSAfb} CMS \cite{Chatrchyan:2012dc}: The unfolded and combined  measurement of $A_{FB}$ at the Born level with $1.5<|y|<2.4$ and $\pT > 20\,\GeV$. The data points are shown with both statistical and combined statistical and systematic error bars. The error bars on the MC prediction points are due to PDF uncertainties. The MC prediction statistical errors are of the same order of magnitude as the PDF uncertainties. }
\end{minipage}
\end{figure*}

The measurement of $\sin^2 \theta_W$ by the CMS experiment is based exclusively 
on the muon decay channel with an integrated luminosity of $1.1\,\ifb$ \cite{Chatrchyan:2011ya}. 
CMS focuses on the muon decay channel because of the smaller background 
uncertainties and a better understanding of the corresponding detector 
performance. 
The actual methodology for the $\sin^2\theta_W$ measurement is different from ATLAS. 
CMS uses an unbinned extended maximum likelihood 
function which is fitted to data in order to extract the effective weak mixing 
angle. The likelihood function is evaluated on an event-by-event basis and 
depends on the number of signal and background events and the expected event 
the probability density functions for the signal and background processes. 
These probability densities are parameterised as a function of the di-lepton 
rapidity, the di-lepton invariant mass, their decay angle $\cos\theta^*$ and the 
weak mixing angle and rely on leading order predictions of \Pythia, leading order PDF set and a full 
detector simulation. The impact of NLO effects has been estimated. The 
minimisation of the likelihood functions leads to a measured value of

\begin{eqnarray}
\nonumber
sin^2 \theta_{eff} &=& 0.2287 \pm 0.0020(stat) \pm 0.0025(syst.) \\
\nonumber
&=& 0.2287 \pm 0.003\, ,
\end{eqnarray}

\noindent also in agreement with the current world average \cite{ALEPH:2005ab}. A summary of the systematic uncertainties 
of the ATLAS and CMS measurements is shown in Table \ref{tab:SinThetaSys}. The dominating 
uncertainty for ATLAS comes from PDF uncertainties. 
Also at CMS, theoretical uncertainties due to PDF, FSR and NLO corrections dominate. The 
remaining experimental uncertainties could be in principle reduced by 
future studies. Hence a competitive measurement of $\sin^2\theta_W$ at the LHC 
relies on a significant improvement of the proton PDFs.

\begin{table}
\caption{Summary of the ATLAS and CMS measurements of the weak mixing angle $sin^2 \theta_{eff}$, together with the associated uncertainties and the current world average}
\label{tab:SinThetaSys}       
\begin{tabular}{p{3cm}llll }
\hline
 					& ATLAS			& CMS		& World Average  		\\		
\hline
$sin^2 \theta_{eff}$		&0.2297			&0.2287		& 0.23153				\\
Total uncertainty		&0.0010			&0.0032		& 0.00016				\\
\hline
Stat. uncertainty		&0.0004			&0.0020		& -					\\
Sys. uncertainty		&0.0009			&0.0025		& -					\\
\hline
PDF					&0.0007			&0.0013		& -					\\
Modelling+FSR		&0.0005			&0.0016		& -					\\
Detector effects		&0.0005			&0.0013		& -					\\
\hline
\end{tabular}
\end{table}

\subsection{\label{sec:WPol}Polarisation measurement of \Wboson bosons}

The measurement of the angular distribution of $W\rightarrow \mu\nu$ and 
$W\rightarrow e\nu$ events allows for the determination of the \Wboson boson polarisation 
in proton-proton collisions. The theoretical basis of this measurement was 
introduced in Section \ref{sec:QCDDynamics} and relies on Equation \ref{EQN:DECOMP}. This can be 
rewritten in terms of the fractions of left-handed, right-handed and 
longitudinal polarised \Wboson bosons, $f_L$, $f_R$ and $f_0$, which is given by

\[\frac{1}{\sigma} \frac{d\sigma}{d\cos\theta} = \frac{3}{8} f_L (1\mp 
\cos\theta)^2 + \frac{3}{8} f_R (1\pm \cos\theta)^2 + \frac{3}{4} f_0 
\sin^2\theta\]

\noindent where the superscripts $\pm$ relate to the charge of the \Wboson boson. By definition, 
$f_i>0$ and $f_L+f_R+f_0 = 1$ must hold. The parameters $f_i$ are not expected 
to be the same for $W^+$ and $W^-$ in proton-proton collisions, as the ratio of 
valence $u$ quarks to sea quarks is higher than for valence $d$ quarks. Hence 
different angular distributions are expected. 

The choice of an appropriate reference frame is not trivial, as the \Wboson 
boson rest frame can be only defined by the full four-momentum information of 
both decay leptons. While the four-momentum information of the muon is present, 
only the transverse momentum of the neutrino can be measured. In 
principle, its longitudinal component can be determined through the \Wboson mass 
constraint. However, the corresponding equations lead to two possible solutions 
thus an unambiguous choice is not possible. Therefore, the measurement is 
based on a highly correlated variable $\cos\theta_{2D}$, defined as

\[\cos\theta_{2D} = \frac{\vec \pT^{l*} . \vec \pT^W}{|\vec p^{l*}_T| \cdot 
|\vec \pT^W|}\, ,\]

\noindent where $\vec \pT^{l*}$ is the transverse momentum of the lepton in the 
transverse \Wboson boson rest frame and $\vec \pT^W$ is the transverse momentum of 
the \Wboson boson in the laboratory frame. The observable $\cos\theta_{2D}$ can be 
interpreted as the 2-dimensional projection of  $\cos\theta$ on the 
transverse plane. The helicity fractions can then be determined by 
fitting the measured $\cos\theta_{2D}$ distributions with a weighted sum 
of templates obtained from the simulations. Each template corresponds to one 
helicity state and is weighted by the corresponding $f_i$ value. 

The ATLAS measurement of the polarisation of \Wboson bosons is based on $\IntLumi = 
35\ipb$ of the 2010 data sample \cite{ATLAS:2012au}. The selection of signal events and the 
background determination methods are similar to the inclusive measurement. In 
addition it is required that the transverse \Wboson boson mass is within a range of 
$50\,\GeV < \mT^W < 110 \,\GeV$. The upper cut was chosen to reject badly 
reconstructed jets. In order to enhance the polarisation effects, two regions 
for events with a large transverse momentum of \Wboson boson have been defined as 
$35<\pT^W<50\,\GeV$ and $\pT^W>50\,\GeV$. The MC prediction of the 
$\cos\theta_{2D}$ templates have been obtained independently from \MCAtNLO~and 
\Powheg. The actual fit was performed using a binned maximum likelihood fit 
based on \MCAtNLO~templates. The templates predicted by \Powheg\ have been used 
for estimating systematic uncertainties. Since only two of the three parameters 
$f_i$ are independent, it was chosen to measure $f_0$ and $f_L-f_R$. The 
results for $\pT^W>50\,\GeV$ averaged over both decay channels and both 
charges are shown in Figure \ref{fig:ATLASWPol} together with the expectations from simulations, which agree 
with the measurement. The systematic uncertainties are dominated by experimental
uncertainties on the measured $\pT^W$.

The CMS measurement is also based on the 2010 data sample \cite{Chatrchyan:2011ig} and relies on a 
binned maximum likelihood for to the $\cos\theta_{2D}$ variable. The 
fitting templates for the different helicity states are based on the \MadGraph\  
generator. In contrast to the inclusive measurements, CMS requires 
in this case a cut on the transverse mass of $\mT>50\,\GeV$ for the electrons and of 
$\mT>30\,\GeV$ for the muon channel, in order to reduce the QCD multi-jet
background. In addition, also a minimal $\pT^W>50\,\GeV$ is required, similar 
to the ATLAS analysis. The background from \ttbar~is reduced by vetoing events with more 
than three reconstructed jets with a $\pT>30\,\GeV$. The resulting fit values 
for $f_0$ and $f_L-f_R$ are shown in Figure \ref{fig:CMSWPol}, independently for both \Wboson 
boson charges. 

\begin{figure*}
\begin{minipage}{0.53\textwidth}
\resizebox{0.5\textwidth}{!}{\includegraphics{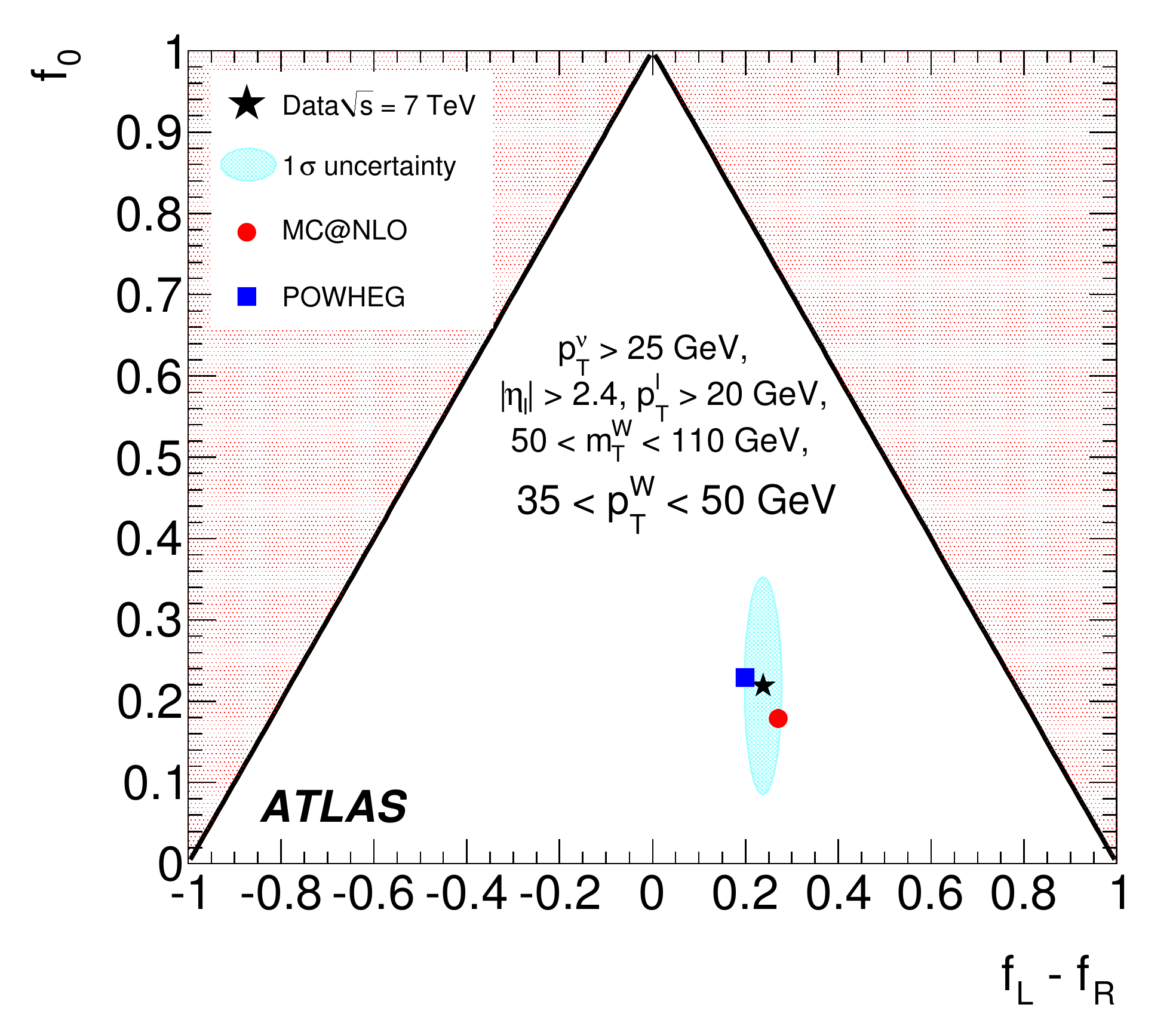}}
\resizebox{0.5\textwidth}{!}{\includegraphics{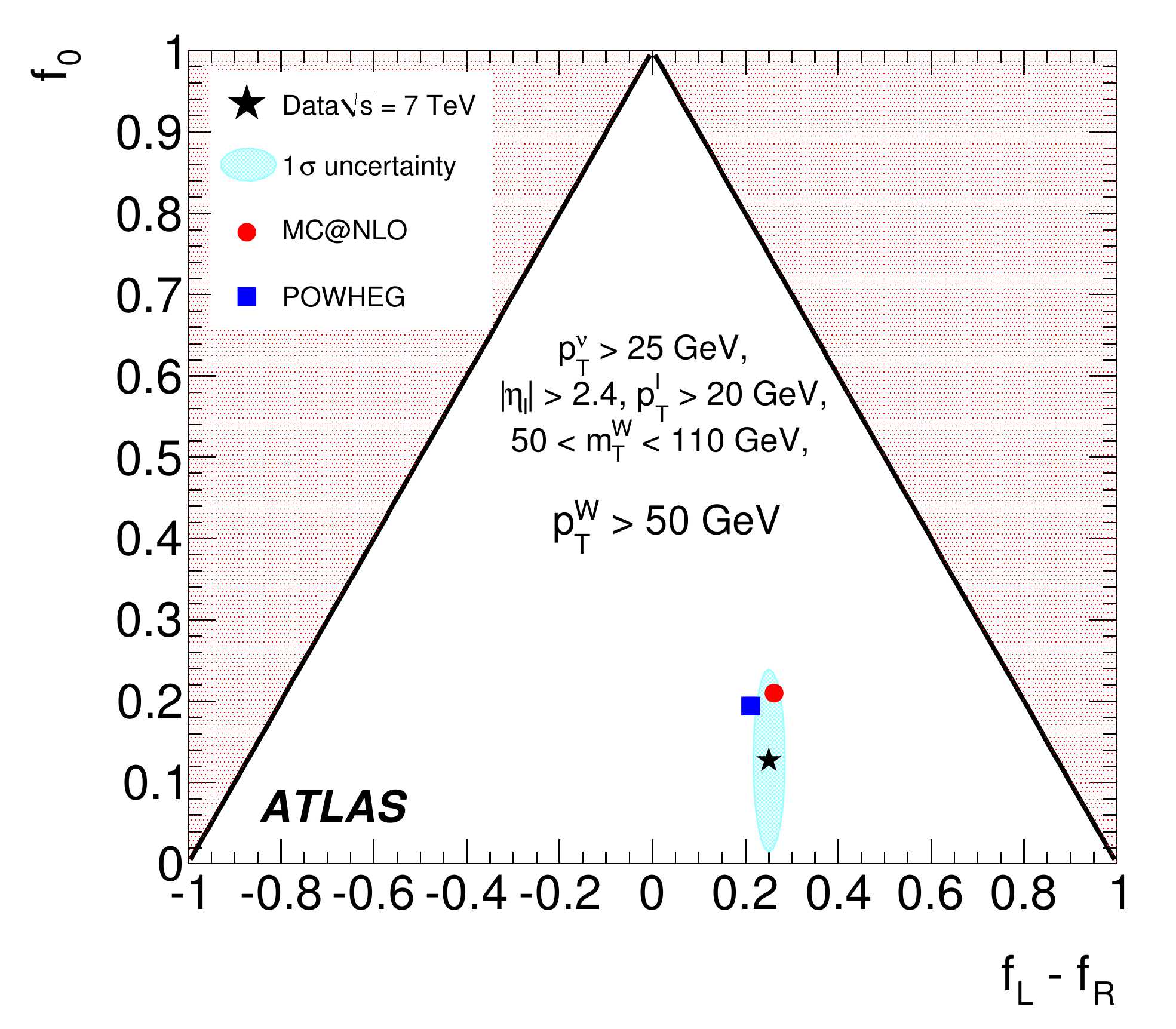}}
\caption{\label{fig:ATLASWPol} ATLAS \cite{ATLAS:2012au}: Measured values of $f_0$ and $f_L - f_R$, after corrections, within acceptance cuts for $35 < \pT^W < 50 \GeV$ (left) and $50\,\GeV < \pT^W$ (right) compared with the predictions corresponding to \MCAtNLO~ and \Powheg. The ellipses around the data points correspond to one standard deviation, summing quadratically the statistical and systematic uncertainties. The forbidden region is hatched. }
\end{minipage}
\hspace{0.2cm}
\begin{minipage}{0.45\textwidth}
\resizebox{0.5\textwidth}{!}{\includegraphics{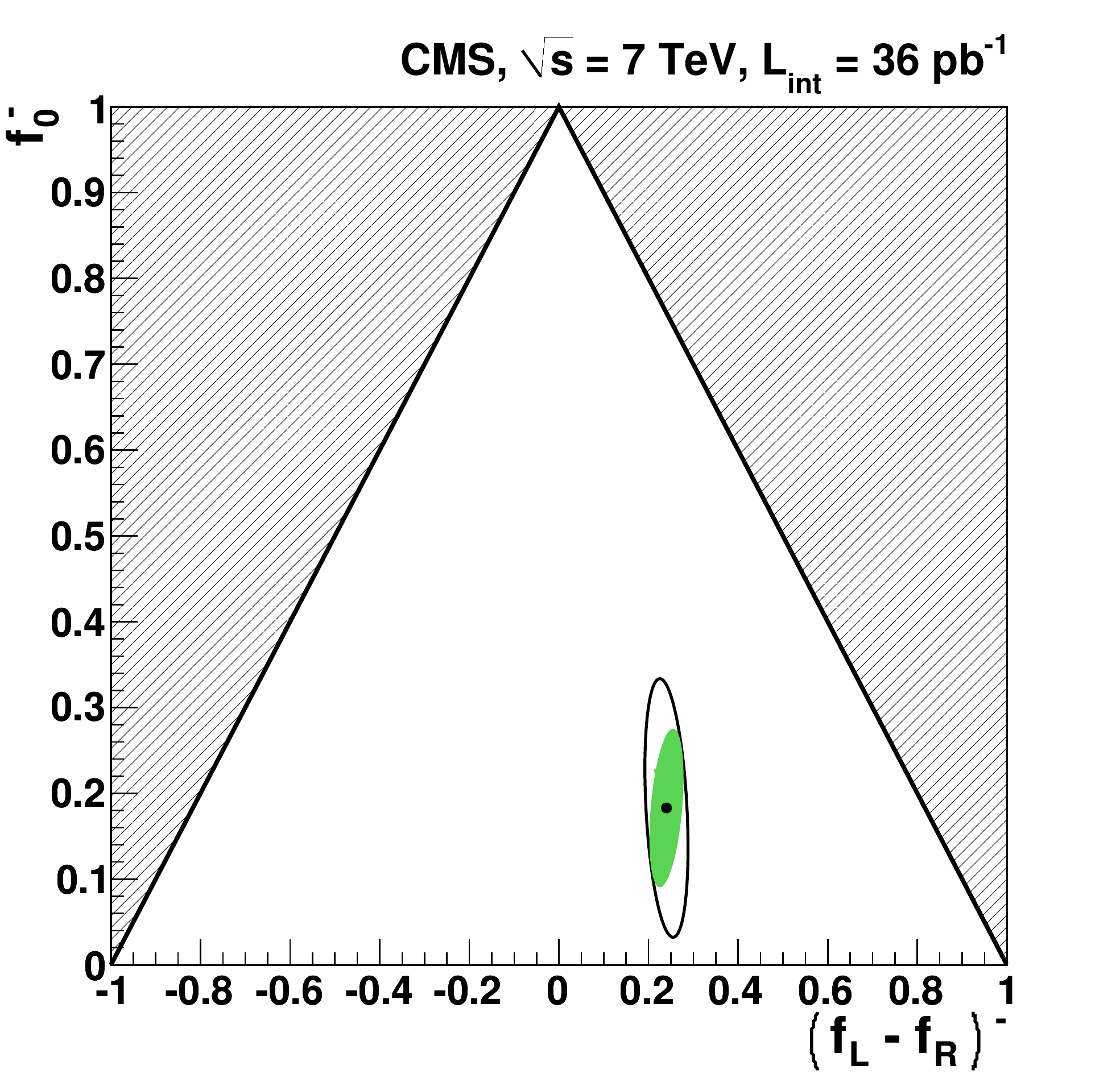}}
\resizebox{0.5\textwidth}{!}{\includegraphics{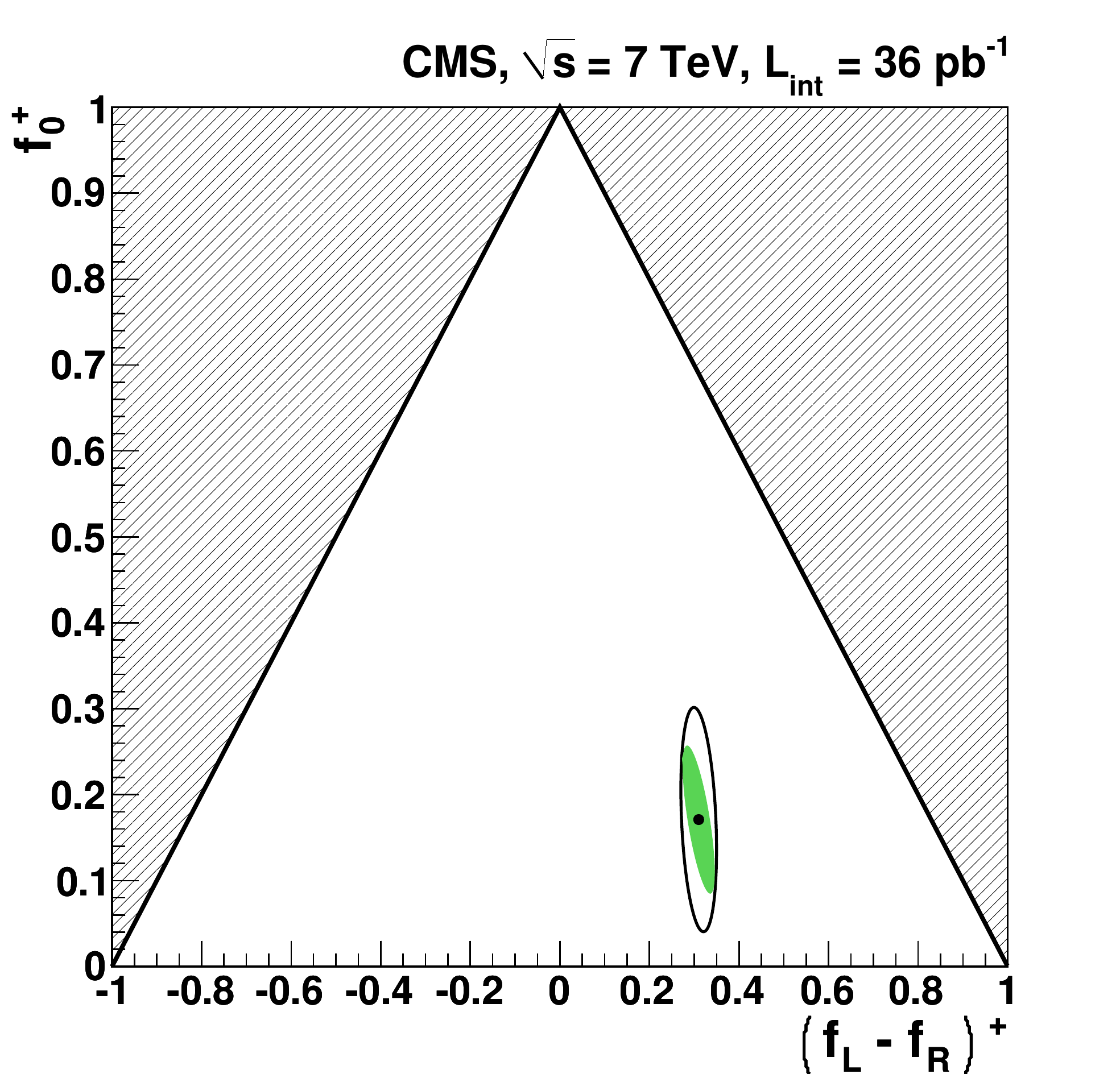}}
\caption{\label{fig:CMSWPol}CMS \cite{Chatrchyan:2011ig}: The muon fit result (black dot) in the $[(f_L-f_R), f_0]$ plane for negatively charged leptons (left) and positively charge leptons (right). The 68\% confidence level contours for the statistical and total uncertainties are shown by the green shaded region and the black contour, respectively. The forbidden region is hatched.}
\end{minipage}
\end{figure*}

Both measurements show a clear difference between the left- and right-handed 
polarisation parameters in proton-proton collisions and are compatible with the 
Standard Model expectations.

\section{Vector Boson Production in Association with Jets}
\label{sec:VjetsMeasurements}

The cross-sections of \Wboson and \Zboson production for different jet multiplicities are an important measure for NLO and MC predictions but the exploratory 
power of measurements of $W$+jets and $Z$+jets production today lies in the precision tests of differential distributions. 
The cross-section as a function of the jet \pT~for example is a sensitive test of the scale used in $\alpha_s$ calculations whereas the 
rapidity distributions of the jets is a sensitive test to different PDF sets. Studying the angular distributions between the jets, 
such as the rapidity differences between two jets, tests hard parton radiation at large angles. Previous publications from CDF and D0
have measured the differential cross-sections as a function of the jet \pT, the boson \pT, the angular separation between jets 
as well as other observables~\cite{Abazov:2009av, Aaltonen:2007ae, Abazov:2009pp, Aaltonen:2007ip, Abazov:2013gpa}. However, the kinematic reach of these 
measurements compared to that at the LHC is limited. For example, at the Tevatron using an integrated luminosity of $\IntLumi \approx 1.7 \ifb$, 
jets up to a \pT~of 400 \GeV~can be measured in $Z$+jets events, while at the LHC using of integrated luminosity of $\IntLumi \approx 4.6 \ifb$, jets 
up to a \pT~of 700 \GeV~can be studied. 
These large data samples at the LHC, allow us to make precision measurements over large
regions of the phase space. In addition, we can measure in detail specific topologies, like $Z$+jets production 
where the $\pT^Z$ is greater than 100 \GeV~or measure differential cross-sections for rare processes, like $Z+ \bbbar$ production.  

In addition, the last five years have been an `NLO revolution'. At the time of the LHC turn-on, NLO calculations up to two 
associated jets were available, while today an NLO calculation up to six associated jets can be achieved~\cite{Berger:2008sj}. 
In addition NLO calculations for \Wboson and \Zboson  
production associated with heavy-flavour jets have also expanded greatly in recent years. 

\subsection{Measurements of $W$+jets and $Z$+jets cross-sections}

At the LHC, $W$+jets and $Z$+jets production is dominated by quark-gluon interactions making these measurements different from 
measurements of the QCD multi-jet process, which is dominated by gluon-gluon interactions. In MC calculations, 
associated jets to \Wboson and \Zboson production can arise either from the matrix-element calculation itself or from quarks or 
gluons in the parton showering. Jets from the matrix element tend to have a higher \pT~compared to those from the parton shower and 
therefore not including multiple partons in the matrix element will result in an underestimate of the jet multiplicity cross-sections. 
For an excellent review on jets and their properties at hadron colliders, see~\cite{Salam:2009jx}.

Both ATLAS and CMS have performed measurements of \Zboson production in association with jets. The ATLAS and CMS results are based 
on an integrated luminosity of $\IntLumi \approx 4.6 \ifb$~\cite{Aad:2013ysa} and $\IntLumi \approx 4.9 \ifb$~\cite{CMS:new-zjets}, respectively. 
Both results measure the jet multiplicity cross-sections up to seven  
associated jets. Shown in Figure~\ref{fig:ATLAS-zjets-njet} for the ATLAS measurement, several theory predictions are compared to the data. The 
\Alpgen~and \Sherpa~generators both include matrix-element calculations that cover up to five partons, with additional jets coming 
from the parton showering. \MCAtNLO\ generates the Drell-Yan process at NLO and includes the real emission of one 
additional parton and any additional jets from the parton showering. The \Blackhat~results provide fixed-order 
calculations at NLO for up to four jets. The data is in excellent agreement with \Blackhat~and \Sherpa~predictions. \Alpgen~
tends to underestimate the data above five jets, as expected, since above five partons all additional jets in the predictions 
originate from the parton shower. While \MCAtNLO~agrees with the data for zero-jet and one-jet events, its parton shower model 
underestimates the observed jet rate by a factor of two. The CMS results are also in good agreement with the \Blackhat~predictions as well as the \MadGraph~predictions, 
which includes matrix-element calculations up to four partons and the \PowhegBox~predictions, which is a NLO calculation for one jet.
The dominant experimental uncertainty in both of these measurements is the uncertainty on the jet energy scale, while the 
dominant theory uncertainty is due to the scale uncertainties. 

\begin{figure}
\resizebox{0.5\textwidth}{!}{\includegraphics{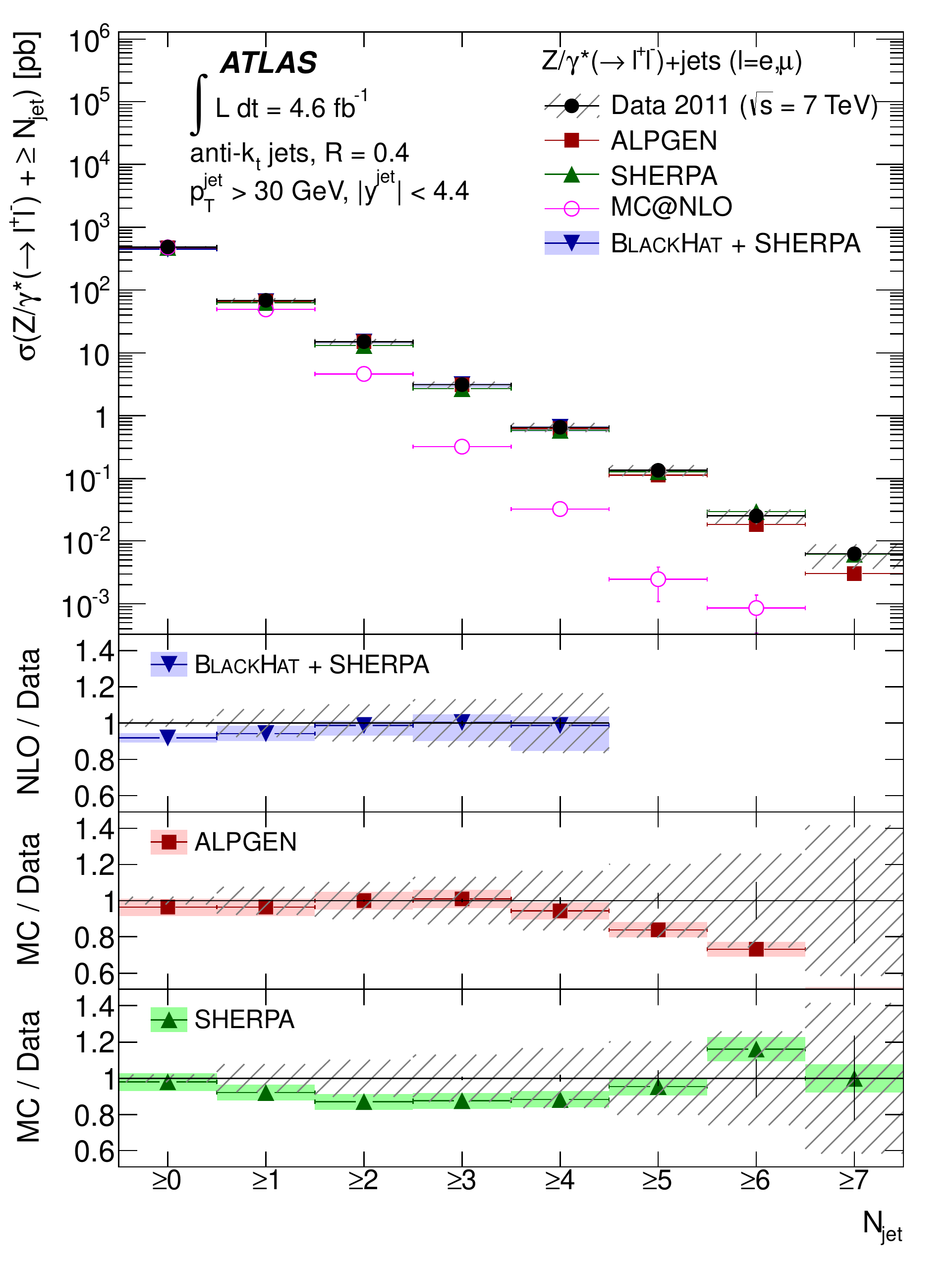}}
\caption{ATLAS~\cite{Aad:2013ysa}:  Measured cross-section for $Z$+jets as a function of the number of jets. The data are compared to 
\Blackhat, \Alpgen, \Sherpa~and \MCAtNLO.}
\label{fig:ATLAS-zjets-njet}
\end{figure}

Taking the ratio of jet multiplicity cross-section allows for many of the experimental systematic uncertainties to cancel, 
thereby improving the precision of the measurement. The ratios, $R_{(n+1)/n}$, shown in Figure~\ref{fig:ATLAS-zjets-njetratio} 
for the ATLAS measurement, exhibit a constant or staircase scaling pattern, as derived in Equation~\ref{eqn:starcaise}. 
As described in Section~\ref{sec:QCDCorrections}, basic quantum field theory would predict a Poisson scaling of 
$R_{(n+1)/n}$ due to successive 
gluon radiation from an energetic quark. At higher jet multiplicities, though, a constant value of $R_{(n+1)/n}$ is 
expected due to the non-abelian nature of QCD final-state radiation, i.e. a final-state gluon can radiate an additional gluon. At low 
multiplicities this constant value is due to a combination of Poisson scaling and parton density suppression, where the 
emission of the first parton has a stronger suppression than any additional parton. This Poisson scaling can be 
recovered if the scale difference between the main process (such as Z+1-jet events) and the $\pT$ of the second leading jet is large~\cite{Englert:2011pq}.
When requiring the leading jet to have $\pT>150\,\GeV$ 
and all other additional jets to have a $\pT>30\,\GeV$, the ratio changes dramatically (Figure~\ref{fig:ATLAS-zjets-njetratio}, 
right) and the Poisson scaling is clearly seen. The theory predictions track this trend and are all in good 
agreement with the data. 

\begin{figure*}
\resizebox{0.5\textwidth}{!}{\includegraphics{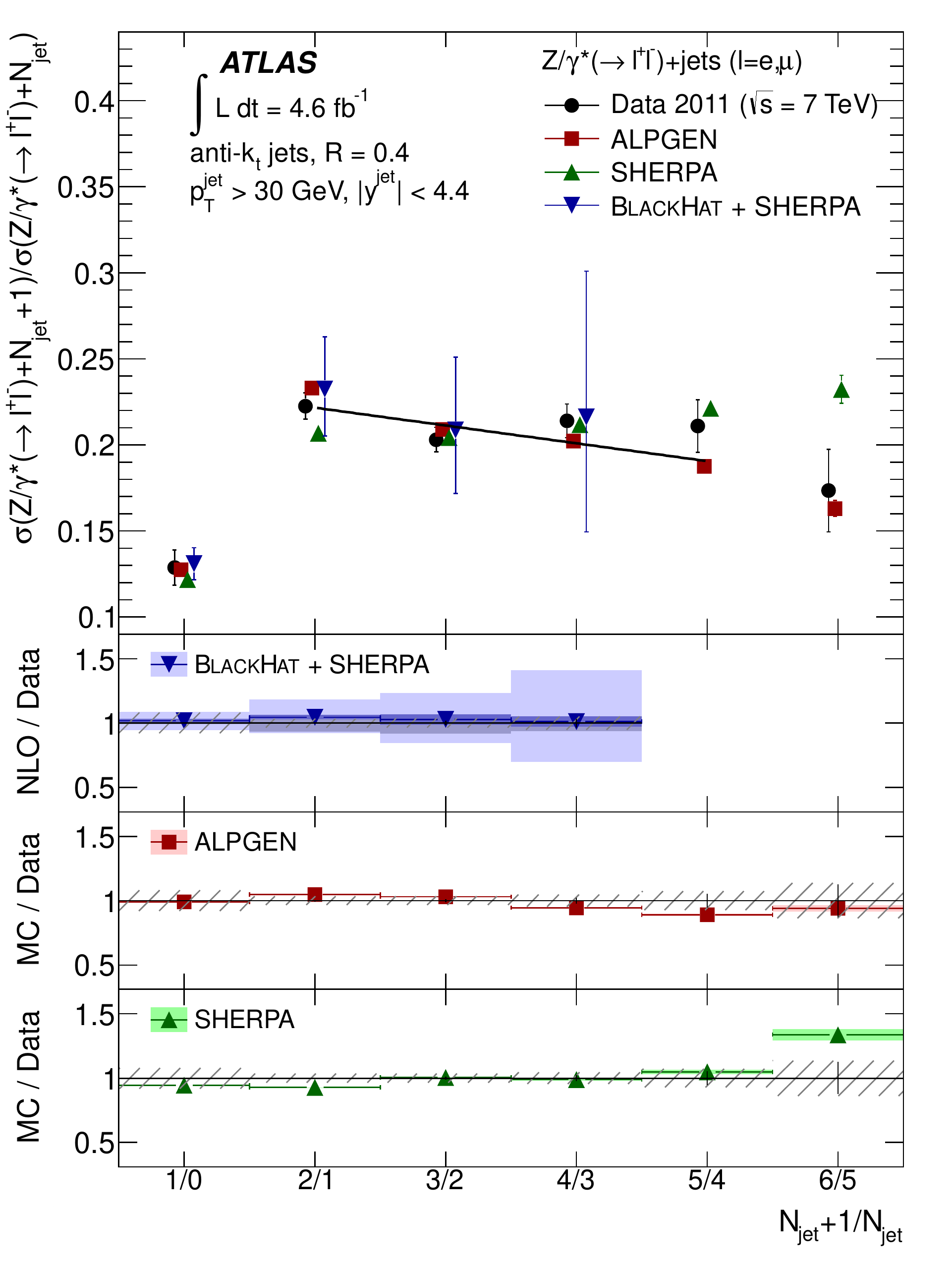}}
\resizebox{0.5\textwidth}{!}{\includegraphics{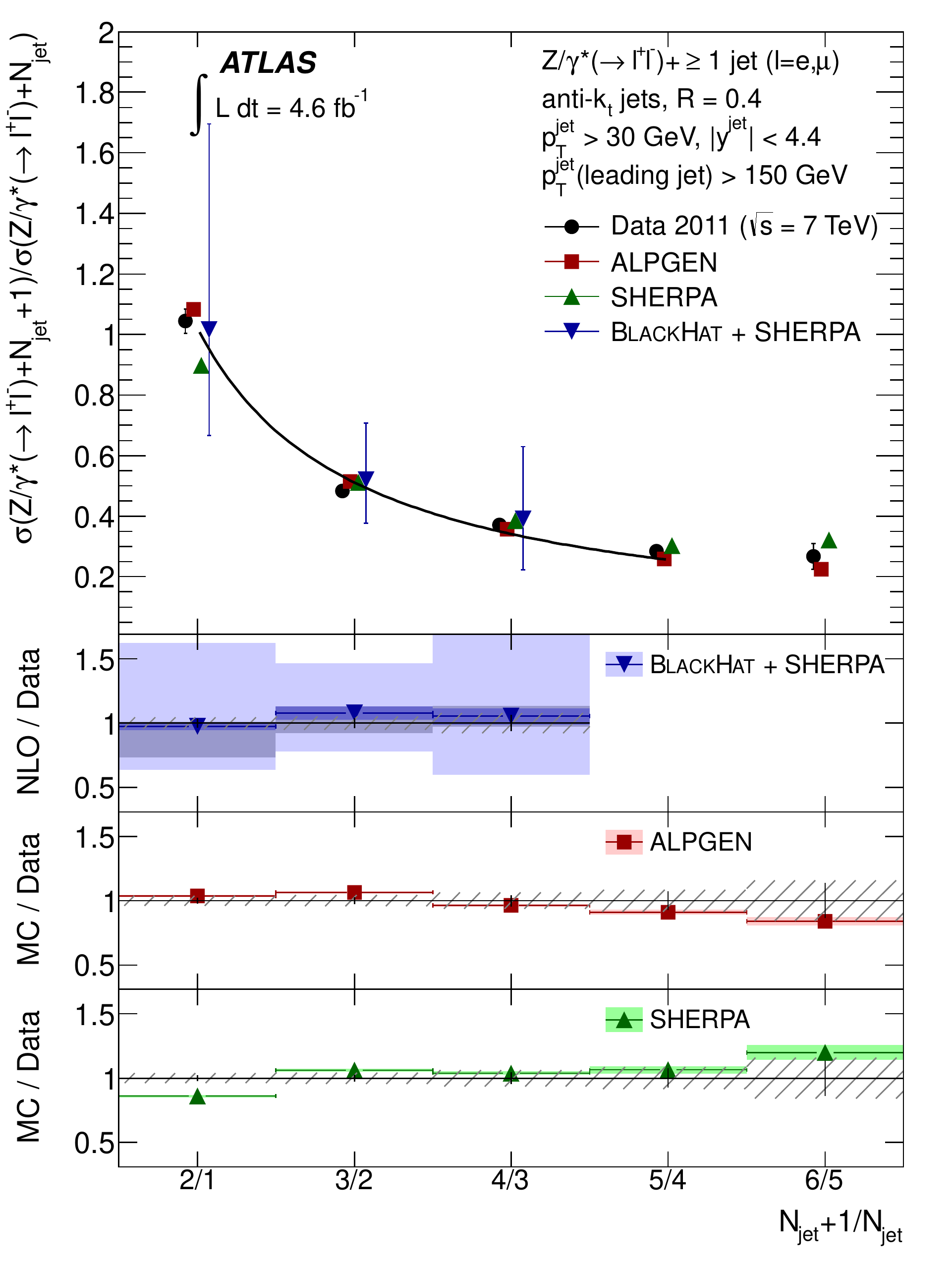}}
\caption{ATLAS~\cite{Aad:2013ysa}:  Measurements of the ratio of cross-sections for successive exclusive jet multiplicities. The data are compared 
to \Blackhat, \Alpgen~and \Sherpa~predictions. Left: The standard jet selection of $\pT>30\,\GeV$ is used. 
Right: The leading jet is required to have $\pT>150\,\GeV$ while all other jets must have $\pT>30\,\GeV$. }
\label{fig:ATLAS-zjets-njetratio}
\end{figure*}

Similarly, using a smaller data sample of $\IntLumi \approx 35 \ipb$, a CMS analysis~\cite{Chatrchyan:2011ne} tested the Berends-Giele scaling hypothesis which similarly states that the ratio can be 
described as a constant. Since phase space effects can modify this ansatz slightly, a linear function is used $C_n= \alpha 
+ \beta n$. The Berends-Giele scaling is confirmed to describe the data for events with up to four jets. 

Measurements of \Wboson production in association with jets has also been performed by ATLAS~\cite{Aad:2012en} using an integrated 
luminosity of $\IntLumi \approx 35 \ipb$ and by CMS~\cite{CMS:new-wjets} using an integrated luminosity of $\IntLumi \approx 5.0 \ifb$. The CMS results, 
preformed using only $W \rightarrow \mu \nu$~events, measured the cross-section with up to 6 associated jets (Figure~\ref{fig:ATLAS-wjets-njet}). 
Similar to the $Z$+jets production, the data is in good agreement with the \Blackhat, \MadGraph\ and \Powheg\ predictions. The Berends-Giele scaling, 
measured by both ATLAS and CMS~\cite{Chatrchyan:2011ne} at $\IntLumi \approx 35 \ipb$, also describes well the $W$+jets data up to 4-jets.


\begin{figure}
\resizebox{0.5\textwidth}{!}{\includegraphics{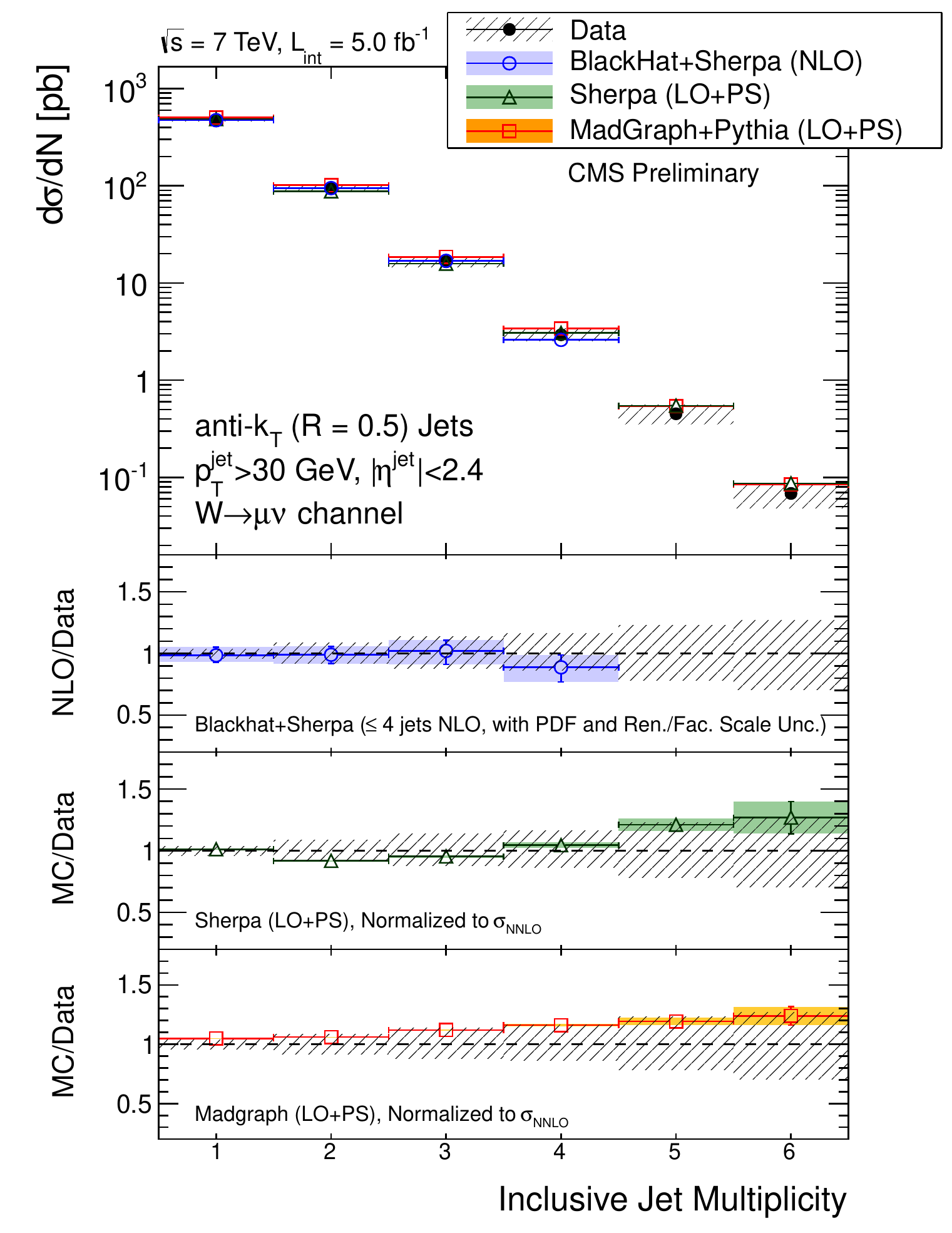}}
\caption{CMS ~\cite{CMS:new-wjets}:  Measured cross-section for $W$+jets as a function of the number of jets. The data are compared to 
\Blackhat, \Sherpa~and \MadGraph.}
\label{fig:ATLAS-wjets-njet}
\end{figure}

Using a common phase space, we summarise in Figure~\ref{fig:comp-wjet-zjet} the $W$+jets and $Z$+jets results from ATLAS and CMS.
For the $W$+jets results the common phase space is one lepton with a \pT $>20 \GeV$ and $|\eta| < 2.5$, a neutrino from the \Wboson decay with \pT$>25$ and the transverse mass of the \Wboson of greater than 40 \GeV. For the $Z$+jets results the phase space is defined as two leptons each with a \pT $>20 \GeV$ and $|\eta| < 2.5$ and invariant mass of $ 66 \GeV < m_{ll} < 116 \GeV$. For both \Wboson and \Zboson production, the jets are defined using an anti-$k_T$ algorithm with a distance parameter of $R=0.5$, a $\pT > 30 \GeV$ and a rapidity less than 2.4. The correction factors applied to both the ATLAS and CMS results are derived from \Sherpa~and range from 1\% to 23\%. For all jet multiplicities, the correction factors are smaller than the experimental systematic uncertainties.  

\begin{figure*}
\resizebox{0.5\textwidth}{!}{\includegraphics{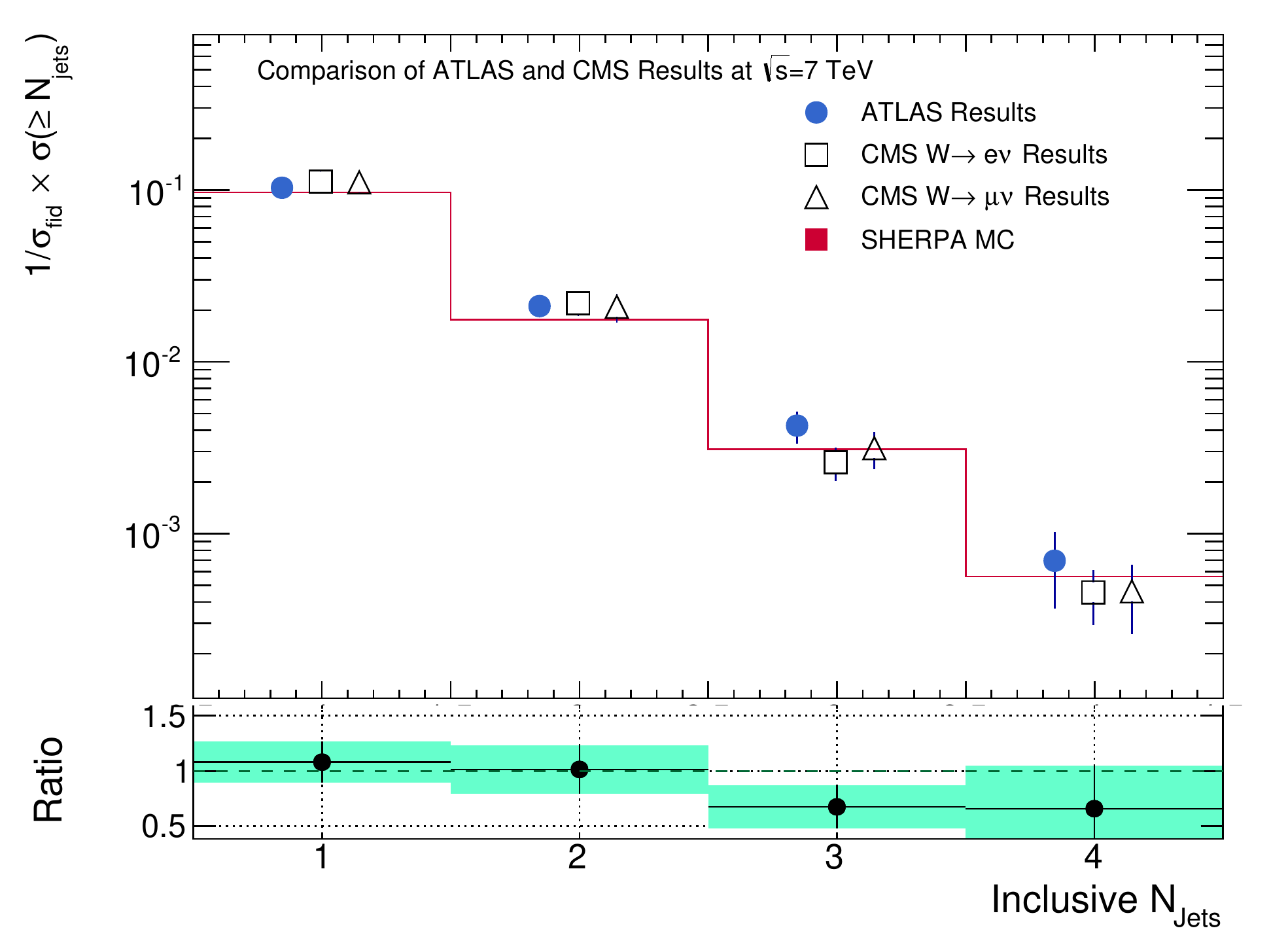}}
\resizebox{0.5\textwidth}{!}{\includegraphics{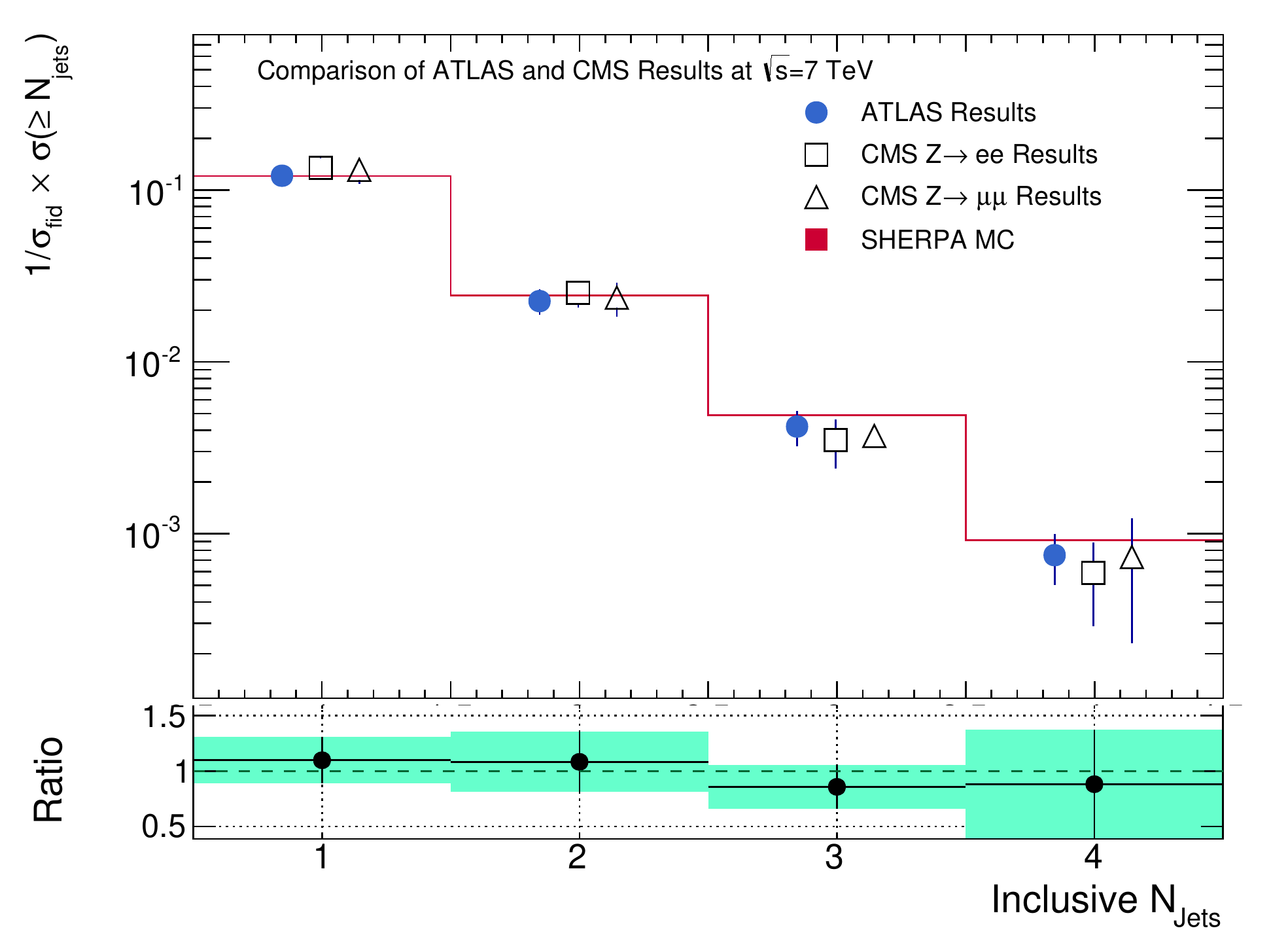}}
\caption{Summary of the $W$+jets (left) and $Z$+jets (right) cross-sections as a function of the number of jets. Both the ATLAS~\cite{Aad:2012en, Aad:2013ysa} and CMS~\cite{Chatrchyan:2011ne} results have been corrected to a common phase space as described in the text. The cross-sections for each jet multiplicity have been normalised by the inclusive \Wboson or \Zboson cross-section. The CMS results shown here use an integrated luminosity of $\IntLumi \approx 35 \ipb$, as the most recent CMS results are not yet published. The ratio between the ATLAS and combined CMS results are shown below. Predictions from \Sherpa~are also shown. }
\label{fig:comp-wjet-zjet}
\end{figure*}

In summary measurements of the cross-sections of \Wboson and \Zboson production in association with jets are in excellent agreement with the predictions. For measurements of the cross-section ratios of $R_{(n+1)/n}$ the experimental uncertainties are much smaller compared to those from the theory predictions. Future measurements should therefore focus on differential measurements of the cross-section to further test perturbative QCD theory. 

\subsection{Differential $W$+jets and $Z$+jets measurements}
\label{sec:wz-jet-diff}

Differential measurements of the properties of the jets in $W$+jets and $Z$+jets events probe not only perturbative QCD theory, 
but are also sensitive to renormalisation scales, PDFs and hard parton radiation at large angles. For these measurements, ATLAS and CMS have 
two major advantages. First, with large data samples very high jet \pT~and scalar sum scales can be probed. Second, the 
detectors can measure jets at large rapidities. Both the high \pT~and large rapidity jet phase spaces 
have not been extensively measured in the past. 

The differential cross-section of both \Wboson or \Zboson events as a function of $H_T$ is of particular interest. In the CMS 
measurements $H_T$ is defined as the scale sum of all jets passing the selection criteria. 
In many fixed-order calculations, $H_T$ is often used as the value of the renormalisation and 
factorisation scales. It is also an observable that is very sensitive to missing higher-order terms in theoretical predictions 
as well as an observable which is often used in searches for new physics. The effect of missing higher-order terms in the 
predictions can readily be seen in Figure~\ref{fig:CMS-wjets-ht}, which shows the differential cross-section as a function 
of $H_T$ for $W+\ge 1$-jet events. At large values of $H_T$, the NLO \Blackhat~predictions underestimate the data. 
This is because of the limited order of the \Blackhat~calculations which do not include matrix-element 
calculations of three or more real emissions. Modifying \Blackhat~to include higher-order NLO terms to the 
$N_{jet} \ge 1$ predictions yields good agreement to the data~\cite{Aad:2012en}. The \Sherpa~predictions shows better agreement to the data, 
compared to \MadGraph. 


\begin{figure}
\resizebox{0.5\textwidth}{!}{\includegraphics{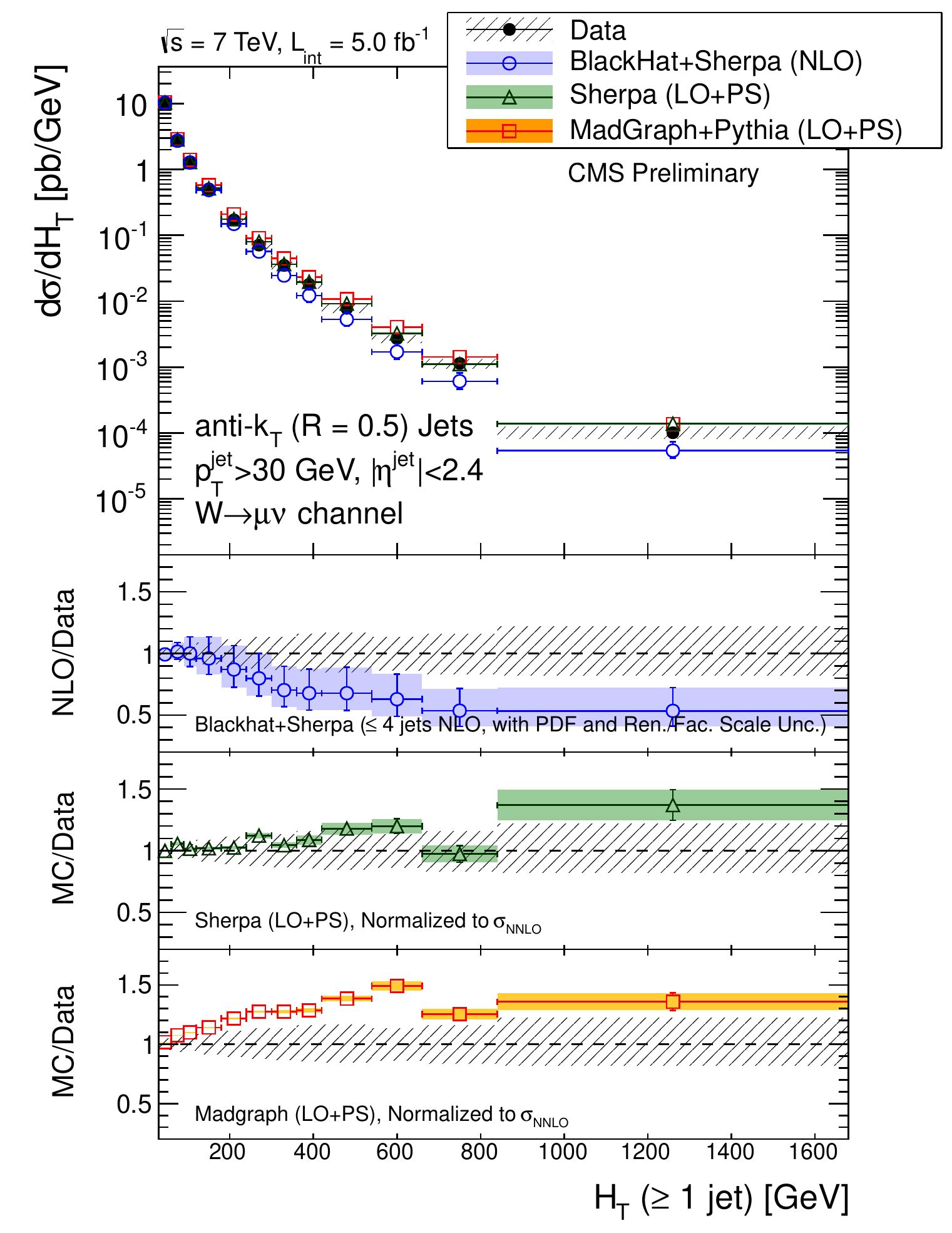}}
\caption{CMS ~\cite{CMS:new-wjets}:  Differential cross-section as a function of $H_T$ for $W+ \ge 1$-jet events. Predictions are shown for \Sherpa, \MadGraph~and \Blackhat. }
\label{fig:CMS-wjets-ht}
\end{figure}

The discrepancy between data and predictions in the $H_T$ distribution, which is attributed to missing higher jet 
multiplicities in fixed-order calculations, can be further investigated by comparing the average jet multiplicity as a function 
of the $H_T$. Figure~\ref{fig:ATLAS-zjets-ht} shows that for higher values of $H_T$\footnote{ATLAS measurements define $H_T$ as the scalar sum of all jets and leptons in the event.}, the average jet multiplicity increases. 
Therefore at large values of $H_T$, a fixed-order calculation for only $N_{jet}=1$, will not model correctly the data and 
agreement to the data can only be restored when including higher jet multiplicities. This conclusion is especially important 
for searches for new physics which rely on simulations to predict the number of \Wboson and \Zboson background events with large values of 
$H_T$. Using simulations with an insufficient number of partons in the final state will lead to an underestimate the 
number of \Wboson and \Zboson events at high values of $H_T$.

\begin{figure}
\resizebox{0.5\textwidth}{!}{\includegraphics{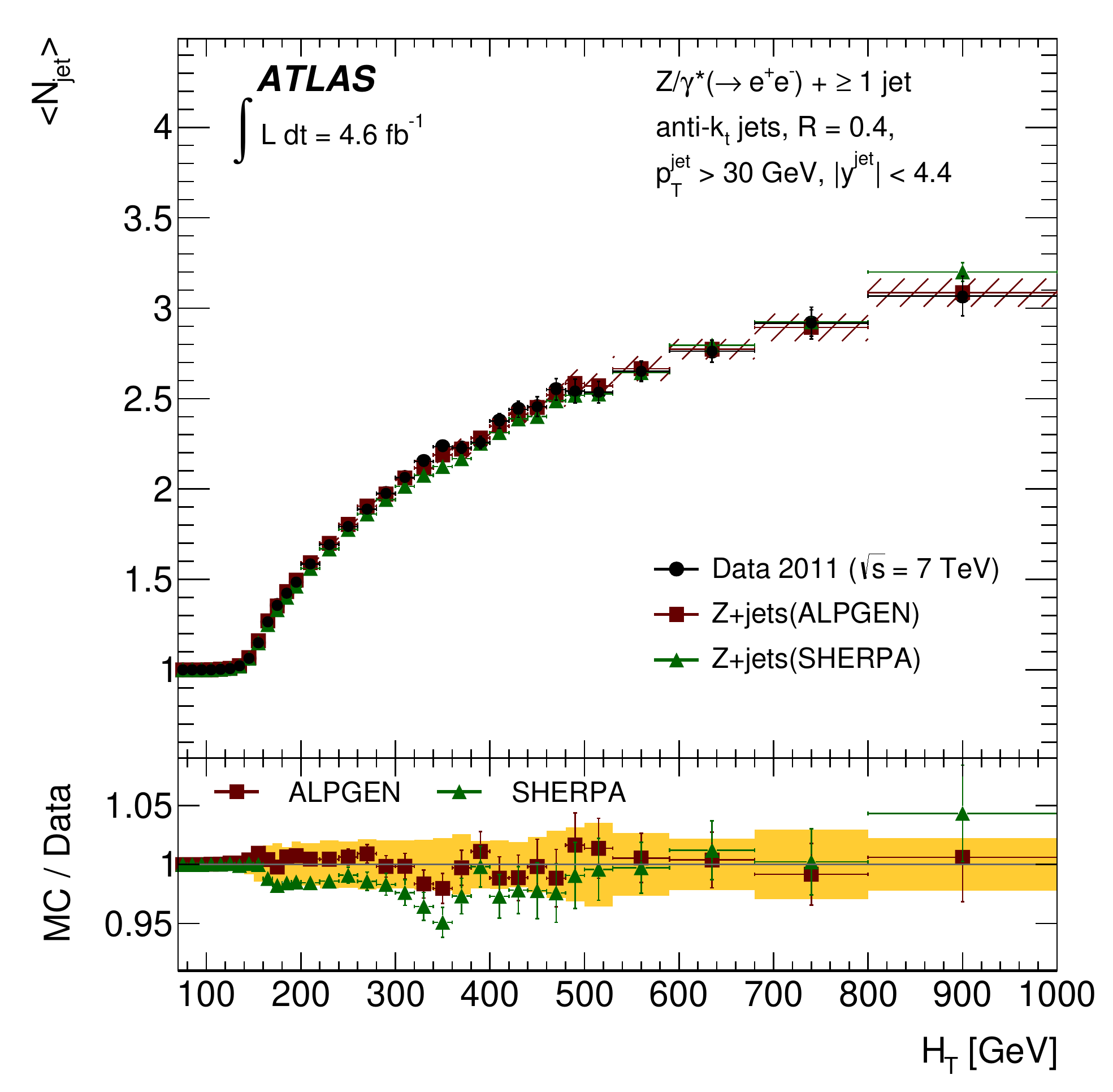}}
\caption{ATLAS~\cite{Aad:2013ysa}:  The average number of jets, $<N_{jet}>$, in $Z \rightarrow ee$ events with jets as a function of the $H_T$. 
\Alpgen~and \Sherpa~predictions are also shown.}
\label{fig:ATLAS-zjets-ht}
\end{figure}

When the \pT~of the jet is larger than the mass of the \Zboson or \Wboson boson, the NLO to 
leading-order correction factors become large due to QCD corrections, which are of the order $\alpha_s \ln^2 (\pT/m_Z)$. 
Also at high values of the jet \pT, the electroweak corrections, which are usually small compared to QCD corrections, can also reduce the 
cross-section by 5-20\% for $100\,\GeV < \pT^{ll} < 500\,\GeV$~\cite{Denner:2011vu}. The differential cross-section as a function of the 
leading jet \pT~for \Zboson events with $N_{jet} \ge 1$ measured by the CMS experiment is shown in Figure~\ref{fig:CMS-zjets-pt}. 
The experimental systematic uncertainties are smaller than those of the theoretical predictions. The NLO 
predictions from \Blackhat~are consistent with the data, while \Powheg\ tends to predict slightly harder jet spectra. 
\MadGraph~also slightly models incorrectly the shape of the data. 


\begin{figure}
\resizebox{0.5\textwidth}{!}{\includegraphics{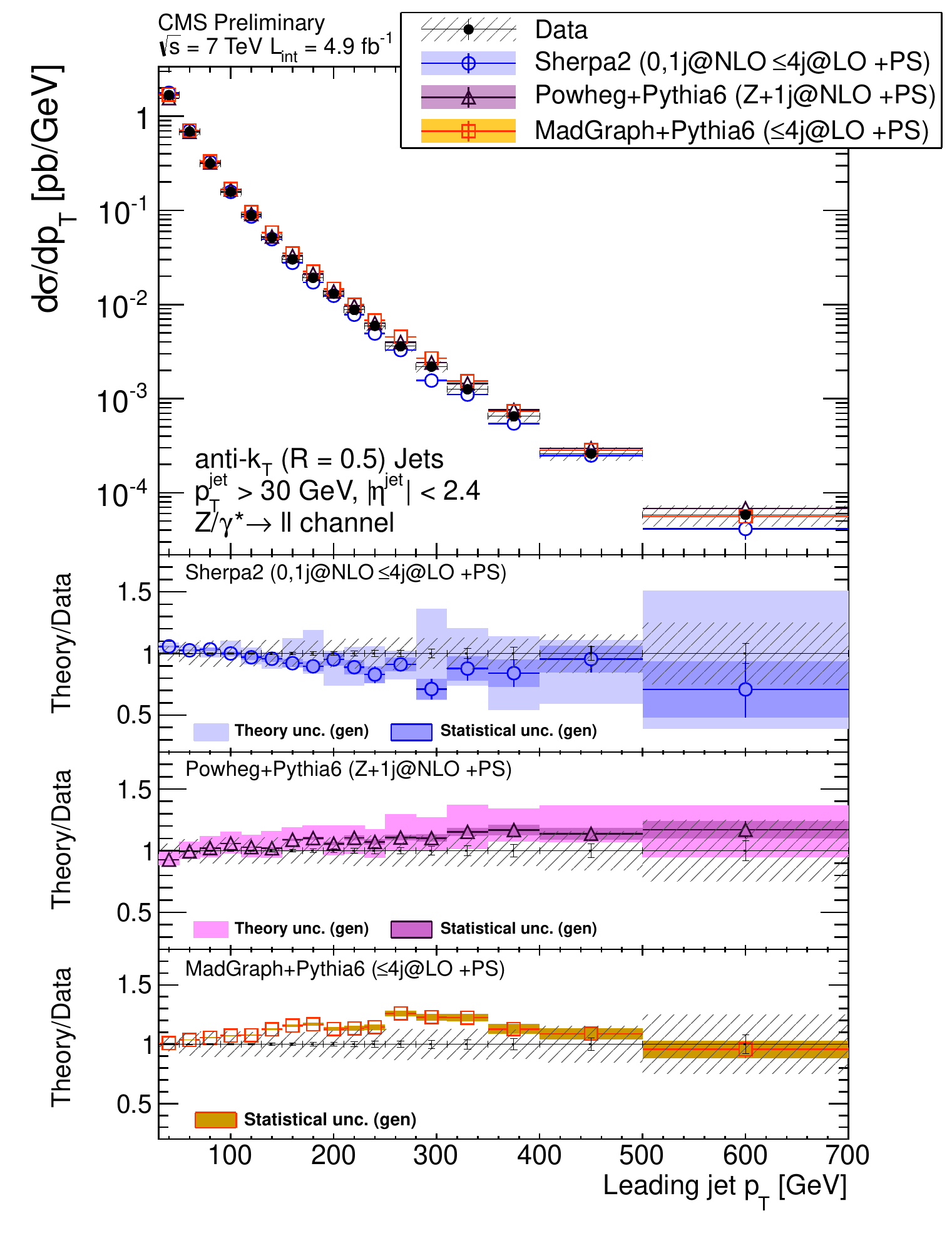}}
\caption{CMS~\cite{CMS:new-zjets}:  Differential cross-section as a function of the leading jet \pT~for $Z$+jets events. Predictions are shown for \Powheg, \MadGraph~and \Blackhat.}
\label{fig:CMS-zjets-pt}
\end{figure}

The differential cross-sections as a function of the \Zboson boson and jet rapidities for Z+1 jet events was performed by the CMS 
collaboration~\cite{Chatrchyan:2013oda}. Since $Z$+jets production involves a relatively high momentum valence quark and a low momentum gluon or 
quark, the \Zboson boson and jet are usually produced in the same end of the detector, which implies that the rapidity of the jet 
and the \Zboson boson in one-jet events is highly correlated. Measuring the rapidity sum, $y_{sum} = |y_z + y_{jet}|$, between the 
jet and the \Zboson boson is therefore sensitive to the PDFs while the rapidity difference $y_{dif} = |y_z - y_{jet}|$ is sensitive to 
the leading-order parton differential cross-section. These results are shown in Figure~\ref{fig:CMS-zjets-rap}. \Sherpa~models the 
data well, whereas the \MadGraph~and \MCFM~predictions are less consistent with the data. The differences between the 
different predictions is most clearly seen in the $y_{dif}$ distribution where \Sherpa~best models the data.  

\begin{figure*}
\resizebox{0.5\textwidth}{!}{\includegraphics{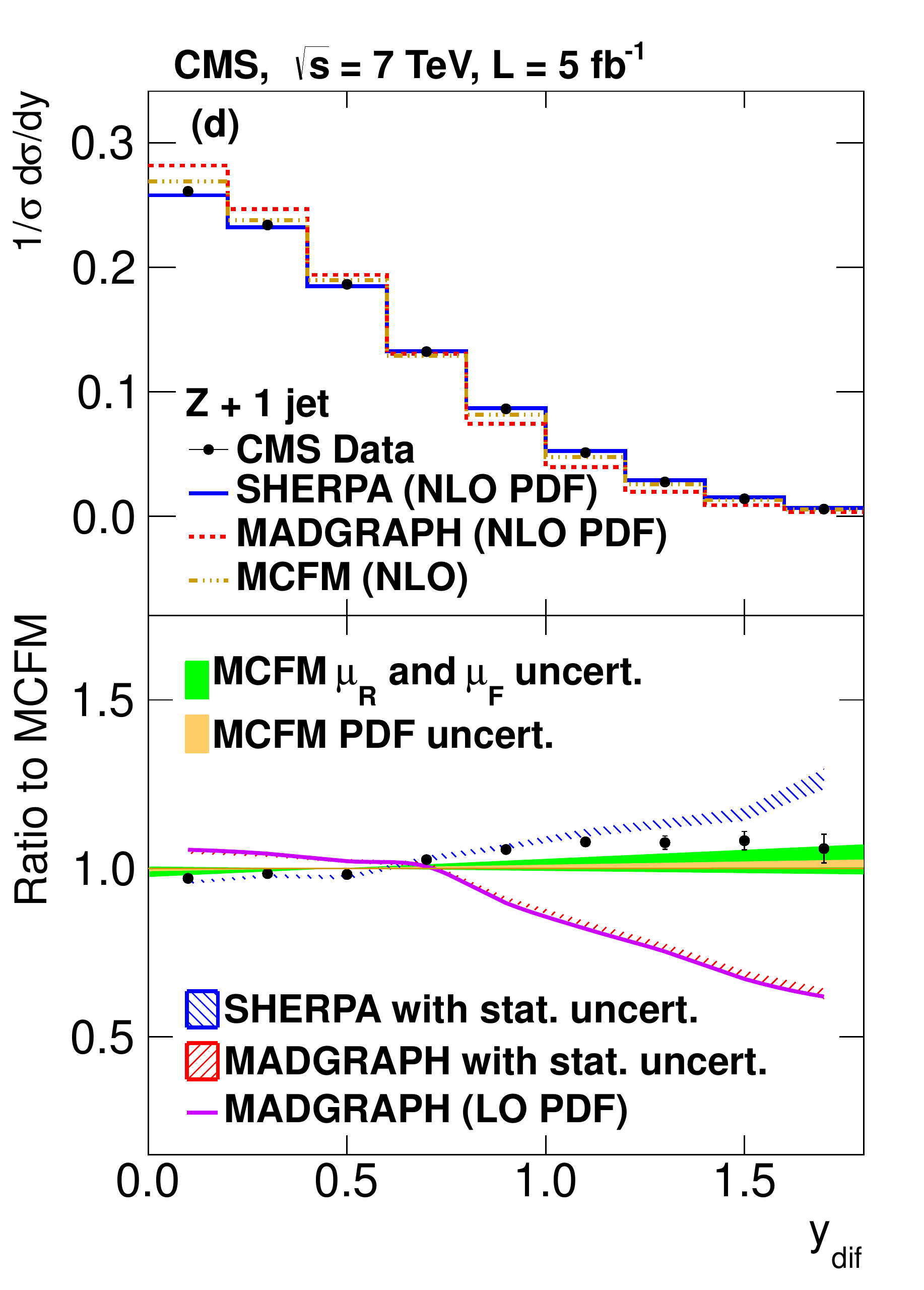}}
\resizebox{0.5\textwidth}{!}{\includegraphics{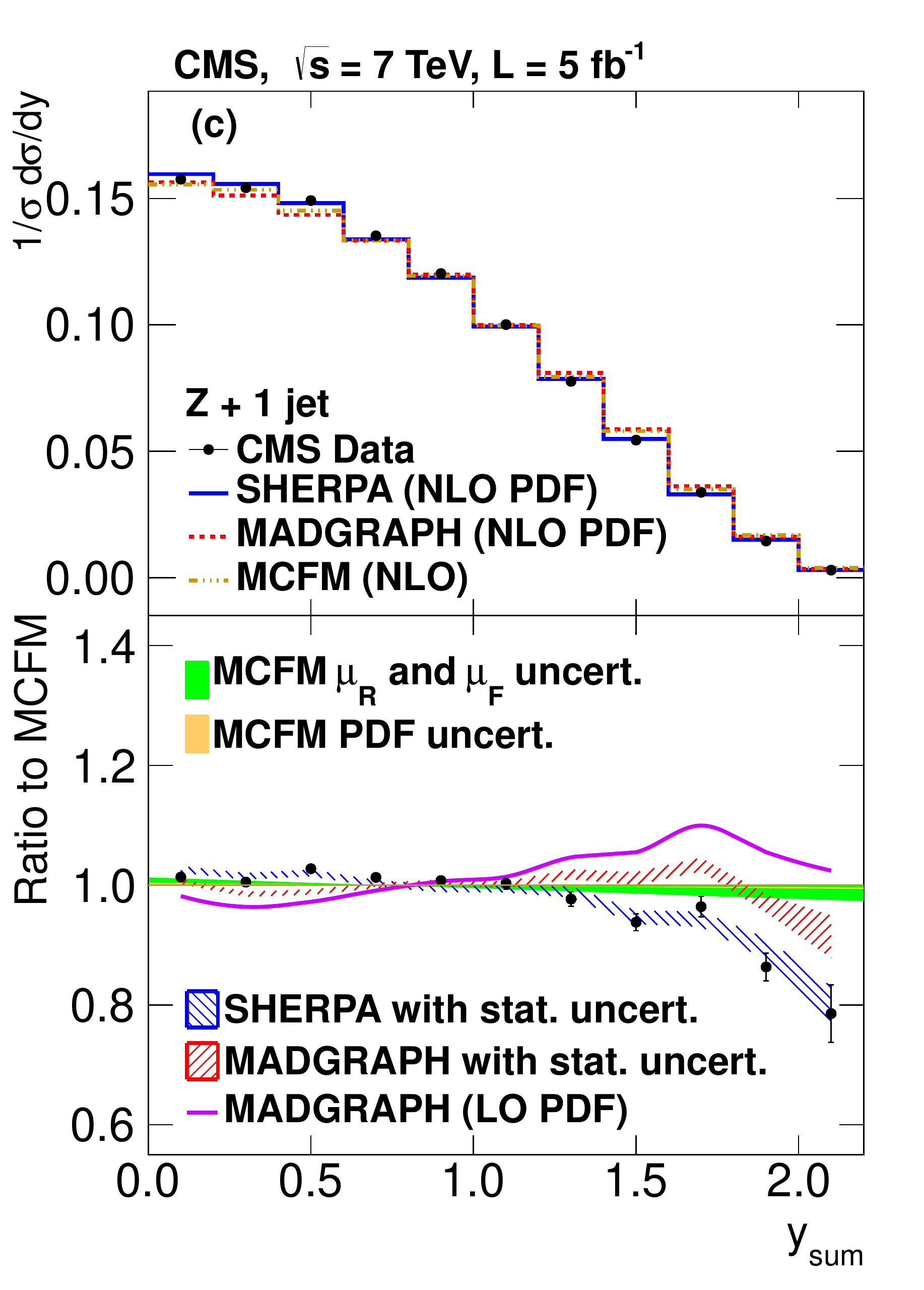}}
\caption{CMS~\cite{Chatrchyan:2013oda}:  Differential cross-sections as a function of rapidity sum, $y_{sum}$ (right) and rapidity difference, $y_{dif}$ 
between the \Zboson boson and the leading jet. The distributions are normalised to unity. The lower panel gives the ratios of data 
and the \Sherpa~and \MadGraph~simulations to the \MCFM~predictions. }
\label{fig:CMS-zjets-rap}
\end{figure*}

The large data samples of the LHC also allow for 
precision measurements at high scales. One topological observable of interest in many searches for new physics is the event 
thrust, which is defined as 

\[ \tau_T \equiv 1 - \rm{max} \frac{\sum_i | \vec{p_{T,i} } \cdot \vec{n_{\tau}} |}{\sum_i p_{T,i}} \, ,\]

\noindent where the index $i$ is over all jets and the \Zboson boson, $\vec{p_{T,i}}$ is the transverse momentum of object $i$ 
and $\vec{n_\tau}$ is the unit vector that maximises the sum. In events where the \Zboson and the jet are back-to-back, the 
thrust is zero. For events with additional jets that are isotropically distributed, the value of the thrust becomes larger. 
Traditionally the results are presented as $\ln \tau_T$, so that back-to-back events have a value approaching infinite and 
isotropic events have a value of $-1$. CMS measured the differential cross-section as a function of the thrust in two different 
phase spaces~\cite{Chatrchyan:2013tna}: an inclusive phase space with $\pT^Z > 0\, \GeV$ and a phase space region with $\pT^Z > 150\, \GeV$. 
Similar to measurements of the jet \pT, applying a cut on the $\pT^Z$ tests perturbative QCD theory in a region of phase space where the QCD corrections 
can be large. As seen in Figure~\ref{fig:CMS-zjets-thrust}, the predictions are within 10-15\% of the data in the inclusive 
phase space, except for \Pythia~which shows large deviations. In the phase space region with large $\pT^Z$, the agreement with \Pythia~
improves but both \Pythia~and \Sherpa~tend to predict more events in back-to-back topologies compared to the data. 

\begin{figure*}
\resizebox{0.5\textwidth}{!}{\includegraphics{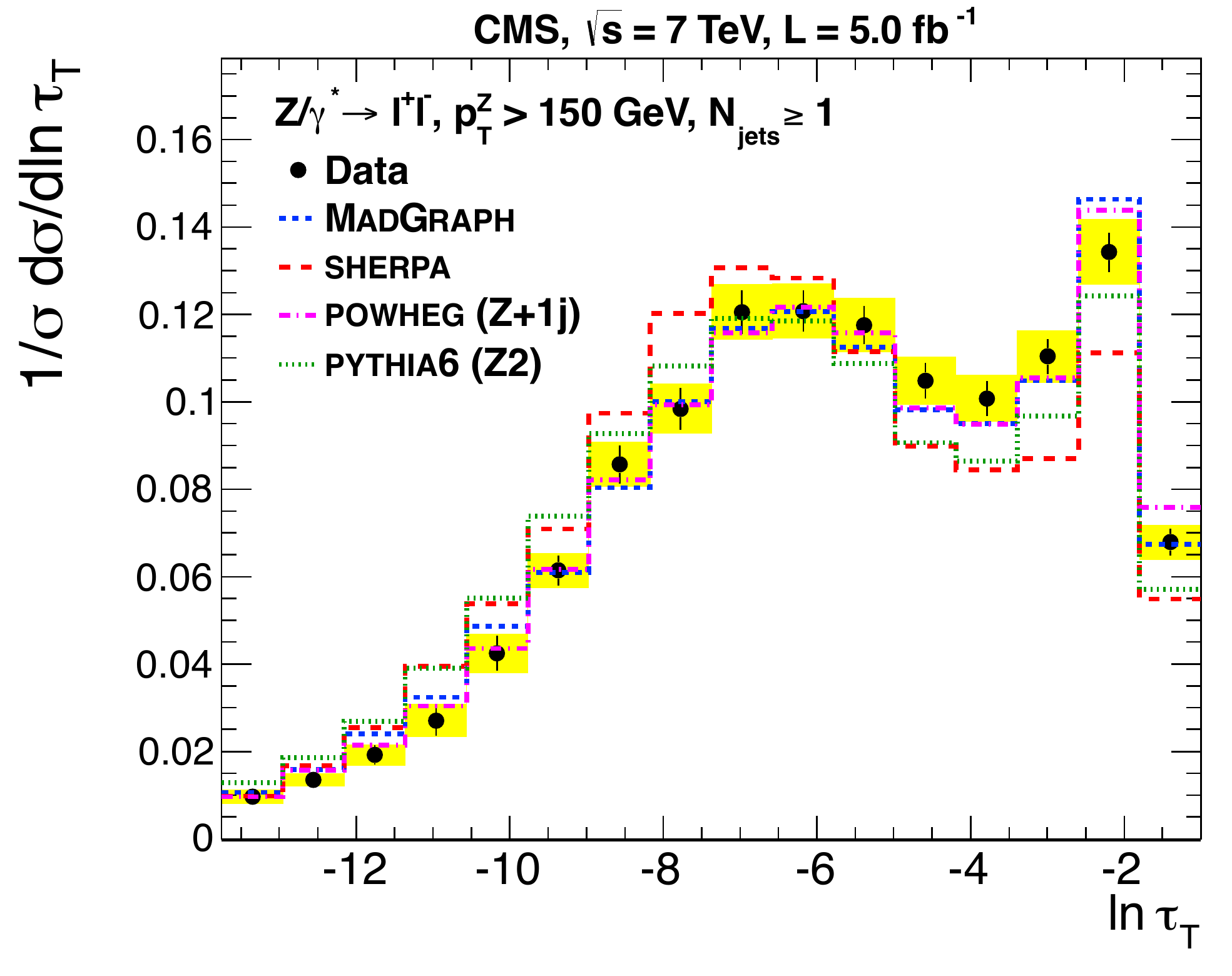}}
\resizebox{0.5\textwidth}{!}{\includegraphics{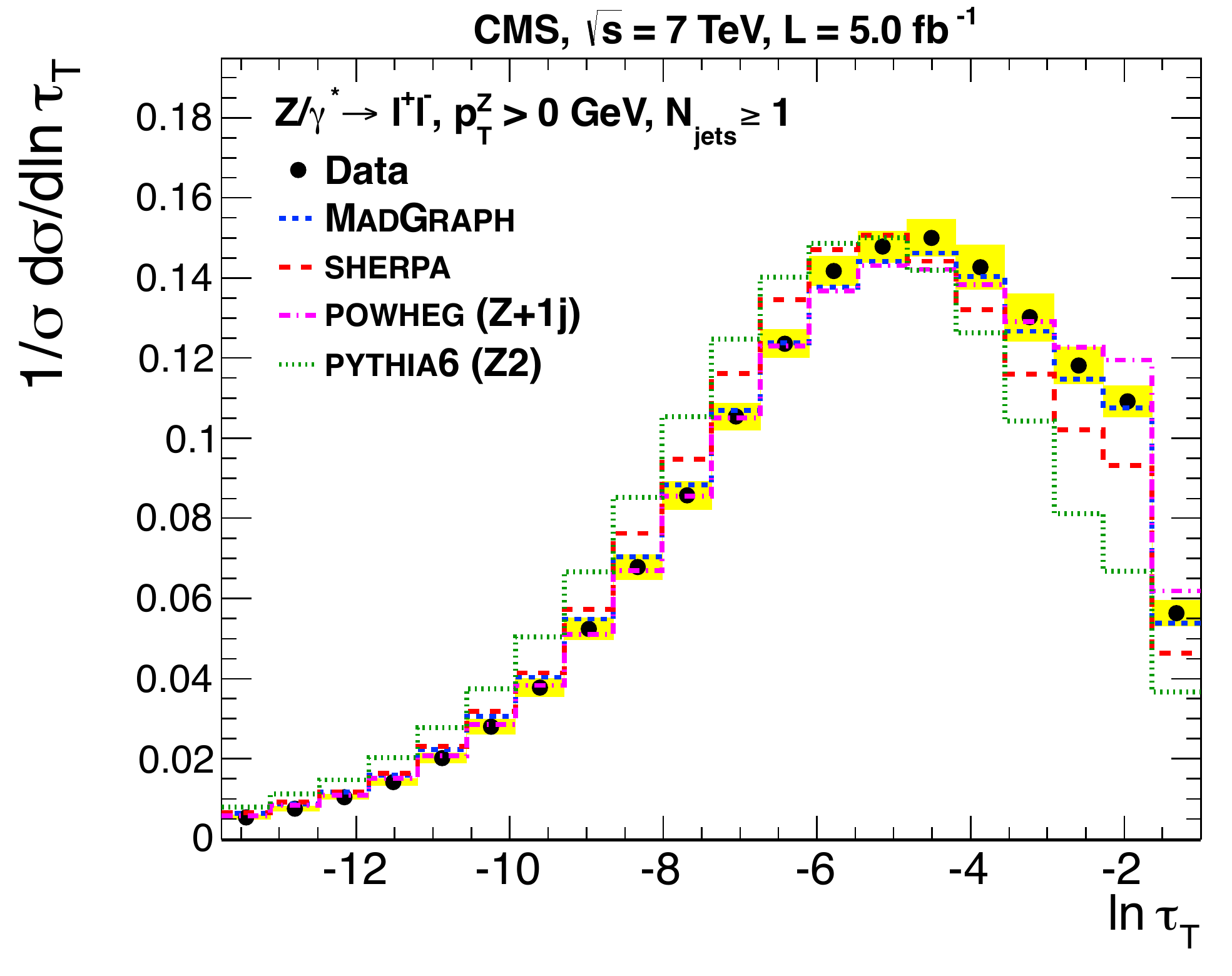}}
\caption{CMS~\cite{Chatrchyan:2013tna}: Differential cross-sections as a function of the $\ln \tau_T$ for $Z$+jets events. The distributions are 
normalised by the inclusive $Z \rightarrow ll$ cross-section. \MadGraph, \Sherpa, \Powheg~and \Pythia~predictions 
are shown. }
\label{fig:CMS-zjets-thrust}
\end{figure*}

A measure of the hadronic activity accompanying \Wboson production can be investigated by studying the 
\textit{splitting scales} in the $k_T$ cluster sequence~\cite{Aad:2013ueu}. These splitting scales are determined by the clustering of objects, either calorimeter energy deposits or particle-level hadrons, 
according to their distance from each other. The final splitting scale in the clustering sequence, called $d_0$, is the hardest scale and corresponds to the \pT~of the jet.
Studying the hardest splitting scales is therefore like studying the $k_T$ jet algorithm clustering in reverse. Since this algorithm clusters the soft 
and collinear branchings first, this clustering sequence is akin to studying the QCD evolution in reverse\footnote{This is not case for the anti-$k_T$ algorithm which clusters 
the collinear branchings first but not the soft emissions.}. The results of the hardest splitting scale, $d_0$ are shown in Figures~\ref{fig:ATLAS-ktscale} and 
compared to \MCAtNLO, \Alpgen, \Sherpa~and \Powheg\ predictions. Although there is reasonable agreement to the data, the NLO predictions 
do not describe well the high tail even though the leading-order accuracy for all of these
generators should be the same. This study, which also includes measurements that are sensitive to the hadronisation effects and multiple parton interactions, 
can be used to help tune these generators in the future.

\begin{figure}
\resizebox{0.5\textwidth}{!}{\includegraphics{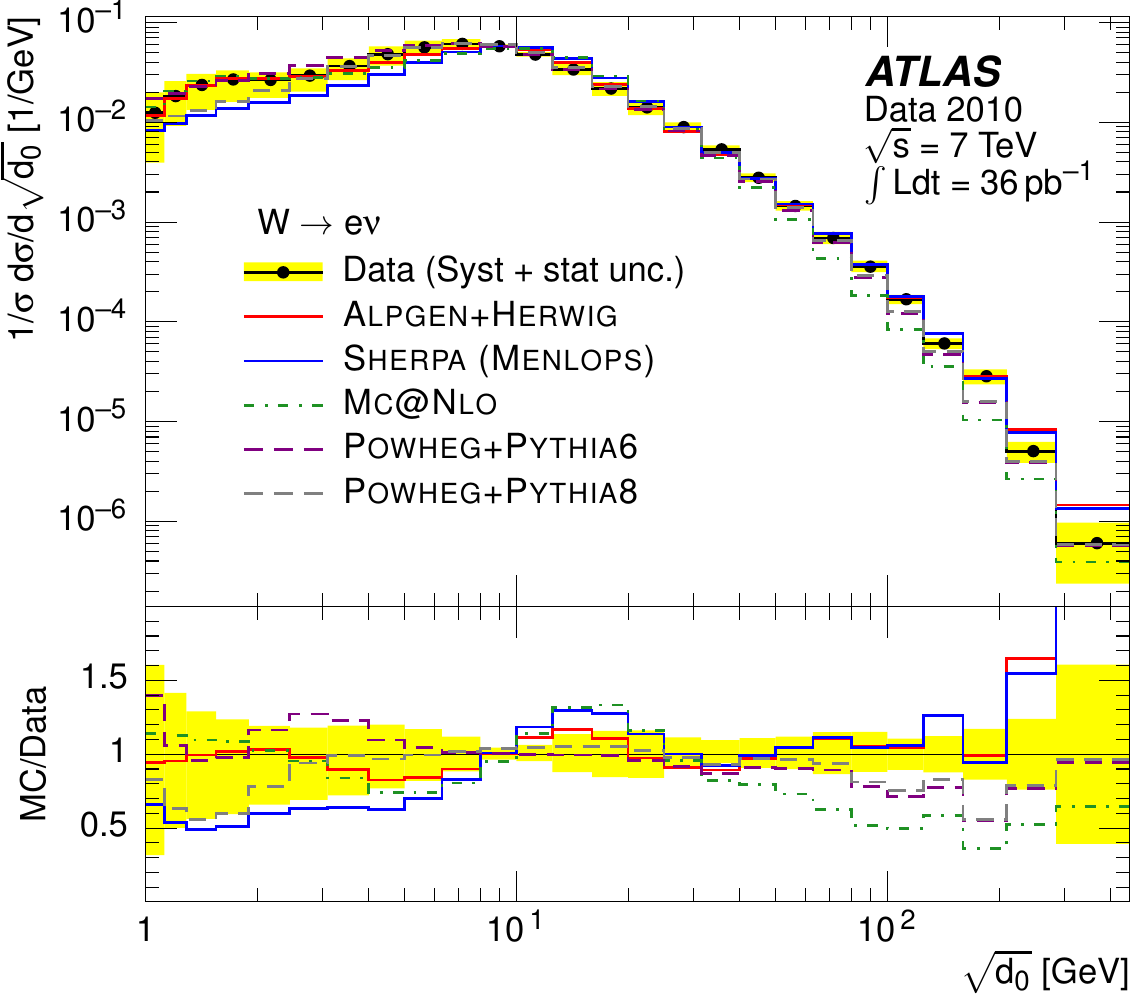}}
\caption{ATLAS~\cite{Aad:2013ueu}:  Distribution of the hardest splitting scale ($d_0$) in the $k_T$ clustering sequence for $W \rightarrow e\nu$ events. The data are compared to the \MCAtNLO, \Alpgen, \Sherpa~
and \Powheg~predictions. The yellow bands represent the combined statistical and systematic uncertainties. The histograms are normalised to unit area.}
\label{fig:ATLAS-ktscale}
\end{figure}

With the large data sets available from the LHC, both the ATLAS and CMS collaborations have studied extensively differential cross-sections for \Wboson and \Zboson production in association with jets. These measurements have highlighted a few features. First, at large values of $H_T$ the predictions must include a sufficient number of partons in the matrix-element calculation in order to model the data correctly, even at low jet multiplicities. Second, the experimental precision of measurements such as the \Zboson boson differential cross-section as a function of leading jet \pT~can test not only QCD corrections but for the first time become sensitive to the QED corrections in one-jet events. Future measurements with higher transverse momenta of the jets and the boson will be able to better probe these large QCD corrections at these high values of \pT~as well as be able to make qualitative statements about the accuracy of the QED corrections. 

\subsection{Measurements of the ratio of $W^+$ to $W^-$ in association with jets}

As discussed in Section~\ref{sec:DifferentialCS}, the $W^\pm$ rapidity distribution is sensitive to the $u\dbar$ and $d\ubar
$ quark distributions. In addition the number of $W^\pm$ events depends on the number of associated jets because the 
fraction of $u$ and $d$ quarks contributing to the different jet multiplicity processes changes. CMS measured the charge 
asymmetry defined as $A_W = \frac{\sigma(W^+) - \sigma(W^-)}{\sigma(W^+) + \sigma(W^-)}$ for different numbers of associated
jets~\cite{Chatrchyan:2011ne}. Since many experimental systematic uncertainties especially the dominant uncertainties due to the jet energy scale 
cancel in this ratio, the charge asymmetry is a sensitive test even at large jet multiplicities. Figure~\ref{fig:CMS-vjets-charge} 
shows the charge asymmetry for $W \rightarrow \mu\nu$ events. The \MadGraph~predictions agree well with the 
data, while \Pythia~fails to model the data even for one-jet events. 

In order to place constraints on the PDFs, measurements of the \Wboson charge asymmetry as a function of the number of jets and as a function of the \pT~of the \Wboson boson are needed. Such measurements at large values of the boson \pT~could constrain the PDFs at larger momentum fractions $x$ compared to inclusive measurements~\cite{Malik:2013kba}. 

\begin{figure}
\resizebox{0.5\textwidth}{!}{\includegraphics{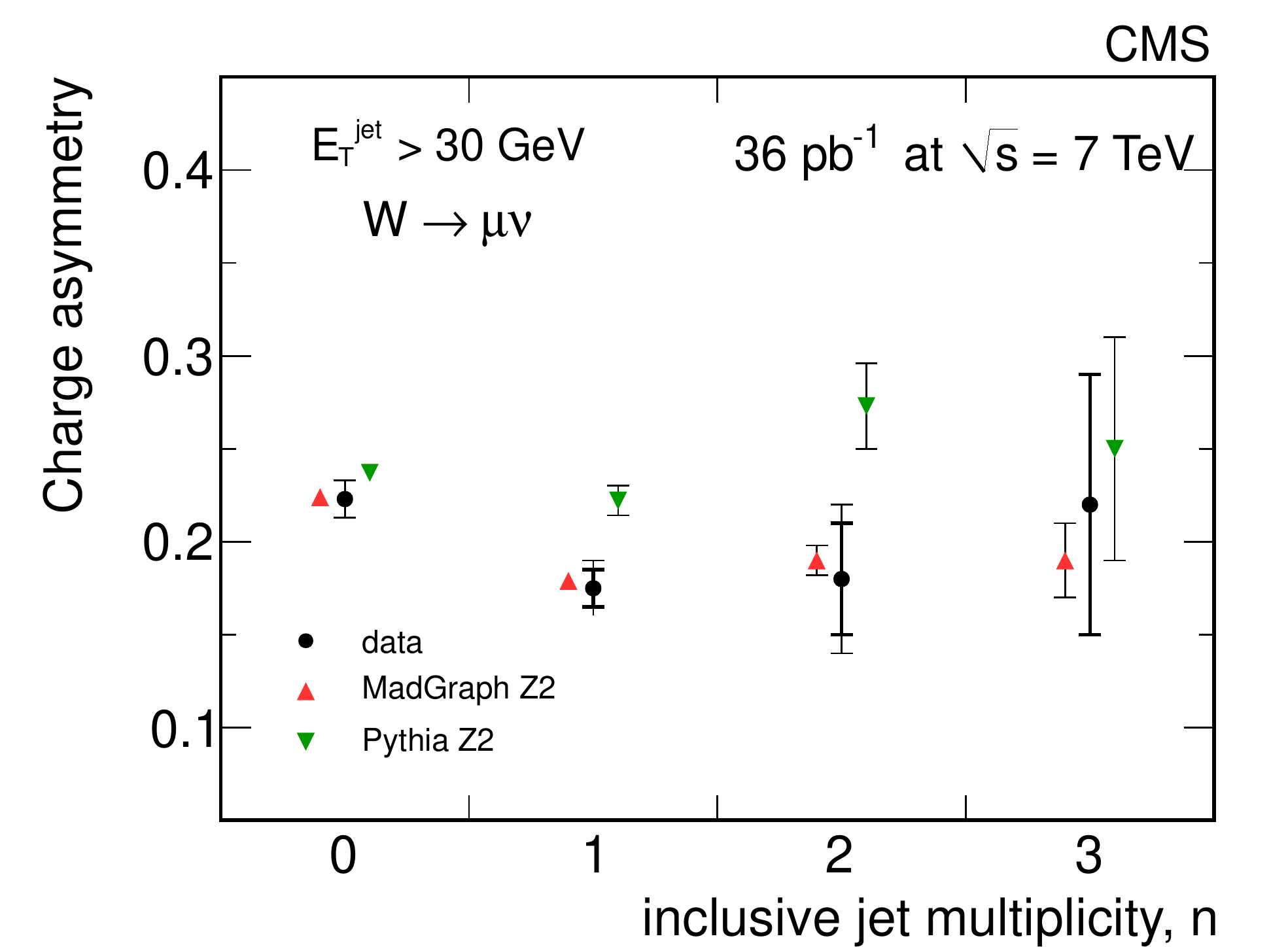}}
\caption{CMS~\cite{Chatrchyan:2011ne}:  \Wboson charge asymmetry $A_W$ as a function of the jet multiplicity for $W \rightarrow \mu\nu$ events. 
\MadGraph~and \Pythia~predictions are also included.}
\label{fig:CMS-vjets-charge}
\end{figure}

\subsection{Measurements of the ratio of $W$+jets to $Z$+jets}

Measurements of the \Wboson and \Zboson cross-sections in association with jets are plagued by dominant uncertainties from the jet 
energy scale. Although both CMS and ATLAS have achieved excellent understanding of the jet energy scale, these
uncertainties still dominate especially for jets at high rapidities. However, in other cases, such as the measurement of the jet 
\pT~(Section~\ref{sec:wz-jet-diff}), the experimental measurement is more precise than the theory predictions. Both the theory and experimental  
uncertainties can be reduced through a measurement of the cross-section ratio between $W$+jets and $Z$+jets processes. For 
example, when comparing ATLAS measurements of $W$+jets production to the ratio of $W$+jets to $Z$+jets production for events with one associated jet, 
the jet energy scale uncertainty is roughly a factor of two smaller in the ratio measurement. 

CMS measured the ratio of $W$+jets to $Z$+jets for up to four associated jets~\cite{Chatrchyan:2011ne}. In both the electron and muon channels, the 
data were in good agreement with the \MadGraph~and \Pythia~predictions. ATLAS measured the ratio for exactly one 
associated jet but for different thresholds of the jet \pT~\cite{Aad:2011xn}. The combined results from the electron and muon channels are 
shown in Figures~\ref{fig:ATLAS-rjets-pt} and compared to predictions from \Pythia, \Alpgen~and NLO predictions from 
\MCFM. The ratio, which is not constant as a function of the jet \pT~threshold, decreases because at large jet momenta the difference 
in the boson masses is small compared to the effective scale of the interaction. All of the predictions model 
this trend well.

\begin{figure}
\resizebox{0.5\textwidth}{!}{\includegraphics{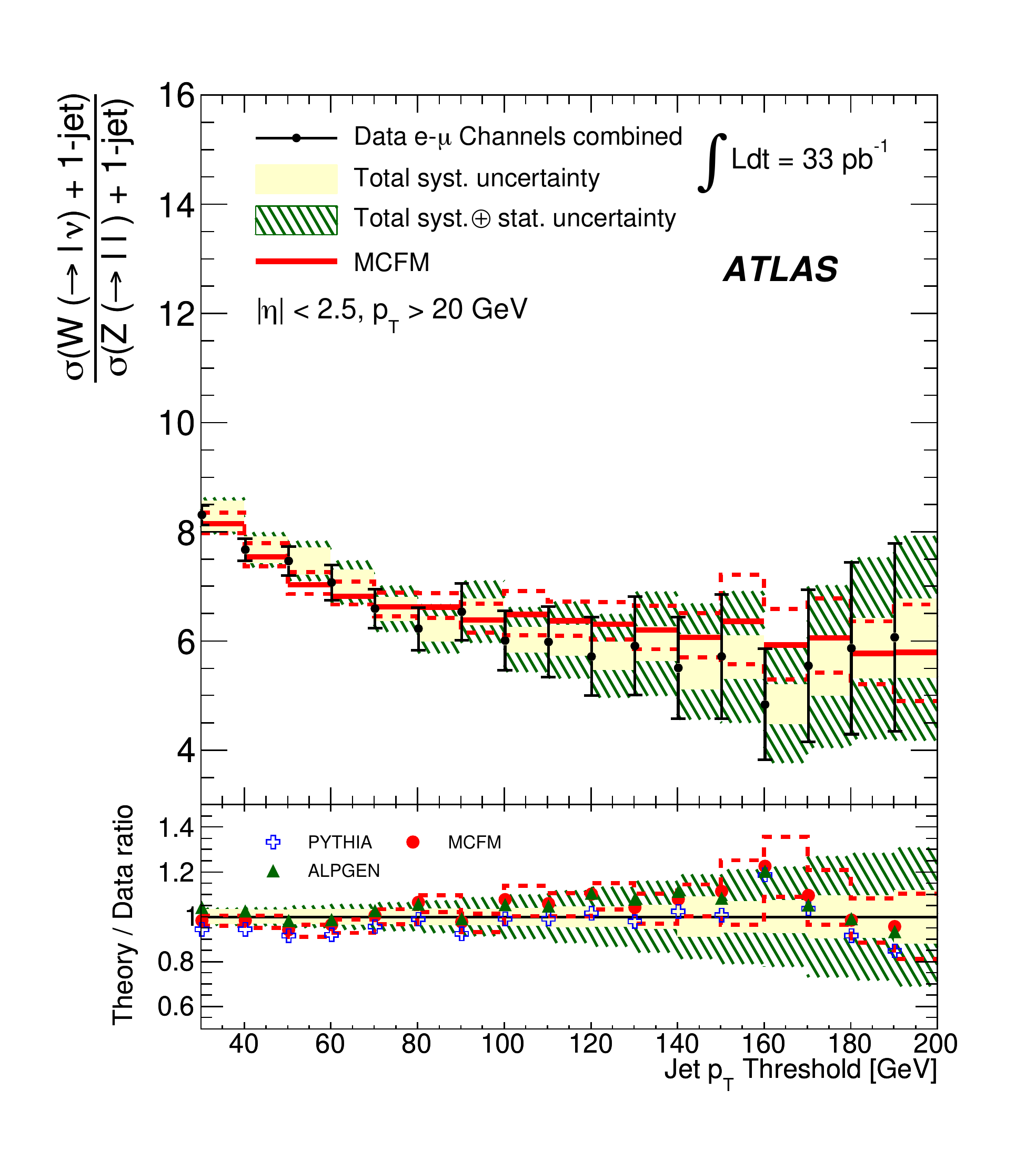}}
\caption{ATLAS~\cite{Aad:2011xn}: Ratio of $W$+jets to $Z$+jets in $N_{jet}=1$ events for varying thresholds of the jet \pT. NLO predictions from 
\MCFM~and predictions from \Alpgen~and \Pythia~are also shown. }
\label{fig:ATLAS-rjets-pt}
\end{figure}

The measurement of the ratio of \Wboson to \Zboson production in association with jets is one of the most precise measurements of perturbative QCD. 
Future measurements, using the full 2011 data set from the LHC will be able to measure this ratio for higher jet multiplicities and as a function of the
\pT~and rapidity of the jets and the $H_T$. This ratio is sensitive to new physics models, especially if the new particles decay preferentially either to the \Wboson vs. the \Zboson final state.

\subsection{Cross-section measurements of \Wboson and \Zboson bosons in association with heavy-flavour quarks}

The study of \Wboson and \Zboson production in association with heavy-flavour quarks is of particular importance today. First, the 
theoretical predictions are less well known compared to the inclusive $W$+jets and $Z$+jets predictions. Second, precision 
measurements of these processes are critical since they are a dominant background in Higgs measurements of $WH$ 
production with $H  \rightarrow bb$ decays and new physics searches involving heavy-flavour production. 

\begin{figure}
\resizebox{0.5\textwidth}{!}{\includegraphics{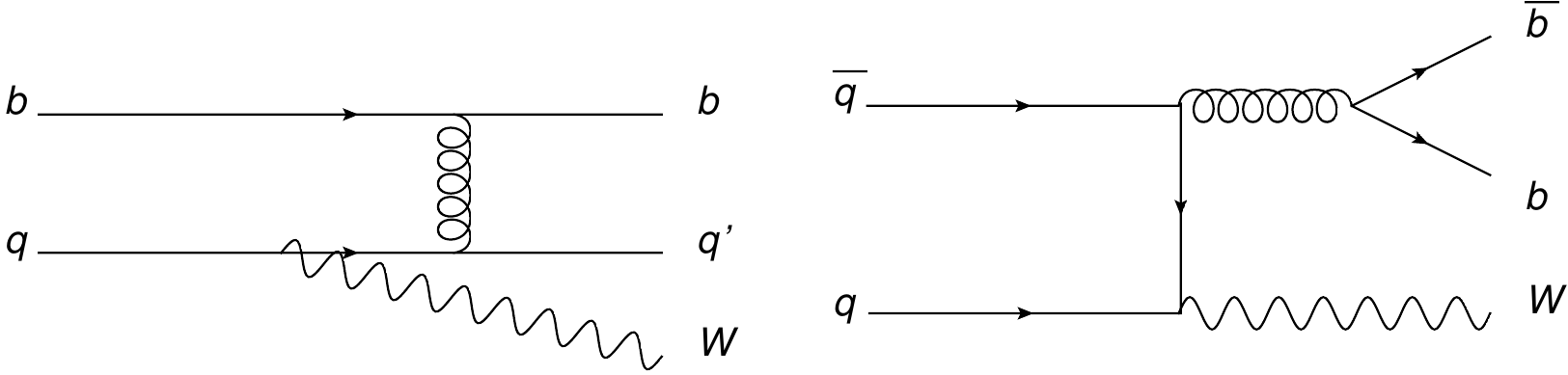}}
\caption{Leading-order Feynman diagram for the production of \Wboson bosons in association with a b-quark.}
\label{fig:WBProd}
\end{figure}

The production of $W+b$-jet events has two main diagrams at leading order (Figure \ref{fig:WBProd}): 
\Wboson plus a gluon in the final state where the gluon splits to a 
\bbbar~pair, and $b$-quark in the initial state where the \Wboson is produced from radiation from a quark. The former can be 
produced both by the matrix-element calculation and by the parton showering model, whereas the latter diagram can only 
be modelled by including a $b$-quark in the initial state from the PDF. 

The predictions and measurements of $W+b$-jet cross-section have a long history. The first measurements by CDF \cite{Aaltonen:2009qi} 
indicated that the measured cross-section was too large by 2.8 standard deviations compared to the predictions. A 
measurement by ATLAS, using only an integrated luminosity of $\IntLumi \approx 35 \ipb$, also reported a larger cross-section 
by 1.5 standard deviations, while the D0 measurement was consistent with the predictions~\cite{D0:2012qt}. 
An updated ATLAS measurement~\cite{Aad:2013vka} of 
the $W+b$-jet cross-section using an integrated luminosity of $\IntLumi \approx 4.5 \ifb$ offers the statistical and 
systematic precision to definitively close this debate. The measurement presented cross-sections for the exclusive one-jet 
and two-jet final states with a fiducial phase space requirement of at least one $b$-jet, defined by the presence of a weakly 
decaying $b$-hadron with $\pT > 5\, \GeV$ and within a cone radius of $\Delta R = 0.3$ of the jet axis. The cross-section results are 
summarised in Figure~\ref{fig:ATLAS-wb-pt} and compared to calculations from \MCFM, \Powheg~and \Alpgen. The 
\MCFM~predictions are calculated using the 5-flavour scheme (5FNS) which accounts for the presence of $b$-quarks in the PDF. The \Alpgen~and \Powheg~predictions use the four-flavour scheme (4FNS). For one-jet 
events, the measured cross-section is consistent with 1.5 standard deviations to the NLO \MCFM~predictions, while for two-jet 
events the measured cross-section is in good agreement with the predictions. In the one-jet case, the difference between 
data and the predictions can be more clearly understood in the differential cross-section measurement as a function of the 
jet \pT~shown in Figure~\ref{fig:ATLAS-wb-pt}. The \MCFM~and \Alpgen~predict a softer jet \pT~spectrum with respect to 
the data. 

In a complementary result, the $W+bb$ final state was measured 
by CMS~\cite{Chatrchyan:2013uza}, by requiring events with only two jets, both of which 
must originate from a $b$-hadron. The measured cross-section of 

$0.53 \pm 0.05 (stat) \pm 0.09 (syst) \pm 0.06 (th) \pm 0.02 (lumi) \pb$ 

\noindent is in excellent agreement with the \MCFM~prediction of $0.52 \pm 0.03 \pb$

\begin{figure*}
\resizebox{0.5\textwidth}{!}{\includegraphics{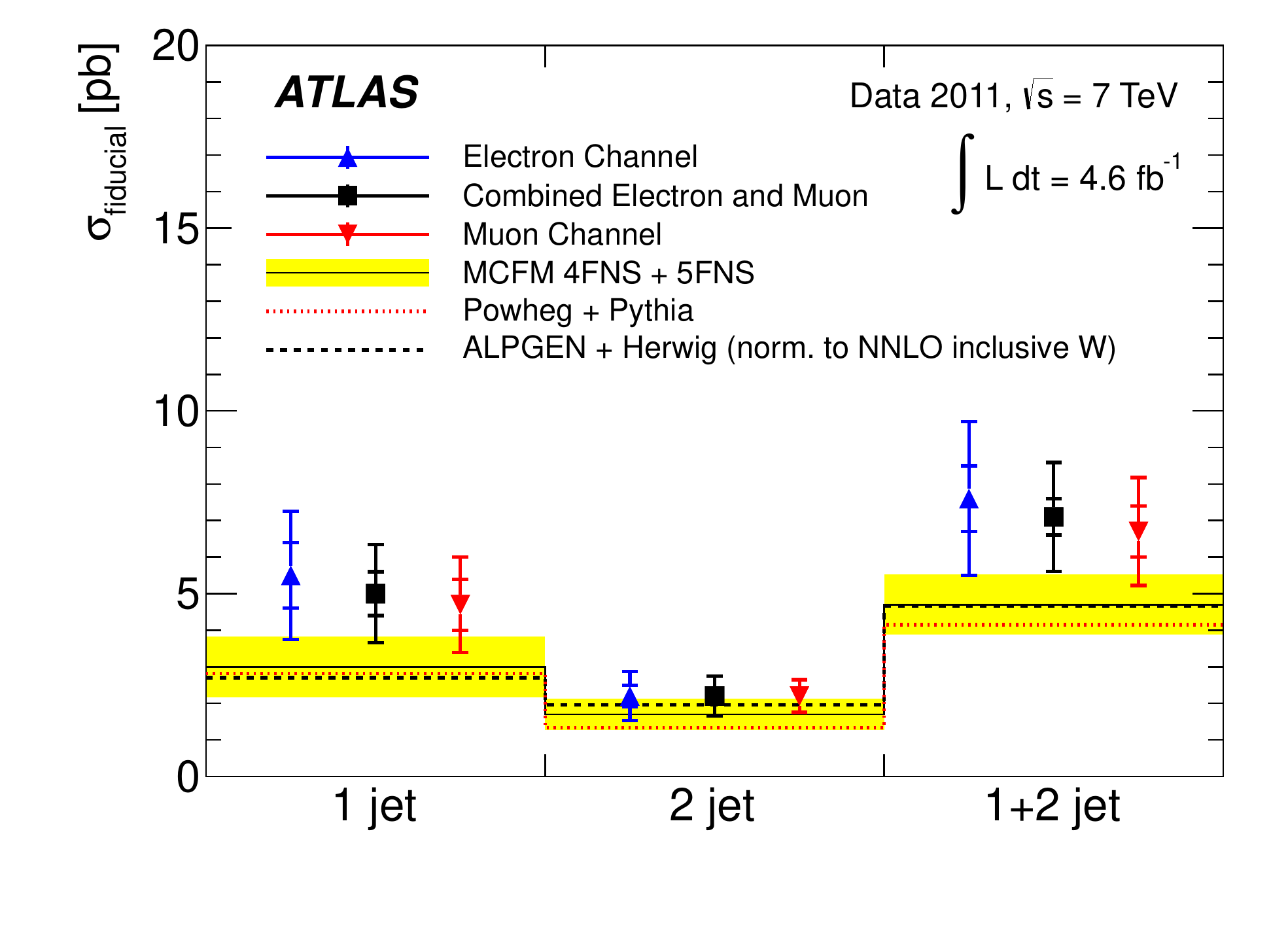}}
\resizebox{0.5\textwidth}{!}{\includegraphics{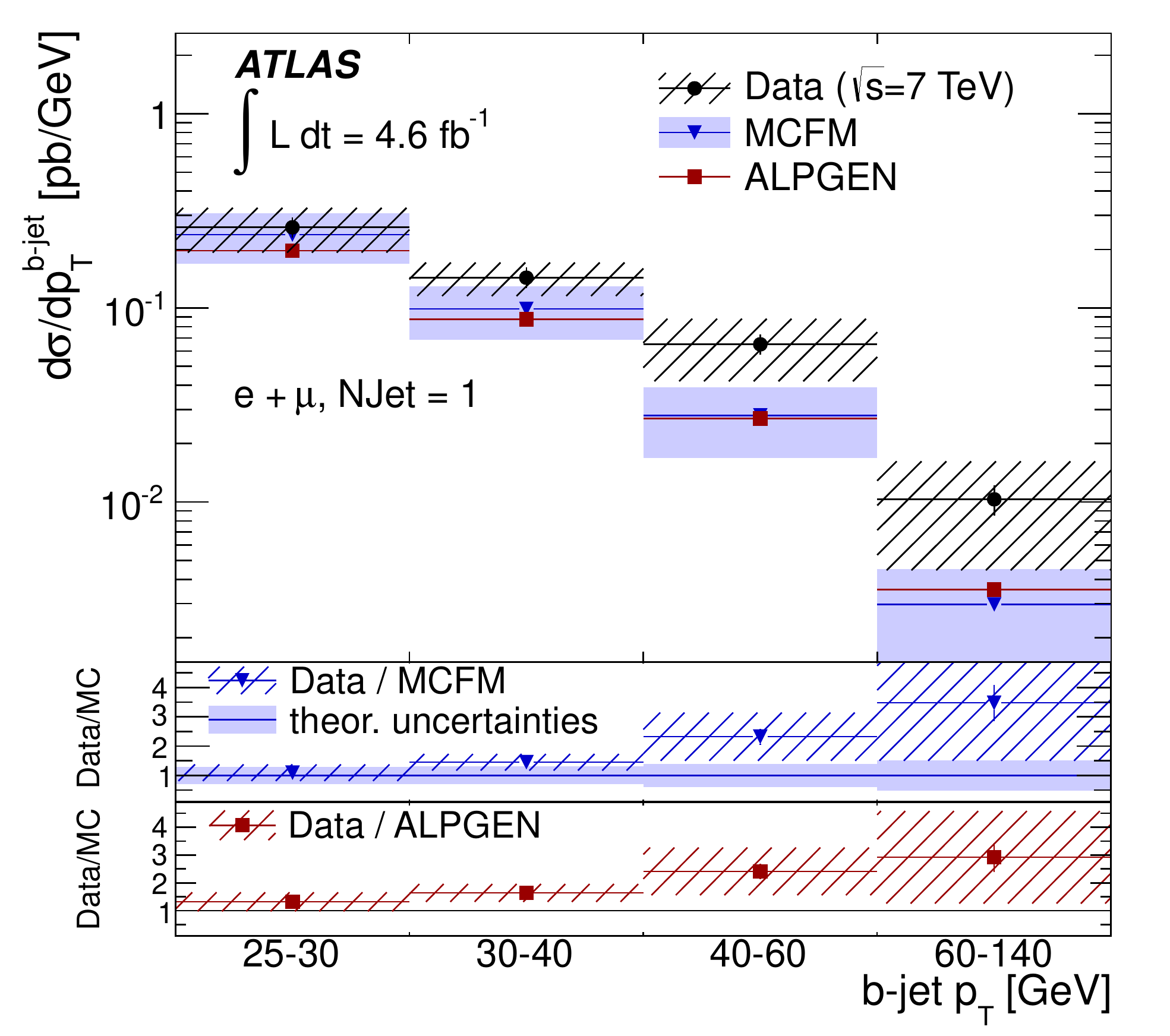}}
\caption{ATLAS~\cite{Aad:2013vka}: Measurement of the cross-section for $W+b$-jet events for 1-jet, 2-jet, and 1+2-jet events (left). NLO 
\MCFM~predictions and \Alpgen~and \Powheg~predictions are also shown. Differential cross-section as a function of the $b
$-jet \pT~for 1-jet events (right). NLO \MCFM~predictions and \Alpgen~predictions are also shown.}
\label{fig:ATLAS-wb-pt}
\end{figure*}

Similar to measurements of $W+b$-jet, the 
$Z+b$-jet cross-section is much less studied compared to the inclusive $Z$ measurements. Unlike $W+b$-jet which only 
has two main contributing diagrams at leading order, $Z+b$-jet production includes additional leading-order diagrams of \Zboson radiation from an 
initial-state $b$-quark and \Zboson radiation from a final-state \bbbar~pair (Figure \ref{fig:ZBProd}). Predictions which include diagrams with an initial-state $b$ 
are therefore necessary. CMS presented a measurement of the $Z+b$-jet cross-section with exactly one b-jet, with at least one b-jet and with at least two b-jets, 
using an integrated luminosity of 
$\IntLumi \approx 5\,\ifb$~\cite{Chatrchyan:2014dha}. The results were compared to \MadGraph, \aMCAtNLO~and \MCFM~predictions. 
While the measured cross-sections
were found to be in fair agreement with \MadGraph~and \aMCAtNLO, the \MCFM~results differ by approximately two standard deviations from the data. 
ATLAS also measured the $Z+b$-jet cross-section using a smaller data sample with an 
integrated luminosity of $\IntLumi \approx 35 \ipb$~\cite{Aad:2011jn}. The NLO \MCFM~predictions as well as \Alpgen~and \Sherpa~were 
found to be consistent with the data but there were signs of tension especially between the \Alpgen~and \Sherpa~predictions themselves. These results are
summarised in Figure~\ref{fig:SummaryCrossSection}.
Updated results with better statistical precision and predictions with massive quark models are needed here to help resolve these differences. 

\begin{figure}
\resizebox{0.5\textwidth}{!}{\includegraphics{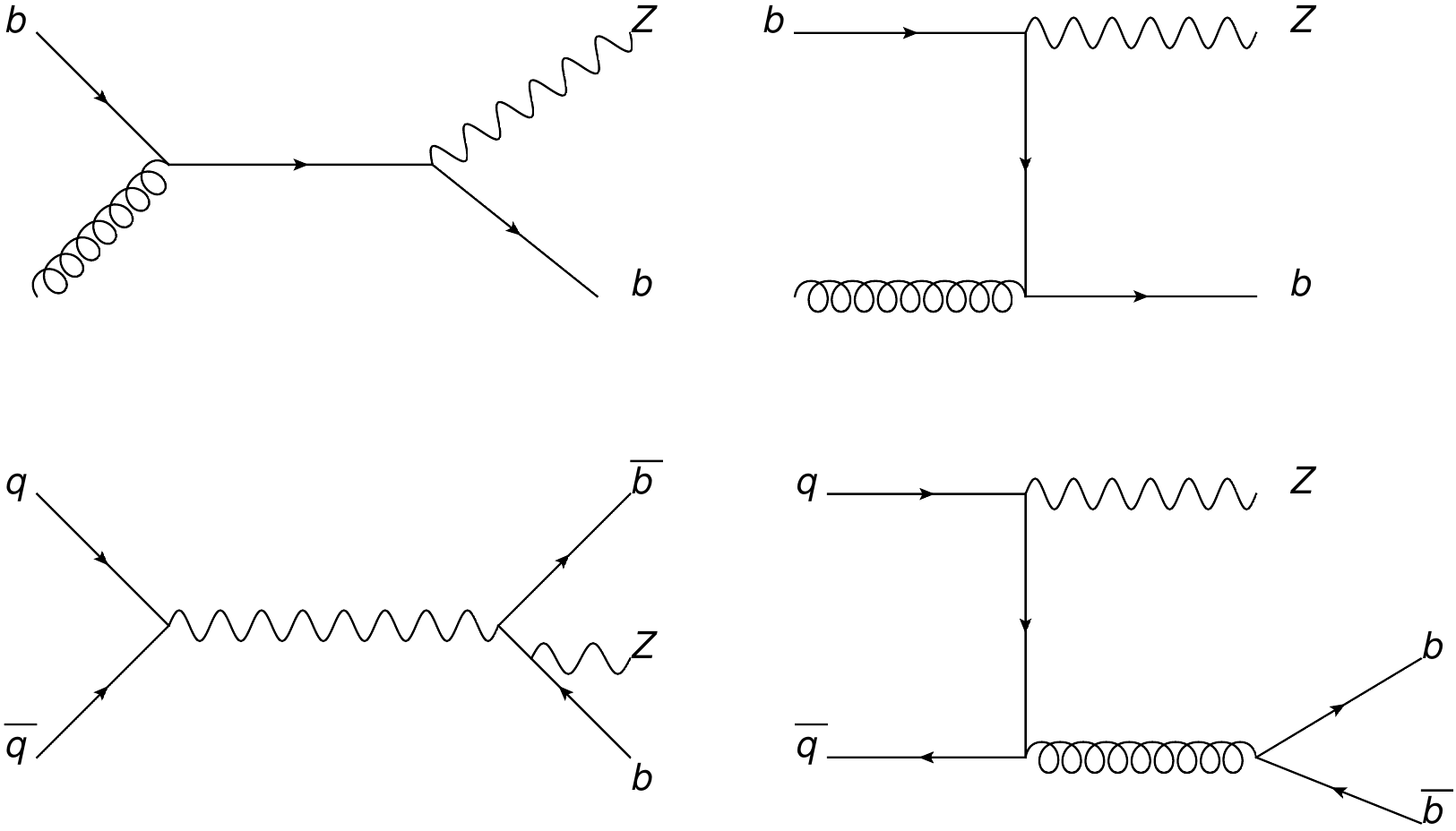}}
\caption{Leading-order Feynman diagrams for the production of \Zboson bosons in association with a $b$-quark.}
\label{fig:ZBProd}
\end{figure}

In both $W+b$-jet and $Z+b$-jet production, diagrams with a gluon splitting to a \bbbar~pair contribute to the matrix-element 
calculations and in the parton showering model. In a generator like \Alpgen, this overlap is removed by applying a $\Delta 
R$ cut, so that at small values of $\Delta R$ the gluon splitting is handled by the parton shower, while at large values it is 
predicted from the matrix element. The theoretical uncertainties describing collinear $b$-quark production are large. To test this 
transition from parton shower to matrix-element calculations, measurements of the gluon splitting at small values of $\Delta R$ is an 
important topic at the LHC today. This is especially important for new physics searches and Higgs measurements that 
select $b$-jets, since high \pT~$b$-jets tend to be produced via gluon splitting. 

While measurements of gluon splitting at small values of $\Delta R$ are interesting, it is experimentally challenging to 
measure since the two $b$-quarks are often reconstructed within the same jet. CMS presented a new approach for 
this measurement by measuring $Z$ events with two $b$-hadrons~\cite{Chatrchyan:2013zja}. As the $b$-hadrons can be reconstructed from 
displaced secondary vertices, only tracking information is needed and there is no dependence on a jet algorithm. The 
angular resolution is $\Delta R \approx 0.02$ between the two $b$ hadrons. The differential cross-section measurement 
as a function of $\Delta R$ of the two $b$-hadrons is shown in Figure~\ref{fig:CMS-zbhad-dr}. The collinear region ($\Delta 
R < 0.5$) is best described by \Alpgen, while \MadGraph~and a\MCAtNLO~predictions tend to underestimate the data. In 
addition, the differential cross-section was measured for a phase space region where $\pT^Z > 50\, \GeV$ (Figure~
\ref{fig:CMS-zbhad-dr}). In this phase space, the relative fraction of events with collinear $b$-quarks increases. Again, 
\Alpgen~gives the best description of the data. 

\begin{figure*}
\resizebox{0.5\textwidth}{!}{\includegraphics{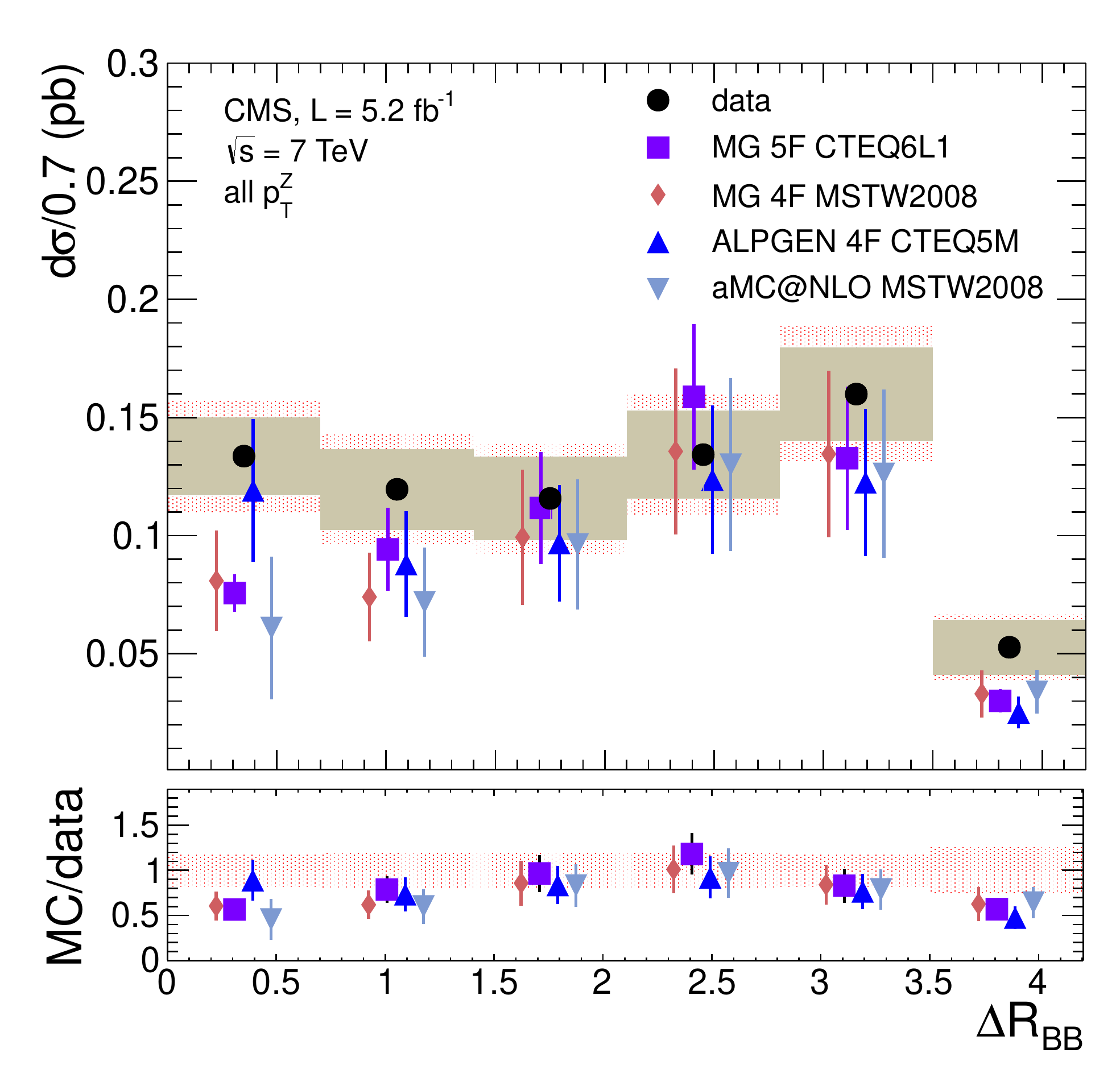}}
\resizebox{0.5\textwidth}{!}{\includegraphics{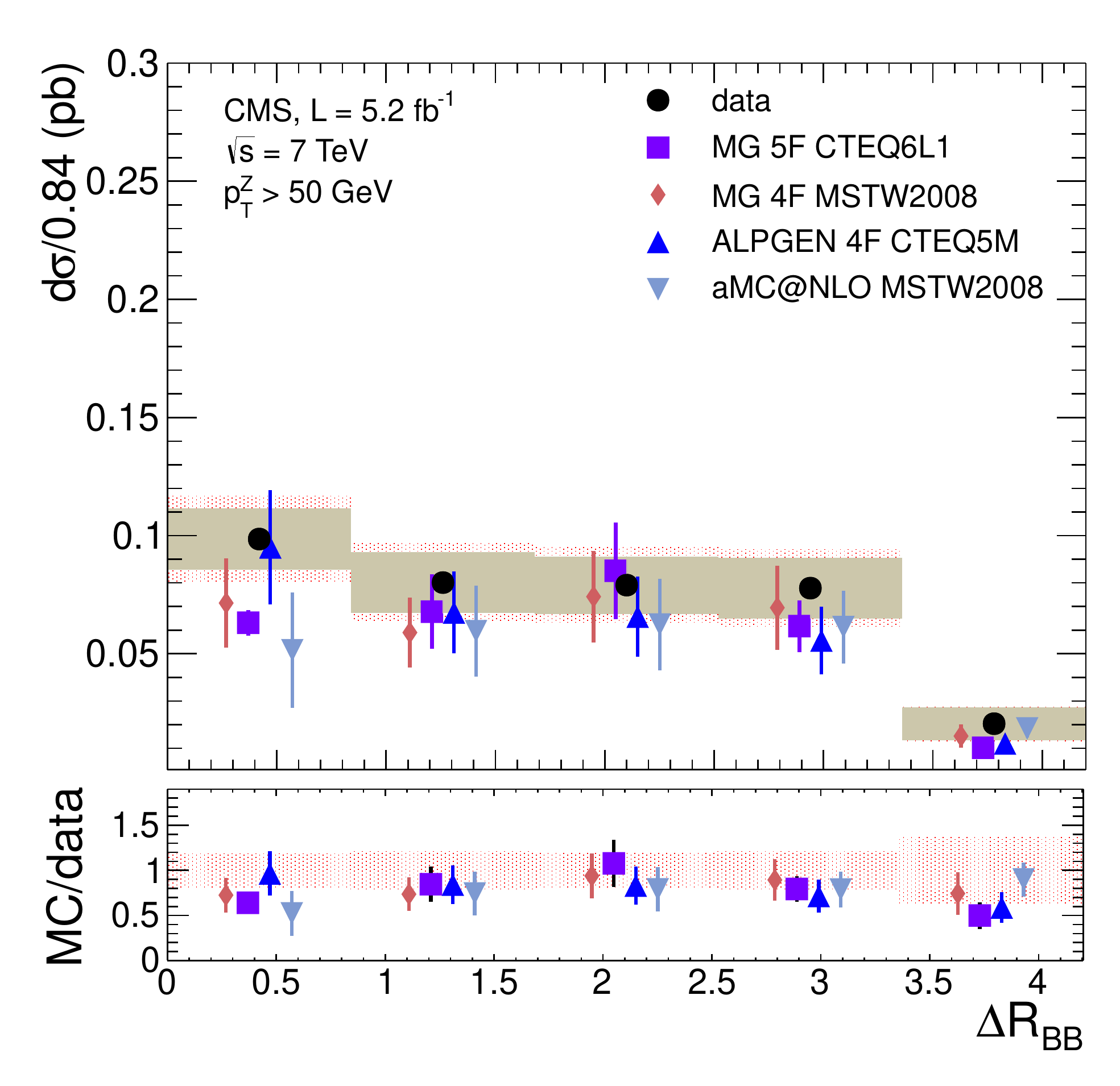}}
\caption{CMS~\cite{Chatrchyan:2013zja}: Differential cross-section as a function of $\Delta R$ of the two $b$ hadrons for $Z+\bbbar$ events. The right plot 
shows the inclusive phase space of $\pT^Z > 0\, \GeV$. The left plot shows the phase space region of $\pT^Z > 50\, \GeV$. 
Also shown are the predictions from \MadGraph~using the 5- and 4-flavour schemes, \Alpgen~and \MCAtNLO. }
\label{fig:CMS-zbhad-dr}
\end{figure*}

\begin{figure}
\resizebox{0.5\textwidth}{!}{\includegraphics{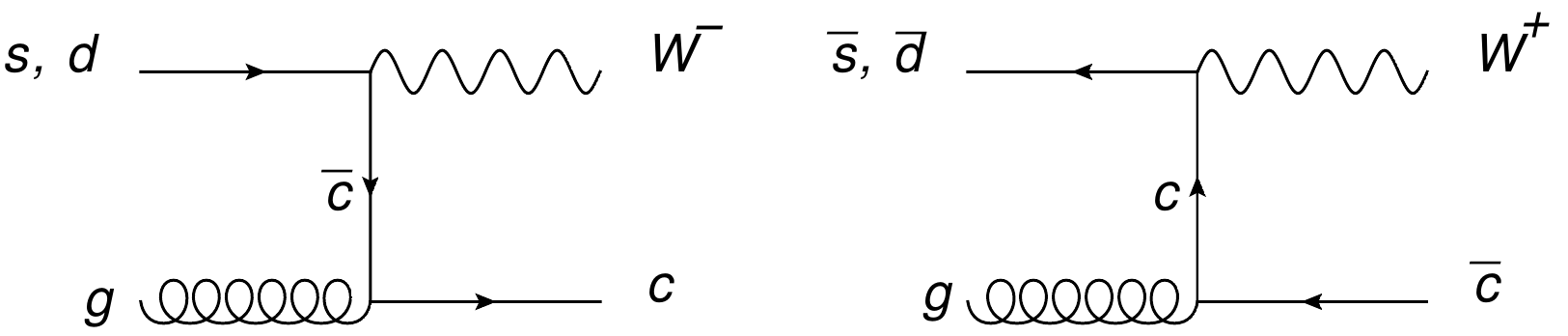}}
\caption{Leading-order Feynman diagrams for the production of the \Wboson\ boson in association with a $c$-quark.}
\label{fig:WcProd}
\end{figure}

The cross-section of the \Wboson boson in association with $c$-quarks, which have sensitivity to the $s$-quark contribution in PDFs, have not been experimentally well measured in the past. 
At the LHC, the dominant W+$c$ production takes place via the reaction  
$sg \rightarrow W^- + c$ and $\sbar g \rightarrow W^+ + \cbar$, as illustrated in Figure~\ref{fig:WcProd}. Due to the high production rates at the LHC, 
measurements of the $W+c$ cross-sections have for the first time, sufficient precision to constrain the $s$-quark
in the PDFs at $x \approx 0.01$. In the future, measurements of $W+c$ may also be able to help determine if there is a 
asymmetry between 
the $s$ and $\sbar$ sea\footnote{In practice this is challenging since the 
cross-section asymmetry in the ratio of $W^+ +c$ to $W^- +c$ production comes mainly from the $d-\bar{d}$ PDF asymmetry.} as suggested by the NuTeV measurements~\cite{Goncharov:2001qe, Martin:2009iq, Ball:2009mk}.
Since the $c$-quark and the $W$ have opposite charge, the W+$c$ production can be measured by subtracting events 
with the same-sign charge from events with opposite-signed charge. This subtraction will have 
no effect on the $W^- + c$ process, but all other background such as $W^- + c\cbar$ and $W^- + b\bbar$ are 
symmetric in same-sign and opposite-sign events and will be removed. In the CMS analysis~\cite{Chatrchyan:2013uja}, the jets originating 
from a $c$-quark are selected in 
one of three ways: a selection of a $D^\pm \rightarrow K^\mp \pi^\pm \pi^\pm$ decay by requiring a displaced secondary 
vertex with three tracks and an invariant mass which is consistent with the $D^\pm$, a selection of a $D^0 \rightarrow K^
\mp \pi^\pm$ decay by requiring a displaced secondary vertex with two tracks which is consistent with the $D^0$, and 
semileptonic $c$-quark decay by requiring a muon matched to a jet. The ATLAS analysis~\cite{Aad:2014xca} selects 
$W+c$ events by reconstructing the $D^\pm \rightarrow K^\mp \pi^\pm \pi^\pm$ and $D^{*\pm} \rightarrow D^0 \pi^\pm$ 
decay modes or by identifying jets with a semileptonic $c$-quark decay.

The ATLAS and CMS results both presented measurements of the $W+c$ cross-sections, the cross-section ratio of $W^{+}+c$ to $W^{-}+c$ as well as the cross-sections as a function of the lepton $\eta$. As summarised in Figure~\ref{fig:wc-total}, the measured $W+c$ cross-section in the ATLAS results is most consistent with PDF sets with a relatively higher $s$-quark density, while the CMS results are most consistent with PDF sets with a relatively lower density. However, the precision of the measurements is not sufficient to make any definitive conclusions. Overall for both the ATLAS and CMS measurements, there is good agreement between the experimental results and the predictions. 

In the ATLAS results, the ratio of the strange-to-down sea-quark distribution, $r_s = 0.5\, (s+\sbar)/\dbar$, as a function of $x$ is treated as free parameter in the HERAPDF1.5 PDF fits and all other eigenvectors in the fit are constrained within the uncertainties from the HERA data. As seen in Figure~\ref{fig:ATLAS-sd-sea}, the ATLAS results support the hypothesis of an SU(3)-symmetric light-quark sea and are consistent with the results from the ATLAS-epWZ12 PDF fits~\cite{Aad:2012sb} where the ATLAS \Wboson and \Zboson cross-section measurements are included in addition to the HERA data (see also Section~\ref{sec:DifferentialCS}).

\begin{figure*}
\resizebox{0.5\textwidth}{!}{\includegraphics{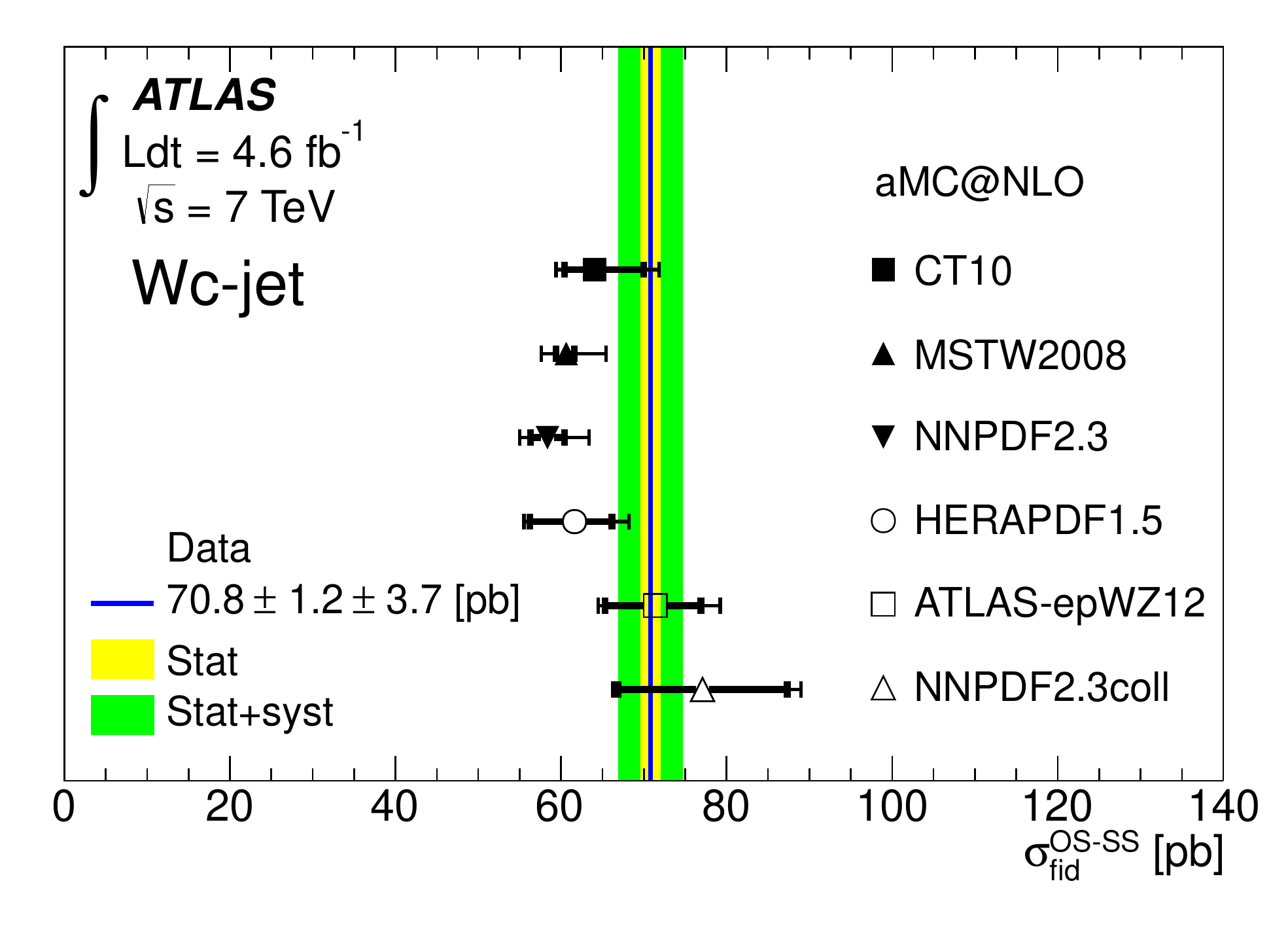}}
\resizebox{0.5\textwidth}{!}{\includegraphics{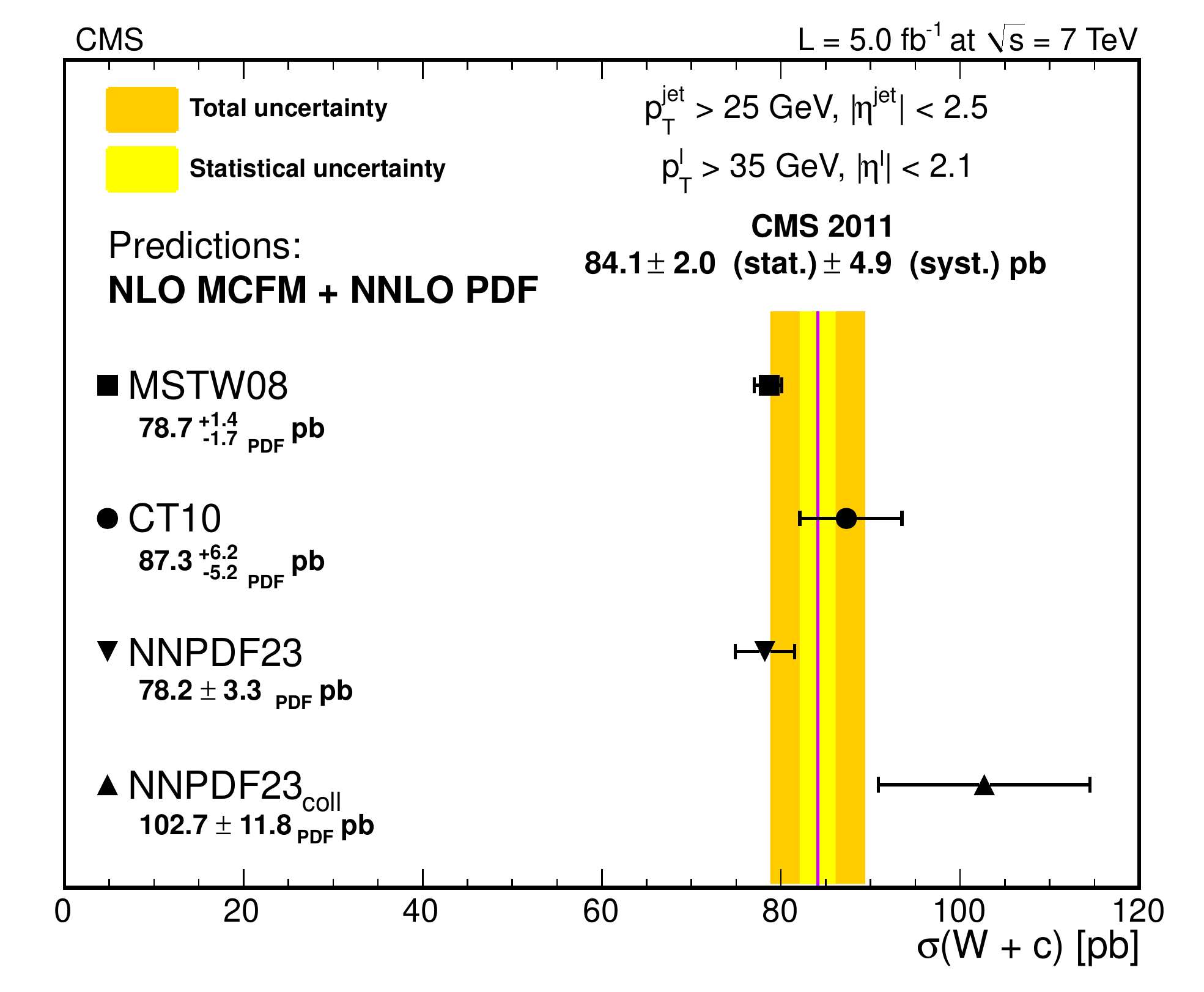}}
\caption{Cross-section of $W+c$ for ATLAS~\cite{Aad:2014xca} (left) and CMS~\cite{Chatrchyan:2013uja} (right), compared to four different PDF sets: \MSTW 2008, \CTEQ 10, \NNPDF 23 and \NNPDF 23coll. The ATLAS results are also compared to \HeraPDF1.5~and \HeraPDF1.5~but including the ATLAS \Wboson\ and \Zboson\ data, called {\sc ATLAS-epWZ12}. The \NNPDF 23coll PDF set is like \NNPDF 23 but excludes all fixed target data. In the ATLAS figure, "OS-SS" refers to the subtraction of events with opposite-signed charges and same-signed charges. The predictions in the ATLAS results are made using the \aMCAtNLO\ generator, while the predictions for the CMS results use the \MCFM\ generator.}
\label{fig:wc-total}
\end{figure*}

\begin{figure}
\resizebox{0.45\textwidth}{!}{\includegraphics{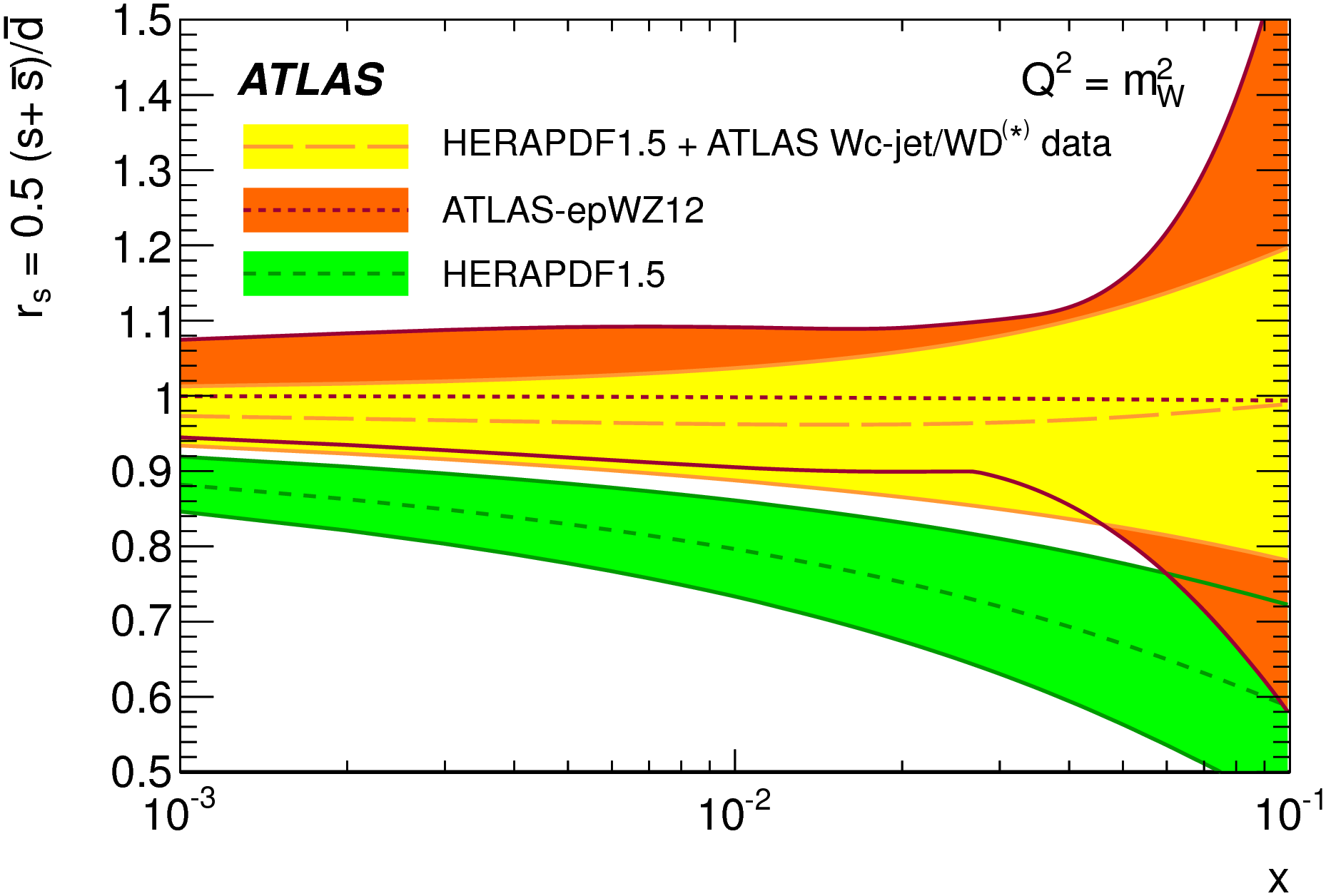}}
\caption{ATLAS~\cite{Aad:2014xca}:  Ratio of strange-to-down sea-quark distributions as a function of $x$ as obtained from the ATLAS-epWZ12 PDF set, the ratio as assumed in the HERAPDF1.5 PDF set and the ratio obtained from the HERAPDF1.5 PDF set but including the ATLAS $W+c$ measurements. The ATLAS-epWZ12 PDF set includes the ATLAS \Wboson and \Zboson cross-section measurements in addition to the HERA data. The error band on the ratio including the ATLAS measurements represents the total uncertainty. }
\label{fig:ATLAS-sd-sea}
\end{figure}


In summary with the large LHC data sets, precision measurements of differential cross-sections of \Wboson and \Zboson production in association with heavy-flavour quarks can be made for the first time. Measurements of the $W+b$-jet cross-sections have indicated that the jet \pT~spectra is not well modelled by the  predictions. In addition the predictions for the $Z+b$-jet cross-section are in tension both with the data and with each other. Future measurements of the differential cross-sections are needed to resolve this. Finally, both measurements of the $W+c$-jet production from ATLAS and CMS agree with a wide range of PDF sets. Although the ATLAS and CMS results tend to prefer PDFs with a different $s$-quark density, additional measurements with more data are needed to study this in greater detail. 

\subsection{Electroweak production of \Zboson bosons}

\Zboson bosons in associations with jets can be produced not only via the Drell-Yan process, but also via electroweak processes, as illustrated in Figure \ref{fig:VBF-diagram}. Electroweak processes are here defined as all processes which lead to a final state of two leptons, two quarks and involve the exchange of electroweak bosons in the t-channel. Of special importance is the vector boson fusion process, shown in the first diagram, as it is an important input for Higgs boson studies and the study of electroweak gauge couplings. In a full calculation of the production cross-section of process, which involves all diagrams, large negative interference exists between the pure vector boson fusion process, the bremsstrahlung and the non-resonant (or multi peripheral) processes.  

\begin{figure}[h]
\resizebox{0.5\textwidth}{!}{\includegraphics{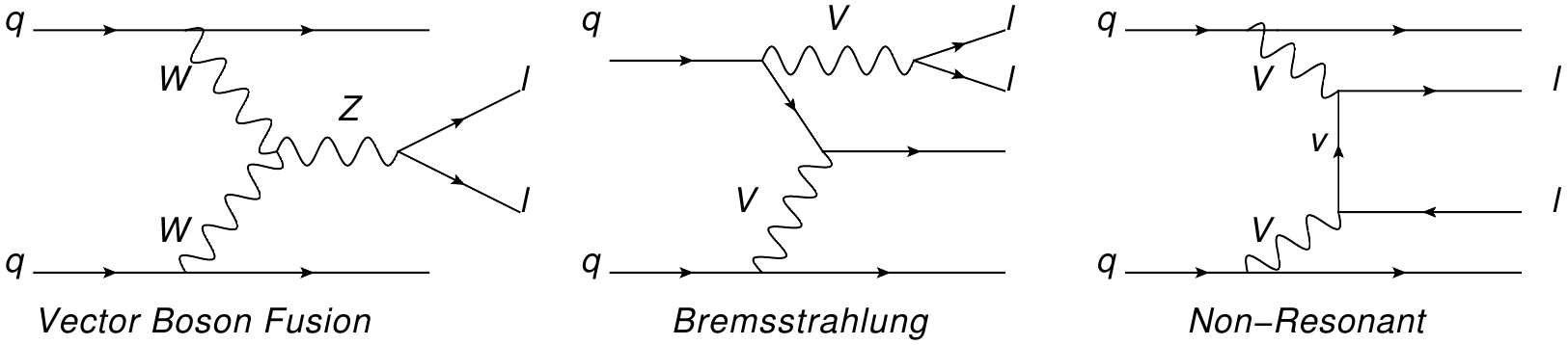}}
\caption{Three contribution Feynman diagrams for the electroweak production of an $l^+l^-q\bar q$ final state: Vector Boson Fusion (left), Bremsstrahlung (middle), non-resonant production (left)}
\label{fig:VBF-diagram}
\end{figure}

The experimental signature of the electroweak production described above, is the typical \Zboson boson topology of two oppositely charged leptons close to the \Zboson boson mass and in addition two high energetic jets. As the momentum transfer to the interacting initial partons caused by the electroweak bosons in the $t$-channel is relatively small, these jets tend to be produced in the forward region of the detector. The dominant background is due to Drell-Yan production of the \Zboson boson in association with jets. Significantly smaller background contributions are expected from the top-pair production and di-boson processes such as $WW$, $WZ$ and $ZZ$. 

The CMS experiment has analysed the full 2011 data set to measure the cross-section of the electroweak production of \Zboson bosons \cite{Chatrchyan:2013jya}. In addition to a standard \Zboson boson selection in the electron and muon decay channels, further cuts on two reconstructed jets within $|\eta|<3.6$ are imposed in order to reject the Drell-Yan background: The transverse momenta of the two jets are required to be $p_T>65\,\GeV$ and $p_T>40\,\GeV$, respectively. In addition, the \Zboson boson rapidity in the rest frame of the two jets has to fulfil $|y^*|<1.2$. These requirements lead to a signal efficiency of 0.06 and to a signal over background ratio of $S/B \approx 0.1$. This low $S/B$ ratio requires the usage of maximum likelihood fits or multivariate techniques for the cross-section determination. Both approaches have been used in \cite{Chatrchyan:2013jya}. The distribution of the invariant mass of both jets is shown in Figure \ref{fig:CMS2JetZ} for the signal-and-background processes. It is the basis of a likelihood fit based on Poisson statistics where the normalisation of the background distributions and the signal distributions are kept as free parameters. The signal extraction via multivariate techniques uses a boosted decision tree approach, which is based on kinematic variables of the jets and leptons and their combinations. Both approaches lead to consistent results, where the multivariate approach provides lower uncertainties. The final measured cross-section within a fiducial phase space\footnote{defined by $m_{ll}>50\,\GeV$, $p_{T,jet}>25\,\GeV$, $|\eta_{jet}|<4.0$, $m_{jj}>120\,\GeV$} is

$\sigma_{EW,Z}^{meas.} = 154 \pm 24 (stat) \pm 46 (syst) \pm 26 (th) \pm 3 (lumi)\,\fb$

\noindent and is in good agreement with the theoretical expectation of $\sigma_{EW,Z} = 166\fb$. The dominating uncertainties are due to the modelling of the background distributions and the jet energy scale. A possible electroweak \Zboson production can therefore be observed with a $2.6\sigma$ significance already within the 2011 data set. The ATLAS experiment has also published a measurement with improved systematic uncertainties based on the 2012 data set at $\sqrt{s} = 8\, \TeV$ \cite{Aad:2014dta} which confirms the observation of the electroweak production of \Zboson above a $5\sigma$ confidence level. Measurements at a higher center-of-mass energy and even larger data samples than the $\sqrt{s} = 8\, \TeV$ sample will be needed to establish this cross-section measurement with higher precision. 

\begin{figure}[h]
\resizebox{0.5\textwidth}{!}{\includegraphics{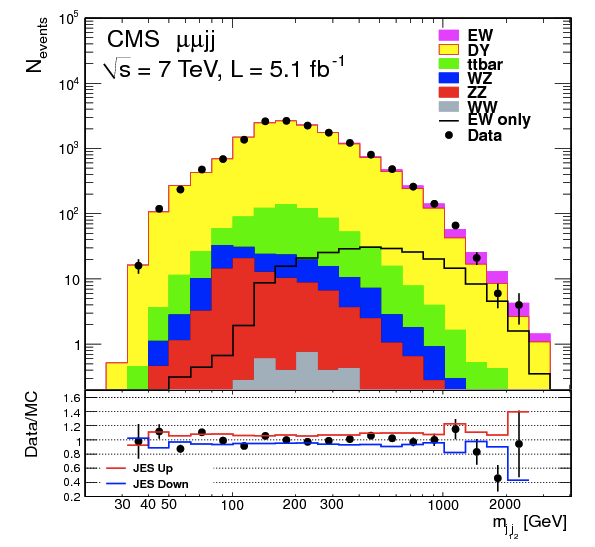}}
\caption{CMS \cite{Chatrchyan:2013jya}: The $m_{jj}$ distribution after selection cuts. The expected contributions from the dominant Drell-Yan (labeled DY) background and the electroweak (labeled EW) signal processes are evaluated from a fit, while the contributions from the small \ttbar~and di-boson backgrounds are estimated from simulation. The solid line with the label "EW only'" shows the distribution for the signal alone.}
\label{fig:CMS2JetZ}
\end{figure}


\section{Summary and Outlook}
\label{sec:SummaryOutlook}

In the first two years of the LHC physics program, a new  
energy regime was investigated with high precision; in some measurements, the experimental systematic uncertainties are 
at the percent level and the statistical uncertainties are even smaller. 
A similar precision was reached at previous colliders 
usually after many years of running. Not only the highest available collision energies 
and large luminosities provided by the LHC, but also the remarkable performance of the 
two general purpose detectors, ATLAS and CMS, are the basis of this success.

This article is the first comprehensive review of all major results regarding the production of
single heavy gauge bosons at a collision energy of $\sqrt{s}=7\,\TeV$. Summaries, comparisons and 
interpretations of the available results have been presented. 
All results are in agreement between the two main experiments at the LHC and are 
furthermore consistent with the presently available Standard Model predictions.  This is illustrated in 
Figure \ref{fig:SummaryCrossSection}, where the ratio between theoretical predictions and measured observables 
for both experiments are shown. 

\begin{figure}
\resizebox{0.5\textwidth}{!}{\includegraphics{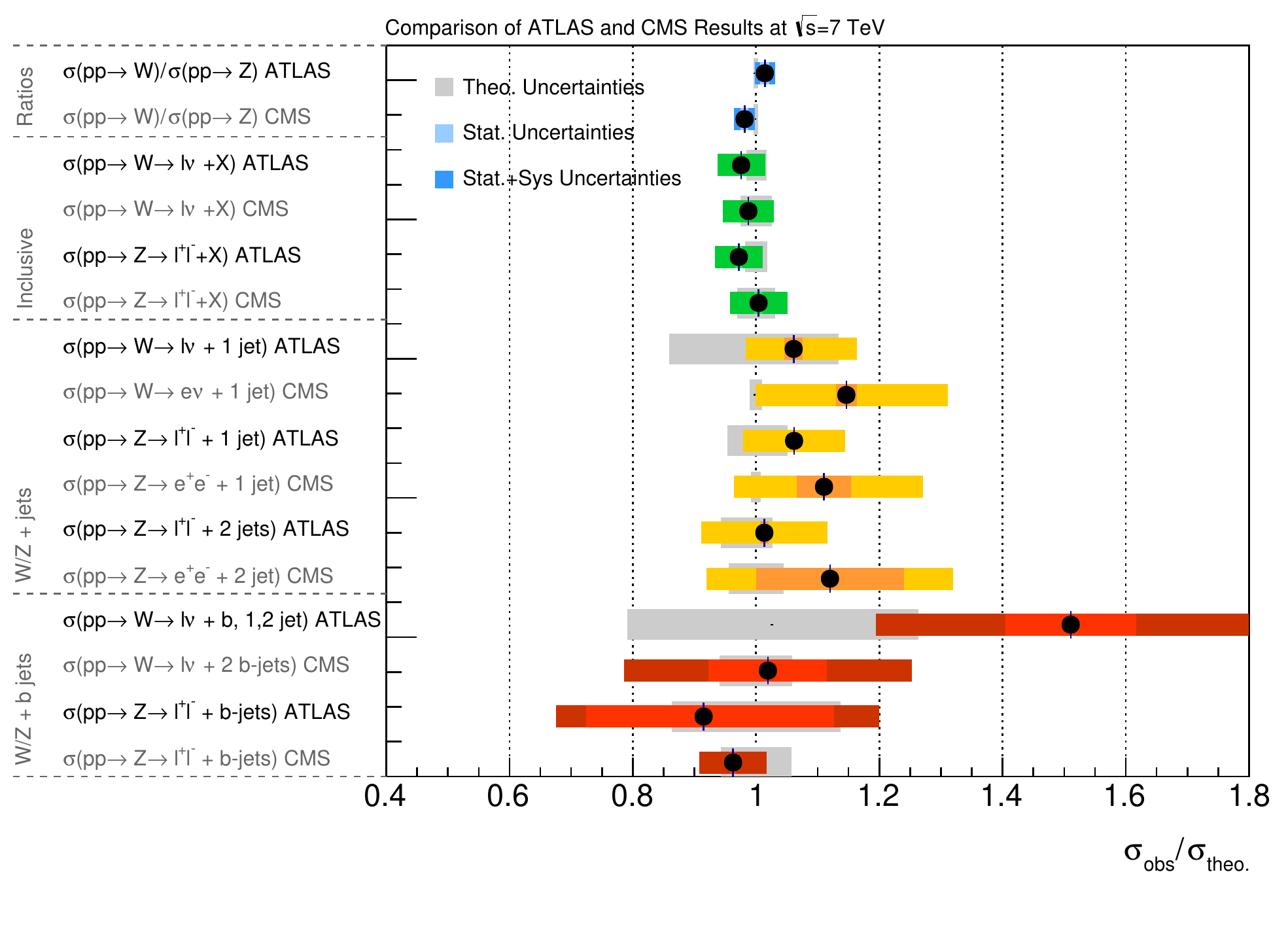}}
\caption{Summary of cross-sections for inclusive \Wboson and \Zboson production and with associated jets. Shown is the ratio between data and predictions. The grey bands represent the theory uncertainties, the light coloured bands are the statistical uncertainties and the dark coloured bands the combined statistical and systematic uncertainties. Note that the CMS $W$+jets and $Z$+jets results are compared to \Sherpa~predictions here. }
\label{fig:SummaryCrossSection}
\end{figure}

Among the numerous measurements which have been presented in this review article, a few should be emphasised.

The inclusive production cross-sections of \Wboson and \Zboson bosons were among the first measurements 
of the LHC which made use of the full detector potential and built the basis for many subsequent physics analyses. 
By now, the experimental precision of the fully inclusive cross-sections of both experiments is below 2\% and hence 
comparable to the NNLO QCD prediction uncertainties. The differential cross-section measurement led to an improved 
understanding of the proton structure functions. In particular, the strange-quark content appears to be 
comparable to the $\bar u$ and $\bar d$-quark content even at low scales, i.e. before the QCD evolution.

Using the large data sets available from the LHC, our understanding of \Wboson and \Zboson production in association with heavy-flavour jets, 
in particular, has greatly improved. Recent measurements of the $W+b$-jet cross-section, which indicate that generators such as 
\MCFM~ and \Alpgen~predict a jet \pT~spectrum that is too soft compared to the data, are the first steps in resolving a long debate over the
source of the disagreement between the measured and predicted cross-section. Measurements of $W+c$ production 
have for the first time the precision to constrain the $s$-quark in the PDFs at $x \approx 0.01$ and 
can help determine if there is an asymmetry between the $s$ and $\sbar$ sea.

With the proton-proton data recorded in the year 2012 at a center-of-mass energy of $\sqrt{s} = 8\, \TeV$, 
the available statistics increased by a factor of four, corresponding to an integrated luminosity of $\IntLumi dt 
\approx 20-25\ifb$. While the increase in center-of-mass energy has only a mild effect on the 
expected cross-section,\footnote{The Standard Model production cross-sections for \Wboson and \Zboson bosons 
are expected to increase by roughly $20\,\%$.} the 
increased data sample allows for calibration of the detector to a higher 
precision. This opens the possibility for the precision measurement of electroweak observables such as the \Wboson boson mass.
In addition, the measurements of multi-differential cross-sections and rare processes such as \Wboson and \Zboson production in association with heavy
flavour quarks will become available.

The next significant step forward will be the LHC run in the years 2015 to 2018 at a center-of-mass energy of $\sqrt{s}=13-14\,\TeV$.
With this, the Standard Model predictions will once again be tested in a new energy regime with the ultimate hope to find signs of `new physics'.

\newpage

\section*{Acknowledgements}
It is our privilege to participate as members of the ATLAS collaboration in the endeavour to investigate the new energy regime opened by the LHC and to present the work of a collaborative effort of both the ATLAS and CMS experiments. We thank Alessandro Tricoli for lively discussions over this text and Lutz Koepke and Joey Houston for the detailed feedback on the first drafts of this review article. In addition, we thank the ATLAS and CMS working group conveners,  Sasha Glazov, Alessandro Tricoli, Jeffrey Berryhill and Maxime Gouzevitch for their input. Special thanks goes to Voica Radescu and Elzbieta Richter-Was, who helped us in the prospects of PDF Fits and the implementation of the studies regarding the QCD angular coefficients, respectively. Last, but not least, we would like to express our gratitude to Dieter Haidt for his guidance and comments during the writing of this review article. The contribution by Matthias Schott was supported by the Volkswagen Foundation and the German Research Foundation (DFG). The contribution by Monica Dunford was supported by the Bundesministerium f\"ur Bildung und Forschung (BMBF).

\bibliography{SingleVectorBoson7TeV}{}
\bibliographystyle{unsrt}

\clearpage

\end{document}